\documentclass[amssymb,usenatbib,useAMS,usedcolumns]{mn2e}
\usepackage{times} 
\usepackage{latexsym,amssymb,amsmath,amsfonts}
\usepackage{rotfloat} 
\usepackage{rotating} 
\usepackage{float}
\usepackage{graphicx}
\usepackage{natbib}
\usepackage{supertabular}
\usepackage{longtable}
\usepackage{url}
\usepackage{dcolumn}
\usepackage{placeins}
\usepackage{multirow} 
\usepackage{textcomp}
\usepackage{appendix}
\usepackage{multicol}
\usepackage{lipsum}
\usepackage{afterpage}
\usepackage{lscape}
\usepackage[caption=false]{subfig}
\usepackage{alphalph}

\begin{document}
\newcommand{\hi}{\mbox{H\,{\sc i}}}
\newcommand{\caii}{\mbox{Ca\,{\sc ii}}}
\newcommand{\nai}{\mbox{Na\,{\sc i}}}
\newcommand{\mgii}{\mbox{Mg\,{\sc ii}}}
\newcommand{\lapp}{\mbox{\raisebox{-0.3em}{$\stackrel{\textstyle <}{\sim}$}}}
\newcommand{\gapp}{\mbox{\raisebox{-0.3em}{$\stackrel{\textstyle >}{\sim}$}}}
\def\h2{$\rm H_2$}
\def\Nh2{$N$(H${_2}$)}
\def\chin{$\chi^2_{\nu}$}
\def\chiu{$\chi_{\rm UV}$}
\def\lya{\ensuremath{{\rm Ly}\alpha}}
\def\lymana{\ensuremath{{\rm Lyman}-\alpha}}
\def\kms{km\,s$^{-1}$}
\def\cms{cm$^{-2}$}
\def\cc{cm$^{-3}$}
\def\zabs{$z_{\rm abs}$}
\def\zem{$z_{\rm em}$}
\def\zrad{$z_{\rm rad}$}
\def\zgal{$z_{\rm gal}$}
\def\nhi{$N$($\hi$)}
\def\ln{log~$N$}
\def\nh{$n_{\rm H}$}
\def\ne{$n_{e}$}
\def\21{21-cm}
\def\ts{T$_{s}$}
\def\th{T$_{01}$}
\def\ll{$\lambda\lambda$}
\def\l{$\lambda$}
\def\fc{$C_{f}$}
\def\mjb{mJy~beam$^{-1}$}
\def\taudv{$\int\tau dv$}
\def\ha{H\,$\alpha$}
\def\taudvl{$\int\tau dv^{3\sigma}_{10}$}
\def\taudv{$\int\tau dv$}
\def\c21{$C_{21}$}
\def\t0{$\tau_0$}
%
%=============================================================================================
%
\voffset=-0.5in
\hoffset=0.2in
\title[Distribution of cold gas around $z<0.4$ galaxies]{
{\hi\ \21\ absorption survey of quasar-galaxy pairs: Distribution of cold gas around $z<0.4$ galaxies
}  
\author[Dutta et al.]{R. Dutta$^1$\thanks{E-mail: rdutta@iucaa.in}, R. Srianand$^{1}$, N. Gupta$^1$, E. Momjian$^2$, P. Noterdaeme $^3$, P. Petitjean $^3$, 
\newauthor{H. Rahmani$^4$} \\ 
$^1$ Inter-University Centre for Astronomy and Astrophysics, Post Bag 4, Ganeshkhind, Pune 411007, India \\
$^2$ National Radio Astronomy Observatory, 1003 Lopezville Road, Socorro, NM 87801, USA \\   
$^3$ Institut d'Astrophysique de Paris, CNRS and UPMC Paris 6, UMR7095, 98bis boulevard Arago, 75014 Paris, France \\
$^4$ Aix Marseille Universit\'{e}, CNRS, LAM (Laboratoire d'Astrophysique de Marseille) UMR 7326, 13388, Marseille, France
} 
}
\date{Accepted. Received; in original form }
\maketitle
\label{firstpage}
%
%============================== ABSTRACT =================================================================================
%
\begin {abstract}  
\par\noindent 
We present the results from our survey of \hi\ \21\ absorption, using GMRT, VLA and WSRT, in a sample of 55 $z<0.4$ galaxies towards 
radio sources with impact parameters ($b$) in the range $\sim$0$-$35 kpc. In our primary sample (defined for statistical analyses) of 
40 quasar-galaxy-pairs (QGPs), probed by 45 sightlines, we have found seven \hi\ \21\ absorption detections, two of which are reported 
here for the first time. Combining our primary sample with measurements having similar optical depth sensitivity (\taudv\ $\le$0.3\,\kms) 
from the literature, we find a weak anti-correlation (rank correlation coefficient = $-$0.20 at 2.42$\sigma$ level) between \taudv\ and $b$,
consistent with previous literature results. The covering factor of \hi\ \21\ absorbers (\c21) is estimated to be 0.24$^{+0.12}_{-0.08}$ at 
$b\le$ 15 kpc and 0.06$^{+0.09}_{-0.04}$ at $b=$ 15$-$35 kpc. \taudv\ and \c21\ show similar declining trend with radial distance along 
the galaxy's major axis and distances scaled with the effective \hi\ radius. There is also tentative indication that most of the \hi\ \21\ 
absorbers could be co-planar with the extended \hi\ discs. No significant dependence of \taudv\ and \c21\ on galaxy luminosity, stellar mass, 
colour and star formation rate is found, though the \hi\ \21\ absorbing gas cross-section may be larger for the luminous galaxies. The higher 
detection rate (by a factor of $\sim$4) of \hi\ \21\ absorption in $z<1$ DLAs compared to the QGPs indicates towards small covering factor 
and patchy distribution of cold gas clouds around low-$z$ galaxies. 
\end {abstract}  
%
%=========================== KEY WORDS ===================================================================================== 
%
%
%=========================== INTRODUCTION ================================================================================== 
%
\section{Introduction} 
\label{sec_introduction}  
The \hi\ gas in a galaxy's extended disc plays an indispensable role in the galaxy formation and evolution by acting as the intermediary phase 
between the accreting ionized gas from the intergalactic medium (IGM) and the molecular gas phase in the stellar disc that gets converted to 
stars. Being more extended than the stellar disc (typically by more than a factor of 2), the \hi\ disc is also the component that gets affected 
the most by tidal interactions and merger events \citep{haynes1979,rosenberg2002,oosterloo2007,sancisi2008,chung2009,mihos2012}. Therefore, 
the \hi\ cross-section of galaxies and its evolution with redshift is expected to have an imprint on galaxy formation in the hierarchical structure 
formation models. 

In the local Universe ($z\lesssim$ 0.2), high spatial resolution \hi\ \21\ emission studies have been used to map the distribution and kinematics of 
the atomic gas around galaxies \citep{vanderhulst2001,zwaan2001,verheijen2007,deblok2008,walter2008,begum2008,catinella2015}. 
However, sensitivities of present day radio telescopes make it difficult to directly map \hi\ \21\ emission from $z\gtrsim$ 0.2 galaxies. 
The highest redshift ($z$ = 0.376) \hi\ \21\ emission detection to date \citep{fernandez2016} has been possible due to very long integration (178 hours) using the Karl G. Jansky Very Large Array (VLA).
Alternatively, absorption lines seen in the spectra of background quasars whose sightline happen to pass through the discs or halos of 
foreground galaxies (we refer to such fortuitous associations as quasar-galaxy-pairs or QGPs from hereon), allow us to probe the physical, 
chemical and ionization state of gas in different environments such as the stellar discs, extended \hi\ discs, high velocity clouds, outflows, 
accreting streams and tidal structures. The relationship between absorption strength and impact parameter ($b$) obtained from a large number 
of quasar sightlines passing near foreground galaxies can then be used to statistically determine the gas distribution in and around galaxies 
\citep[as pioneered by][in case of \mgii\ systems]{bergeron1991,steidel1995}. 

Considerable progress has been made in mapping the distribution of gas in the circumgalactic medium (CGM), that extends upto $\sim$100s of kpc, 
using absorption from \mgii\ \citep{chen2010,kacprzak2011,churchill2013,nielsen2013} as well as \lymana\ \citep{chen2001,prochaska2011,tumlinson2013,stocke2013,borthakur2015}.
On the other hand, the high column density \hi\ gas around galaxies can be traced using Damped \lymana\ absorption-line systems (DLAs) 
\citep[absorbers with neutral hydrogen column density, log~\nhi\ $\ge$ 20.3; see for a review][]{wolfe2005}. 
Observations of host galaxies of DLAs have found anti-correlation between \nhi\ and $b$, at $z<1$ \citep{rao2011}, at $1<z<2$ 
\citep{peroux2012,rahmani2016}, and at $z>2$ \citep{krogager2012}. 
At $z=$ 0, \citet{zwaan2005}, using \hi\ \21\ emission maps of local galaxies, have found that \nhi\ decreases with galactocentric radius. 

\hi\ \21\ emission line observations of nearby dwarf and spiral galaxies indicate that the properties of the cold neutral medium (CNM) phase of the 
\hi\ gas and small-scale structures detected in the \hi\ gas are closely linked with the in-situ star formation in galaxies \citep[e.g.][]{tamburro2009,bagetakos2011,ianjamasimanana2012}. 
However, identification of CNM gas through \hi\ \21\ emission is not straightforward as it depends on Gaussian decomposition of the line profiles and
associating components exhibiting smaller line widths with the CNM. Moreover, \hi\ \21\ emission studies usually do not have sufficient spatial resolution
to detect parsec-scale structures. On the other hand, \hi\ \21\ absorption is an excellent tracer of the CNM phase \citep{kulkarni1988}, and can be 
used to study parsec-scale structures in the \hi\ gas using sub-arcsecond-scale spectroscopy \citep[e.g.][]{srianand2013,biggs2016}. The optical depth 
of the \hi\ \21\ absorption line depends on \nhi\ and inversely on the spin temperature (\ts), which is known to follow the gas kinetic temperature in
the CNM \citep{field1959,bahcall1969,mckee1977,wolfire1995a,roy2006}, and can be coupled to the gas kinetic temperature via the \lymana\ radiation 
field in the warm neutral medium (WNM) \citep{liszt2001}.
This along with its very low transition probability and its resonance frequency falling in the radio wavelengths make the \hi\ \21\ absorption line a good 
tracer of high column density cold \hi\ gas without being affected by dust and luminosity biases.

Systematic searches of \hi\ \21\ absorption in samples of intervening \mgii\ systems and DLAs towards radio-loud quasars have estimated the CNM filling factor 
of galaxies as 10$-$20\% \citep{briggs1983,kanekar2003,curran2005,gupta2009,curran2010,srianand2012,gupta2012,kanekar2014}. 
However, such studies have led to degenerate interpretations of the evolution of cold atomic gas fraction of galaxies, since there is a degeneracy between \ts\ 
and the fraction \fc\ of the background radio source covered by the absorbing gas. Moreover, selection based on optical/ultraviolet (UV) spectra leads to bias 
against dusty sightlines. 

An alternative technique, that connects the \hi\ \21\ absorption with the galaxy properties, has been to search for \hi\ \21\ absorption from low-$z$ ($z<$0.4) 
QGPs \citep[e.g.][]{carilli1992,gupta2010,borthakur2011,borthakur2016,zwaan2015,reeves2016}. Such studies have revealed a weak anti-correlation between the \hi\ 
\21\ optical depth and impact parameter \citep{gupta2013,zwaan2015,curran2016}. However, the number of low-$z$ QGPs that have been searched for \hi\ \21\ absorption, 
with a good optical depth sensitivity (i.e., 3$\sigma$ integrated optical depth, for a line width of 10\,\kms, is $\le$ 0.3\,\kms, which is sensitive to detect 
100 K gas with \nhi\ of $\le$ 5$\times$10$^{19}$\,\cms), till date is still small. As \nhi\ of few times 10$^{19}$ \cms\ is observed in the outer disks of galaxies 
in \hi\ \21\ emission maps \citep[e.g.][]{vanderhulst2001,walter2008}, any survey aiming to detect CNM gas in galaxies via \hi\ \21\ absorption should reach such 
limiting column densities. Further, the connections between the distribution of \hi\ \21\ absorbers around low-$z$ galaxies and different parameters of galaxies, 
in particular, those that are governed by the ongoing star formation in these galaxies, geometry and stellar distributions of galaxies, are not yet well-established. 

Here we present the results from our systematic survey of \hi\ \21\ absorption in a large homogeneous sample of $z<$ 0.4 galaxies. Our survey has resulted in seven 
\hi\ \21\ detections, the largest from any single survey of low-$z$ QGPs. The measurements from our survey increase the existing number of sensitive \hi\ \21\ optical depth 
measurements (as defined above) at low-$z$ by a factor of three, and the number of \hi\ \21\ absorption detections from QGPs by almost a factor of two. In addition, while most 
studies of QGPs in the literature have focused at $z\lesssim$ 0.1, our sample probes galaxies upto $z\sim$ 0.4, and has produced four \hi\ \21\ detections at $z\ge$ 0.1.
In this work, combining the \hi\ \21\ absorption measurements with the host galaxy properties, we attempt to statistically determine the distribution of \hi\ \21\ absorbers 
and hence that of cold \hi\ gas around low-$z$ galaxies.  

Identifying how the \hi\ \21\ absorption line properties depend on the locations of the radio lines-of-sight with respect to the foreground galaxies
is essential to make progress in using \hi\ \21\ absorption measurements in future blind \hi\ \21\ absorption surveys \citep{morganti2015}, with the 
Square Kilometre Array (SKA) pre-cursors like MeerKAT \citep{jonas2009} and ASKAP \citep{deboer2009} and pathfinders like AperTIF \citep{verheijen2008} 
and uGMRT, to probe galaxy evolution. These surveys are expected to detect several hundred \hi\ \21\ absorbers. Hence, we expect that the results from 
our survey would play a crucial role in interpreting the huge volume of data expected from these upcoming surveys. In addition, our study will help to 
bridge the missing link between the Galactic interstellar medium (ISM) studies and high redshift ($z>1$) \hi\ \21\ absorber studies for which host galaxy 
properties are difficult to study using present day telescopes.

This paper is structured as follows. In Section \ref{sec_sample}, we describe the sample selection and various properties of the galaxies in our sample. 
In Section \ref{sec_observations}, we describe our radio observations using the Giant Metrewave Radio Telescope (GMRT), VLA and Westerbork Radio Synthesis 
Telescope (WSRT), and optical observations using the South African Large Telescope (SALT). In Section \ref{sec_individual}, we discuss the physical conditions 
in a few individual systems from our sample. Next, we study the distribution of \hi\ \21\ absorbers around low-$z$ galaxies and their dependence on host galaxy 
properties in Section \ref{sec_results}. We discuss our results in Section \ref{sec_discussion} and conclude by summarizing our results in Section \ref{sec_summary}. 
Throughout this work we have adopted a flat $\Lambda$-CDM cosmology with $H_{\rm 0}$ = 70\,\kms~Mpc$^{-1}$ and $\Omega_M$ = 0.30.
%
%==================================================== Sample Selction ======================================================
%
\section{Sample Description}  
\label{sec_sample}
\subsection{Sample Selection}
\label{sec_samplesel}
At first we constructed a parent sample of QGPs using the Sloan Digital Sky Survey \citep[SDSS;][]{york2000} and the Faint Images of the Radio Sky at Twenty-Centimeters 
\citep[FIRST;][]{white1997} databases. To do this we cross-matched SDSS-Data Release 9 (DR9) galaxies with photometric redshift, $z_{phot}$ $\le$0.4, and FIRST radio 
sources with peak flux density at 1.4 GHz $\ge$50 mJy within a 40$''$ search radius. The redshift range was chosen to ensure that the redshifted \hi\ \21\ absorption 
frequency is uniformly covered at the most sensitive radio interferometers (i.e. GMRT, VLA and WSRT), and to maximize the possibility of sub-arcsecond-scale spectroscopic 
and \hi\ \21\ emission follow-ups. The search radius was chosen to restrict to radio sightlines that would probe the foreground galaxies at low impact parameters (40$''$ 
corresponds to $b$ = 16$-$215 kpc for the range of $z_{phot}$ = 0.02$-$0.4 of the sample, with $b$ = 93 kpc at median $z_{phot}$ = 0.13). The flux density cut on the 
radio sources was imposed in order to obtain a \hi\ \21\ optical depth sensitivity, defined as 3$\sigma$ upper limit on the integrated optical depth from spectra smoothed 
to 10\,\kms, \taudvl\ $\le$0.3\,\kms. This corresponds to a sensitivity of \nhi\ $\le$ 5$\times$10$^{19}$\,\cms\ for \ts\ = 100 K and \fc\ = 1.
Note that 10\,\kms\ is the typical width of the individual \hi\ \21\ absorption line components detected from QGPs (see Section~\ref{sec_radioobs}). We then visually inspected 
the SDSS images of the resultant pairs to confirm that the galaxies have been correctly identified. This selection technique resulted in 106 QGPs which we refer to as the SDSS 
parent sample. Note that at this stage we did not place any condition on the nature of the radio source except for a flux density cut. For simplicity, we use the term QGP to refer 
to all pairs of radio sources and galaxies throughout this work, irrespective of the nature of the radio sources. 

We carried out \hi\ \21\ absorption line searches for 43 of the QGPs in the SDSS parent sample. To select which sources to observe from the parent sample, 
preference was given to those QGPs: (a) which had spectroscopic redshift of the galaxy already available from SDSS or the literature, (b) whose optical 
spectra we were able to obtain with the optical telescopes accessible to us, and (c) which had impact parameters (based on spectroscopic redshift) 
$\lesssim$30 kpc. We restricted to QGPs with $b\lesssim$30 kpc, as ground-based imaging studies of galaxies responsible for DLAs/sub-DLAs at $0.1\le z\le 1$
have found that DLA and sub-DLA column densities occur at median impact parameters of 17 and 33 kpc, respectively \citep{rao2011}. Further, the expected 
median $b$ for systems with log~\nhi\ $>$ 20.3 is 8 kpc, based on conditional probability distribution of \nhi\ and $b$ from \hi\ \21\ emission maps of 
galaxies at $z$ = 0 \citep{zwaan2005}.  

In addition to the above QGPs from the parent sample, we searched for \hi\ \21\ absorption from five `galaxy on top of quasars' \citep[GOTOQs; see][]{noterdaeme2010,york2012,straka2013}, 
in order to sample sightlines at small impact parameters passing close to the stellar discs of galaxies. The GOTOQs were selected by visually identifying intervening 
\ha\ and other emission lines superimposed in the SDSS spectra of the background quasars (with FIRST peak flux density $\ge$50 mJy). Additionally, we searched 
for \hi\ \21\ absorption towards four quasars where faint optical continuum emission is seen around the quasars in the SDSS images and weak intervening \caii\ 
absorption is detected in the background quasar's spectrum, but no corresponding emission lines are detected in the SDSS spectrum. Further, in order to include 
very low-$z$ ($z \lesssim$ 0.01) galaxies not present in the SDSS parent sample, we carried out \hi\ \21\ absorption line search of three QGPs identified from 
literature \citep{boisse1988,corbelli1990,womble1993}. At such low-$z$, sub-arcsecond-scale spectroscopy and \hi\ \21\ emission observations become feasible.
For example, joint analysis of \hi\ \21\ emission and absorption was possible in case of the QGP, J0041$-$0143, from our sample \citep{dutta2016}. In total, 
we searched for \hi\ \21\ absorption from 55 QGPs. 

\begin{figure*}
\subfloat[]{ \includegraphics[width=0.4\textwidth, angle=90]{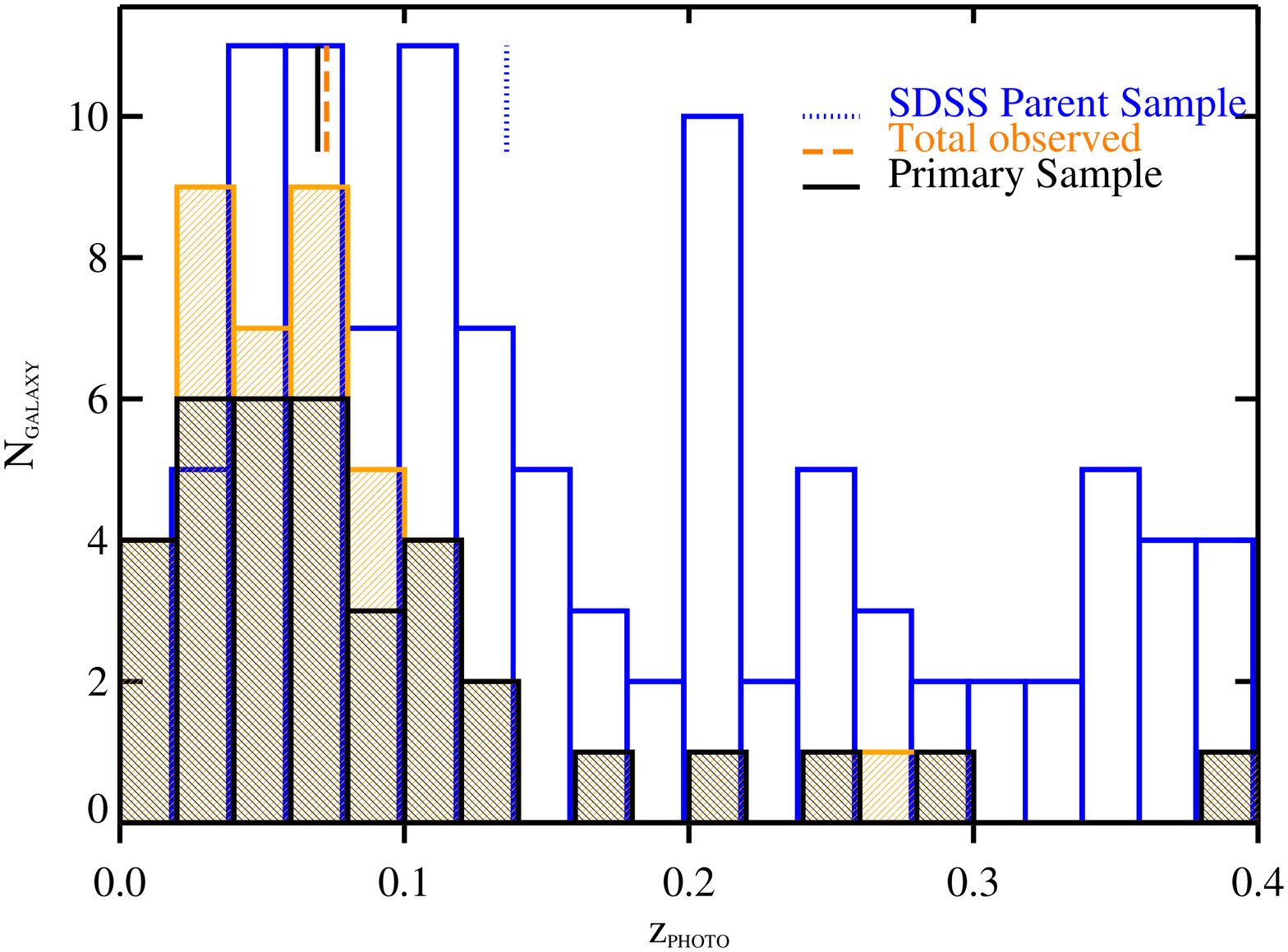} }
\subfloat[]{ \includegraphics[width=0.4\textwidth, angle=90]{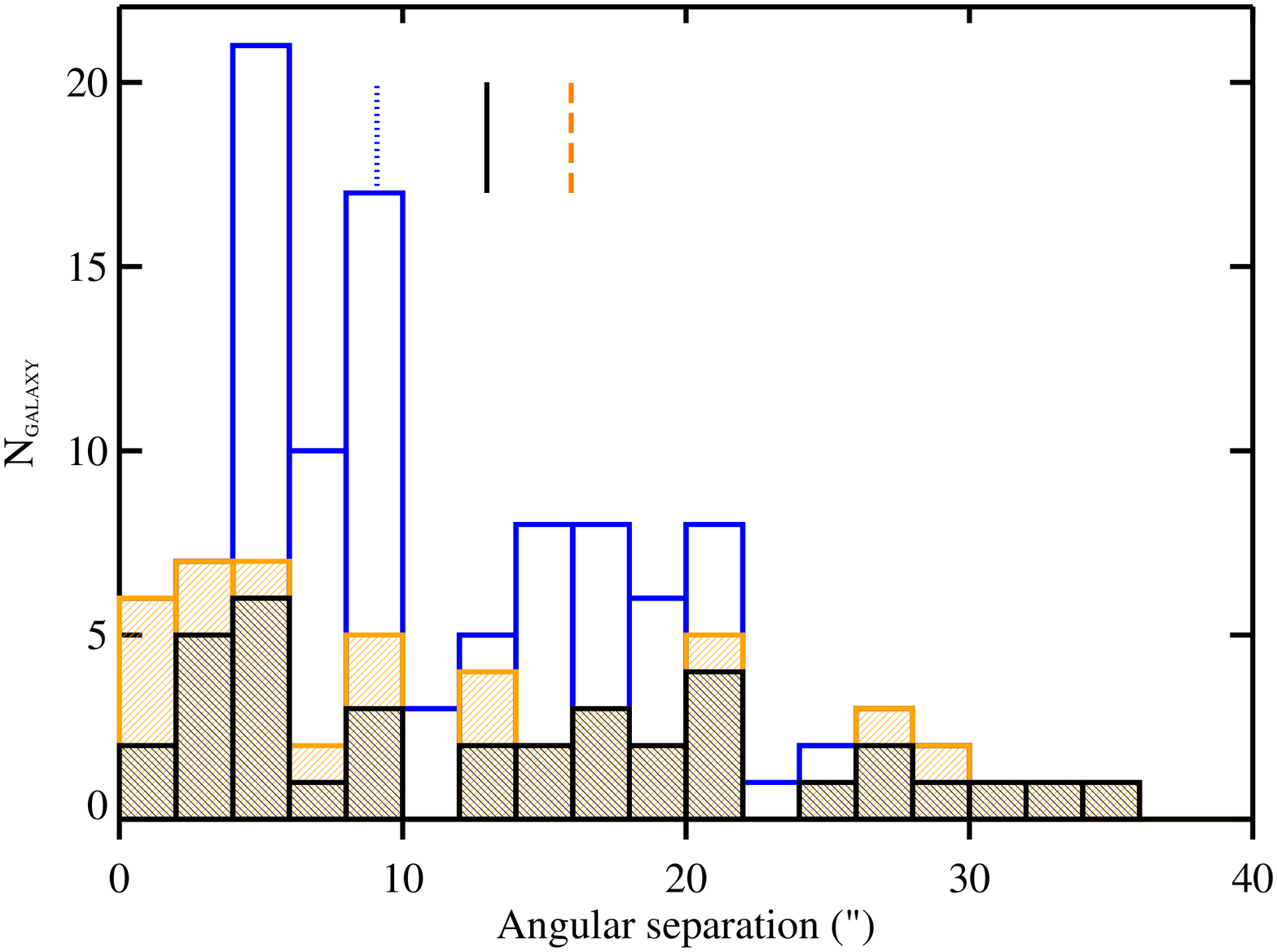} } \hspace{0.1cm}
\subfloat[]{ \includegraphics[width=0.4\textwidth, angle=90]{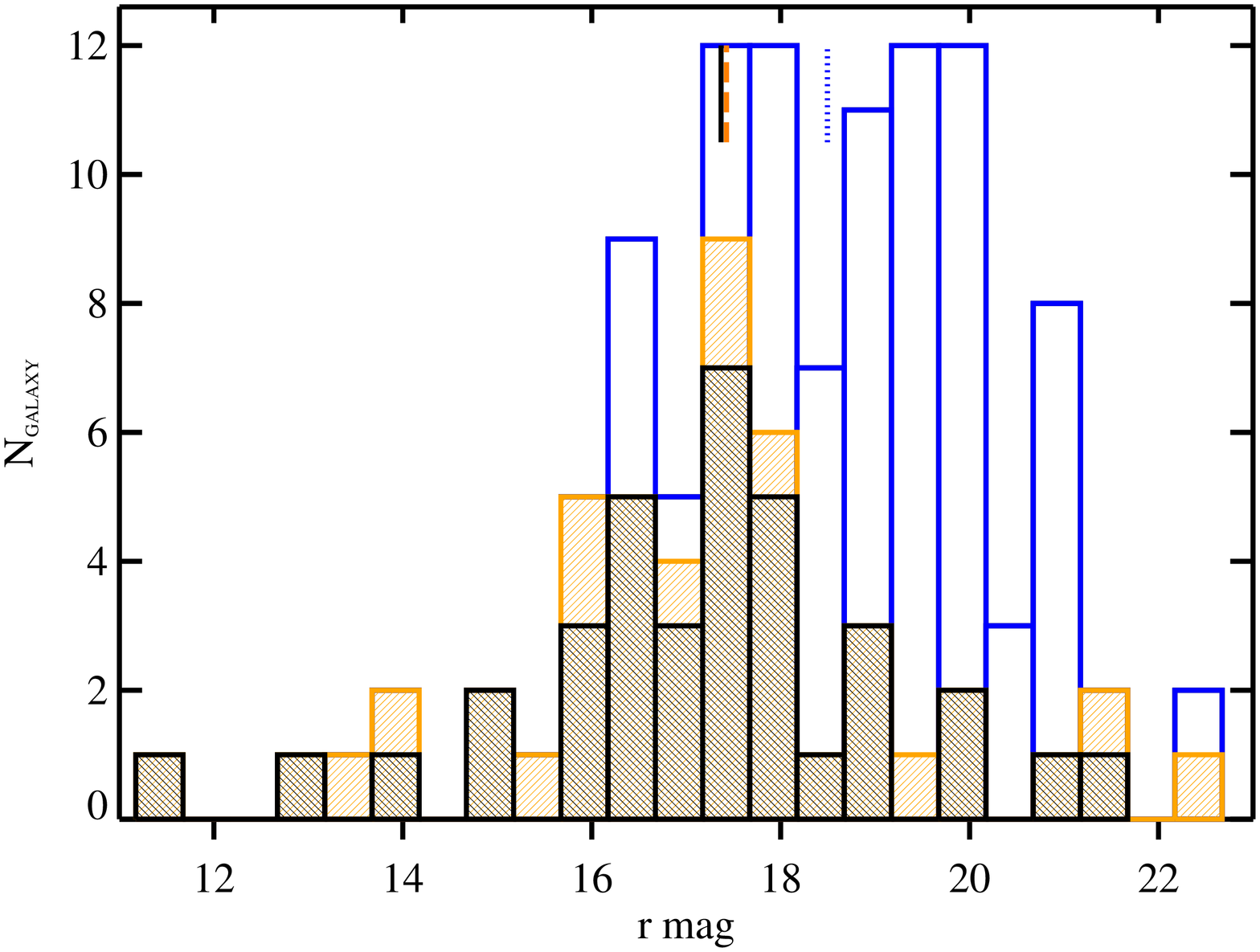} } 
\subfloat[]{ \includegraphics[width=0.4\textwidth, angle=90]{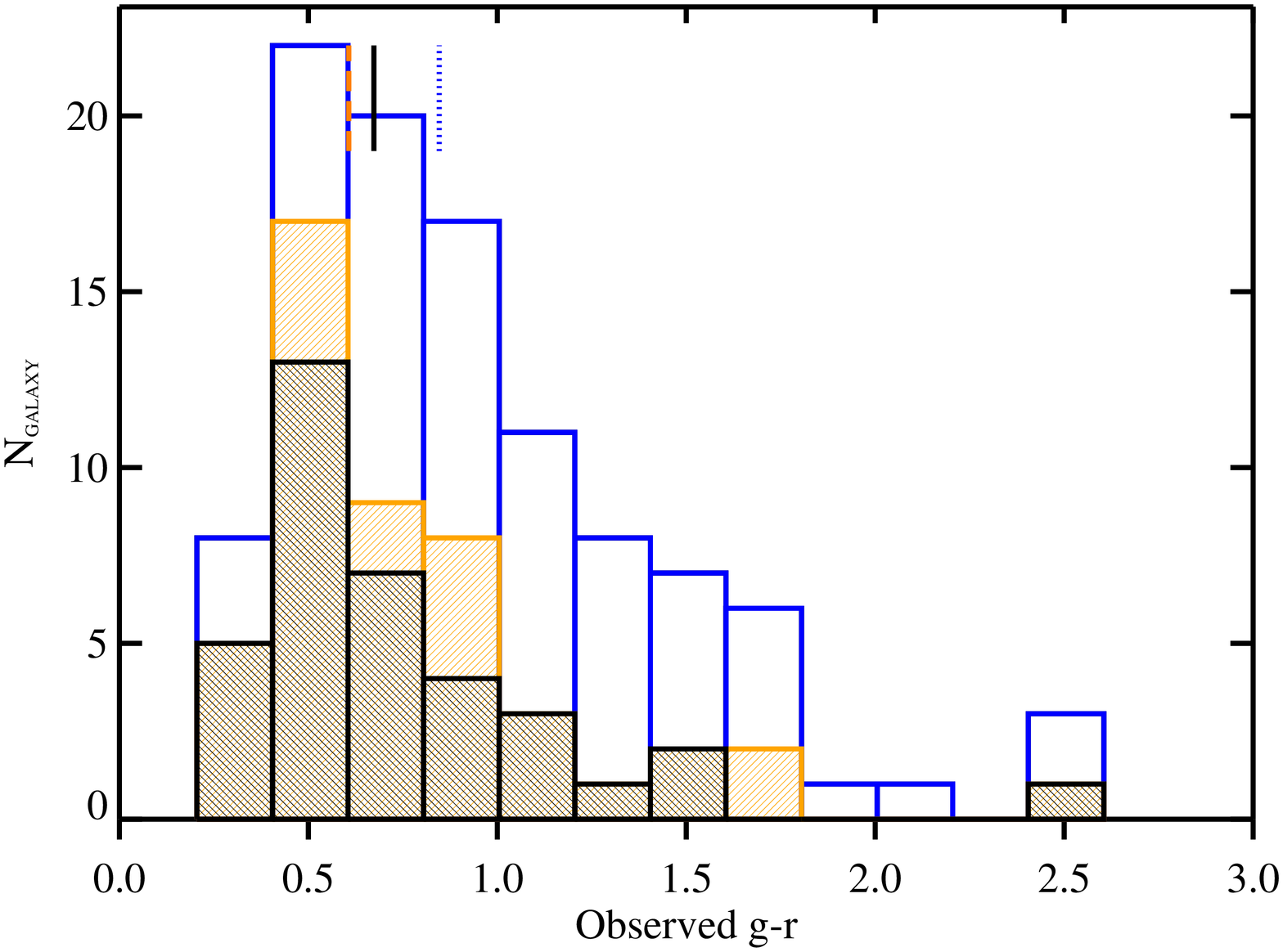} }
\caption{Distributions of: (a) the photometric redshift, (b) angular separation from the background radio source, (c) $r$-band magnitude, and (d) the observed $g-r$ colour of the galaxies in the 
SDSS parent sample are shown in open blue histograms. The shaded orange histograms (with hashes at 45$^\circ$) are for the galaxies searched for \hi\ \21\ absorption by us, and the shaded black histograms 
(with hashes at 135$^\circ$) are for those that are in the primary sample. The median values of the parameters for the parent sample, the observed sample (primary $+$ supplementary $+$ miscellaneous) 
and the primary sample are marked by dotted, dashed and solid ticks, respectively, in each plot. Note that the angular separation from the background radio source of the three $z\lesssim$ 0.01 galaxies 
(70$''$, 220$''$, 320$''$) are  not shown in panel (b).}
\label{fig:master} 
\end{figure*}

The distribution of the photometric redshift, the angular separation from the background radio source, the SDSS $r$-band magnitude, and the observed SDSS $g-r$ 
colour for the galaxies in the parent sample are shown in Fig.~\ref{fig:master} (open blue histograms). We also show the same for the galaxies searched for \hi\ 
\21\ absorption by us in orange shaded histograms. Note that SDSS photometric catalog contains objects to a magnitude limit of $r \sim$ 22.5 mag, which corresponds 
to $r$-band limiting luminosity of $5.5\times 10^6-3.4\times 10^9 L_\odot$ for the range of $z_{phot}$ = 0.02$-$0.4 of the parent sample, and $2.7\times 10^8 L_\odot$ 
at median $z_{phot}$ = 0.13. It can be seen from Fig.~\ref{fig:master}, that due to the requirement of spectroscopic redshifts of galaxies for carrying out \hi\ \21\ 
spectral line search, our observed sample is biased towards the brighter and the bluer galaxies from the parent sample, and hence, most of the galaxies observed by us 
are at $z_{phot} \le$ 0.2. 

In order to carryout statistical studies, we categorize the QGPs observed by us as `primary', `supplementary' and `miscellaneous'. The primary sample consists of
QGPs where we are confident that the radio source is a background source. The supplementary sample consists of QGPs where we are not sure about the radio source
being in the background or about the galaxy redshift. The miscellaneous sample consists of QGPs where we are sure that the radio source is not a background source.  

\noindent{\it (i) Primary Sample:} 
Our primary sample consists of 40 QGPs. In all except eight pairs, spectroscopic redshifts of the background radio sources are available in SDSS or in the 
literature, ensuring that these are background sources. We measured the redshift of one radio source using IUCAA Girawali Observatory (IGO). Out of the seven 
radio sources without spectroscopic redshifts, four have photometric redshifts available in SDSS which ensure that the probability of \zrad\ $>$ \zgal\ is high, 
even after taking into account the redshift errors. The remaining three radio sources also have high probability of being $z\ge$0.5 quasars and hence background 
sources, based on their SDSS and WISE colours \citep{wu2012}. Hence, we take these seven radio sources to be background sources. We discuss in Section~\ref{sec_results} 
the effects of excluding these sources in our statistical analyses. Our primary sample consists of 32 QGPs from SDSS, five GOTOQs, and three $z\lesssim$ 0.01 QGPs 
selected from literature as mentioned above. Three background radio sources in the primary sample show multiple radio emitting components at arcsecond-scales, 
that can be used for searching \hi\ \21\ absorption. Because of this, in total we have 45 radio sightlines around 40 galaxies in our primary sample where we have 
searched for \hi\ \21\ absorption. \hi\ \21\ absorption line searches towards these sources have resulted in seven \hi\ \21\ detections. The distribution of the 
photometric redshift, the angular separation from the background radio source, the SDSS $r$-band magnitude, and the observed SDSS $g-r$ colour for the galaxies 
in the primary sample are shown in Fig.~\ref{fig:master} (black shaded histograms).\\

\noindent{\it (ii) Supplementary Sample:} 
Our supplementary sample consists of 10 QGPs. In five cases, the radio sources have no optical and infrared counterparts detected in the SDSS and WISE images, 
respectively. One of the radio sources is detected in the SDSS images but we cannot place any constraints on its redshift. These six sources are included in 
the supplementary sample, since it cannot be confirmed with the existing data whether these radio sources are behind the galaxies. From the Combined EIS-NVSS 
Survey Of Radio Source \citep{best2003,brookes2006,brookes2008}, we see that 21 out of 24, i.e. 88$^{+12}_{-19}$\%, of the radio sources brighter than 
50 mJy at 1.4 GHz have $z>$ 0.4 (the quoted errors are 1$\sigma$ Gaussian confidence intervals computed using tables 1 and 2 of \citet{gehrels1986} assuming a Poisson distribution).
However, to remove additional sources of uncertainty in our analyses we do not include radio sources without optical/infrared counterparts in our primary sample. 
Note that \hi\ \21\ absorption has not been detected towards such radio sources till date. The other four systems in the supplementary sample were identified as 
potential GOTOQs, i.e. faint optical emission is seen around the quasars in the SDSS images and weak \caii\ absorption is detected in the quasar's SDSS spectra. 
However, no corresponding galactic emission lines are detected in the SDSS spectra. Further, we do not detect any continuum or line emission from foreground 
galaxies in two of these cases, using SALT long-slit observations with the slit aligned along the extension seen around the quasars. We carried out \hi\ \21\ 
absorption line search of these sources at the redshift of the \caii\ absorption. We do not include these four sources in the primary sample as the galaxy 
redshifts are not certain and we cannot rule out the possibility that the optical emission seen around the quasars in SDSS images could be from the quasar 
host galaxies. No \hi\ \21\ absorption was detected towards these 10 sources. The measurements from these systems are shown in the plots presented in this 
work, but they are not included in any of the statistical analyses.\\

\noindent{\it (iii) Miscellaneous Sample:} 
We report the \hi\ \21\ absorption line measurements towards 5 QGPs as part of a miscellaneous sample in this work, but do not include them in any analysis/discussion.
Spectroscopic redshifts of the background radio sources were not available in SDSS or in the literature for these systems at the time of their radio observations. 
We measured the redshift of two radio sources using IGO and SALT and that of three sources subsequently became available in SDSS or in the literature. These radio 
sources turned out to be either at the same redshift as or slightly lower (within $\sim$200\,\kms) than that of the foreground galaxy. Three of these sources are 
part of merging systems while two appear to be part of group of nearby galaxies at similar redshifts. \hi\ \21\ absorption has been detected towards two of the merging 
galaxy pairs. One of them has already been reported in \citet{srianand2015}. We report for the first time here the detection of a broad ($\sim$150\,\kms) and strong \hi\
\21\ absorption (\nhi\ (\fc$/$\ts) $\sim$ 5.5 $\times$ $10^{19}$\,\cms\,K$^{-1}$) at $z$ = 0.2 towards the pair J205449.61$+$004153.0$/$J205449.64$+$004149.8 (see Fig.~F3). 
Such a strong \hi\ \21\ absorption is rare and usually seen in merging galaxy pairs \citep{srianand2015}. Details of \hi\ \21\ absorption towards this system will be presented 
in a future work. 

The summary of all the sources observed by us, split into primary, supplementary and miscellaneous samples as described above, are provided in Table~\ref{tab:samples}. 
This table also gives the number of radio sightlines probed and \hi\ \21\ absorption detected in each category. The properties of all the sources observed by us are provided 
in Table~A1 and discussed in Section~\ref{sec_sampleprop}.
\begin{table}  
\centering
\caption{Summary of different samples discussed in this work.}
\begin{tabular}{lccc}
\hline
Sample & No. of   & No. of radio & \hi\ \21\  \\
       & galaxies & sightlines   & detections \\
\hline
Total Observed & 55 & 62          & 9          \\
\hline
Primary        & 40 & 45 (7$^*$)  & 7 (0$^*$)  \\
Supplementary  & 10 & 12          & 0          \\
Miscellaneous  & 5  & 5           & 2          \\
\hline
Literature     & 24 & 24 (3$^*$)  & 9 (3$^*$)  \\
\hline
Combined       & 64 & 69 (10$^*$) & 16 (3$^*$) \\
(Primary $+$ Literature) &   &    &            \\
\hline
\end{tabular}
\label{tab:samples}
\begin{flushleft}
$^*$ Number of radio sources without spectroscopic redshifts albeit having identified optical/infrared counterparts.
\end{flushleft}
\end{table}
\subsection{Literature Sample}
\label{sec_litsample}
In order to interpret the nature of \hi\ \21\ absorbers around low-$z$ galaxies, we combine the \hi\ \21\ measurements of our primary sample with those of QGPs from literature. 
Since the existing studies of QGPs in literature have been carried out with various selection techniques and sensitivities, we define here a sample of QGPs searched for \hi\
\21\ absorption in literature that is relevant for our analyses and consistent with the definition of our primary sample. We consider the literature measurements 
from the compilation presented in Table 5 of \citet{gupta2010}, as well as more recent measurements from \citet{borthakur2011,borthakur2014,borthakur2016,zwaan2015,reeves2015,reeves2016}, 
subject to the following conditions.
(i) We consider only systems selected on the basis of presence of a radio sightline near the galaxy, without any prior detection of absorption lines in the quasar's optical/UV spectra. 
This is to obtain an unbiased estimate of the covering factor of \hi\ \21\ absorbers around low-$z$ galaxies (see Section~\ref{sec_results}).
(ii) Most of the \hi\ \21\ measurements obtained in the literature are less sensitive than ours. Hence, in order to have uniform optical depth sensitivity across all the measurements,
we include only measurements of \hi\ \21\ non-detections in literature with \taudvl\ $\le$ 0.3 \kms\ (see Section~\ref{sec_radioobs}).
(iii) We restrict to measurements which are at impact parameters $\lesssim$30 kpc. 
(iv) We exclude measurements towards associated systems, or where \zrad\ is within $\sim$200\,\kms\ of \zgal. 
(v) We exclude radio sources without optical and/or infrared counterparts and hence whose redshifts cannot be constrained.
(vi) In case of low spatial resolution single dish measurements of \citet{borthakur2016}, we do not consider the \hi\ \21\ non-detections with \hi\ \21\ emission detected in the spectrum, 
as the absorption could be filled-in by the emission \citep[see e.g.][]{borthakur2011}.

The details of the literature sample so defined are presented in Table~B1. Note that seven of the QGPs in the SDSS parent sample are part of the sample of \citet{borthakur2010,borthakur2014,borthakur2016}. 
We observed three of these QGPs, while the other four are part of the literature sample. All except three of the background radio sources in the literature sample are classified as quasars based on their 
optical spectra. For one of the radio sources only photometric redshift is available and two others no redshifts are available. We include these sources in the sample since their optical and infrared 
counterparts are present and \hi\ \21\ absorption have been detected towards them. We discuss in Section~\ref{sec_results} the effects of excluding these sources in the statistical analyses. In total 
there are 24 QGPs in the literature sample, out of which \hi\ \21\ absorption have been detected in 9 cases (see Table~\ref{tab:samples}).  We refer to our primary sample plus the literature sample as 
the ``Combined sample''.
\subsection{Sample Properties}
\label{sec_sampleprop}
In this section, we discuss various properties of the primary and literature samples as listed in Tables~A1 and B1, respectively.

\noindent{\it (i) Nature of background radio sources:}
As mentioned in Section~\ref{sec_samplesel}, the optical counterparts of all except seven radio sources in the primary sample have spectroscopic redshifts available from SDSS, 
NASA/IPAC Extragalactic Database (NED), or our IGO observations. The ones without spectroscopic redshifts are taken as background sources based on their SDSS and WISE photometry. 
All the radio sources with spectroscopic redshifts available are classified as quasars based on their optical spectra. For the rest either no classification is available or classification 
based on photometry indicates that they are radio galaxies. We provide the optical classification from literature and morphology of the radio sources based on their FIRST images in 
Table~A1. Of the 38 background radio sources in our primary sample, 28 are compact, 7 are resolved and 3 show double peaks in the FIRST images \citep[quasars with deconvolved sizes 
$\le$2$''$ are classified as compact;][]{white1997}. The three sources which are double in the FIRST images are resolved into a central core (cospatial with the optical core) and two 
lobes in our GMRT images. In total there are 45 radio sightlines (with peak flux density $\ge$50 mJy) in the primary sample.

In case of the radio sources in the literature sample, as mentioned in the previous section, all except three have spectroscopic redshifts available which classify them as background 
quasars. For the remaining three sources, optical/infrared counterparts are present but no classification is available. Radio morphology at arcsecond resolution is available for 20 of 
the radio sources from FIRST ($\sim$5$''$) or other VLA images ($\sim$1$-$5$''$) \citep{stanghellini1998,reid1999,keeney2011}. Among these sources seven are resolved while rest are compact.
 
\begin{figure*}
\subfloat[]{ \includegraphics[width=0.4\textwidth, angle=90]{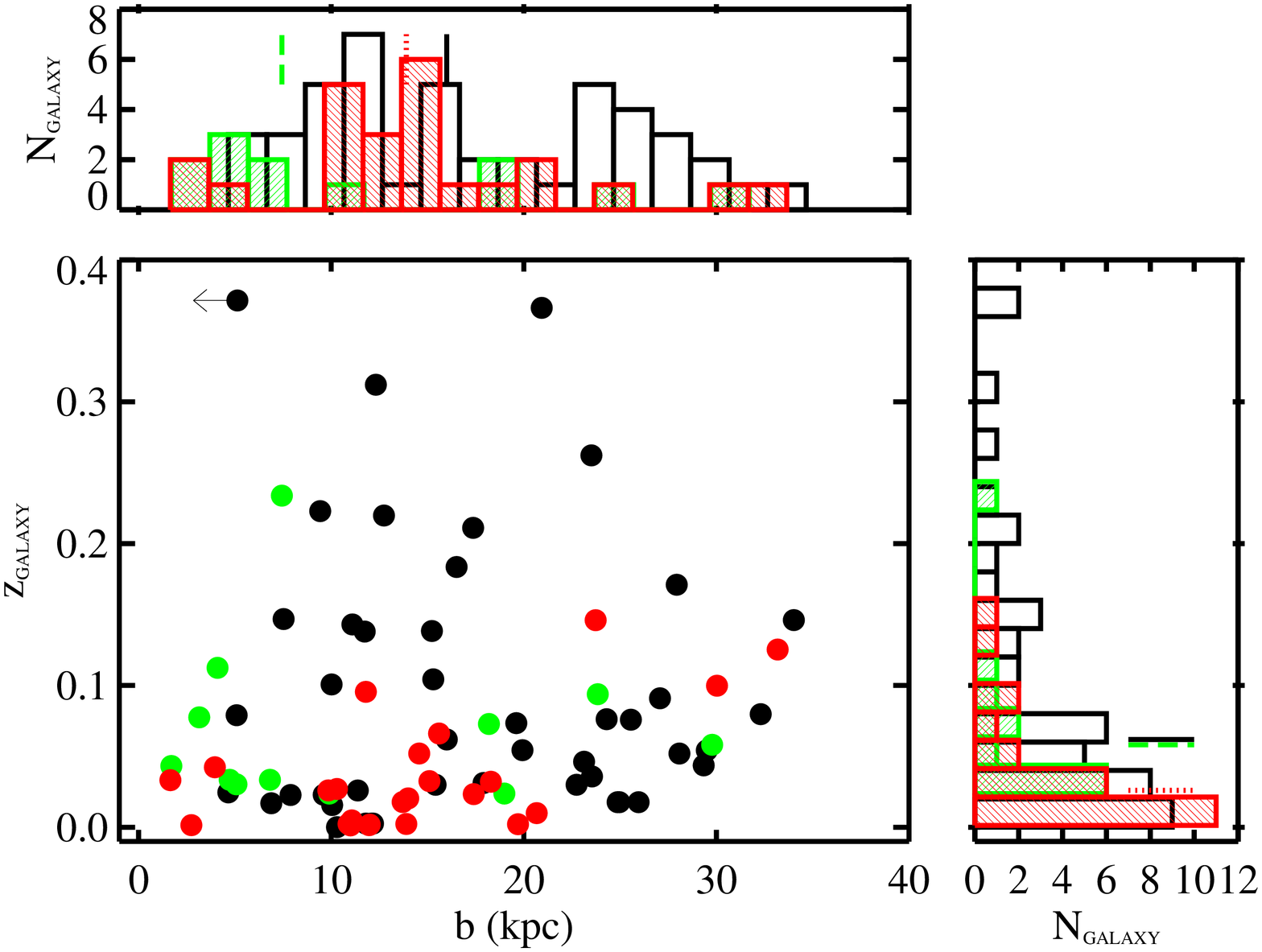} }
\subfloat[]{ \includegraphics[width=0.4\textwidth, angle=90]{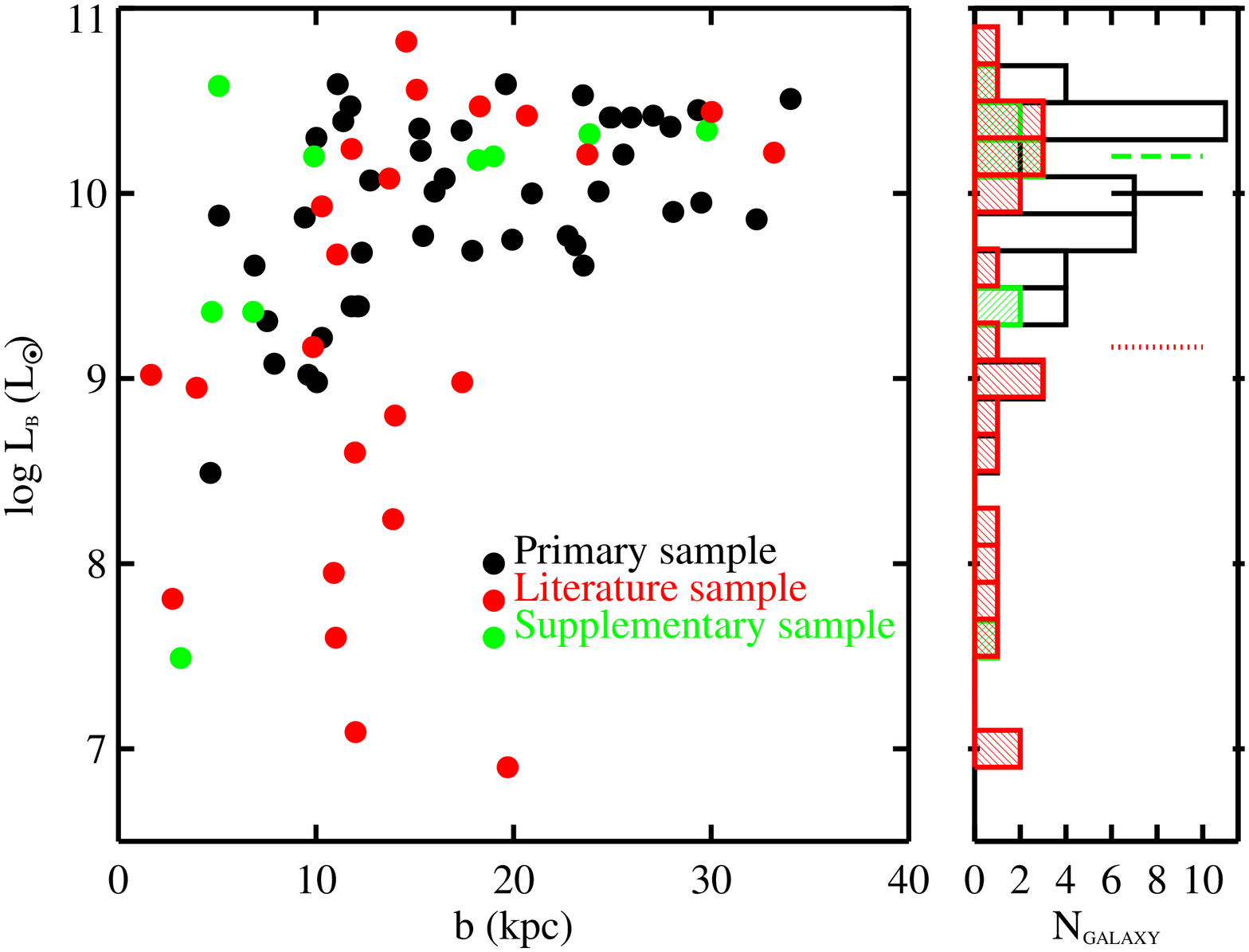} } \hspace{0.1cm}
\subfloat[]{ \includegraphics[width=0.4\textwidth, angle=90]{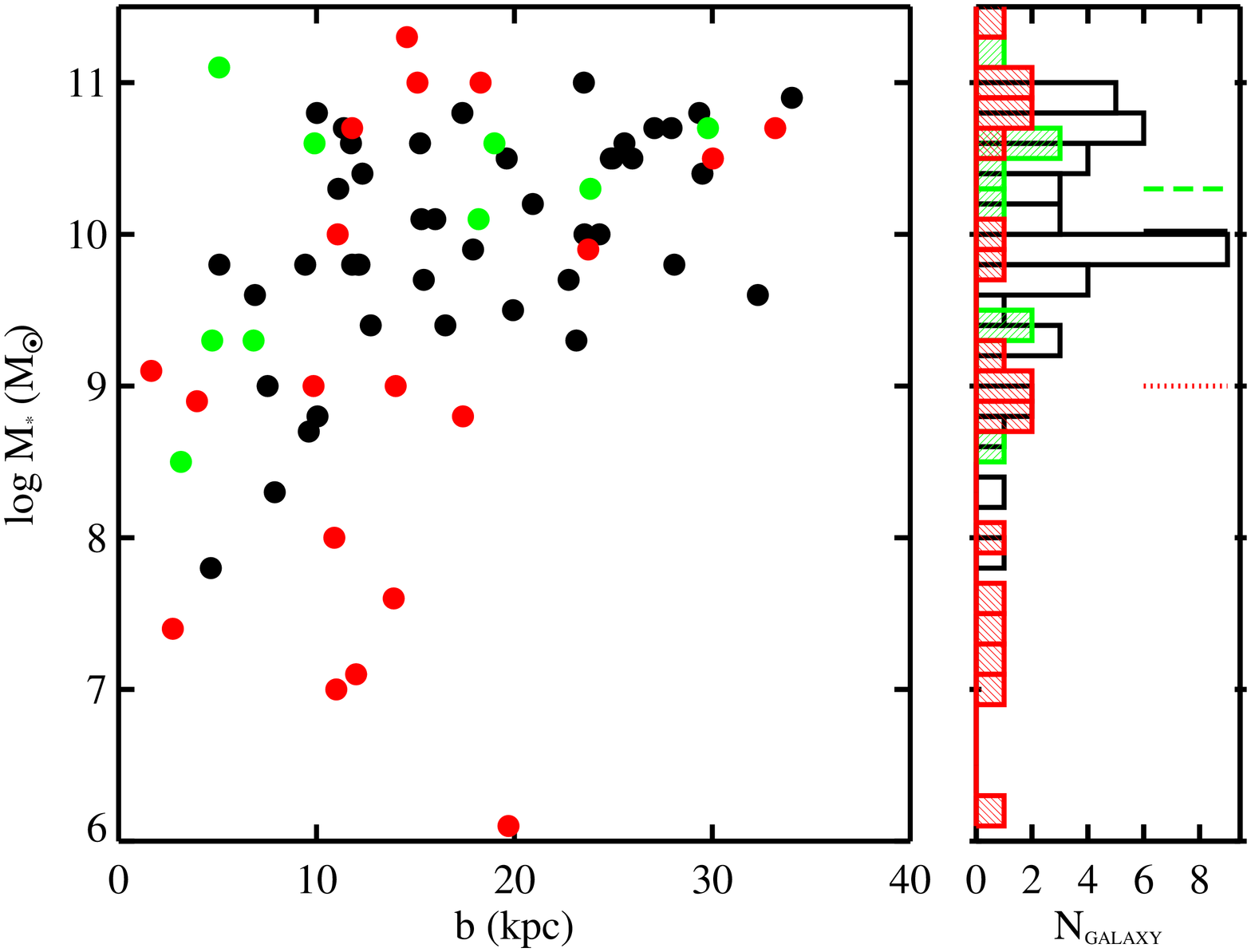} } 
\subfloat[]{ \includegraphics[width=0.4\textwidth, angle=90]{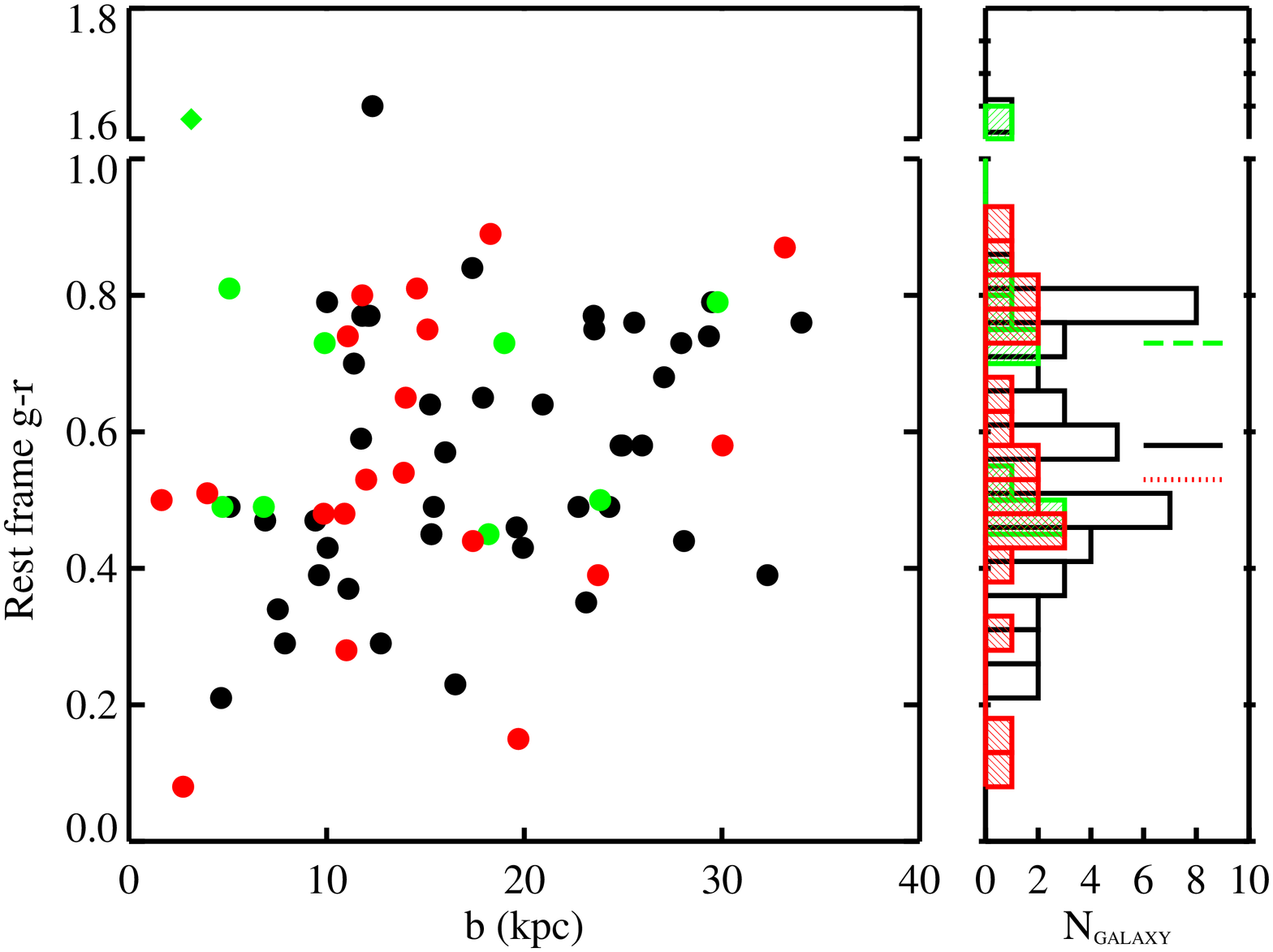} }
\caption{The distributions of: (a) galaxy redshifts, (b) $B$-band luminosity, (c) stellar mass, and (d) rest-frame SDSS $g-r$ colour as a function of impact parameter. 
The black, red and green circles are for the galaxies in the primary, literature and supplementary samples, respectively. The histograms of the impact parameters, galaxy 
redshifts, luminosities, stellar masses and colours are also shown next to the plots, following the same colour convention. The solid, dotted and dashed ticks mark the 
median of the distributions for the primary, literature and supplementary samples, respectively.}
\label{fig:hist} 
\end{figure*}

\noindent{\it (ii) Galaxy redshifts:}
The spectroscopic redshifts of 21 galaxies in our primary sample are taken from SDSS, while that of five galaxies are taken from NED (the detailed references are provided in Table~A1). 
We measured the redshifts of two galaxies using the Apache Point Observatory (APO) 3.5-m telescope, and that of seven galaxies using SALT. The details of the APO observations are provided in 
\citet{gupta2010} and that of the SALT observations are provided in the next section. There are five galaxies in our primary sample that are identified as GOTOQs based on emission in the 
background quasar's SDSS spectra. The galaxies in our primary sample have redshifts in the range 0.00008 $\le z\le$ 0.37, with a median redshift of 0.08. The distribution of galaxy redshifts 
as a function of impact parameter is shown in panel (a) of Fig.~\ref{fig:hist}. It can be seen that galaxies in the literature sample have $z<$ 0.15 with median $z\sim$ 0.03. 

\noindent{\it (iii) Impact parameters:}
The impact parameters ($b$) are calculated based on the projected angular separation between the centre of the galaxy and the radio sightline. The GOTOQ J1443$+$0214 is not detected in the 
SDSS images even after subtracting the quasar contribution, and the galactic line emission probably originates in a low surface brightness region within 5 kpc to the quasar sightline \citep{gupta2013}.
The impact parameter for this GOTOQ is taken to be $<$5 kpc. The 45 radio sightlines in the primary sample probe a range of impact parameters $\sim$5 $\le b$ (kpc) $\le$ 34, with a median 
$b$ $\sim$ 16 kpc. From panel (a) of Fig.~\ref{fig:hist}, it can be seen that the range of impact parameters (median $b\sim$ 13.9 kpc) probed by the literature sample is similar to that 
by the primary sample.

\noindent{\it (iv) Orientation:}
The galaxy inclination ($i$) and azimuthal orientation ($\phi$) of the radio sightline with respect to the galaxy are calculated from the minor-to-major axis ratio and the position angle of the 
major axis of the galaxy, respectively, from SDSS-DR12 $r$-band photometry \citep{alam2015}. The orientation is defined as the projected angle that the radio sightline makes with the projected 
major axis of the galaxy, with $\phi$ = 0\textdegree\ for sightlines along the major axis and $\phi$ = 90\textdegree\ for sightlines along the minor axis. We use the parameters from the best fit 
between the exponential and the de Vacouleurs fit provided by SDSS and also visually verify the parameters. The errors in $i$ range from $\sim$2$-$10\textdegree. We treat galaxies with 
$i<$30\textdegree\ as face-on galaxies and do not estimate $\phi$ in these cases. 

For the GOTOQ, J1300$-$2830, we estimate $i$ and $\phi$ from the SDSS $r$-band image (after masking the quasar) based on isophotal ellipses using {\tt STSDAS} package in {\tt IRAF}. For the GOTOQ, 
J1438$+$1758, we estimate $i$ and $\phi$ from visual inspection of the SDSS $r$-band image, since the galaxy emission could not to be well fit by ellipses. One of the galaxies in our primary 
sample (J1748$+$7005) and six of the galaxies in the literature sample are not covered in the SDSS footprint. Since these are very low-$z$ galaxies their optical photometry are present in NED and 
we use their $B$-band photometric measurements to estimate $i$ and $\phi$. For the galaxy J104257.74$+$074751.3, we use the photometric measurements provided in \citet{borthakur2010}. Note that excluding 
the galaxies for which photometric measurements are not available in SDSS does not change the statistical results obtained in this work beyond the quoted statistical uncertainties.

\noindent{\it (v) Galaxy properties:}
We estimate the absolute magnitudes of the galaxies using the {\it kcorrect} algorithm (v\_4.2) by \citet{blanton2007a}. Specifically, we use the IDL based {\it sdss\_kcorrect} 
routine, the SDSS-DR12 Galactic extinction corrected asinh {\it ugriz} model magnitudes \citep{alam2015}, and the galaxy redshifts to obtain the $k$-corrections. For two of the GOTOQs 
(J1300$+$2830 and J1438$+$1758), we estimate the {\it ugriz} magnitudes from the total counts due to the galaxy (after masking the quasar and subtracting the sky contribution) within 
an aperture of 1.5$''$ radius in the respective SDSS images. Recall that the GOTOQ, J1443$+$0214, is not detected in the SDSS images and hence photometric measurements of this 
galaxy are not possible. The stellar masses are obtained from the {\it kcorrect} output. The typical error in the SDSS magnitudes is 0.01 and the typical uncertainty in the stellar 
masses derived from SDSS photometry is 50\% \citep{blanton2007a}. 

The stellar masses of the galaxies in the primary sample are in the range of 10$^{7.8}-$10$^{11.0}$ $M_\odot$ with a median of 10$^{10.0}$ $M_\odot$, while the $B$-band luminosities 
are in the range of 10$^{8.5}-$10$^{10.6}$ $L_\odot$ with a median of 10$^{10.0}$ $L_\odot$. The median $g-r$ colour is 0.58, with most of the galaxies having $g-r <$1.0, except for 
one with $g-r\sim$1.6, which is identified as a luminous red galaxy \citep{gupta2013}. Note that the galaxies in the literature sample have lower median stellar mass (10$^{9.0}$ $M_\odot$) 
and $B$-band luminosity (10$^{9.2}$ $L_\odot$).

The distribution of the galaxy luminosities, stellar masses and colours as function of impact parameter are shown in Fig.~\ref{fig:hist}. Considering the primary sample,
it can be seen that at $b<$10 kpc, our sample consists mostly of low luminosity and low stellar mass galaxies. This is most probably because the number density of fainter 
and smaller galaxies is higher and hence the probability of selecting such galaxies is higher. From the same figure, we see that at $b\ge$10 kpc, our sample probes mostly 
galaxies with high luminosity and high stellar mass. It seems that we have preferentially selected the brightest and largest galaxy that is closest to the radio sightline. 
Hence, it is to be noted that the distributions of galaxy luminosities and stellar masses as function of impact parameter are not uniform by construct of our sample. On 
the other hand, these distributions are more uniform for the literature sample, which probes more low luminosity and low stellar mass galaxies at $b\ge$10 kpc than our 
primary sample.

The SDSS local galaxies show a bimodal colour distribution, i.e. they fall on either the blue or the red sequence. To check how our galaxies compare with the general galaxy 
population at similar redshifts, we used a magnitude-dependent colour cut, $^{0.1}$($g-r$) = 0.80 $-$ 0.30($^{0.1}M_{r} +$ 20) \citep{blanton2007b}. Here $^{0.1}M_{r}$ 
indicates the absolute magnitude in the $r$-band $k$-corrected to $z$ = 0.1. We find that $\sim$70\% of the galaxies in the primary sample are blue and rest are red,
as also indicated by panel (d) of Fig.~\ref{fig:master}, and this holds on combining our primary sample with the literature sample.

In the following section, we present the details of our optical and radio observations.
%
%==================================================== OBSERVATIONS =========================================================
%
\section{Observations and Data Reduction}  
\label{sec_observations}  
\subsection{Optical Observations}
\label{sec_opticalobs}
We obtained the optical spectra of seven galaxies in our sample using the Robert Stobie Spectrograph (RSS) on SALT in long-slit mode, under the programs
2013-2-IUCAA-001, 2014-1-IUCAA-001, 2014-2-SCI-017 (PI: Dutta). The observations were carried out with a 1.$''$5 slit and the PG0900 grating (resolution 
of $\sim$5\,\AA\ at 5000\,\AA). The slit positions were chosen to align both the galaxy and the quasar in the slit. Each pair was observed for a total $\sim$4800 
sec, split into two blocks with spectral coverage of $\sim$4000$-$7000\,\AA\ and $\sim$5000$-$8000\,\AA. The wavelength ranges were chosen to maximize the 
coverage of the expected wavelength ranges of the \ha, H$\beta$ and O\,{\sc iii} emission lines and the \caii\ absorption lines (within the photometric 
redshift errors). We used data reduced by the SALT science pipeline \citep{crawford2010}, i.e. {\tt PYSALT}\footnote{{\tt PYSALT} user package is the primary 
reduction and analysis software tools for the SALT telescope (http://pysalt.salt.ac.za/).} tasks were used to prepare the image headers for the pipeline, 
apply CCD amplifier gain and crosstalk corrections, subtract bias frames and create mosaic images. We subsequently flat-fielded the data, applied a wavelength 
solution using arc lamp spectra, performed cosmic ray removal, applied background subtraction, combined the two-dimensional spectra and extracted the 
one-dimensional spectra using standard {\tt IRAF} tasks\footnote{IRAF is distributed by the National Optical Astronomy Observatories, which are operated by the 
Association of Universities for Research in Astronomy, Inc., under cooperative agreement with the National Science Foundation.}. The emission and/or 
absorption lines used to determine the galaxy redshifts and the galaxy spectra are given in Appendix~\ref{appendix_saltspectra}.
\subsection{Radio Observations}
\label{sec_radioobs}
Of the 55 sources that were searched for \hi\ \21\ absorption by us, 50 were observed with the GMRT, two with the VLA, two with the WSRT, and one was observed with 
both the GMRT and the WSRT. The GMRT observations were carried out using the L-band receiver. Prior to 2011 the sources were observed using the 2 MHz baseband 
bandwidth split into 128 channels (spectral resolution $\sim$4\,\kms\ per channel, velocity coverage $\sim$500\,\kms). The later observations were carried 
out using the 4.17 MHz baseband bandwidth split into 512 frequency channels (spectral resolution $\sim$2\,\kms\ per channel, velocity coverage $\sim$1000\,\kms). 
The VLA observations (Proposal ID: 15A$-$176) were carried out in A-configuration (maximum baseline of 36.4 km) using a single 8 MHz sub-band in dual 
polarization split into 4096 channels (spectral resolution $\sim$0.5\,\kms\ per channel, velocity coverage $\sim$ 2000\,\kms). 
In the case of WSRT observations a total bandwidth of 10 MHz split into 2048 spectral channels was used to acquire data in the dual polarization mode. 
This corresponds to a total bandwidth of $\sim$2400 \kms\ and a channel width of $\sim$1.2 \kms\ before Hanning smoothing. 

In all the observations, the pointing centre was at the quasar coordinates and the band was centred at the redshifted \hi\ \21\ frequency estimated from the 
galaxy redshifts. Data were acquired in the two linear polarization products, XX and YY (circular polarization products, RR and LL, in the case of the VLA). 
Standard calibrators were regularly observed during the observations for flux density, bandpass, and phase calibrations (no phase calibrators were observed
for WSRT observations). The details of the radio observations of the sources are given in Table~\ref{tab:radiolog1}.

The data were reduced using the National Radio Astronomy Observatory (NRAO) Astronomical Image Processing System ({\tt AIPS}) following standard procedures as outlined in \citet{gupta2010}.
The results obtained from the \hi\ \21\ absorption line searches of the QGPs are provided in Table~\ref{tab:radioparam1}. 
During the course of our observations we have reported the results of \hi\ \21\ absorption studies of 10 sources \citep[9 from the primary sample,][]{gupta2010,gupta2013,srianand2013,srianand2015,dutta2016}. 
The results of our \hi\ \21\ absorption searches of the remaining 45 sources (31 from the primary sample) are presented for the first time in this work. 

We provide the standard deviation in the optical depth at the observed spectral resolution ($\sigma_\tau$), and 3$\sigma$ upper limit on the integrated 
optical depth from spectra smoothed to 10\,\kms\ (\taudvl). In case of \hi\ \21\ absorption detections, we also provide the peak optical depth ($\tau_p$) 
and the total integrated optical depth (\taudv), both at the observed spectral resolution. The choice of 10\,\kms\ for estimating the optical depth sensitivity 
is motivated by the fact that the median Gaussian full-width-at-half-maximum (FWHM) of all the detected \hi\ \21\ absorption components in the primary sample 
is $\sim$10\,\kms. The \hi\ \21\ absorption spectra in the primary sample have \taudvl\ in the range 0.015$-$0.380\,\kms, with median of 0.131\,\kms\ and 
90\% have \taudvl\ $\le$ 0.3\,\kms. Note that the \hi\ \21\ measurements in our primary sample have increased the existing number of sensitive (i.e. \taudvl\ 
$\le$ 0.3\,\kms) \hi\ \21\ optical depth measurements of QGPs by a factor of $\sim$3.

The \hi\ \21\ absorption spectra of the QGPs and the radio continuum contours of the radio sources overlaid on the optical images of the galaxies are provided in 
the online version, in Appendix~F and Appendix~G, respectively. The typical spatial resolution of the GMRT images is $\sim$2$-$3 arcsec, while that of the VLA 
images is $\sim$1$-$2 arcsec, and that of the WSRT images is $\sim$20 arcsec (see Table~D1).
%
%=================================================== RESULTS =============================================================== 
%
\section{Individual systems}
\label{sec_individual}
We have detected \hi\ \21\ absorption from seven QGPs in our primary sample. Five of these have been reported in \citet{gupta2010,gupta2013,srianand2013,dutta2016}.
In this work we report two new \hi\ \21\ absorption detections from the QGPs, J1243$+$4043 and J1438$+$1758. We discuss the physical conditions in the QGP, J1438$+$1758, 
in Section~\ref{sec_J1438}. Detailed study of the QGP, J1243$+$4043, including Global Very Long Baseline Interferometry \hi\ \21\ absorption spectra and optical long-slit 
spectra, will be presented in a future work. In Section~\ref{sec_J1243}, we discuss the implications of our measurements of the QGP, J1243$+$1622, from which \citet{schneider1993} 
have reported tentative \hi\ \21\ absorption. In Sections~\ref{sec_J1300} and \ref{sec_J1551}, we discuss about two QGPs, J1300$+$2830 and J1551$+$0713, respectively.
These two are intriguing since these are the only QGPs in our sample where the quasar sightlines appear to pass through the optical disc of the galaxies in the 
SDSS images, but no significant \hi\ \21\ absorption is detected towards them (though tentative absorption features are present). Readers interested in only the 
statistical results may go directly to Section~\ref{sec_results}.
\subsection{QGP J1438$+$1758}
\label{sec_J1438}
\begin{figure*}
\subfloat[]{ \includegraphics[width=0.32\textwidth, bb= 50 30 740 745, clip=true]{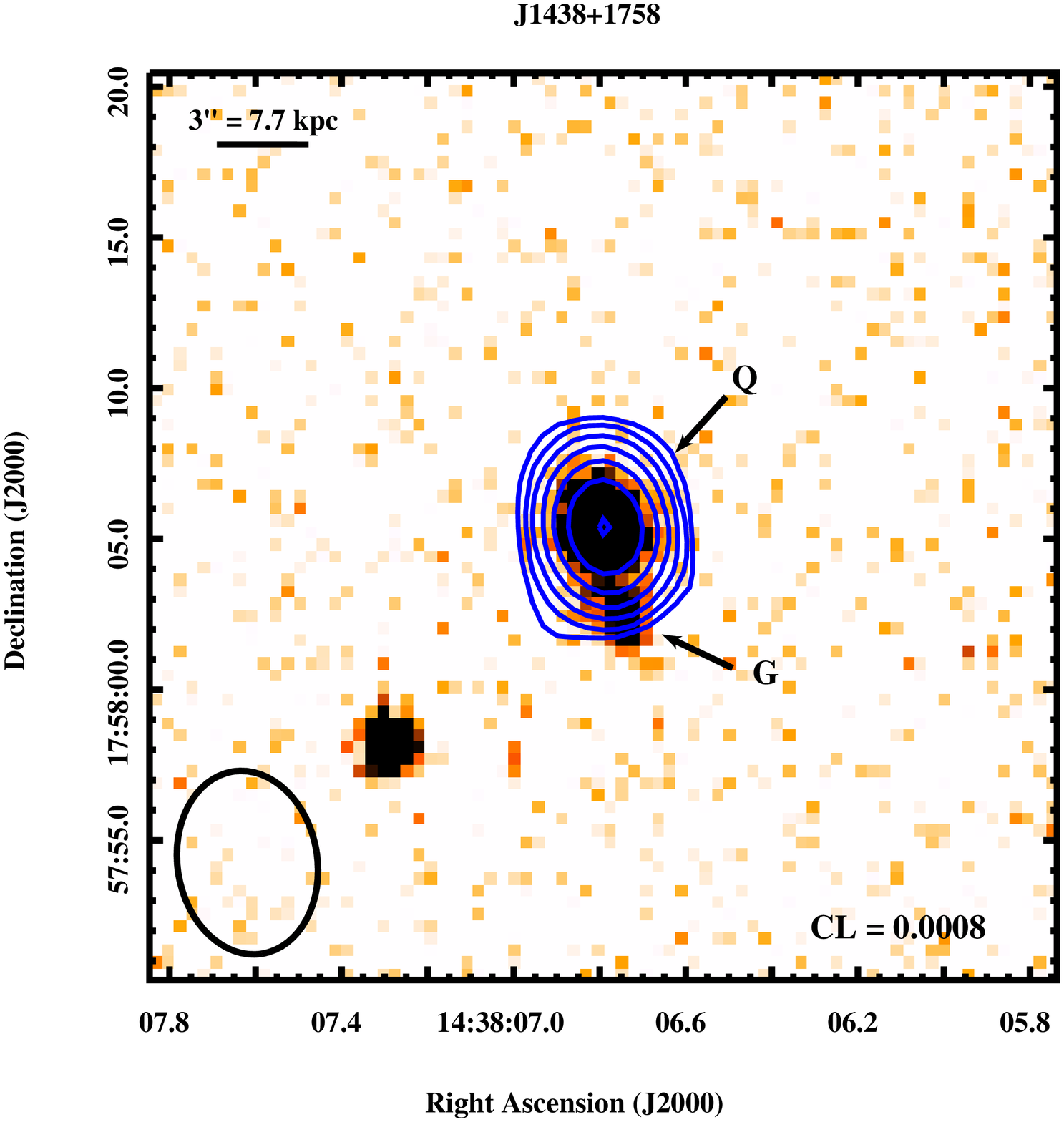} } 
\subfloat[]{ \includegraphics[width=0.33\textwidth, angle=90]{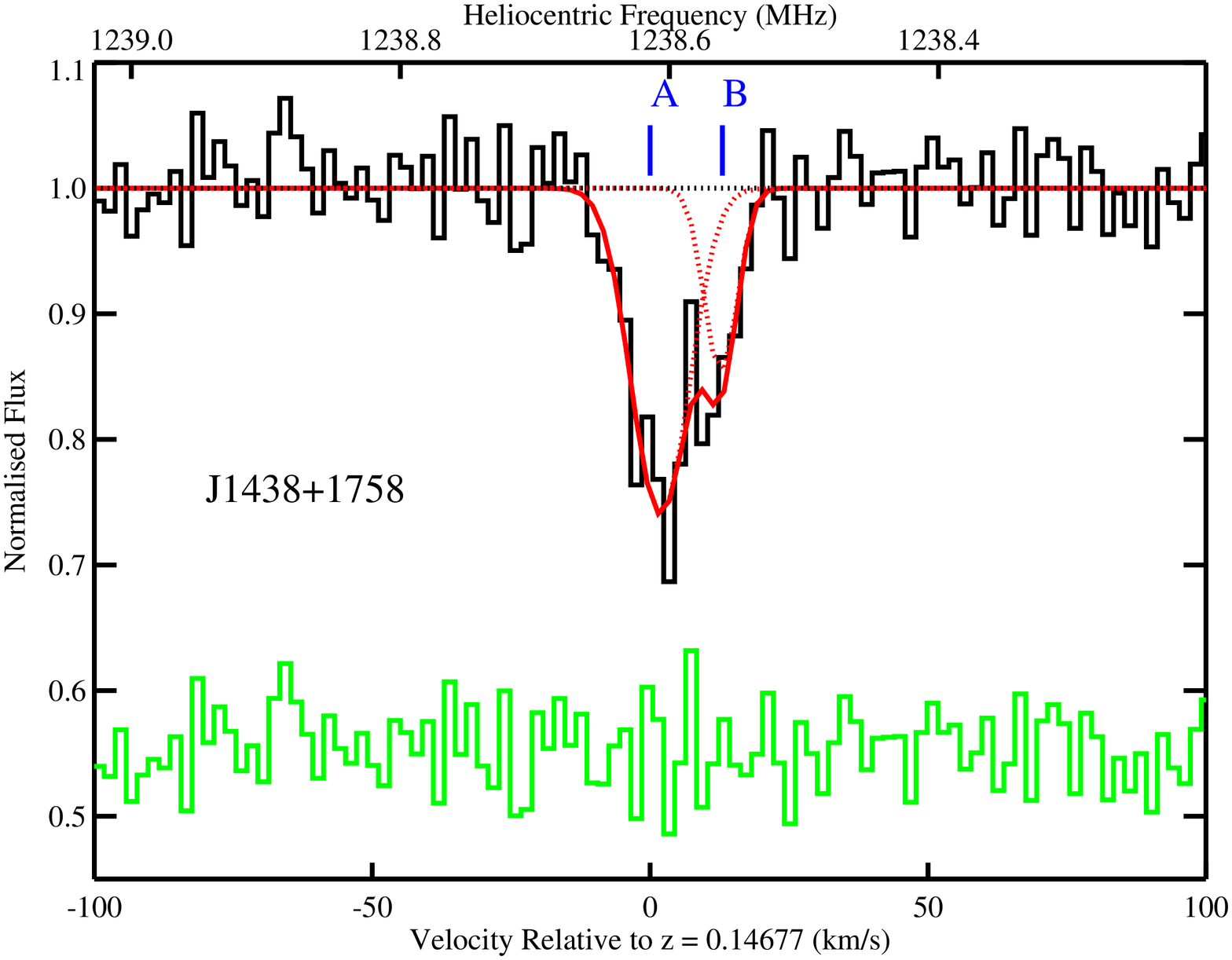} } \hspace{0.01cm}
\subfloat[]{ \includegraphics[width=0.32\textwidth, bb= 55 40 735 735, clip=true]{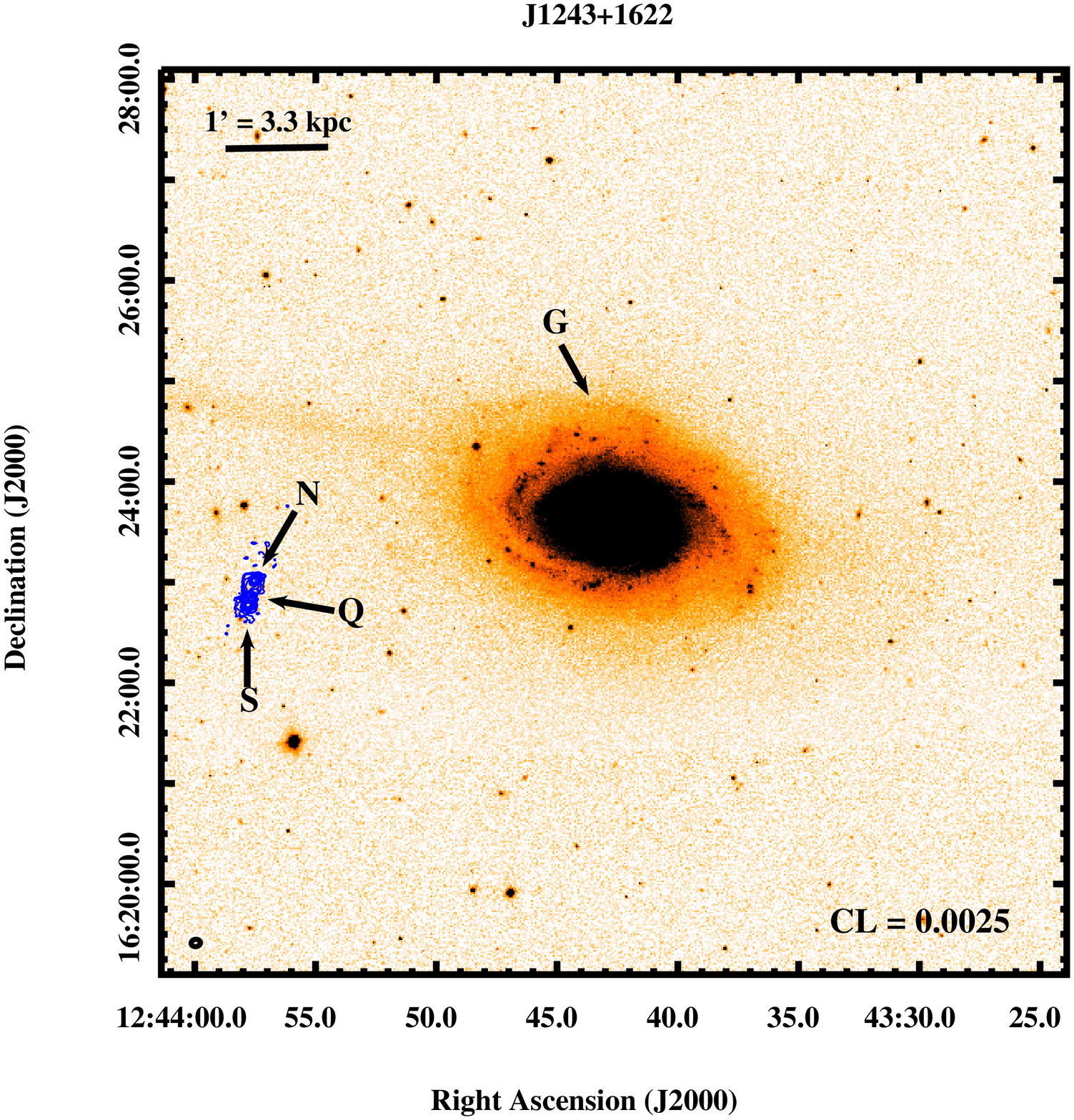} }
\subfloat[]{ \includegraphics[width=0.33\textwidth, angle=90]{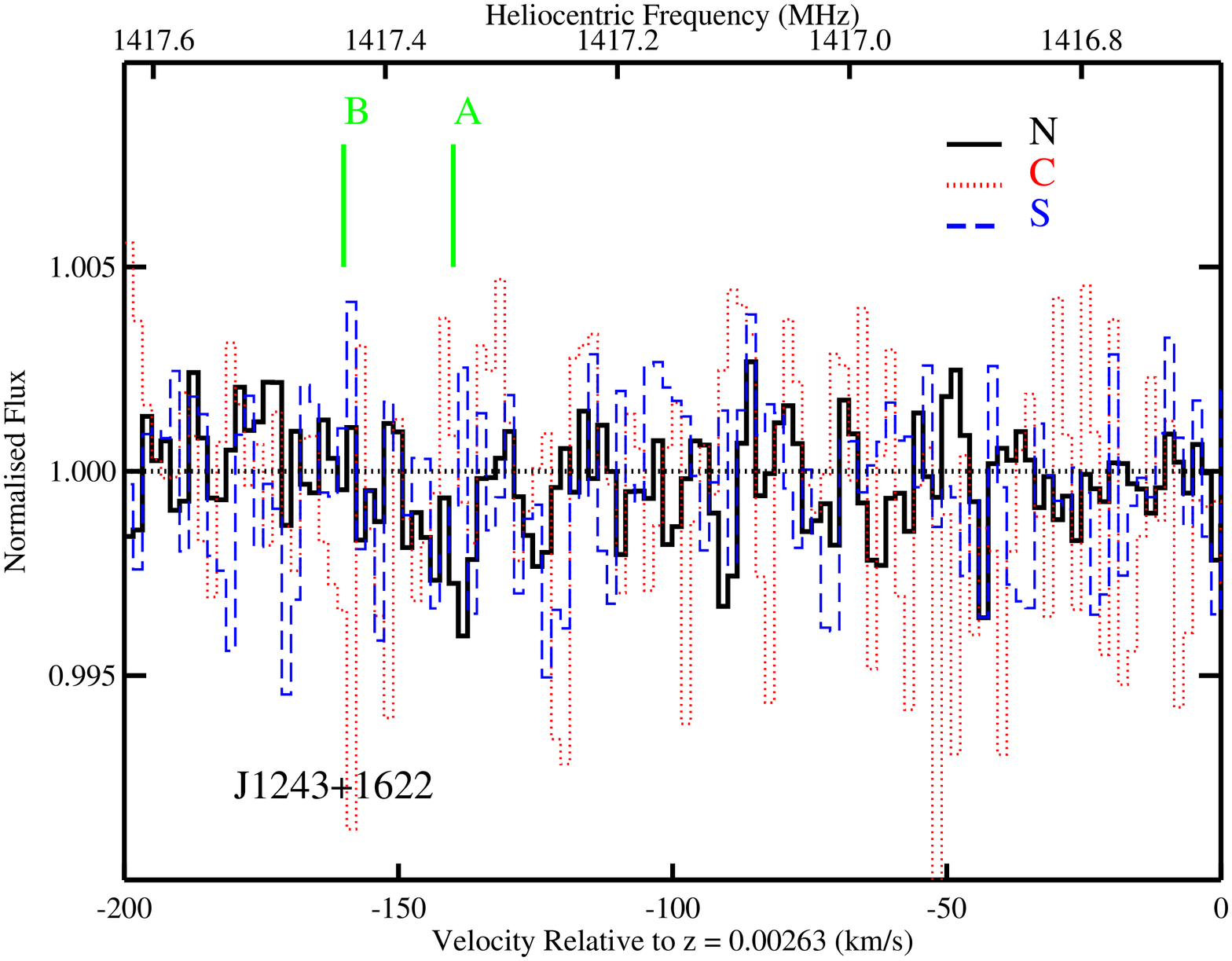} } \hspace{0.01cm}
\caption{{\it Left:} SDSS $r$-band images overlaid with the 1.4 GHz continuum contours of the QGPs (a) J1438$+$1758 and (c) J1243$+$1622. 
In each image the quasar is marked by `Q' and the galaxy is marked by `G'. In case of J1243$+$1622 the north and south lobes are marked by `N' and `S', respectively. 
The restoring beam of the continuum map is shown at the bottom left corner. The spatial scale is indicated at the top left corner. The contour levels are plotted as 
CL $\times$ ($-$1,1,2,4,8,...)~Jy~beam$^{-1}$, where CL is given in the bottom right of each image. The rms in the images (a) and (b) are 0.2 and 0.5 mJy~beam$^{-1}$, respectively.
{\it Right:} \21\ absorption spectra towards (b) J1438$+$1758 and (d) J1243$+$1622. In case of J1438$+$1758, individual Gaussian components and the resultant fits 
to the absorption profile are plotted as dotted and continuous lines, respectively. Residuals, shifted by an arbitrary offset for clarity, are also shown. Locations 
of the peak optical depth of the individual components are marked by vertical ticks. In case of J1243$+$1622, the \hi\ \21\ absorption spectra towards the core and southern 
lobe are overplotted on the spectrum towards the northern lobe (solid line), in dotted and dashed lines, respectively. Locations of the weak tentative absorption features 
suggested by \citet{schneider1993} are marked by vertical lines. We do see weak albeit statistically insignificant ($\le3\sigma$) features at these locations even in our spectra.
}
\label{fig:gmrtcontours1} 
\end{figure*}

\begin{figure*}
\subfloat[]{ \includegraphics[width=0.25\textwidth, angle=90]{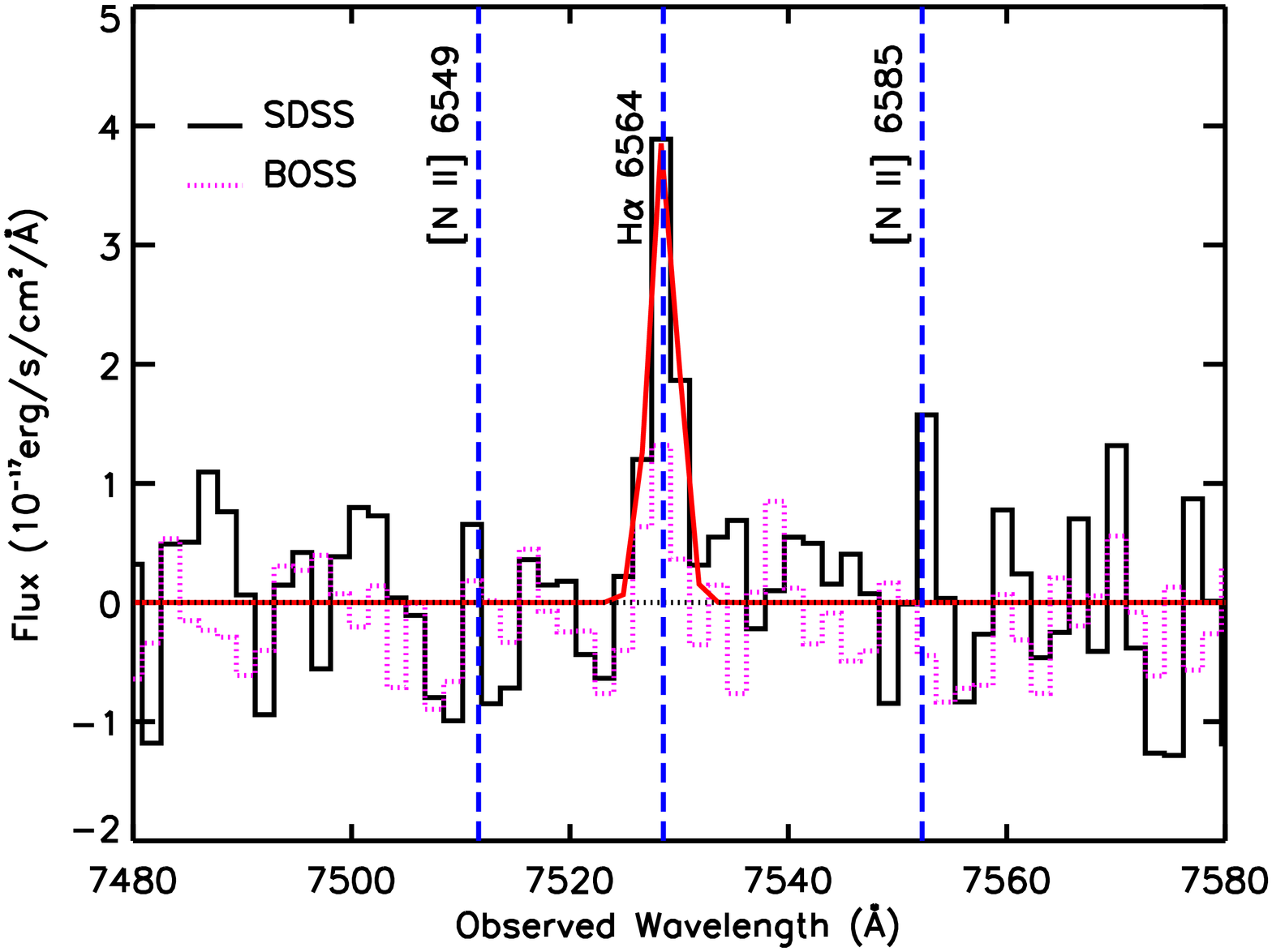} }
\subfloat[]{ \includegraphics[width=0.25\textwidth, angle=90]{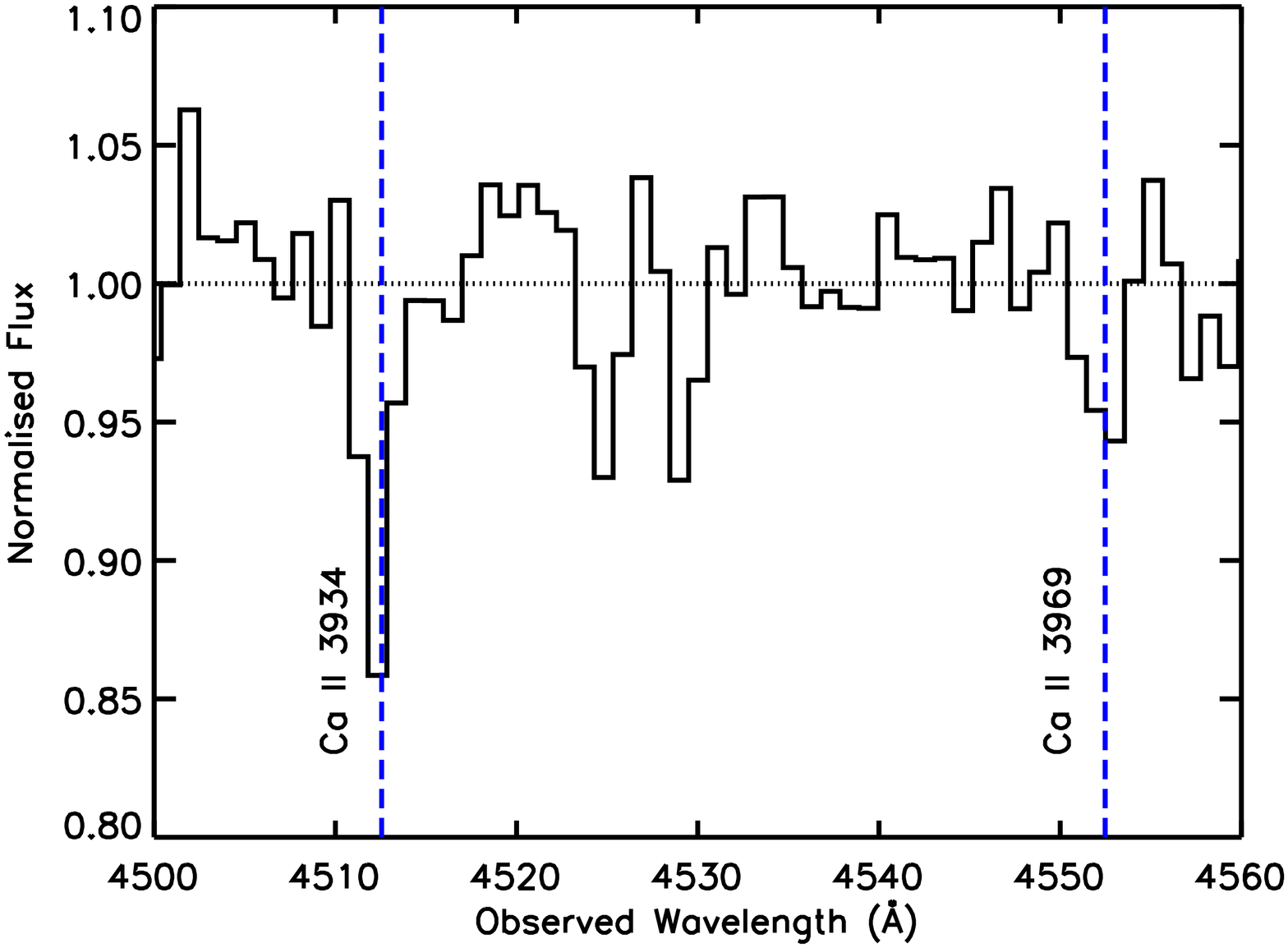} }
\subfloat[]{ \includegraphics[width=0.25\textwidth, angle=90]{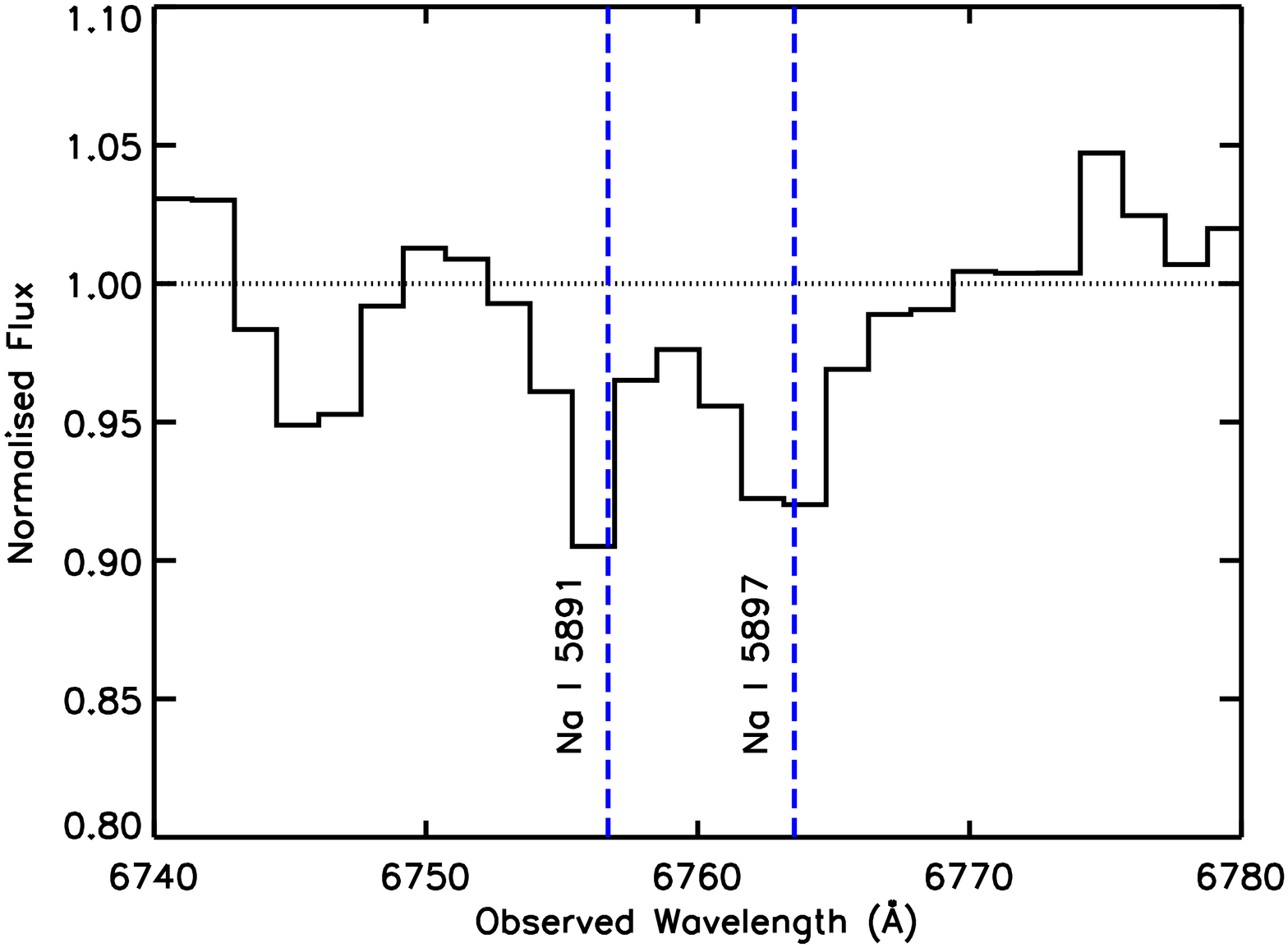} }
\caption{Gaussian fit to the \ha\ emission at \zgal\ = 0.1468 towards J1438$+$1758 is overplotted on the SDSS-DR7 spectrum, after subtracting the quasar continuum, in panel (a). 
The SDSS-BOSS spectrum is also shown in dotted histogram for comparison. The flux normalized SDSS-BOSS spectrum towards J1438$+$1758 with the \caii\ and \nai\ absorption detected at 
\zgal\ = 0.1468, is shown in panels (b) and (c), respectively. The vertical ticks mark the position of the \ha\ and [N\,{\sc ii}] emission lines and absorption lines of \caii\ and \nai\ doublet.}
\label{fig:j1438_sdssspec}
\end{figure*}

The QGP J1438$+$1758 is a GOTOQ, i.e. the foreground galaxy was identified by us from emission lines in the SDSS spectrum of the background quasar. The foreground galaxy, located 
$\sim$3$''$ ($b$ = 7.5 kpc) south of the quasar (see panel (a) of Fig.~\ref{fig:gmrtcontours1}), is not identified as a separate photometric object in SDSS. A narrow \ha\ emission 
line is seen at \zgal\ = 0.1468 in both the SDSS-DR7 and the SDSS-Baryon Oscillation Spectroscopic Survey (BOSS) spectra (fibre diameters of 3$''$ and 2$''$ or $\sim$8 and $\sim$5 
kpc at \zgal, respectively), with the line strength being a factor of $\sim$3 weaker in the SDSS-BOSS spectrum than in the SDSS-DR7 spectrum. From Gaussian fit to the \ha\ emission 
(see panel (a) of Fig.~\ref{fig:j1438_sdssspec}), we obtain \zgal\ = 0.1468 $\pm$ 0.0001. From the SDSS-DR7 spectrum, we measure flux of the \ha\ line as, $F$(\ha) = (12.6 $\pm$ 2.8) 
$\times$ 10$^{-17}$ erg~s$^{-1}$~\cms. This gives the dust-uncorrected luminosity of $L$(\ha) = (7.3 $\pm$ 1.6) $\times$ 10$^{39}$ erg~s$^{-1}$. Following \citet{argence2009}, we 
estimate the total star formation rate (SFR) in the portion of the galaxy covered by the SDSS fibre as 0.03 $\pm$ 0.01 M$_\odot$\,yr$^{-1}$ (uncorrected for dust attenuation and 
fibre filling factor). For the SDSS fibre size, this leads to an average surface density of the SFR ($\Sigma_{SFR}$) of 0.0006 M$_\odot$\,yr$^{-1}$\,kpc$^{-2}$, for unit fibre 
filling factor. We also detect weak [O\,{\sc ii}] emission in the SDSS-DR7 spectrum and measure $L$([O\,{\sc ii}] $\lambda$3727) = (6.9 $\pm$ 2.5) $\times$ 10$^{39}$ erg~s$^{-1}$ 
(uncorrected for dust). This gives the SFR as 0.05 $\pm$ 0.02 M$_\odot$\,yr$^{-1}$ following \citet{kewley2004}, consistent with the SFR obtained from \ha. 

The SFR obtained above is well below the median SFR ($\sim$2$M_\odot$\,yr$^{-1}$) of [O\,{\sc iii}]-selected \mgii\ absorbers at $0.4<z<0.7$ \citep{noterdaeme2010}. 
The $\Sigma_{SFR}$ is an order of magnitude smaller compared to that measured in $z<0.4$ galaxies which show \hi\ \21\ absorption \citep[see Table 3 of][]{gupta2013}, as 
well as that inferred for the CNM in $z>$2 DLAs by \citet{wolfe2003} using C\,{\sc ii}*. Hence, the quasar sightline seems to be probing the outer, low star-forming disc 
of the galaxy. If we assume that the Kennicutt-Schmidt law is obeyed in the outer disc of the galaxy, then we get log \nhi\ = 20.3 \citep{kennicutt1998a,kennicutt1998b}. 
However, if the star formation efficiency in the outer disc is much lower than that predicted by the Kennicutt-Schmidt law \citep[see e.g.][]{bigiel2010,rafelski2011,elmegreen2015}, 
the \nhi\ will be higher.

We estimate the dust optical depth in the systematic $V$ band of the galaxy ($\tau_V$) and the colour excess ($E(B-V)$) along the quasar sightline by fitting 
the quasar spectral energy distribution (SED) using the SDSS composite quasar spectrum \citep{vandenberk2001}, reddened by the Milky Way, LMC and SMC extinction 
curves \citep[see][for the detailed procedure]{srianand2008,srianand2013,noterdaeme2009,noterdaeme2010}. The best-fitting (\chin\ $\sim$ 1.3) Milky Way extinction 
curve gives $A_V$ = 0.10 $\pm$ 0.01 (where $A_V$ = 1.086$\tau_V$). The errors quoted are mainly statistical. We applied the same procedure to a control sample of 
159 SDSS non-BAL quasars within $\Delta z$ = $\pm$0.01 of \zrad\ and $\Delta r_{mag}$ = $\pm$2.0 of $r_{mag}$ of the quasar and with SNR $\ge$10. The resulting 
distribution of $A_V$ has median value of 0.06 and rms of 0.24. This rms reflects a typical systematic error in the SED-fitting method due to the dispersion of 
the unreddened quasar SED. Hence, the quasar sightline does not show any significant evidence of reddening. If we assume $R_V$ = 3.1 as seen in the Milky Way, 
then $E(B-V)$ = 0.03, which is much lower than what is measured in high-$z$ dusty \mgii\ and CO systems with 2175 \AA\ dust absorption bump \citep{srianand2008,noterdaeme2009,jiang2011} 
as well as in low-$z$ galaxies with \hi\ \21\ absorption \citep{gupta2010,gupta2013,srianand2013}. The small value of extinction is more in line with the values 
generally seen towards intervening DLAs, \mgii\ and \caii\ systems \citep{york2006,khare2012,sardane2015}. Using the observed relation between $A_V$ and \nhi\ 
in the Milky Way \citep{bohlin1978}, we get log \nhi\ = 20.2 (1$/\kappa$), where $\kappa$ is the ratio of dust-to-gas ratio in the absorption system to that in 
the Milky Way. 

Both the SDSS-DR7 and the SDSS-BOSS spectra of the quasar show \caii\ and \nai\ absorption lines at the redshifted wavelengths expected for the foreground galaxy (see panels (b) 
and (c) of Fig.~\ref{fig:j1438_sdssspec}). We use the higher SNR SDSS-BOSS spectrum to obtain the rest equivalent widths of the \caii\ $\lambda$3934 and the \nai\ $\lambda$5891 
lines as 0.22 $\pm$ 0.05 \AA\ and 0.27 $\pm$ 0.09 \AA, respectively. The rest equivalent width of \caii\ $\lambda$3934 measured here is consistent with the weak \caii\ absorber 
population proposed by \citet{sardane2015}, which were shown to have an abundance pattern typical of halo gas with less depletion of the refractory elements and a typical 
$E(B-V)$ $\sim$0.01. Using the empirical relation between $E(B-V)$ and rest equivalent width of \nai\ obtained by \citet{poznanski2012}, we estimate $E(B-V)$ = 0.05$^{+0.05}_{-0.03}$, 
consistent within uncertainties with the value obtained from SED fitting. 

We detect \hi\ \21\ absorption towards the background quasar at the redshift of the galaxy. The \hi\ \21\ absorption profile is best fit with two Gaussian components, 
(i.e. \chin\ = 1.2 for double Gaussian fit compared to \chin\ = 2.0 for single Gaussian fit), A and B, separated by $\sim$13\,\kms. The fit to the absorption is 
shown in panel (b) of Fig.~\ref{fig:gmrtcontours1}, and the details of the fit are provided in Table~\ref{tab:j1438_21cmfit}. The velocity offset between the \hi\ \21\ 
absorption components and the redshift estimated from the \ha\ emission is well within the redshift measurement error. The total integrated optical depth of the \hi\ \21\ 
absorption is 4.89 $\pm$ 0.19 \kms\, which yields \nhi\ (\fc$/$\ts) = (8.9 $\pm$ 0.4) $\times$ 10$^{18}$ \cms\,K$^{-1}$. 90\% of the total optical depth is contained within 
20 \kms. Our GMRT image of the background radio source (panel (a) of Fig.~\ref{fig:gmrtcontours1}) shows that it is compact with a flux density of 53 mJy, consistent with 
that obtained by FIRST. High-resolution (sub-arcsecond-scale) map is not available for this source, and hence \fc\ is uncertain. For \fc\ = 1 and a typical CNM mean temperature 
of 100 K, as observed in the Milky Way \citep{heiles2003}, log \nhi\ = 20.95 $\pm$ 0.02. Comparing this with the \nhi\ obtained from SED fitting above gives $\kappa$ = 0.2, 
consistent with the absorption arising from cold gas in the outer disc of the galaxy, where the dust-to-gas ratio is expected to be lower.
\begin{table}
\caption{Details of the Gaussian fit to the 21-cm absorption line detected towards J1438$+$1758.}
\centering
\begin{tabular}{cccccc}
\hline
ID  & \zabs & FWHM   & $\tau_{p}$ & $T_k$ & \nhi\ (\fc$/$\ts)           \\
    &       & (\kms) &            & (K)   & ($10^{19}$\,\cms\,K$^{-1}$) \\
(1) & (2)   & (3)    & (4)        & (5)   & (6)                         \\
\hline
A & 0.14677 & 12 $\pm$ 1 & 0.33 $\pm$ 0.02 & $<$3150 & 0.76 $\pm$ 0.11 \\
B & 0.14682 & 7 $\pm$ 1  & 0.15 $\pm$ 0.03 & $<$1070 & 0.20 $\pm$ 0.07 \\
\hline
\end{tabular}
\label{tab:j1438_21cmfit}
\begin{flushleft}
Column 1: Absorption component identification. Column 2: Redshift of absorption component. Column 3: Full width at half maximum (FWHM; \kms). 
Column 4: Peak optical depth. Column 5: Upper limit on the gas kinetic temperature ($T_k$; K), obtained assuming the line width is purely due to thermal motions. 
Column 6: $N$(H\,{\sc i}) in terms of the covering factor, \fc, and the spin temperature, \ts, in units of $10^{19}$\,\cms\,K$^{-1}$.
\end{flushleft}
\end{table}
\subsection{QGP J1243$+$1622}
\label{sec_J1243}
The QGP J1243$+$1622 (also known as 3C 275.1$/$NGC 4651; \zrad\ = 0.56; \zgal\ = 0.0026) has been studied in \hi\ \21\ absorption and emission by \citet{corbelli1990,schneider1993}. 
A tentative \hi\ \21\ absorption feature, with peak optical depth of 0.0067 $\pm$ 0.0012, was found in their VLA C-configuration observations at heliocentric velocity 643\,\kms.
In their subsequent VLA B-configuration observations, the radio source (3C 275.1) resolves into two components. However, the absorption feature appeared weaker in the spectra towards 
both these components, as well as in the spectra extracted from data smoothed to the resolution of the C-configuration. Another tentative feature was identified at heliocentric velocity 
636\,\kms, with peak optical depth of 0.0059 $\pm$ 0.0016 and integrated optical depth of 0.03\,\kms, towards the southern component. The authors also report \hi\ \21\ emission along the 
quasar sightline from the outer disk of NGC 4651 at the level of \nhi\ $\sim$ 3$\times$ 10$^{19}$ \cms\ from VLA D-configuration observations. Combining this with their optical depth 
measurement they constrain the \hi\ gas in the outer \hi\ disk of NGC 4651 to be warmer than $\sim$500 K.

The radio source resolves into a core (cospatial with the optical quasar) and two lobes separated by $\sim$15$''$ in our GMRT image (panel (c) of Fig.~\ref{fig:gmrtcontours1}). 
There appears to be tentative weak absorption features (marked by `A' and `B' in panel (d) of Fig.~\ref{fig:gmrtcontours1}), near the velocities of the tentative features reported 
by \citet{schneider1993}, in the spectra towards the continuum peaks of the northern lobe and the core. 
However, the integrated optical depth of these features (0.03 $\pm$ 0.01 \kms\ and 0.04 $\pm$ 0.02 \kms\ for A and B, respectively) are not significant ($\le$3$\sigma$). 
We obtain \taudvl\ $\le$ 0.030 \kms\ towards the two lobes and $\le$ 0.060 \kms\ towards the core, consistent with the measurement of \citet{schneider1993}. All three radio 
sightlines are at $b\sim$ 12 kpc from NGC 4651. \citet{chung2009} have estimated the isophotal \hi\ diameter (where the azimuthally averaged \hi\ surface density falls to 1 
$M_\odot$pc$^{-2}$) and the effective \hi\ diameter (encompassing 50\% of the total flux) of NGC 4651 to be 30.3 kpc and 15.6 kpc, respectively. From the high resolution and 
high sensitivity \hi\ emission map of NGC 4651 presented in \citet{chung2009}, the \hi\ distribution is asymmetric and extended to the west, i.e. away from the radio sightlines. 
In the opposite side of the \hi\ tail i.e. to the north-east, a stellar tail is found in the optical images (see Fig.~\ref{fig:gmrtcontours1}). The quasar sightlines lie just 
at the edge of the \hi\ disk of NGC 4651 in this map, where \nhi\ $\le$ 6$\times$ 10$^{19}$ \cms, consistent with \citet{schneider1993}. It is important to note that cold gas, 
if present in the outer disk of NGC 4651, could be clumpy and the \nhi\ obtained from \hi\ emission represents the average \hi\ surface density in that region. The \hi\ \21\ 
optical depth sensitivity achieved in this case will be good enough to detect \ts\ $\sim$ 100 K gas for \nhi\ = $10^{19}$ \cms\ when \fc\ = 1. The lack of \hi\ \21\ absorption 
is consistent with either \nhi\ being very small or \ts$/$\fc\ $\gtrsim$ 550 K towards the core and $\gtrsim$ 1100 K towards the two lobes if we assume \nhi\ = 6$\times$ 10$^{19}$ 
\cms\ as seen in the outer regions of the emission map.
\subsection{QGP J1300$+$2830}
\label{sec_J1300}
\begin{figure*}
\subfloat[]{ \includegraphics[width=0.32\textwidth, bb= 70 20 755 730, clip=true]{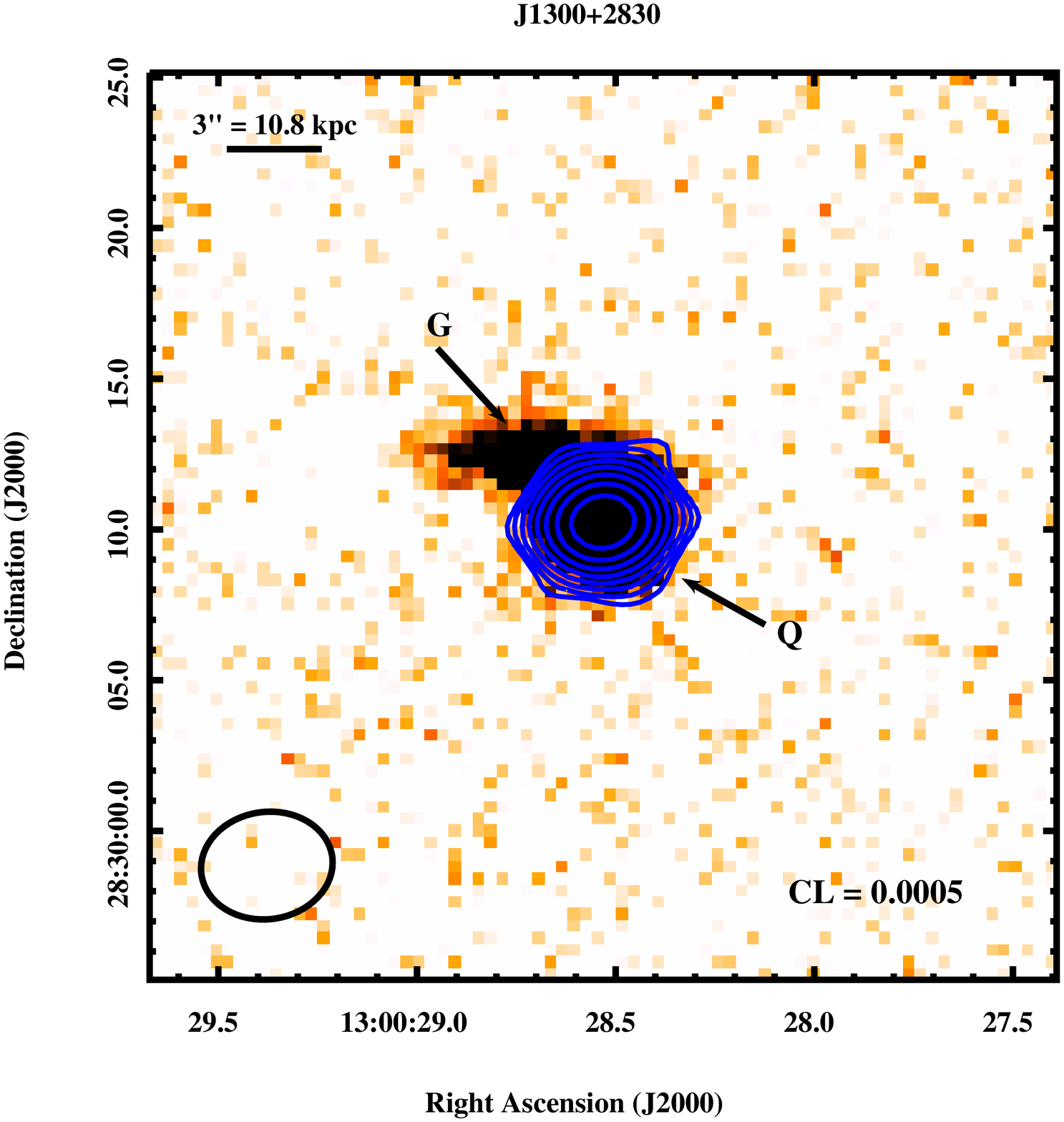} } 
\subfloat[]{ \includegraphics[width=0.33\textwidth, angle=90]{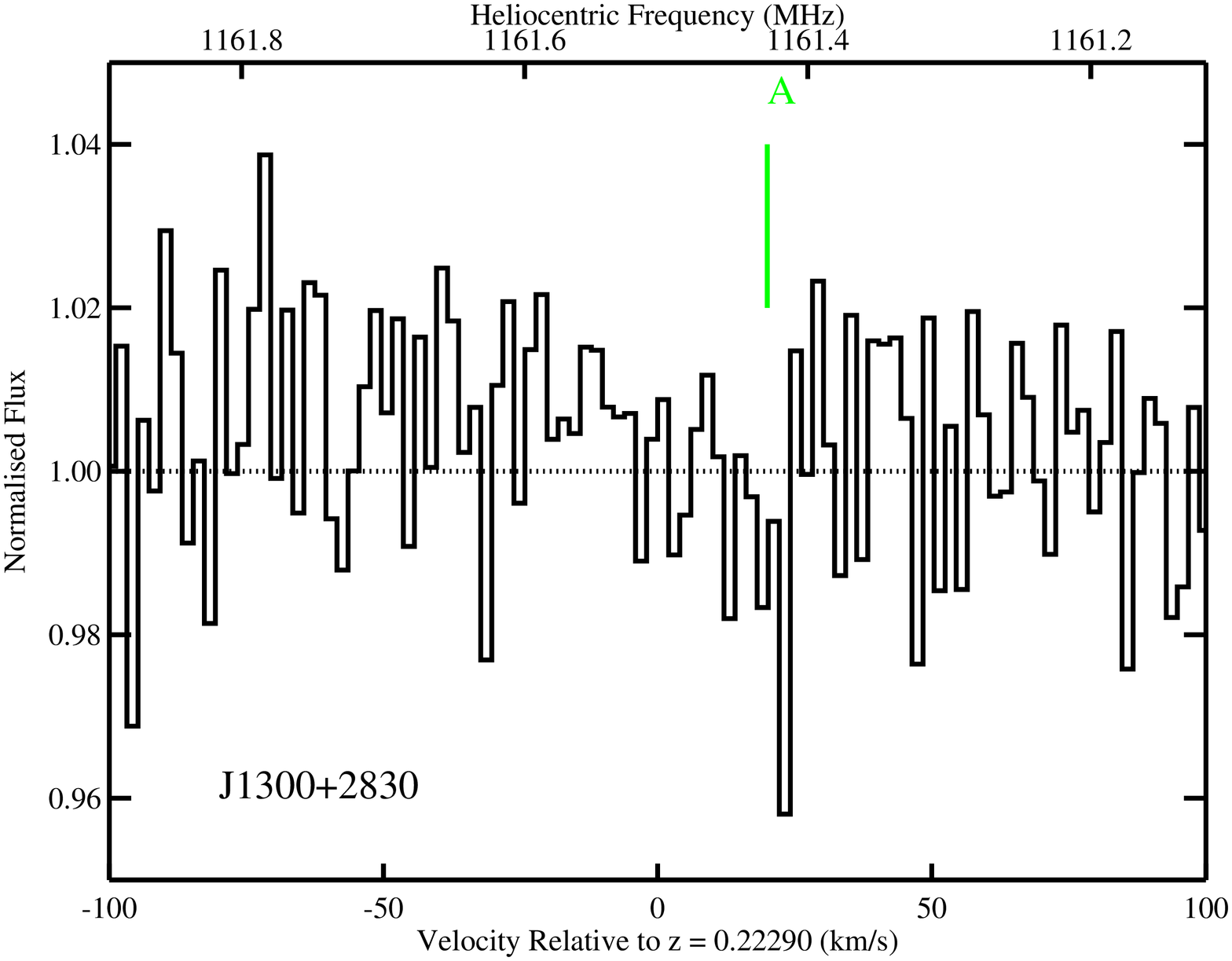} } \hspace{0.01cm}
\subfloat[]{ \includegraphics[width=0.32\textwidth, bb= 55 10 765 745, clip=true]{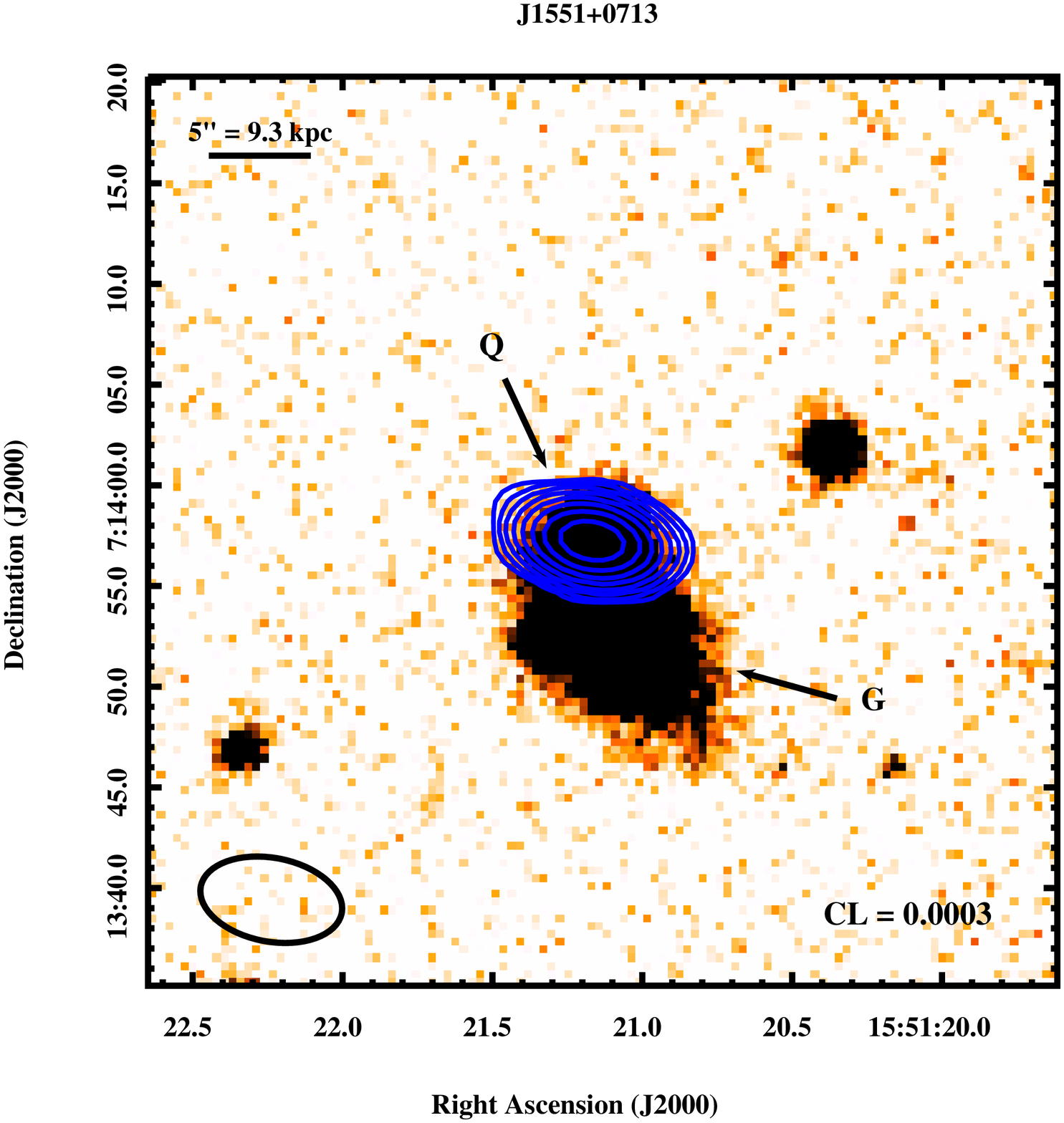} } 
\subfloat[]{ \includegraphics[width=0.33\textwidth, angle=90]{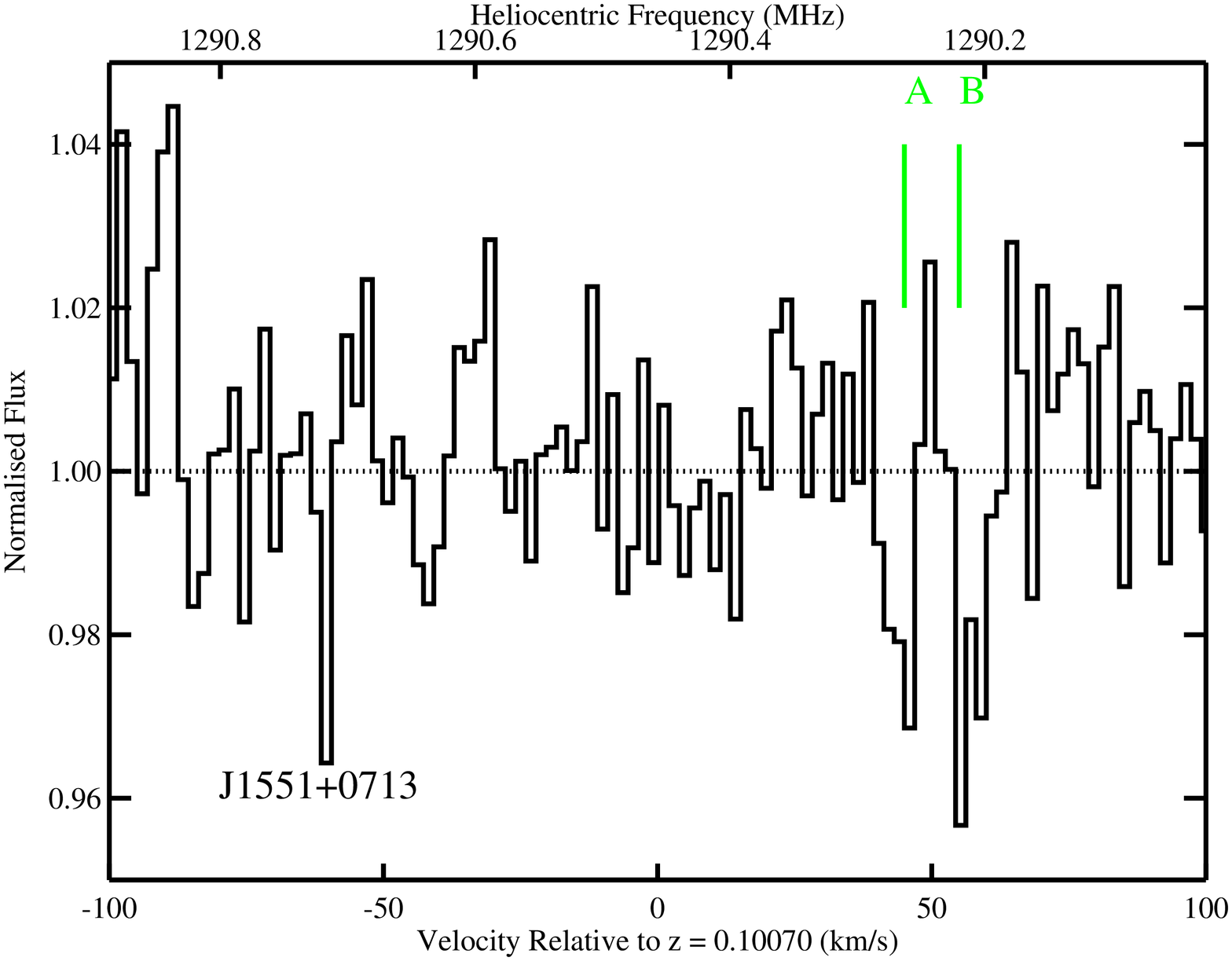} }
\caption{{\it Left:} SDSS $r$-band images overlaid with the 1.4 GHz continuum contours of the QGPs (a) J1300$+$2830 and (c) J1551$+$0713. 
In each image the quasar is marked by `Q' and the galaxy is marked by `G'. The restoring beam of the continuum map is shown at the bottom left corner. 
The spatial scale is indicated at the top left corner. The contour levels are plotted as CL $\times$ ($-$1,1,2,4,8,...)~Jy~beam$^{-1}$, where CL is given in the bottom right of each image. 
The rms in the images is 0.1 mJy~beam$^{-1}$. {\it Right:} \hi\ \21\ absorption spectra towards (b) J1300$+$2830 and (d) J1551$+$0713. Locations of the weak tentative absorption features present 
in these spectra are identified by vertical ticks.}
\label{fig:gmrtcontours2} 
\end{figure*}

\begin{figure*}
\subfloat[]{ \includegraphics[width=0.25\textwidth, angle=90]{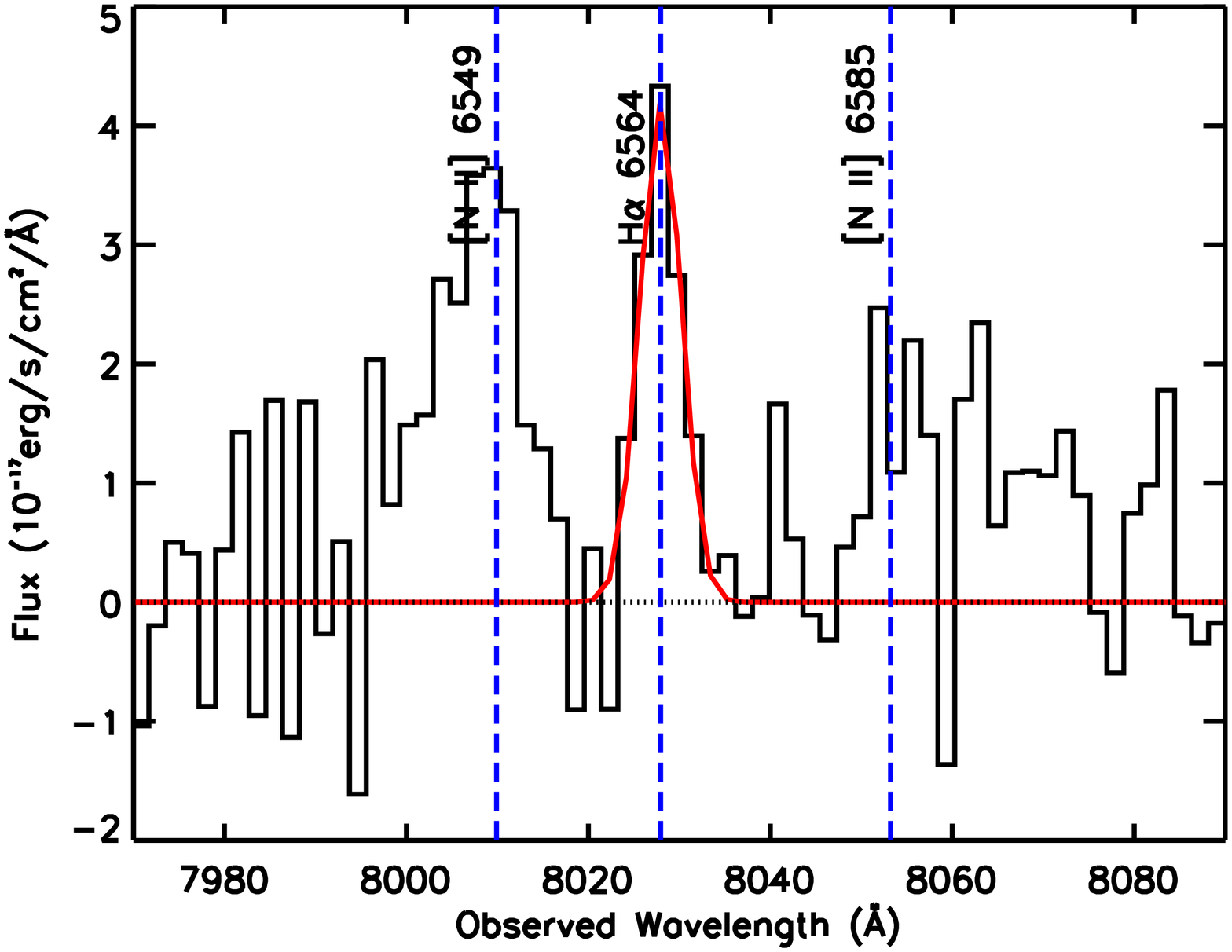} }
\subfloat[]{ \includegraphics[width=0.25\textwidth, angle=90]{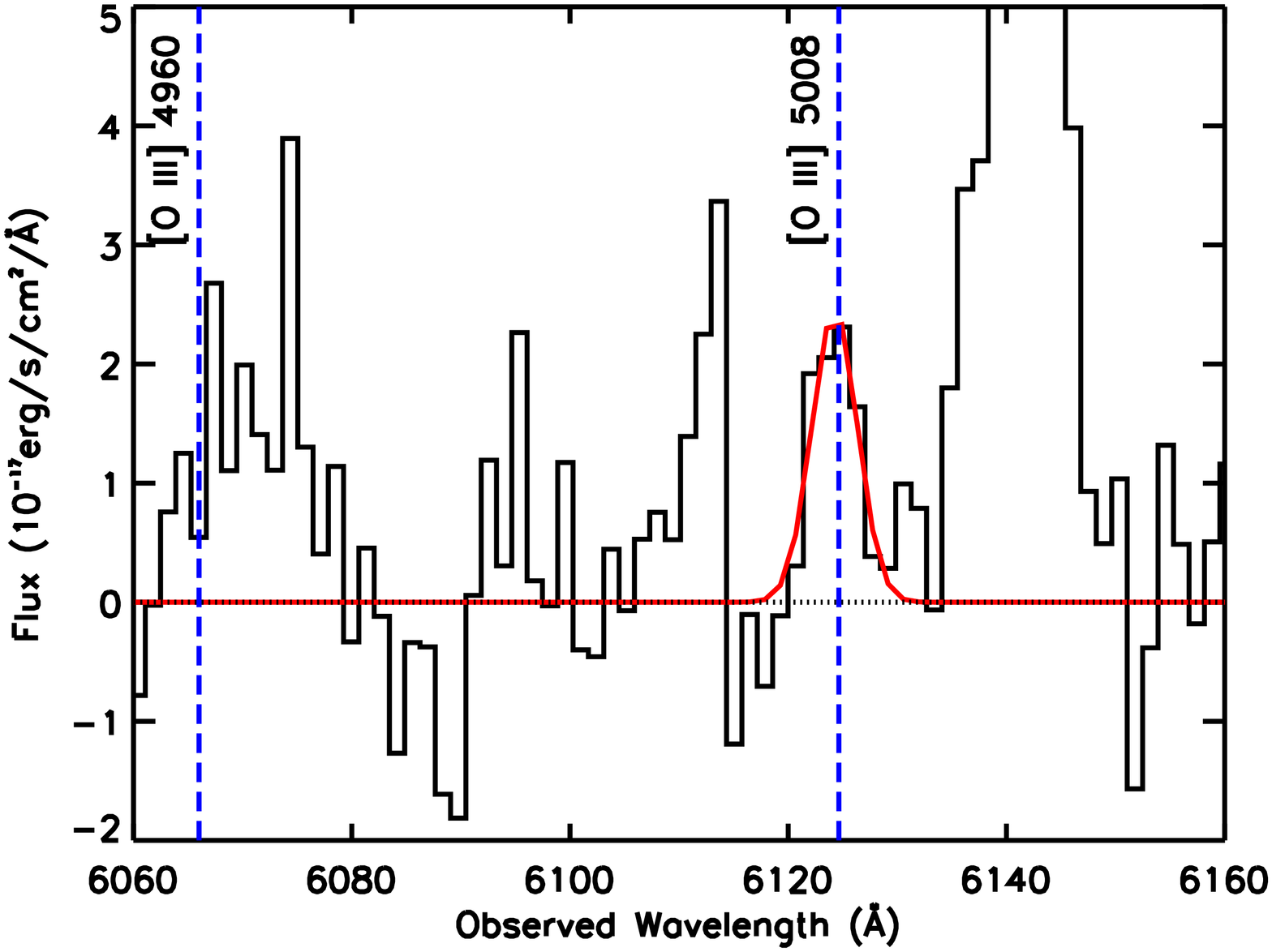} }
\subfloat[]{ \includegraphics[width=0.25\textwidth, angle=90]{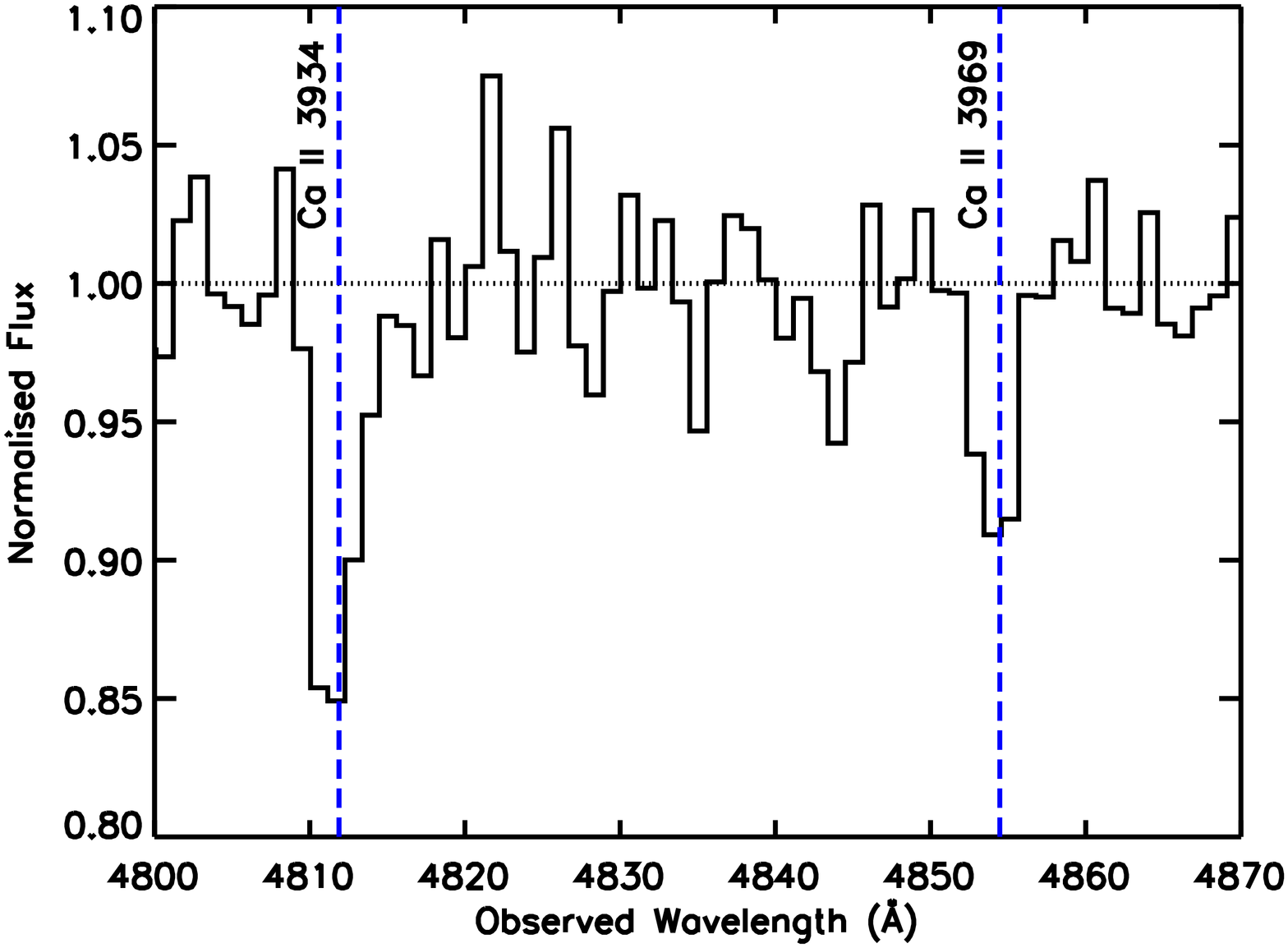} }
\caption{Gaussian fits to the \ha\ and [O\,{\sc iii}] $\lambda$5008 emission lines at \zgal\ = 0.2229 towards J1300$+$2830 are overplotted on the SDSS spectrum, 
after subtracting the quasar continuum, in panels (a) and (b), respectively. 
The flux normalized SDSS spectrum towards J1300$+$2830 with the \caii\ absorption detected at \zgal\ = 0.2229 is shown in panel (c).
The vertical ticks mark the position of the \ha, [N\,{\sc ii}] and [O\,{\sc iii}] emission lines and the absorption from \caii\ doublet.}
\label{fig:j1300_sdssspec}
\end{figure*}

The QGP, J1300$+$2830, is a GOTOQ with the foreground galaxy located $\sim$3$''$ ($b$ = 9.4 kpc) north-east of the quasar (\zrad\ = 0.65) in the SDSS image (see panel (a) of 
Fig.~\ref{fig:gmrtcontours2}). A narrow \ha\ emission line from the foreground galaxy is detected on top of the broad H$\beta$ emission from the background quasar in the SDSS 
spectrum. From Gaussian fit to the \ha\ emission line, we obtain \zgal\ = 0.2229 $\pm$ 0.0001 and $F$(\ha) = (23.52 $\pm$ 6.47) $\times$ 10$^{-17}$ erg~s$^{-1}$~\cms\ (see panel 
(a) of Fig.~\ref{fig:j1300_sdssspec}). This leads to a dust-uncorrected luminosity of $L$(\ha) = (3.45 $\pm$ 0.95) $\times$ 10$^{40}$ erg~s$^{-1}$ and SFR (in the portion of the 
galaxy covered by the SDSS fibre) = 0.14 M$_\odot$\,yr$^{-1}$ \citep{argence2009}. A weak [O\,{\sc iii}] $\lambda$5008 emission line is also detected at a redshift consistent with 
the \ha\ emission, on top of [O\,{\sc ii}] emission from the quasar (see panel (b) of Fig.~\ref{fig:j1300_sdssspec}), which gives $L$([O\,{\sc iii}] $\lambda$5008) = (3.44 $\pm$ 1.06) 
$\times$ 10$^{40}$ erg~s$^{-1}$ (uncorrected for dust).

The $\Sigma_{SFR}$ = 0.003 M$_\odot$\,yr$^{-1}$\,kpc$^{-2}$ is comparable to the values measured in $z<0.4$ galaxies which show \hi\ \21\ absorption \citep[see Table 3 of][]{gupta2013}, 
as well as that inferred for the CNM in $z>$2 DLAs by \citet{wolfe2003} using C\,{\sc ii}*. However, we do not detect \hi\ \21\ absorption from this galaxy (\taudvl\ $\le$0.239\,\kms). 
It is interesting to note that this is the only GOTOQ, out of the six (five in our sample and one from \citet{borthakur2010}) that have been searched for \hi\ \21\ absorption, where \hi\ 
\21\ absorption has not been detected. We show the VLA \hi\ \21\ absorption spectra smoothed to 2\,\kms\ in panel (b) of Fig.~\ref{fig:gmrtcontours2}. We note that a weak absorption feature 
(\taudv\ = 0.10 $\pm$ 0.06) is observed with $\lesssim$2$\sigma$ significance at 20\,\kms\ from the \ha\ redshift. The radio source is compact in our VLA image (panel (a) of Fig.~\ref{fig:gmrtcontours2}) 
with a flux density of 119 mJy, similar to that obtained by \citet{miller2009}. No sub-arcsecond-scale map of the radio source is available at 1.4 GHz.

We measure rest equivalent width of the \caii\ $\lambda$3934 line as 0.43 $\pm$ 0.06 \AA\ (see panel (c) of Fig.~\ref{fig:j1300_sdssspec}). The presence of \caii\ absorption means that there 
is \hi\ gas along the quasar sightline. However, the \nai\ doublet is not detected and we put a 3$\sigma$ upper limit on the rest equivalent width of \nai\ $\lambda$5891 $\le$ 0.11 \AA. In the 
Galactic ISM, $N$($\nai$)/$N$($\caii$) $>$ 1 for the observed values of $N$($\nai$) \citep[][figure~9]{welty1996}, while the present ratio is $\lesssim$0.1 assuming linear part of the curve of 
growth. This could mean that the gas is not arising from the dusty and dense ISM where Ca depletion is high ($\sim$ $-$3 dex). This is discussed further in Section~\ref{sec_metalabs}. 
\subsection{QGP J1551$+$0713}
\label{sec_J1551}
The QGP J1551$+$0713 consists of a quasar (\zrad\ = 0.68) whose sightline appears to pass through the outer optical disc of a foreground galaxy (5$''$ north of the galaxy's centre) in the 
SDSS image (see panel (c) of Fig.~\ref{fig:gmrtcontours2}). We obtained the redshift of the galaxy, \zgal\ = 0.1007 $\pm$ 0.0005, from our SALT long-slit spectrum. The galaxy appears to be 
a non-star-forming early-type galaxy. No emission lines are detected in the SALT spectrum of the galaxy, but stellar absorption from \ha, \caii\ and \nai\ lines are detected (see 
Fig.~\ref{fig:saltspec1}), implying an old stellar population.

The radio emission from the quasar is compact in our GMRT image (panel (c) of Fig.~\ref{fig:gmrtcontours2}) with a flux density of 68 mJy, consistent with that obtained by FIRST. 
No sub-arcsecond-scale image of this quasar is available. Weak tentative absorption features are detected towards this quasar (marked by `A' and `B' in panel (d) of Fig.~\ref{fig:gmrtcontours2}). 
However they are not statistically significant ($\le$2.0$\sigma$), with \taudv\ = 0.10 $\pm$ 0.06 \kms\ and 0.13 $\pm$ 0.07 \kms\ for A and B, respectively. For a typical CNM temperature 
of 100 K, the \taudvl\ $\le$0.265 \kms\ constraint leads to log~\nhi\ $\le$ 19.7. The \hi\ \21\ non-detection in this case can be due to the absence of gas along the sightline or most of 
the gas along the sightline being warm. The first possibility is supported by the non-detection of \caii\ and \nai\ absorption lines at the redshift of the galaxy in the quasar spectrum, 
with 3$\sigma$ upper limits on rest equivalent widths as 0.13\,\AA\ (\caii\ $\lambda$3934) and 0.09\,\AA\ (\nai\ $\lambda$5891), respectively. 
\section{Statistical Analyses}
\label{sec_results}
In this section we study the optical depth and covering factor of \hi\ \21\ absorbers around low-$z$ galaxies as function of their spatial location as well as the host galaxy properties.
To estimate the covering factor of \hi\ \21\ absorbers (\c21) we use a 3$\sigma$ optical depth sensitivity, \t0, i.e. \c21\ is defined as the fraction of systems showing \hi\ \21\ detections 
with \taudvl\ $\le$ \t0\ and \taudv\ $\ge$ \t0. As mentioned in Section~\ref{sec_radioobs}, $\sim$90\% of the QGPs in our primary sample have \taudvl\ $\le$ 0.3\,\kms, corresponding to a 
sensitivity of log~\nhi\ $\le$ 19.7 for \ts\ = 100 K and \fc\ = 1. Hence, we use \t0\ = 0.3 \kms\ throughout this work, unless mentioned otherwise, to estimate \c21. For the statistical 
analyses in this section we include the measurements from our primary sample and the literature sample. In total there are 64 galaxies (69 radio sightlines) in the Combined sample, 
out of which \hi\ \21\ absorption has been detected in 16 sightlines (see Table~\ref{tab:samples}). Note that for the QGPs, J1015$+$1637, J1606$+$2717 and J1748$+$7005, we adopt the
more sensitive optical depth constraints obtained by \citet{borthakur2016}, \citet{borthakur2014} and \citet{greisen2009}, respectively, for the purpose of the statistical analyses here.
\subsection{Radial profile of \hi\ \21\ absorbers}
\label{sec_distance}
\subsubsection{Impact parameter}
\label{sec_impact}
\begin{figure*}
\includegraphics[width=0.35\textwidth, angle=90]{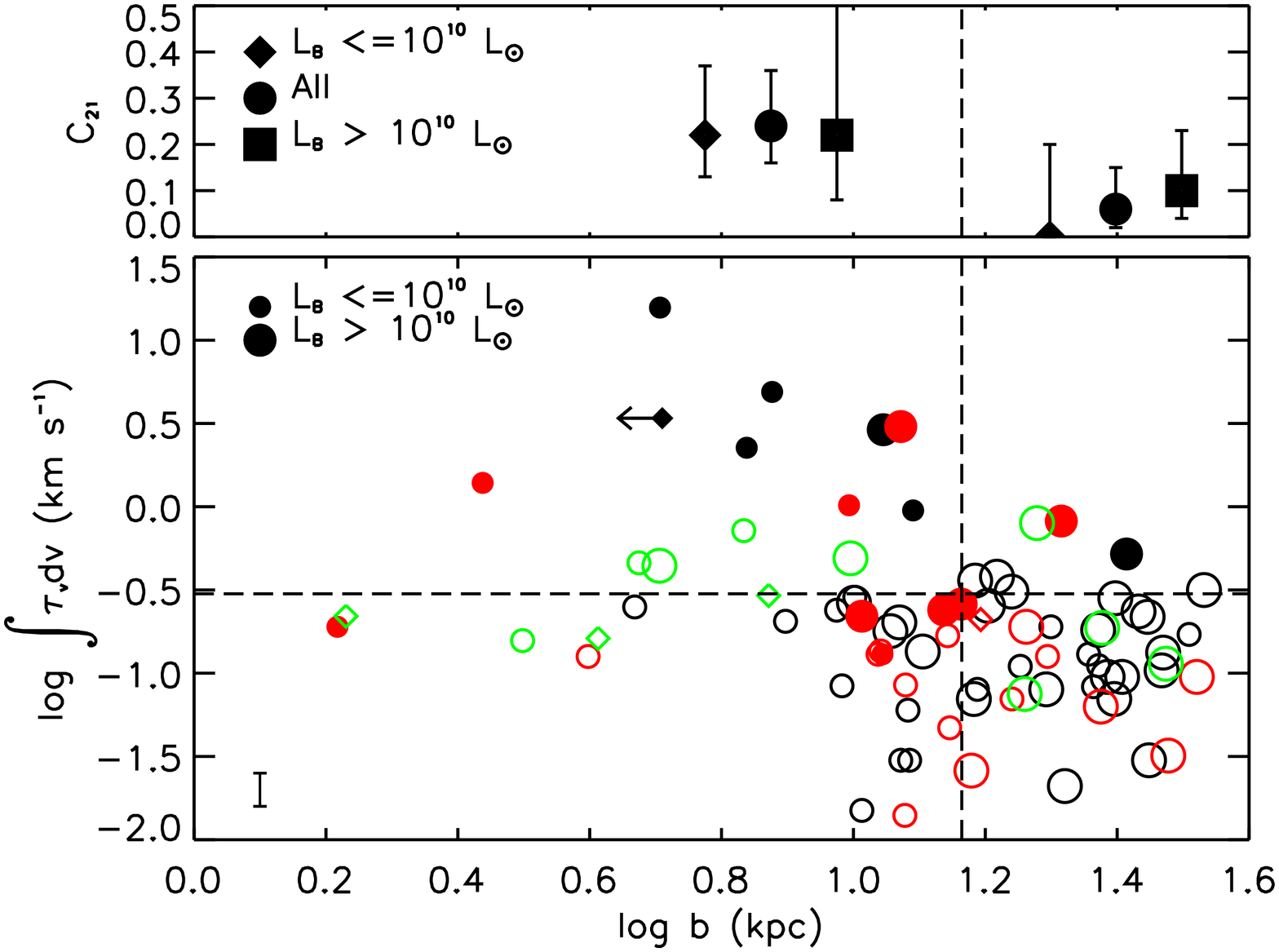} 
\includegraphics[width=0.35\textwidth, angle=90]{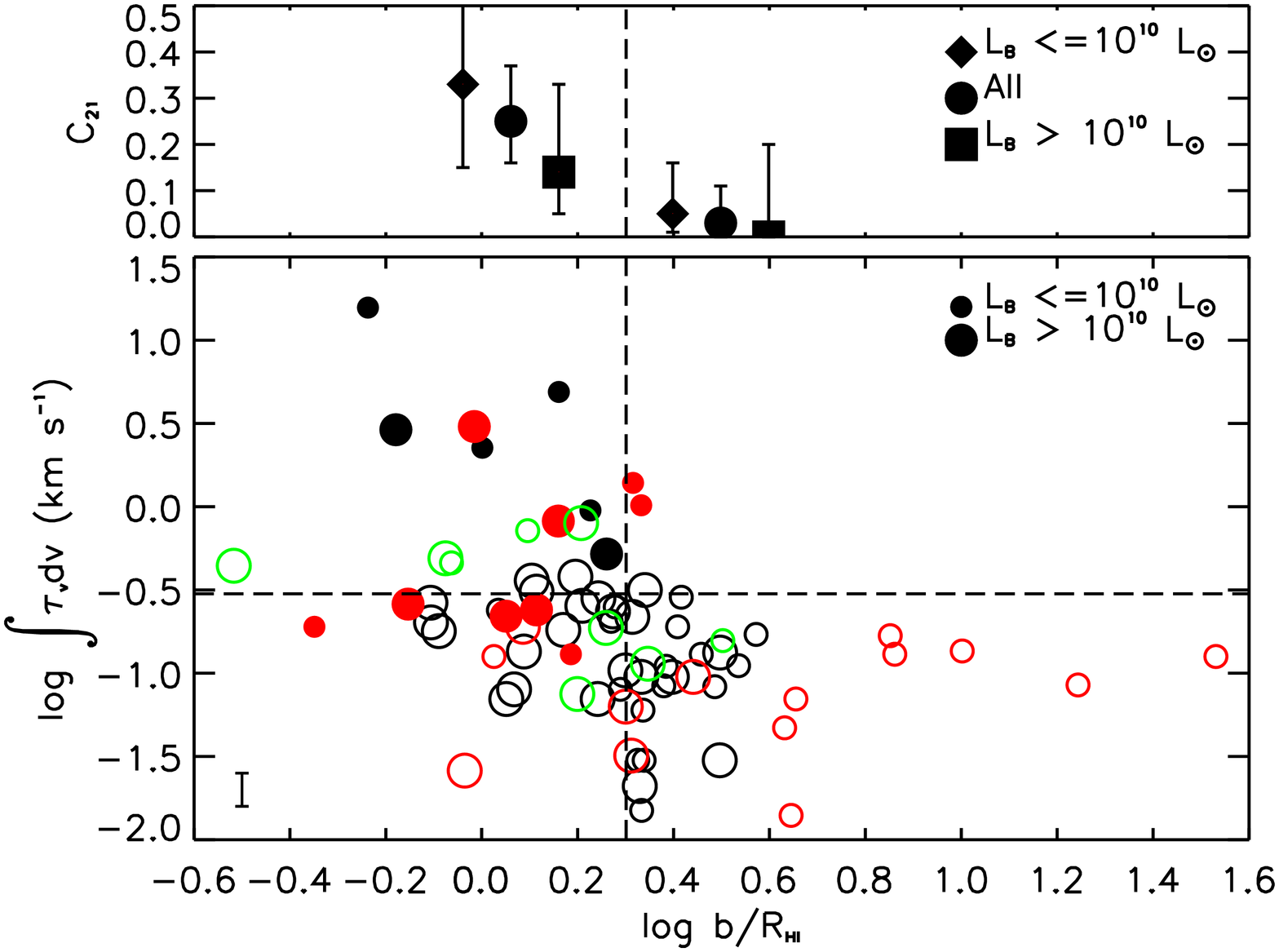} 
\caption{{\it Left:} Integrated \hi\ \21\ optical depth around galaxies as a function of impact parameter. The black, red and green circles are for QGPs in the primary sample, the literature sample and the 
supplementary sample, respectively. The solid and open circles represent \taudv\ and \taudvl\ of the \hi\ \21\ detections and non-detections, respectively. The small circles are for galaxies with $L_B \le$ 10$^{10}$
$L_\odot$ while the large circles are for galaxies with $L_B >$ 10$^{10}$ $L_\odot$. Galaxies without luminosity measurements available are plotted as diamonds. The typical error in the optical depth measurements 
is shown at the bottom left of the plot. To the top of the plot, the covering factor of \hi\ \21\ absorbers (for \t0\ = 0.3 \kms) in the Combined sample is shown as a function of impact parameter in two different 
bins demarcated at the median $b$. The circles, diamonds and squares in the top panels show \c21\ for all the galaxies, galaxies with $L_B \le$ 10$^{10}$ $L_\odot$ and galaxies with $L_B >$ 10$^{10}$ $L_\odot$, 
respectively. {\it Right:} The same as in the figure to the left for impact parameter scaled with effective \hi\ radius (see Section~\ref{sec_scaledimpact} for details). In both panels, the horizontal dotted 
lines mark \taudv\ = 0.3 \kms\ and the vertical dotted lines marks the median $b$ and $b/R_{\hi}$.}
\label{fig:taudvimpact}
\end{figure*}

Left panel in Fig.~\ref{fig:taudvimpact} shows the integrated \hi\ \21\ optical depth as a function of impact parameter around galaxies. The filled symbols correspond to \taudv\ of the \hi\ 
\21\ detections and the open symbols to \taudvl\ of the \hi\ \21\ non-detections. The black, red and green symbols are for QGPs in the primary, literature and supplementary samples, respectively. 
The size of the symbols represents the luminosity of the galaxies, with small size for galaxies with $L_B \le$ 10$^{10}$ $L_\odot$ and large for galaxies with $L_B >$ 10$^{10}$ $L_\odot$. Galaxies
without photometric measurements are plotted with a different symbol (diamond). \hi\ \21\ absorption has been detected over the range 1.7$\le b$(kpc)$\le$25. It can be seen from Fig.~\ref{fig:taudvimpact} 
that the \hi\ \21\ optical depth declines with increasing $b$ albeit with a large scatter. As shown in figure~9 of \citet{zwaan2015}, the distribution of \nhi\ (derived from \taudv) with impact parameter 
of the low-$z$ QGPs follows the general declining trend of \nhi\ with galactocentric radius obtained from emission maps of local galaxies \citep{zwaan2005}. 

A weak anti-correlation between \taudv\ of the \hi\ \21\ detections and $b$ (Spearman rank correlation coefficient of $\sim -$0.3 at a significance level of 0.81) has been reported previously by 
\citet{gupta2013,zwaan2015}. Recently, \citet{curran2016} have reported an anti-correlation using Kendall-tau test (probability of the anti-correlation arising by chance, $P(r_k)$ = 9.39 $\times$ 
10$^{-4}$, which is significant at $S(r_k)$ = 3.31$\sigma$) between the \hi\ \21\ absorption strength and impact parameter using 90 low-$z$ \hi\ \21\ observations in literature and including upper 
limits on the \hi\ \21\ optical depth as censored data points. However, they do not provide the value of the correlation coefficient, $r_k$. We perform a survival analysis on all the measurements 
in the Combined sample, by including the 3$\sigma$ upper limits as censored data points. In particular, we perform Kendall's $\tau$ correlation test using the {\tt `cenken'} function under the 
{\tt `NADA'} package in {\tt R} to obtain $r_k$ = $-$0.20, $P(r_k)$ = 0.01, $S(r_k)$ = 2.42$\sigma$. The results from test for correlation between \taudv\ and various parameters are given in 
Table~\ref{tab:correlations}.

The anti-correlation obtained by us is at a lesser significance than that obtained by \citet{curran2016}. This could be due to the differences between the samples used by them and us, which we list here:
(a) Our Combined sample consists of 31 new measurements from our primary sample and 7 new measurements from \citet{borthakur2016} that were not present in the sample of \citet{curran2016}.
(b) We have considered only measurements upto $b\sim$30 kpc, whereas \citet{curran2016} have considered measurements upto $b\sim$200 kpc.
(c) From table 1 of \citet{curran2016}, we notice that for the upper limits on the integrated optical depth measurements from \citet{borthakur2011}, they have considered a line width of 1 \kms. 
However, for the rest of the upper limits they have considered a line width of 10 \kms. We have used a line width of 10 \kms\ uniformly for all upper limits in the Combined sample.
(d) As per the definition of our primary sample (see Section~\ref{sec_litsample}), we have not included measurements towards radio sources without optical counterparts and measurements towards systems 
that were picked based on the presence of metal absorption in the quasar spectra, whereas \citet{curran2016} have not imposed such conditions on their sample.

To check the dependence of the anti-correlation between \taudv\ and $b$ on galaxy properties, we divide the galaxies into two groups based on the median $L_B =$ 10$^{10}$ $L_\odot$. 
We find that the anti-correlation becomes stronger but not more significant considering only the low-luminosity galaxies ($r_k$ = $-$0.26, $P(r_k)$ = 0.03, $S(r_k)$ = 2.09$\sigma$), 
and weaker and less significant considering only the high-luminosity galaxies ($r_k$ = $-$0.15, $P(r_k)$ = 0.20, $S(r_k)$ = 1.23$\sigma$). Similar results are found when we divide 
the galaxies based on their stellar masses and colours, i.e. the anti-correlation is stronger for low-mass and bluer galaxies. 

\begin{figure}
\includegraphics[width=0.4\textwidth, angle=90]{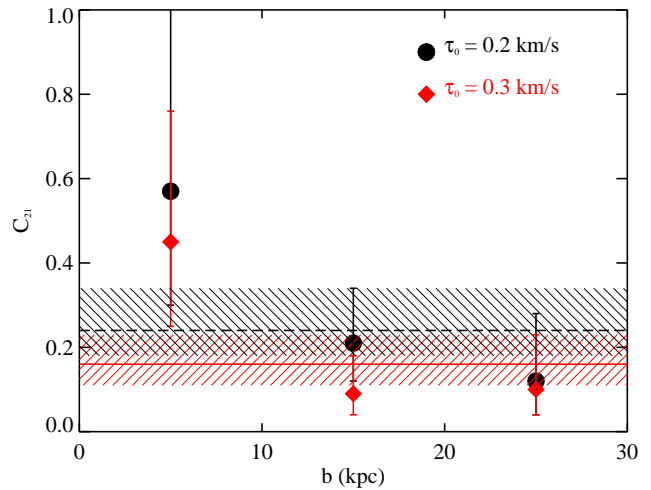} 
\caption{The covering factor of \hi\ \21\ absorbers (\c21) in three different impact parameter bins (0$-$10, 10$-$20 and 20$-$30 kpc) for two different optical depth sensitivities, \t0\ = 0.2 and 0.3 \kms, 
for the combined sample. The horizontal red solid line is the mean \c21\ within 30 kpc for \t0\ = 0.3 \kms, while the dashed black line is for \t0\ = 0.2 \kms. The shaded regions mark the corresponding
1$\sigma$ errors.}
\label{fig:coveringfactor}
\end{figure}

Coming to covering factor of \hi\ \21\ absorbers, top left panel of Fig.~\ref{fig:taudvimpact} shows \c21\ in two bins of $b$ based on the median value ($\sim$15 kpc). \c21\ drops 
from 0.24$^{+0.12}_{-0.08}$ at $b\le$ 15 kpc to 0.06$^{+0.09}_{-0.04}$ at $b=$ 15$-$35 kpc. The dependence of \c21\ on various parameters is given in Table~\ref{tab:coveringfactor}. 
Next, we divide galaxies in each impact parameter bin based on their luminosity into high and low luminosity groups. It can be seen from the above figure that \c21\ shows declining trend 
with $b$ for both luminosity groups. We also find that \c21\ shows similar declining trend with $b$ when we divide galaxies based on their stellar masses and colours. The decrease of 
\c21\ with $b$ is further illustrated in Fig.~\ref{fig:coveringfactor}, where \c21\ is estimated for three different impact parameter bins, $b$ = 0$-$10, 10$-$20 and 20$-$30 kpc, and 
for two different sensitivities, \t0\ = 0.2 and 0.3 \kms. While $C_{21}$ can be 0.45$^{+0.31}_{-0.20}$ (respectively 0.57$^{+0.45}_{-0.27}$) at $b<$ 10 kpc for \t0\ = 0.3 \kms\ 
(respectively 0.2 \kms), the average \c21\ within $b$ = 30 kpc is 0.16$^{+0.07}_{-0.05}$ (respectively 0.24$^{+0.10}_{-0.06}$) for \t0\ = 0.3 \kms\ (respectively 0.2 \kms).
Note that due to the small number of detections and large 1$\sigma$ Poisson errors, the above dependence of \c21\ on $b$ in not statistically significant.

Recall that seven radio sources in the primary sample and three radio sources in the literature sample do not have spectroscopic redshifts available, but are considered as background
sources based on optical and infrared photometry (see Sections~\ref{sec_samplesel} and \ref{sec_litsample}). We find that excluding these 10 systems does not change the anti-correlation
between \taudv\ and $b$. The average \c21\ within $b$ = 30 kpc (0.14$^{+0.07}_{-0.05}$) and the trend of \c21\ with $b$ also remain the same within the uncertainties. 
Similarly, including the sources in the supplementary sample does not change the above results within the uncertainties, with average \c21\ = 0.14$^{+0.06}_{-0.04}$ within $b$ = 30 kpc. 
\begin{table} 
\caption{Results of correlation analysis between the \hi\ \21\ absorption strength and various parameters for the Combined sample.}
\centering
\begin{tabular}{ccccccccc}
\hline
Parameter & N   & $r_k$ & $P(r_k)$ & $S(r_k)$ \\
(1)       & (2) & (3)   & (4)      & (5)      \\
\hline    
$b$               & 69 & $-$0.20 & 0.01 & 2.42$\sigma$ \\
$b/R_{\hi}$       & 67 & $-$0.20 & 0.01 & 2.45$\sigma$ \\
$R$               & 68 & $-$0.16 & 0.05 & 1.90$\sigma$ \\
$R/R_{\hi}$       & 67 & $-$0.18 & 0.03 & 2.13$\sigma$ \\
$i$               & 68 & $+$0.05 & 0.59 & 0.54$\sigma$ \\
$\phi$            & 53 & $-$0.17 & 0.07 & 1.79$\sigma$ \\
$sin(i)cos(\phi)$ & 53 & $+$0.15 & 0.11 & 1.58$\sigma$ \\
$L_B$             & 67 & $-$0.02 & 0.80 & 0.26$\sigma$ \\
$M_*$             & 62 & $-$0.01 & 0.89 & 0.15$\sigma$ \\
$g-r$             & 62 & $-$0.03 & 0.77 & 0.30$\sigma$ \\
$\Sigma_{SFR}$    & 36 & $-$0.03 & 0.83 & 0.22$\sigma$ \\
\hline
\end{tabular}
\label{tab:correlations}
\begin{flushleft}
Column 1: Parameter against which correlation of \taudv\ is tested. Column 2: Number of measurements included in the test. Column 3: Kendall rank correlation coefficient. 
Columns 4: Probability of the correlation arising by chance. Column 5: Significance of the correlation assuming Gaussian statistics. The upper limits (\taudvl) are considered
as censored data points while performing survival analysis using the {\tt `cenken'} function under the {\tt `NADA'} package in {\tt R}.
\end{flushleft}
\end{table}
\begin{table} 
\caption{Covering factor (\c21) of \hi\ \21\ absorbers in the Combined sample for different parameters estimated for \t0\ = 0.3 \kms.}
\centering
\begin{tabular}{cccc}
\hline
Parameter range & $N$ & $N_{21}$ & \c21\ \\
(1)             & (2) & (3)      & (4)      \\
\hline    
0.0$<b$ (kpc) $\le$14.6  & 33 & 8 & 0.24$^{+0.12}_{-0.08}$ \\
14.6$<b$ (kpc) $\le$33.2 & 31 & 2 & 0.06$^{+0.09}_{-0.04}$ \\
\hline
0.0$<b/R_{\hi}$ $\le$2.0  & 32 & 8 & 0.25$^{+0.12}_{-0.09}$ \\
2.0$<b/R_{\hi}$ $\le$34.0 & 30 & 1 & 0.03$^{+0.08}_{-0.03}$ \\
\hline
0.0$\le R$ (kpc) $<$20.6  & 31 & 6 & 0.19$^{+0.12}_{-0.08}$ \\
20.6$\le R$ (kpc) $<$74.0 & 32 & 3 & 0.09$^{+0.09}_{-0.05}$ \\
\hline
0.0$\le R/R_{\hi} <$2.4  & 31 & 7 & 0.23$^{+0.12}_{-0.08}$ \\
2.4$\le R/R_{\hi} <$57.0 & 31 & 2 & 0.06$^{+0.09}_{-0.04}$ \\
\hline
0.0$\le i$ (\textdegree) $<$30.0  & 17 & 3 & 0.18$^{+0.17}_{-0.10}$ \\
30.0$\le i$ (\textdegree) $<$57.9 & 25 & 1 & 0.04$^{+0.09}_{-0.03}$ \\
57.9$\le i$ (\textdegree) $<$90.0 & 26 & 5 & 0.19$^{+0.13}_{-0.08}$ \\
\hline
0.0$\le \phi$ (\textdegree) $<$44.9  & 25 & 5 & 0.20$^{+0.14}_{-0.09}$ \\
44.9$\le \phi$ (\textdegree) $<$90.0 & 26 & 1 & 0.04$^{+0.09}_{-0.03}$ \\
\hline
0.0$\le sin(i)cos(\phi) <$0.5 & 25 & 1 & 0.04$^{+0.09}_{-0.03}$ \\
0.5$\le sin(i)cos(\phi) <$1.0 & 26 & 5 & 0.20$^{+0.14}_{-0.09}$ \\
\hline
\end{tabular}
\label{tab:coveringfactor}
\begin{flushleft}
Column 1: The range of parameter values. Column 2: $N$ is the total number of measurements with \taudvl\ $\le$ \t0. 
Column 3: $N_{21}$ is the number of \hi\ \21\ detections with \taudvl\ $\le$ \t0\ and \taudv\ $\ge$ \t0. Column 4: \c21\ = $N_{21}/N$. \\
Note that for estimating $\phi$, we have considered only galaxies with $i\ge$30\textdegree. Errors represent Gaussian 1$\sigma$ confidence 
intervals computed using tables of \citet{gehrels1986} assuming a Poisson distribution.
\end{flushleft}
\end{table}
\subsubsection{Scaled impact parameter}
\label{sec_scaledimpact}
We scale the impact parameter with an estimate of the \hi\ radius of the galaxy based on its optical luminosity. Since \hi\ emission cannot be mapped from all our galaxies, for the sake of uniformity 
we use an estimate of the probable \hi\ size based on the correlation found between the optical B-band magnitude and \hi\ size of spiral and irregular galaxies in the field at low redshift \citep{broeils1997,lah2009}.
Using this relation we estimate the effective radius ($R_{\hi}$) within which half of the \hi\ mass of the galaxy is contained. This amounts to scaling the impact parameter with the galaxy luminosity 
since $R_{\hi}$ $\propto$ $L_B^\beta$, where $\beta$ $\sim$ 0.4. This is similar to the scaling relation found between optical size and luminosity of nearby galaxies detected by the Arecibo Legacy Fast 
Arecibo L-Band Feed Array (ALFALFA) survey and SDSS \citep{toribio2011}. \hi\ \21\ optical depth as a function of the scaled impact parameter is shown in the right panel of Fig.~\ref{fig:taudvimpact}.

We find an anti-correlation between \taudv\ and $b/R_{\hi}$, similar to that between \taudv\ and $b$ (see Table~\ref{tab:correlations}). As in the previous section, we find that the anti-correlation 
between \taudv\ and $b/R_{\hi}$ is mostly driven by the low-luminosity (alternatively low stellar mass or bluer) galaxies. From top right panel of Fig.~\ref{fig:taudvimpact} and Table~\ref{tab:coveringfactor}, 
it can be seen that \c21\ falls with $b/R_{\hi}$ and the probability of detecting \hi\ \21\ absorption at $b/R_{\hi} >$ 2 is very small ($\le$0.05). From the same figure, there does not appear to be 
any difference in dependence of \c21\ on $b/R_{\hi}$ based on the galaxy luminosity. We note that similarly there is no difference in the trend of \c21\ with $b/R_{\hi}$ based on stellar mass and colour 
of galaxies.

We can treat $\beta$ in the scaling relation between $R_{\hi}$ and $L_B$ as a free parameter, in order to check whether the scatter in the relation between \taudv\ and $b/R_{\hi}$ can be reduced.
Varying $\beta$ over 0$-$0.4 introduces $\sim$15$-$20\% variation in the strength and significance of the anti-correlation between \taudv\ and $b/R_{\hi}$. The strongest ($r_k$ = $-$0.24) and most 
significant ($S(r_{k})$ = 2.82$\sigma$) anti-correlation is obtained for $\beta$ = 0.2. This may suggest that the cold \hi\ gas distribution may not follow the same scaling relation with optical 
luminosity as that of the total \hi\ gas. However, for consistency we quote the results for $\beta$ = 0.4 throughout this work.
\subsubsection{Radial distance}
\label{sec_radialdist}
\begin{figure*}
\includegraphics[width=0.35\textwidth, angle=90]{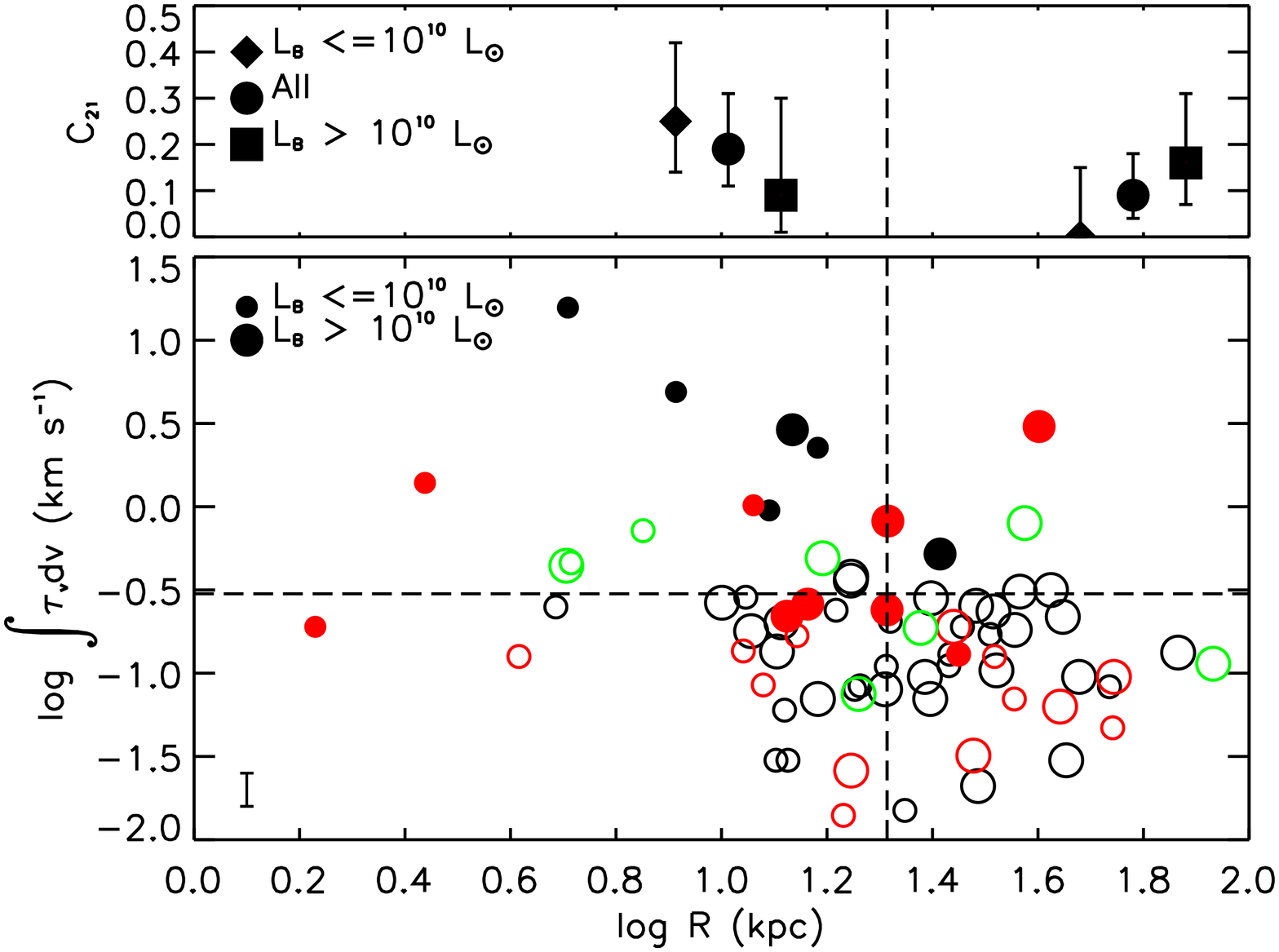}
\includegraphics[width=0.35\textwidth, angle=90]{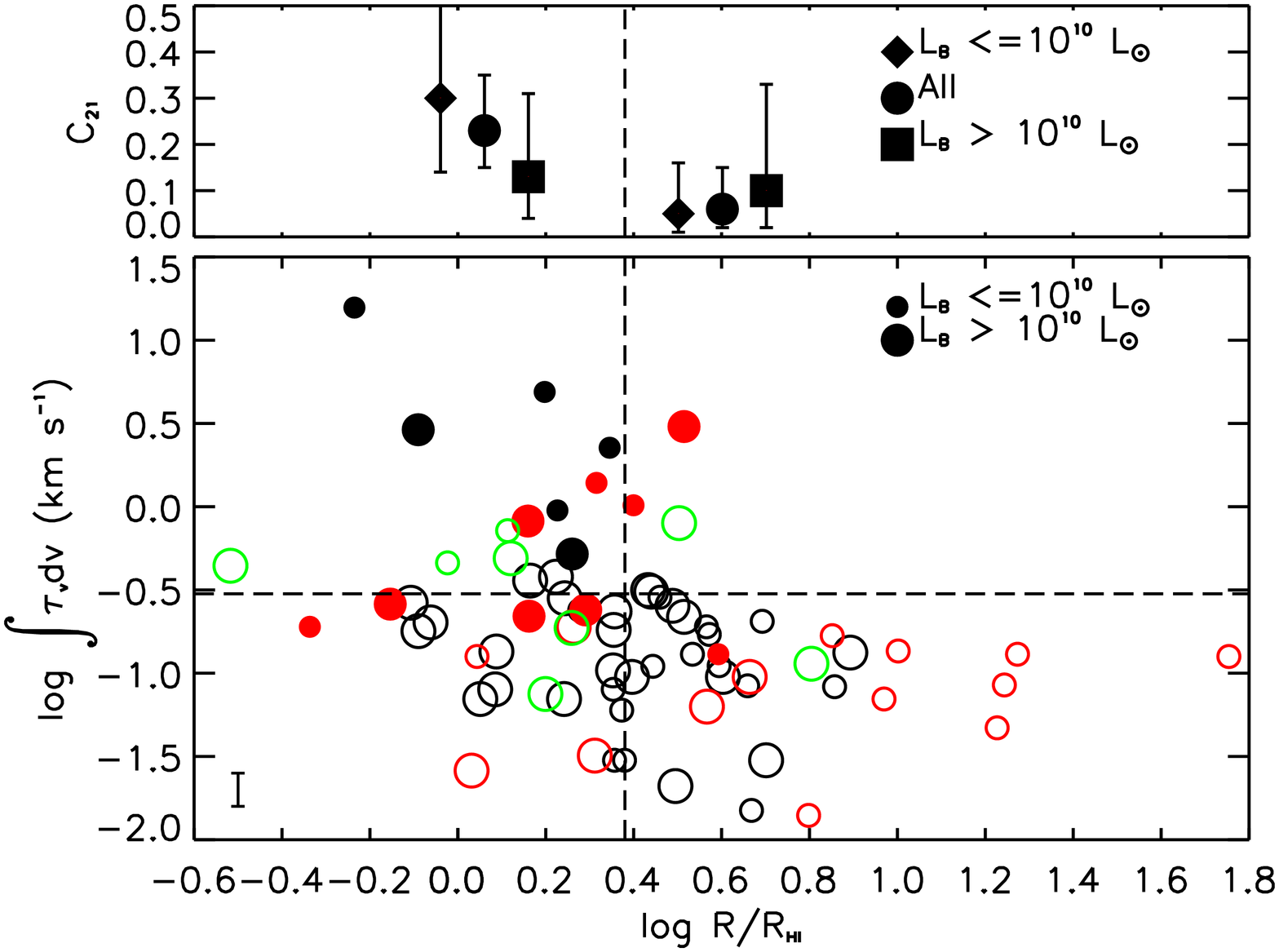} 
\caption{Same as in Fig.~\ref{fig:taudvimpact} for {\it (left)} the radial distance in the plane of the galaxy disk (as defined in Eqn.~1) and {\it (right)} the radial distance scaled by the effective \hi\ radius.}
\label{fig:taudvradius}
\end{figure*}

We define the radial distance from the galaxy's centre to where the radio sightline passes through the plane of the disk as, 
\begin{equation}
 R = b [cos^2 \phi + (sin^2 \phi / cos^2 i)]^{1/2}.
\end{equation}
Here we are assuming that the \hi\ and stellar components of the galaxies have similar morphology. As with $b$, \taudv\ is weakly anti-correlated with $R$, however the anti-correlation is slightly 
weaker and less significant (see Table~\ref{tab:correlations}). The anti-correlation is retained (at $\sim$2$\sigma$) when we scale $R$ with the effective \hi\ radius, $R_{\hi}$, as defined in 
Section~\ref{sec_scaledimpact}. As in the previous section, the strength and significance of the anti-correlation can change by $<$20\% on varying $\beta$ between 0 to 0.4. Again as with $b$, 
the anti-correlations are stronger for the low-luminosity galaxies. The distribution of \hi\ \21\ optical depth as a function of $R$ and $R/R_{\hi}$ is shown in Fig.~\ref{fig:taudvradius}.

From Table~\ref{tab:coveringfactor} and top panels of Fig.~\ref{fig:taudvradius}, we can see that \c21\ declines with increasing $R$ and $R/R_{\hi}$. The two galaxies (NGC~3067 \citep{carilli1992}
and J135400.68$+$565000.3 \citep{zwaan2015}) in which \hi\ \21\ absorption arises at $R/R_{\hi}$ $\ge$ 3 are the ones with the radio sightline oriented close to the minor axis. Hence, in these systems 
the \hi\ \21\ absorption is not likely to arise from the extended \hi\ disks, but due to high-velocity clouds (HVCs) or tidally disturbed/warped \hi\ disks. We discuss such systems in the next section. 
From Fig.~\ref{fig:taudvradius}, it can be seen that \c21\ for galaxies with $L_B \le$ 10$^{10}$ $L_\odot$ follows a similar trend, with \c21\ $\sim$0.3 at $R<$ 20 kpc and $R/R_{\hi} <$ 2, and \c21\ 
$\lesssim$0.05 beyond. On the other hand, \c21\ for galaxies with $L_B >$ 10$^{10}$ $L_\odot$ seems to be more or less constant ($\sim$0.1$-$0.2) with $R$ and $R/R_{\hi}$. The same result holds when we 
divide galaxies based on their stellar masses and colours.

{\it In conclusion: Both the \hi\ \21\ absorption strength (\taudv) and detectability (\c21) around low-$z$ galaxies show a weak declining trend with $b$.
Similar trends are found on converting $b$ to the radial distance along the galaxy major axis as well on scaling the distances with the effective \hi\ radius.}
\subsection{Azimuthal profile of \hi\ \21\ absorbers}
\label{sec_orientation}
\begin{figure*}
\includegraphics[width=0.5\textwidth, angle=90]{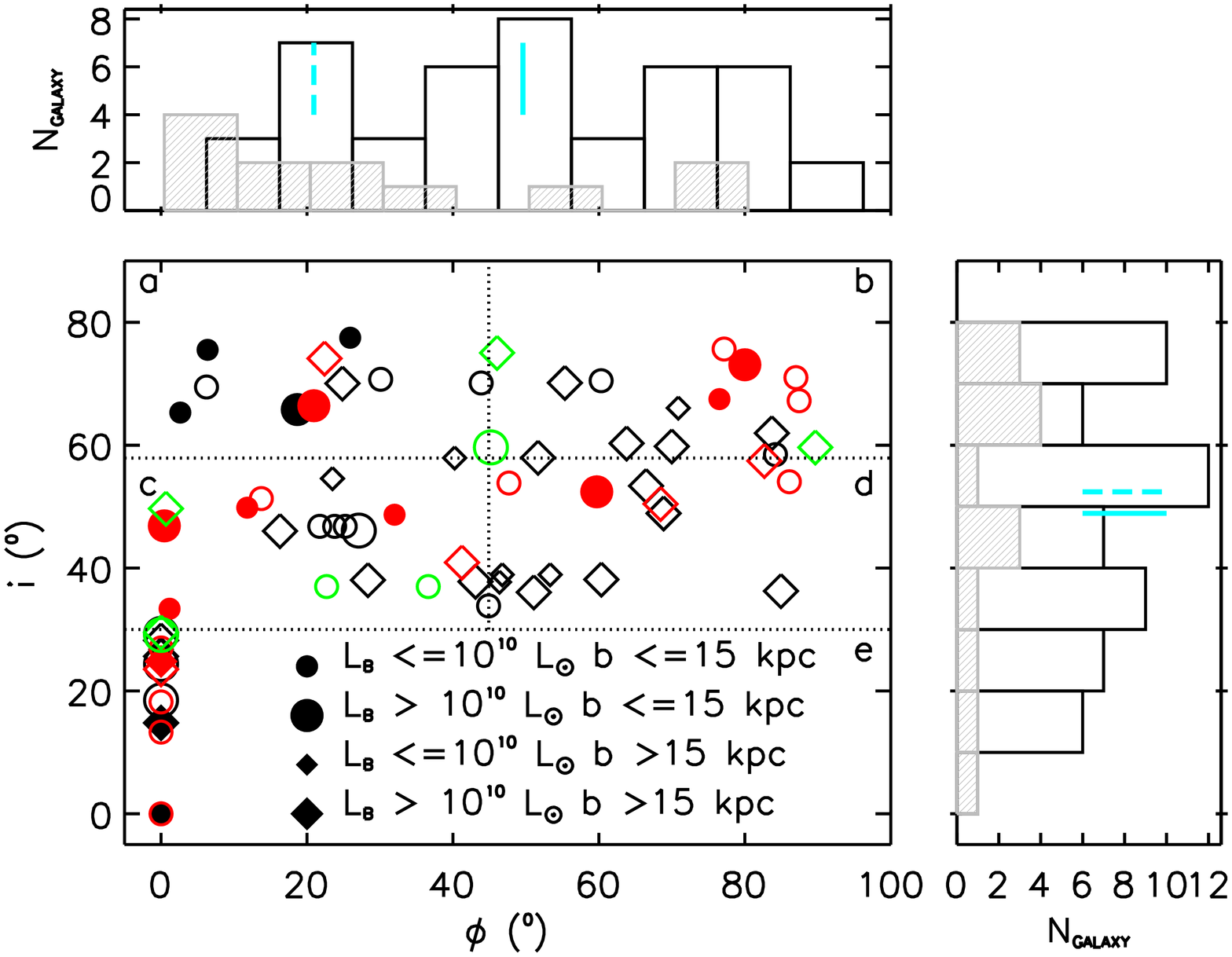}
\subfloat[\c21\ = 0.31$^{+0.24}_{-0.15}$ (0.44$^{+0.35}_{-0.21}$)]{\includegraphics[width=0.25\textwidth, angle=0]{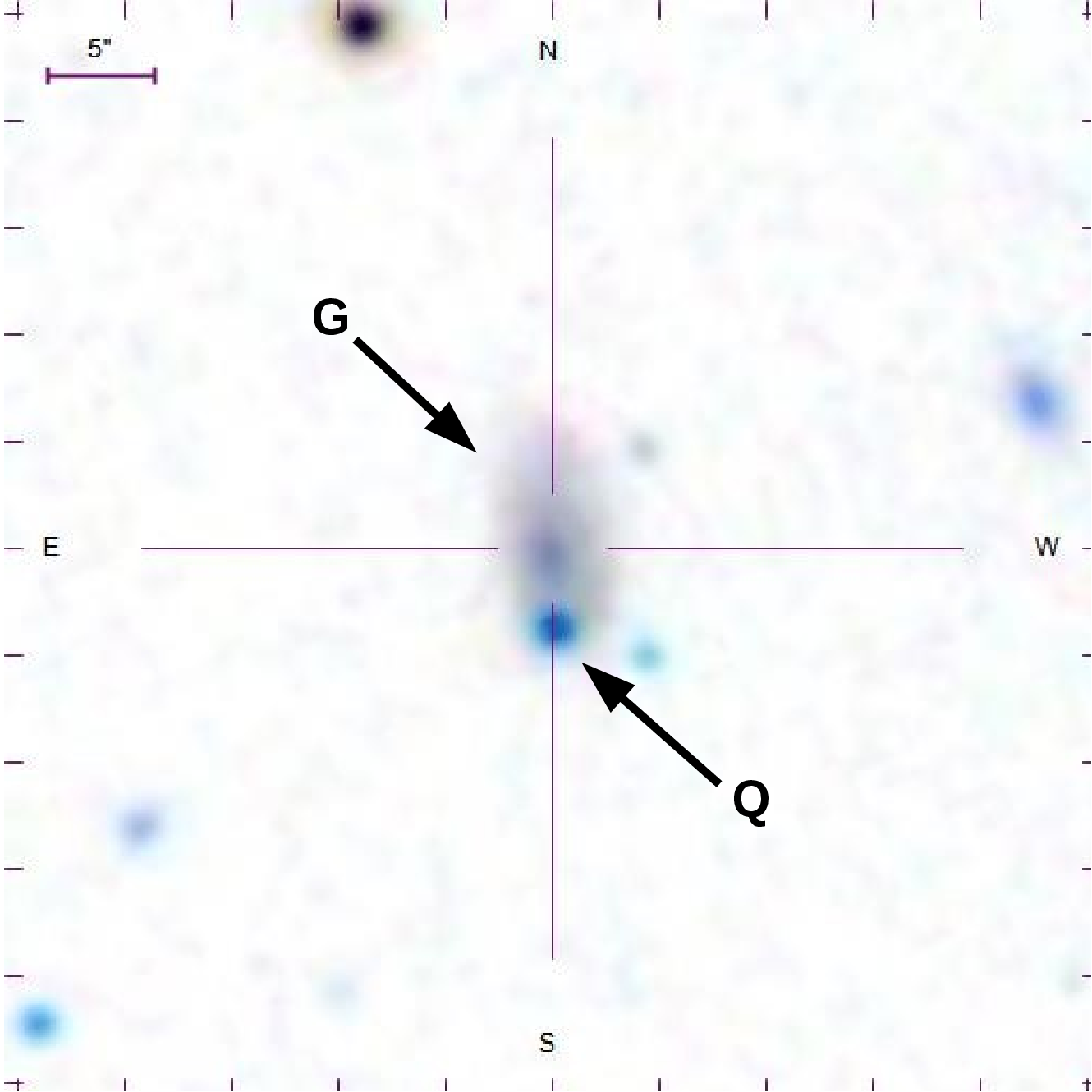} } 
\subfloat[\c21\ = 0.08$^{+0.18}_{-0.06}$ (0.08$^{+0.19}_{-0.07}$)]{\includegraphics[width=0.25\textwidth, angle=0]{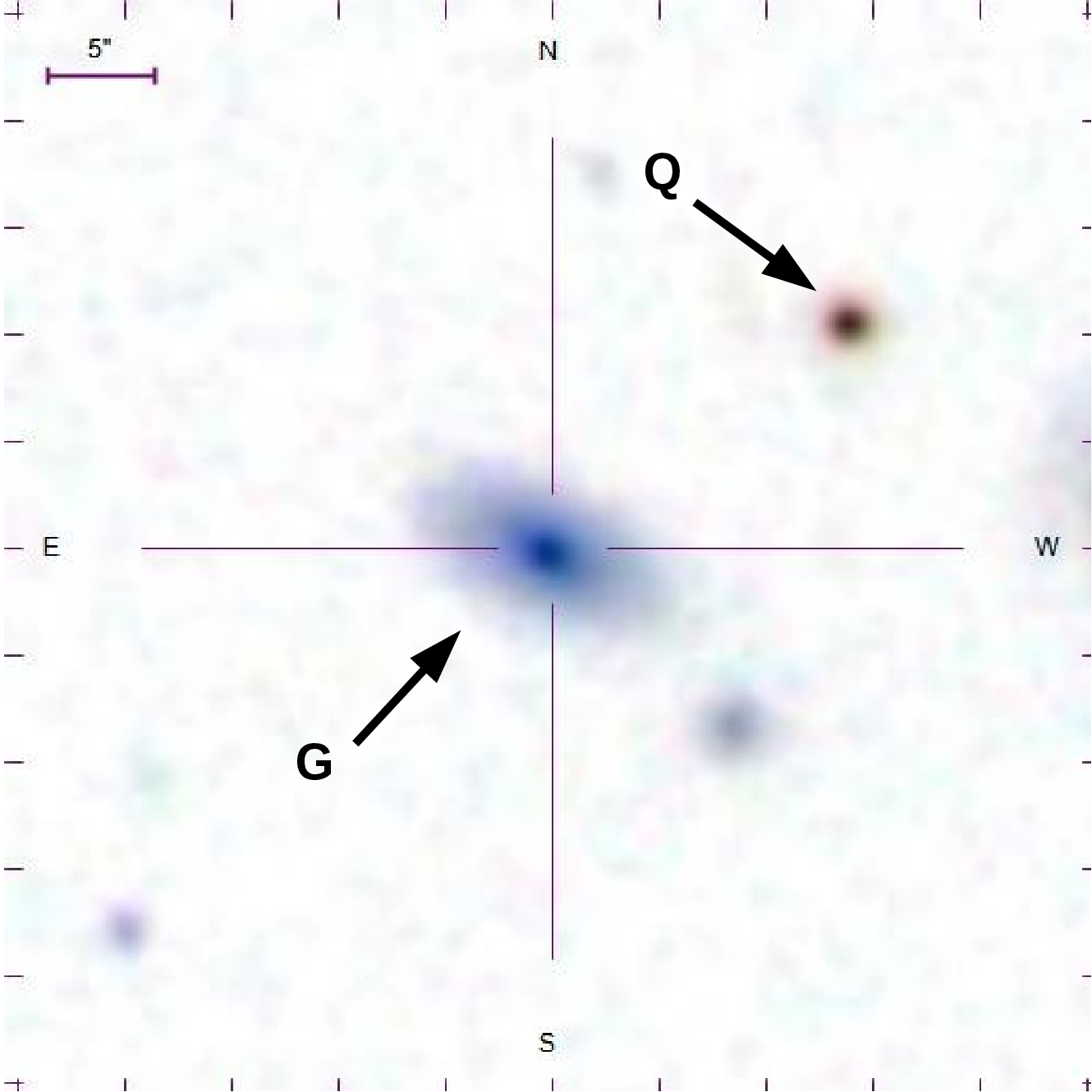} } 
\subfloat[\c21\ = 0.08$^{+0.19}_{-0.07}$ (0.18$^{+0.24}_{-0.12}$)]{\includegraphics[width=0.25\textwidth, angle=0]{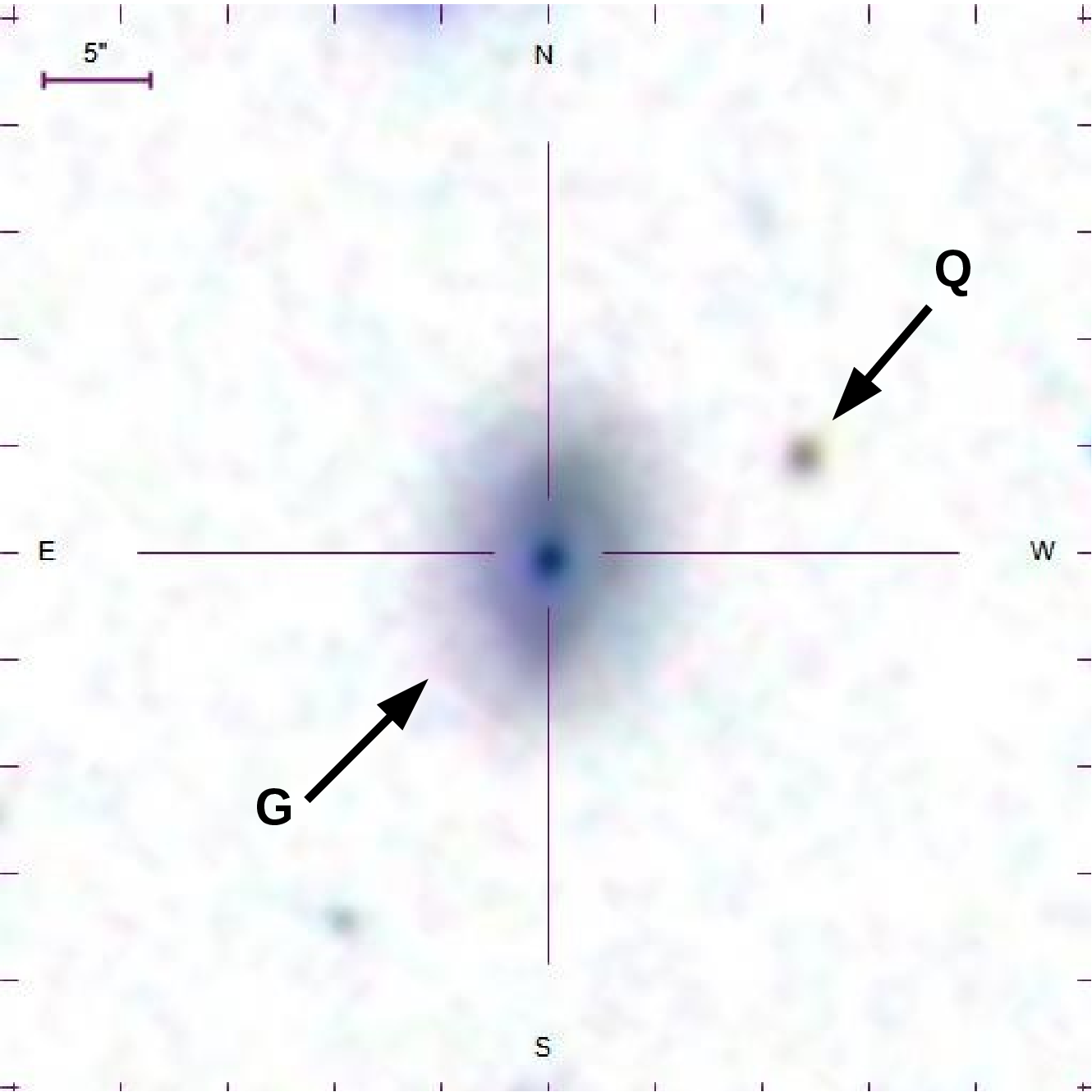} } \hspace{0.1cm}
\subfloat[\c21\ = 0.00$^{+0.14}_{-0.00}$ (0.11$^{+0.26}_{-0.09}$)]{\includegraphics[width=0.25\textwidth, angle=0]{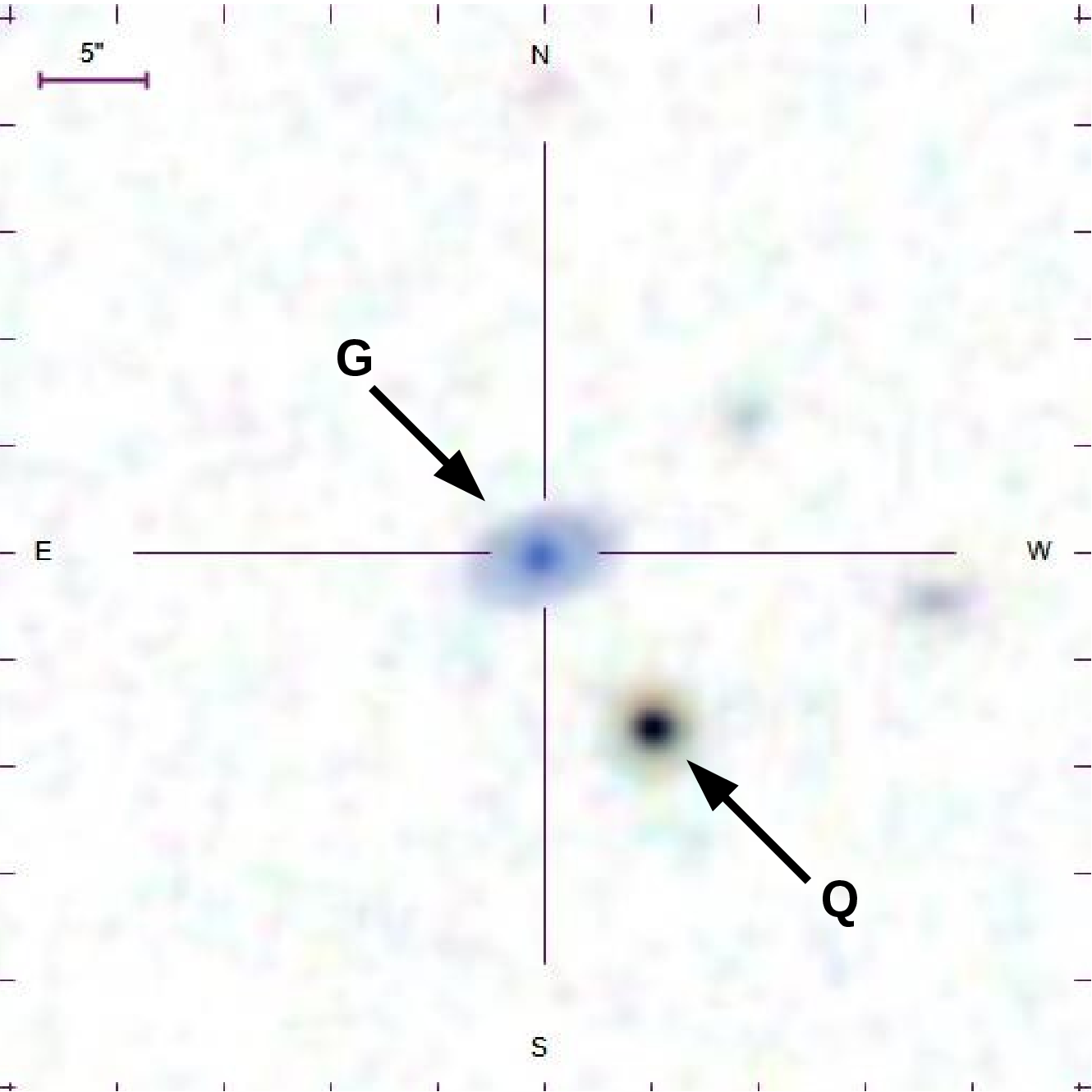} } 
\subfloat[\c21\ = 0.18$^{+0.17}_{-0.10}$ (0.14$^{+0.19}_{-0.09}$)]{\includegraphics[width=0.25\textwidth, angle=0]{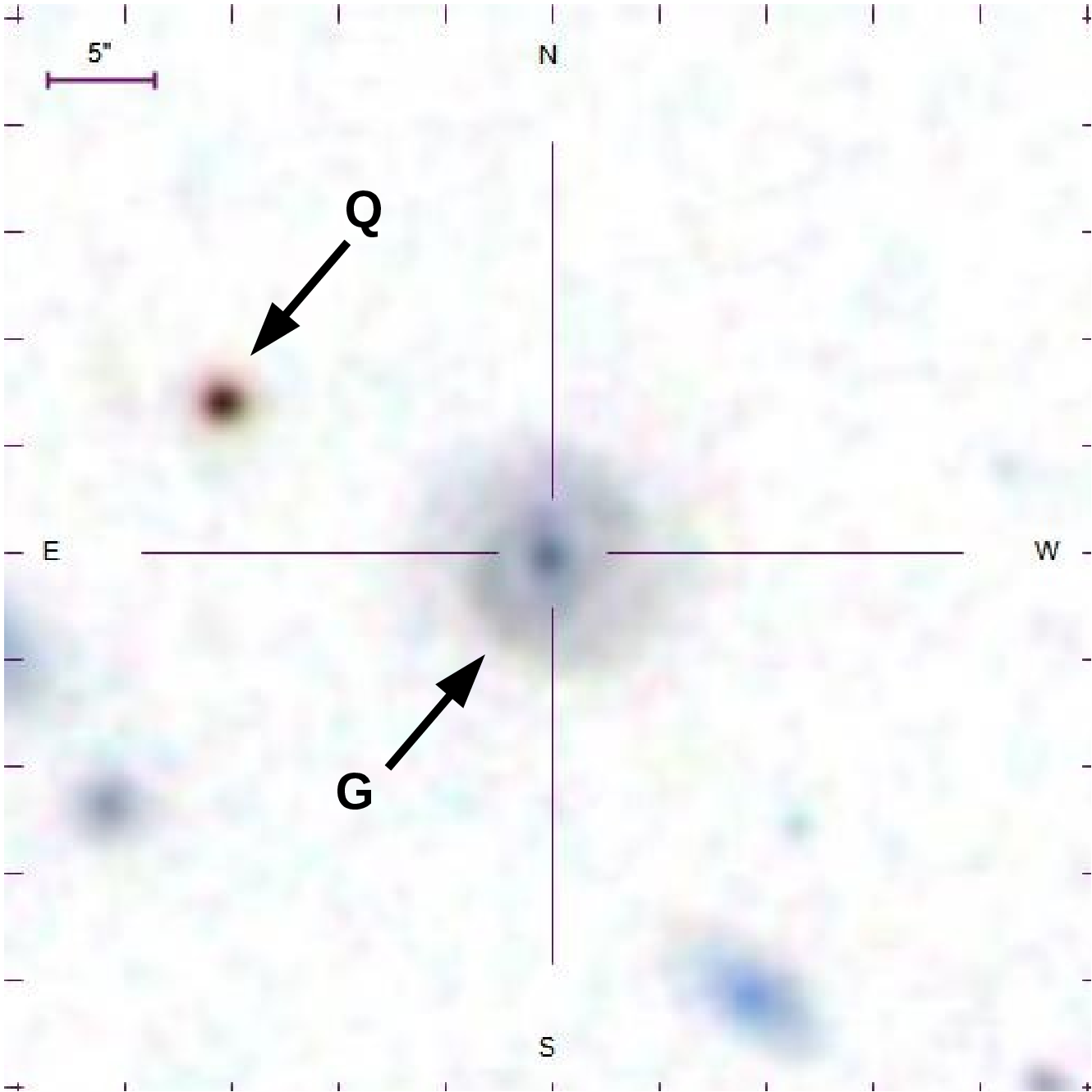} }
\caption{{\it Top:} The distribution of galaxy inclination ($i$) and orientation ($\phi$) angles for the QGPs in the primary sample (black), the literature sample (red), and the supplementary sample (green). 
The filled symbols are for the \hi\ \21\ detections and the open symbols are for the non-detections. The small symbols are for galaxies with $L_B \le$ 10$^{10}$ $L_\odot$ while the large symbols are for 
galaxies with $L_B >$ 10$^{10}$ $L_\odot$. The circles represent QGPs with $b \le$ 15 kpc and the diamonds represent QGPs with $b>$ 15 kpc. Note that for galaxies with $i<$30\textdegree\ (i.e. face-on galaxies) 
we do not estimate $\phi$, but we show them at $\phi$ = 0\textdegree\ in the plot. The horizontal dotted lines mark $i$ = 30\textdegree\ and 57\textdegree\ and the vertical dotted line marks the median 
$\phi$ = 43\textdegree. The five different bins of $i$ and $\phi$ as marked with the dotted lines are identified by a, b, c, d and e. The histogram plots of $i$ and $\phi$ (excluding the supplementary measurements) 
are shown to the right and top, respectively. The open histograms are for the non-detections and the filled histograms are for the detections. The median values of $i$ and $\phi$ for the detections and the 
non-detections are marked by dashed and solid ticks, respectively. 
{\it Bottom:} SDSS composite colour images of five QGPs from our sample are shown as examples of the QGP configurations in the five bins a, b, c, d and e as marked in the figure to the top. 
The quasars and galaxies are identified by `Q' and `G', respectively. The covering factor of \hi\ \21\ absorbers for \t0\ = 0.3 \kms\ for galaxies in the Combined sample in a particular $i$ 
and $\phi$ bin are given below the corresponding image (\c21\ for \t0\ = 0.2 \kms\ are given in parenthesis).}
\label{fig:orientation}
\end{figure*}

We investigate here whether the distribution of \hi\ \21\ absorbers around low-$z$ galaxies is spherically symmetric, or whether there is any dependence on the galaxy inclination and orientation of the radio 
sightline with respect to the galaxy. Due to lack of information about the \hi\ morphology of the galaxies, we assume that the \hi\ disks of the galaxies have the same ellipticity and position angle as 
that of their stellar disks. Additionally, we assume that the \hi\ disks are not warped or twisted. In Fig.~\ref{fig:orientation} we show the distribution of $i$ and $\phi$ of the QGPs, where the sizes 
of the symbols are based on the galaxy luminosities and their shapes are based on the impact parameters. We remind here that $\phi$ is estimated only for galaxies with $i\ge$30\textdegree. The distributions 
of $i$ for galaxies with \hi\ \21\ detections (median $\sim$52\textdegree) and non-detections (median $\sim$47\textdegree) are not very different. A two-sided Kolmogorov-Smirnov (KS) test suggests that the 
maximum deviation between the two cumulative distribution functions is $D_{KS}$ = 0.24 with a probability of $P_{KS}$ = 0.4 (where $P_{KS}$ is the probability of finding this $D_{KS}$ value or lower by 
chance). However, when it comes to $\phi$, the distributions for the systems with \hi\ \21\ detections and the non-detections are quite different ($D_{KS}$ = 0.49 with $P_{KS}$ = 0.02). The \21\ detections 
have median $\phi$ = 21\textdegree\ whereas the non-detections have median $\phi$ = 47\textdegree. This suggests that orientation may play a role in determining properties of \hi\ \21\ absorbers.

From Fig.~\ref{fig:orientation} we can see that \hi\ \21\ absorption is detected only for $\phi$ $\lesssim$30\textdegree\ (i.e. near the galaxy major axis) and for $\phi$ $\gtrsim$60\textdegree\ (i.e. near 
the galaxy minor axis). A bimodality in the azimuthal orientation of \mgii\ absorbing gas around galaxies has been observed, which is believed to be driven by gas accreting along the galaxy major axis and 
outflowing along the galaxy minor axis \citep{bouche2012,kacprzak2012,bordoloi2014,nielsen2015}. However, there are only three \hi\ \21\ detections near the minor axis compared to nine near the major axis. 
These nine absorption are likely to arise from the \hi\ disk of the galaxies. In all three cases where \hi\ \21\ absorption has been detected close to galaxy's minor axis, $b\le$ 15 kpc, i.e. the absorption 
could be arising from a warped/disturbed \hi\ disk. Indeed in two of these galaxies (NGC~3067 and ESO~1327$-$2041), the \hi\ \21\ absorption was found to be arising from the tidally disturbed \hi\ gas around 
the galaxy \citep{carilli1992}. Later studies of these QGPs using Hubble Space Telescope images have attributed the \hi\ \21\ absorbers to HVCs and extended spiral arm of a merging galaxy 
\citep{keeney2005,keeney2011,stocke2010}. 

We point out that in the two cases where \hi\ \21\ absorption have been detected at $b\gtrsim$20 kpc \citep{dutta2016,reeves2016}, the galaxies are face-on (i.e. $i<$30\textdegree), and hence the probability
of intersecting the \hi\ disk is higher as compared to galaxies with inclined disks. This along with larger number of \hi\ \21\ detection near the major axis could imply that \hi\ \21\ absorbers are mostly 
arising from the plane of the \hi\ disk. In that case, the \hi\ \21\ optical depth is expected to correlate with the galaxy inclination and orientation. Specifically, the \hi\ \21\ optical depth should be 
maximum in the configuration where the sightline is aligned along the major axis of the galaxy which is viewed edge-on, since the sightline would traverse the full \hi\ disk in that case. While we do not 
find any significant correlation of \taudv\ with $i$, we find that \taudv\ is weakly anti-correlated with $\phi$ and weakly correlated with $sin(i)cos(\phi)$ (see Table~\ref{tab:correlations}). 

The covering factor of \hi\ \21\ absorbers as a function of $i$, $\phi$ and $sin(i)cos(\phi)$ are given in Table~\ref{tab:coveringfactor}. \c21\ falls from 0.20$^{+0.14}_{-0.09}$ near the major axis 
($\phi <$ 45\textdegree) to 0.04$^{+0.09}_{-0.03}$ near the minor axis ($\phi \ge$ 45\textdegree), and \c21\ increases similarly with $sin(i)cos(\phi)$. To illustrate the dependence of \c21\ on $i$ 
and $\phi$, we show in Fig.~\ref{fig:orientation} the SDSS images of typical QGPs from our sample in five bins of $i$ and $\phi$ as marked in the top figure. Given below each image is \c21\ corresponding 
to that particular bin of $i$ and $\phi$. While the errors are large, it can be seen that \c21\ is maximum for sightlines that pass near the major axis of edge-on galaxies, and minimum for sightlines 
that pass near the minor axis of inclined galaxies. 

However, we find that $\phi$ is correlated with $b$ ($r_k$ = $+$0.28, $P(r_k)$ = 0.004, $S(r_k)$ = 2.91$\sigma$) in the Combined sample. This is due to the dearth of galaxies probed along the minor axis
at small $b$. The correlation is mostly driven by the GOTOQs, where by construct the radio sightlines pass through the optical disks in SDSS images, and hence they are probed both at low $b$ and at low 
$\phi$. We caution that space-based high spatial resolution imaging of these GOTOQs are required to accurately measure their orientation angles. Since the strongest and most significant anti-correlation 
of \taudv\ is found with $b$ (see Table~\ref{tab:correlations}), it is possible that this is driving the weaker and less significant anti-correlation of \taudv\ with $\phi$ (as well as correlation with 
$sin(i)cos(\phi)$). However, if we consider the average \c21\ within 30 kpc (0.16$^{+0.07}_{-0.05}$; see Fig.~\ref{fig:coveringfactor}), we find that it can be enhanced by $\sim50\%$ by restricting to 
sightlines passing close to the disk, i.e. \c21\ = 0.24$^{+0.19}_{-0.11}$ at $\phi \le$ 30\textdegree. Further, when we consider galaxies only at $b<$ 15 kpc (where \c21\ is highest; see 
Table~\ref{tab:coveringfactor}), there is a tendency for more \hi\ \21\ absorbers to be detected closer to the major axis, i.e. \c21\ = 0.31$^{+0.21}_{-0.14}$ at $\phi <$ 45\textdegree compared to 
\c21\ = 0.11$^{+0.26}_{-0.09}$ at $\phi \ge$ 45\textdegree. However, this is not significant considering the 1$\sigma$ Poisson confidence intervals. More \hi\ \21\ detections are required to confirm 
this tentative dependence of \c21\ and \taudv\ on $\phi$, and whether it is independent of that on $b$.

{\it In conclusion: \taudv\ and \c21\ show weak dependence on the galaxy orientation, with tentative indication for most of the \hi\ \21\ absorbers to be co-planar with the \hi\ disks. 
However, the dependence may not be independent from that on impact parameter.}
\subsection{Galaxy properties}
\label{sec_galaxyprop}
In Section~\ref{sec_distance}, we have seen that the anti-correlation between \taudv\ and $b$ (as well as $b/R_{\hi}$, $R$ and $R/R_{\hi}$) becomes more pronounced for the low luminosity, low stellar 
mass, bluer galaxies. However, none of the anti-correlations are at greater than $\sim$2$\sigma$ level of significance. Further, at larger radial distances, \c21\ tends to be higher for the higher 
luminosity, higher stellar mass, redder galaxies (see Fig.~\ref{fig:taudvradius}). While the errors are large due to small number statistics, these results seem to indicate that the probability 
of detecting \hi\ \21\ absorbers at large galactocentric distances is higher for galaxies with high luminosities and masses, in other words, the \hi\ \21\ absorbing gas cross-section may be larger 
for these galaxies. 

However, we do not find any significant correlation of \taudv\ with luminosity, stellar mass and colour of galaxies (see Table~\ref{tab:correlations}). Further, the \hi\ \21\ detections do not 
have different distributions of these galaxy properties from the \hi\ \21\ non-detections. Additionally, we can use the definition of magnitude-dependent $g-r$ colour cut \citep{blanton2007b}, 
as described in Section~\ref{sec_sampleprop}, to classify the galaxies as red and blue. We find that \c21\ is similar for the blue (0.15$^{+0.09}_{-0.06}$) and the red galaxies (0.12$^{+0.16}_{-0.08}$).
However, note that only two of the \hi\ \21\ absorption detections are from early morphological type galaxies \citep{gupta2013,zwaan2015}.

In order to study the dependence of \hi\ \21\ absorption on host galaxy SFR, we consider a subset of the galaxies (30) whose spectra are available from SDSS, since all the galaxies in our sample and 
in the literature do not have homogeneous optical spectra available. We also consider the GOTOQs (6) from which emission lines are detected in the background quasars' SDSS spectra. We estimate the 
surface SFR density, $\Sigma_{SFR}$ (uncorrected for dust attenuation and fibre filling factor) from the \ha\ flux detected in the SDSS fibre using equations 6 and 7 of \citet{argence2009}. The \hi\
\21\ optical depth does not have any significant correlation with $\Sigma_{SFR}$ (see Table \ref{tab:correlations}). Note that the total SFR of the host galaxies in case of the GOTOQs is likely 
to be higher than what is measured from the quasar spectra. In addition, we do not find any significant correlation between \taudv\ and the SFR as well as the specific SFR (SFR per unit stellar mass)
estimated from photometric measurements by the {\it kcorrect} algorithm. However, consistent measurement of SFR for all the galaxies is required to confirm any trend between \taudv\ and star formation. 
Finally we note that, over the redshift range (0$-$0.4) probed by the Combined sample, we do not find any significant ($<1\sigma$) redshift evolution of \taudv\ and \c21.

{\it In conclusion: We do not find any significant dependence of the strength and distribution of \hi\ \21\ absorbers on the host galaxy properties studied here, 
i.e. luminosity, stellar mass, colour, $\Sigma_{SFR}$ and redshift.}
\subsection{Kinematics}
\label{sec_kinematics}
The \hi\ \21\ absorption lines are detected within $\sim \pm$150\,\kms\ of the systematic redshift of the galaxy, with median value $\sim$60\,\kms. Absorption from metal lines at the redshift of the \hi\ \21\ 
absorption have been detected in the SDSS spectra of the quasars in most cases (see Section~\ref{sec_metalabs}). However, high resolution optical spectra of the quasars are required to compare the kinematics 
of the \hi\ \21\ absorption and the metal lines.

Considering the \hi\ \21\ absorption detections where the lines are resolved, the lines have $v_{90}$ in the range $\sim$4$-$70\,\kms, with median $\sim$ 22\,\kms\ ($v_{90}$ is the velocity width 
within which 90\% of the total optical depth is detected). The FWHM, from Gaussian fits of all the \hi\ \21\ absorption components which are resolved, is in the range $\sim$4$-$50\,\kms, with 
median $\sim$10\,\kms. The upper limits on the gas kinetic temperature, obtained assuming the line widths are purely due to thermal motion, range over $\sim$300$-$60000 K. Note that \citet{borthakur2010} 
have detected a narrow, single channel (FWHM $\sim$1.1\,\kms) \hi\ \21\ absorption towards a dwarf galaxy (UGC~7408), which corresponds to $T_K$ $\sim$ 26 K. The typical temperature of the CNM gas 
($\sim$100 K) corresponds to a thermal velocity dispersion of $\sim$1\,\kms (FWHM $\sim$2\,\kms), while that of the WNM gas ($\sim$5000$-$8000 K) corresponds to $\sim$6$-$8\,\kms (FWHM $\sim$14$-$19\,\kms), 
and spectral lines broader than this can be attributed to non-thermal motion in the ISM \citep{wolfire1995a,heiles2003}. However, we note that the some of the large line widths of the \hi\ \21\ absorption 
can also be due to structure of the background radio sources, and multi-wavelength sub-arcsecond-scale images of the radio sources are required to characterize their structure \citep{gupta2012}.

The kinematics of the \hi\ \21\ absorbers are expected to shed light on their spatial distribution. We do not find any significant correlation ($<$1.5$\sigma$) between the kinematical parameters of the \hi\ \21\ 
absorbers and the various geometrical parameters explored in the previous sections. However, we note that we are limited by the small number of detections to infer any general trends. In addition, information
about the velocity field of the galaxies is required to interpret the kinematics of the \hi\ \21\ absorbers. 
%
%=========================== DISCUSSION ====================================================================================
%
\section{Discussion} 
\label{sec_discussion}
In Section~\ref{sec_results}, we have studied the radial and azimuthal profiles of \hi\ \21\ absorption around low-$z$ galaxies. We have quantified the covering factor of \hi\ \21\ absorbers as a 
function of various geometrical parameters. We have found that the covering factor of \hi\ \21\ absorbers decreases with increasing impact parameter/radial distance. Further, the distribution 
of galaxy orientation hints that most of the \hi\ \21\ absorbers could be co-planar with the \hi\ disks. Additionally, we have found that the distribution of \hi\ \21\ absorbers is more sensitive 
to geometrical parameters than physical parameters related to the star formation in galaxies. 

The next step would be to relate the distribution of \hi\ \21\ absorbers with the distribution of cold \hi\ gas around low-$z$ galaxies. However, that is not very straightforward since the \hi\ \21\ 
optical depth depends on \nhi, \ts\ and \fc. It is difficult to quantify \fc\ since sub-arcsecond resolution images of all the radio sources are not available. 
We find \c21\ = 0.18$^{+0.09}_{-0.06}$ towards radio sources that are compact at arcsecond-scale resolution in the Combined sample, as compared to \c21\ = 0.10$^{+0.13}_{-0.06}$ towards those that are resolved.
Note that \citet{gupta2012} have found that the \hi\ \21\ detection rate is higher towards the quasars with flat or inverted spectral index at cm wavelengths and towards the quasars with linear sizes 
less than 100 pc (see their figure 6). 

For the purpose of the discussion here if we consider \fc\ $\sim$1, the non-detection of \hi\ \21\ absorption could mean either that there is lack of \hi\ gas along the sightline or that most of the \hi\ 
gas along the sightline is warm. Independent information about \hi\ gas along the sightline, for example from absorption from \lymana\ or metals in the optical/UV spectra or from \hi\ \21\ emission, can lift 
this degeneracy. 
Stacking the spectra of all the \hi\ \21\ non-detections can help to detect weak absorption that is not detected due to noise in individual spectra. However, we do not detect any \hi\ \21\ absorption
in the stacked spectrum and estimate \taudvl\ $\le$ 0.009\,\kms. Hence, we can place constraint on the average \nhi\ along these sightlines as $\le$ 18.2 (19.2) for \ts\ = 100 (1000) K and \fc\ = 1.

From the results presented in Section~\ref{sec_results}, it appears that majority of the \hi\ \21\ non-detections could be due to the radio sightline not intersecting the \hi\ disk of the galaxy. 
In $\sim$77\% of the non-detections, the sightline passes outside twice the expected radius containing 50\% of the \hi\ mass of the galaxy or away from the major axis ($\phi \ge$ 30\textdegree) 
of the galaxy. In rest of the \hi\ \21\ non-detections where the sightline is likely to pass through the \hi\ disk, the non-detections could be due to the \hi\ gas along the sightline being warm or
small filling factor of cold gas. Radio interferometric \hi\ \21\ emission maps are available for only three of the galaxies with \hi\ \21\ non-detections. One of them has been discussed in Section~\ref{sec_J1243}, 
where the radio sightline passes through the outer regions of the \hi\ disc. In the other two cases, the radio sightline does not pass through the \hi\ disc \citep{greisen2009} or \hi\ \21\ emission 
is not detected from the galaxy \citep{carilli1992}, respectively. 

Next, we discuss about metal absorption towards the quasars in our sample in Section~\ref{sec_metalabs}, and compare our results with \hi\ \21\ absorption searches from low-$z$ DLAs in Section~\ref{sec_dlas}.
Finally, we discuss the implications of our results on the redshift evolution of \hi\ \21\ absorbers in Section~\ref{sec_dndz}. 
\subsection{Metal absorption}
\label{sec_metalabs}
\caii\ and \nai\ absorption are believed to arise from neutral gas near star-forming regions. Measurements of \caii\ and \nai\ rest equivalent widths from the SDSS spectra of the background quasars
are possible for 43 of the QGPs in the Combined sample. \caii\ measurement is available for three other QGPs from \citet{womble1993}. \caii\ and \nai\ measurements from Keck High Resolution Echelle 
Spectrometer (HIRES) spectrum are available towards the QGP J0041$-$0043 \citep{dutta2016}. Out of the 16 sightlines towards which \hi\ \21\ absorption have been detected, metal line absorption information 
is available for 11 cases. Absorption from either \caii\ or \nai\ or both is detected in 9 of these systems. However, in two cases of \hi\ \21\ absorption detection \citep{borthakur2010,borthakur2014}, 
no metal lines are detected in the SDSS spectra (3$\sigma$ upper limits on $W_r$(\caii) $\le$ 1.1 \AA\ and $\le$ 0.27 \AA, and $W_r$(\nai) $\le$ 0.35 \AA\ and $\le$ 0.17 \AA, respectively).

We do not find any correlation of \taudv\ with $W_r$(\caii) and $W_r$(\nai). Furthermore, there does not appear to be a minimum $W_r$(\caii) or $W_r$(\nai) beyond which \hi\ \21\ absorption is 
detected with high probability and vice versa. However, higher resolution optical spectra towards the background quasars are required to study in detail the connection between \caii\ and/or 
\nai\ absorption and \hi\ \21\ absorption. In most cases of \hi\ \21\ non-detections, \nai\ or \caii\ absorption lines are not detected in the SDSS spectra. Stacking of the SDSS spectra without 
metals lines detected at the redshift of the galaxies gives 3$\sigma$ upper limit on $W_r$(\caii) $\le$ 0.094 \AA\ and $W_r$(\nai) $\le$ 0.046 \AA. Assuming linear part of the curve of growth and 
correlation between \nhi\ and $N$(\nai) as in the Milky Way \citep{wakker2000}, gives log~\nhi\ $\le$ 19.5. However, note that unlike in the Milky Way, no correlation is found between $N$(\caii) 
and \nhi\ in DLAs/sub-DLAs over the range of log~\nhi\ = 19.5$-$21.5 \citep{rahmani2016}. 

It is interesting to note that in eight cases \hi\ \21\ absorption is not detected despite \caii\ or \nai\ absorption being present along the quasar sightline. Note that these eight QGPs have a range of 
$b$ ($\sim$10$-$30 kpc) as well as galaxy orientation ($\sim$30\textdegree$-$80\textdegree). In these cases at least one knows that \hi\ gas is present along the sightline. One such case has been discussed 
in Section~\ref{sec_J1300}, where \caii\ absorption has been detected but no \hi\ \21\ or \nai\ absorption. Another such example is J0041$-$0143, where \caii\ and \nai\ absorption have been detected towards 
the core but not \hi\ \21\ absorption \citep{dutta2016}. Interestingly, \hi\ \21\ absorption is detected towards one of the lobes ($\sim$4 kpc away from the core) of this radio source, implying a patchy 
distribution of cold gas in the \hi\ disc. Comparison of the \nhi\ estimated from the metal column densities assuming optically thin gas and correlations seen in the Milky Way \citep{wakker2000}, with the 
\hi\ \21\ optical depth measurements, indicates that the \hi\ gas along these sightlines is likely to be warmer than 100 K.

The ratio of column density of \nai\ to \caii\ can be used to probe the physical conditions in the gas. High values ($\gtrsim$1) of $N$(\nai)$/N$(\caii) are expected in the cold dense ($T$ $\lesssim$ 100 K, $n_H$ 
$\gtrsim$ 100\,\cc) ISM where Ca is usually heavily depleted onto dust grains, whereas low values ($\lesssim$1) are expected in the warmer diffuse gas ($T$ $\gtrsim$ 1000 K, $n_H$ $\lesssim$ 10\,\cc), where most 
of Ca is still in the gas phase \citep{welty1996}. The ratio of $W_r$(\nai)$/W_r$(\caii) is $\lesssim$1 ($N$(\nai)$/N$(\caii) $<$1 under the optically thin approximation) in most ($\sim$90\%) cases of \hi\ 
\21\ absorption, indicating that these sightlines are not likely to be passing through the dusty star-forming disks. Only in one case of \hi\ \21\ detection, it has been demonstrated that the sightline is probing 
the translucent \citep[defined as a region with $1\le A_V \le10$;][]{vandishoeck1989,snow2006}  of the foreground galaxy \citep{srianand2013}. The quasar shows strong reddening along with absorption from 
\nai\ and diffuse interstellar band in this case. The poor SNR of the SDSS spectrum near the wavelength of the \caii\ absorption prevents measurement of $W_r$(\caii) (3$\sigma$ upper limit on $W_r$(\caii) 
$\le$2.3 \AA). Out of the eight \hi\ \21\ absorbers for which $W_r$(\caii) is measured, only two have $W_r$(\caii)$\ge$0.7\AA\ and can be classified as strong absorbers, while the rest can be classified as 
weak absorbers, according to \citet{wild2006,sardane2015}. Based on abundance patterns and dust extinction, the former is shown to have properties intermediate between halo- and disc-type gas, while the latter 
is shown to have properties consistent with halo-type gas \citep{sardane2015}. This is again consistent with the fact that even at low impact parameters, most of the sightlines studied here do not pass through 
the stellar disk of galaxies where there is ongoing star formation.
\subsection{\hi\ \21\ absorption from low-$z$ DLAs}
\label{sec_dlas}
The QGPs discussed in this work have been selected based on the presence of a foreground galaxy at small angular separation from a background quasar, with no prior information about absorption along the quasar 
sightline, i.e. this sample of QGPs is absorption-blind. On the other hand, there have been numerous searches for \hi\ \21\ absorption in literature from samples selected on the basis of absorption lines 
(\mgii\ or \lymana) detected in the optical/UV spectra of the quasar \citep[e.g.][]{briggs1983,kanekar2003,curran2005,gupta2009,curran2010,srianand2012,gupta2012,kanekar2014}. However, most of these have been 
at high-$z$ ($z>1$), where identification of the host galaxy is difficult. In Table~\ref{tab:dlagalaxy}, we list 13 DLAs at $z<$ 1 that have been searched for \hi\ \21\ absorption in literature and whose host 
galaxies have been identified subsequently. Note that in some cases the host galaxy identification is based on photometric redshift or proximity to the quasar sightline. From Table~4 of \citet{kanekar2014}, 
we see that these systems have a range of \ts\ (median $\sim$ 800 K) and metallicities (median $\sim$ $-$0.8). Sufficient information is not available for all the DLA host galaxies in order to study the 
dependence of the \hi\ \21\ optical depth on orientation and other galaxy properties. \hi\ \21\ absorption has been detected in 10 of these DLAs.

\begin{figure}
\includegraphics[width=0.35\textwidth, angle=90]{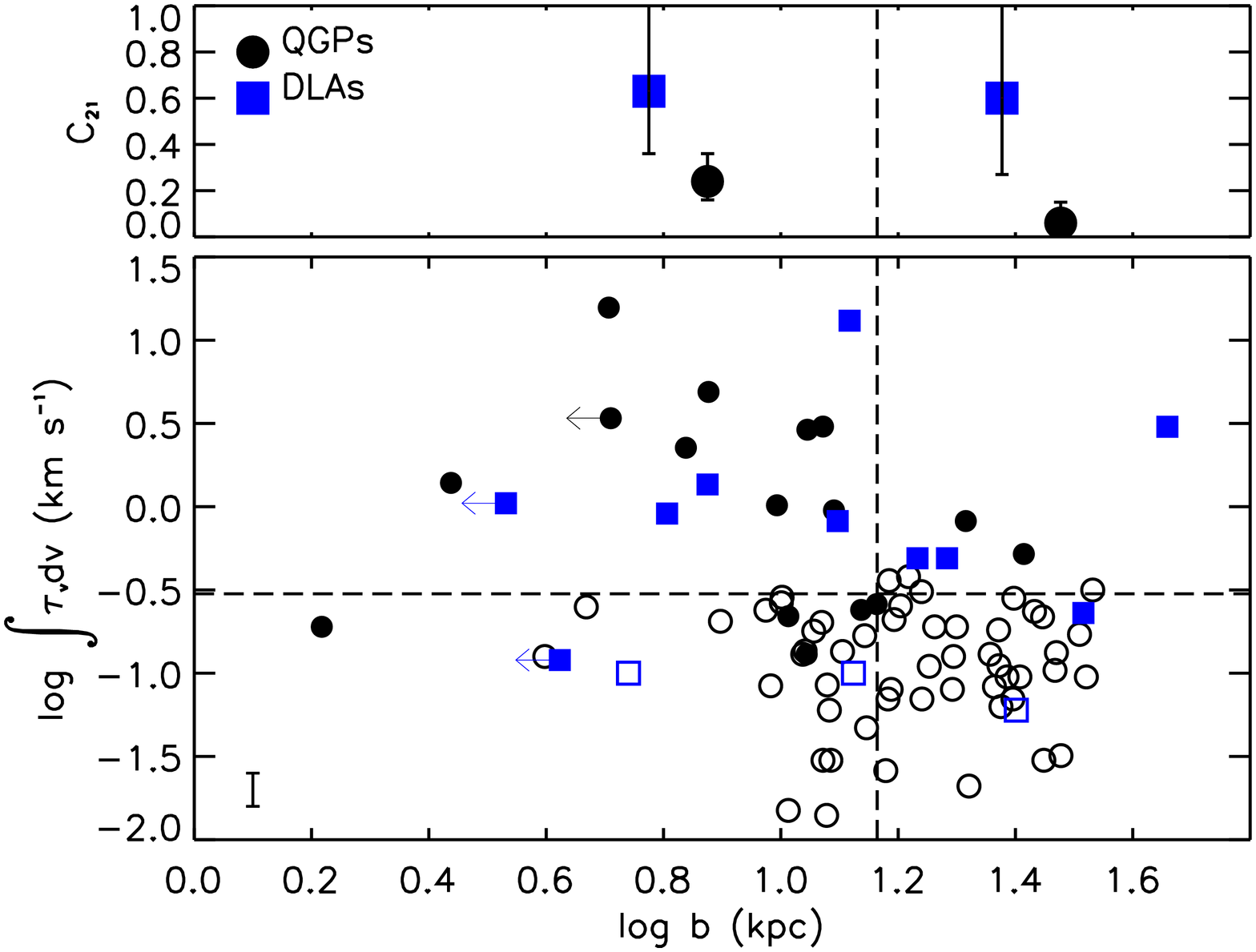}
\caption{Integrated \hi\ \21\ optical depth around galaxies as a function of impact parameter. The black circles and blue squares are for QGPs in the Combined sample and $z<1$ DLAs from the literature, respectively. 
The solid and open symbols represent \taudv\ and \taudvl\ of the \hi\ \21\ detections and non-detections, respectively. The typical error in the optical depth measurements is shown at the bottom 
left of the plot. The horizontal dotted line marks \taudv\ = 0.3\,\kms\ and the vertical dotted line marks $b$ = 15 kpc. To the top of the plot, the covering factor of \hi\ \21\ absorbers (for \t0\ 
= 0.3\,\kms) is shown as a function of impact parameter in two different bins demarcated at the median $b$.}
\label{fig:dlas}
\end{figure}

In Fig.~\ref{fig:dlas}, we show the distribution of \hi\ \21\ optical depth as a function of impact parameter for the QGPs studied in the Combined sample and the DLAs listed in Table~\ref{tab:dlagalaxy}. 
It can be seen that the distribution and range of \taudv\ values for the DLAs are not different from those of the QGPs, and median $b$ = 13 kpc and 15 kpc for the DLAs and the QGPs, respectively. We do not 
find any correlation between \taudv\ and $b$ of the DLAs ($r_k$ = $-$0.05, $P(r_k)$ = 0.85, $S(r_k)$ = 0.24$\sigma$). Unlike in the case of QGPs, the covering factor of \hi\ \21\ absorbers in case of DLAs 
remains more or less constant ($\sim$0.6) with $b$. 
The overall covering factor in DLAs ($C_{21}^{DLA}$ = 0.62$^{+0.30}_{-0.21}$) is higher by a factor of $\sim$4 (at 2$\sigma$ significance) compared to that in the QGPs (\c21\ = 0.16$^{+0.07}_{-0.05}$).
We note that the detection rate of \hi\ \21\ absorption in low-$z$ DLAs is comparable to that of molecular \h2\ absorption in low-$z$ DLAs/sub-DLAs \citep[$\sim$50\%;][]{muzahid2015}. Further, it has been 
found that the covering fraction of \mgii\ absorbers around low-$z$ galaxies drops from unity to $\lesssim$0.5 when going from absorption-selected samples to galaxy-selected samples 
\citep[e.g.][]{bergeron1991,steidel1995,bechtold1992,tripp2005}. This is consistent with cold gas clouds in the extended disks/halos of galaxies having small sizes (parsec to sub-parsec scale) and being patchy 
in distribution. Indeed, variations in the \hi\ \21\ optical depth have been detected from parsec-scales \citep{srianand2013,biggs2016} to kilo-parsec-scales \citep{dutta2016}, indicating varying covering 
fractions over different spatial scales. Further, the \hi\ \21\ optical depth is known to vary as a power law over AU- to pc-scales \citep{stanimirovic2010,roy2012,dutta2014}.

With a simple approach we can try to reconcile the measured covering factors of \hi\ \21\ absorption in DLAs and QGPs.
We can estimate the covering factor of DLAs ($C_{DLA}$) around low-$z$ galaxies using \c21\ = $C_{21}^{DLA}$ $\times$ $C_{DLA}$. We have seen that $C_{21}^{DLA}$ is remaining constant (0.6) with $b$ 
(for $b\lesssim$30 kpc), while \c21\ shows a declining trend with $b$ (Fig.~\ref{fig:dlas}). However, the errors on \c21\ being large, we assume for the sake of simplicity a constant \c21\ of $\sim$0.2 
within $b\le$ 30 kpc (see Fig.~\ref{fig:coveringfactor}), and no luminosity dependence of \c21. Hence, using these we obtain average $C_{DLA}$ $\sim$ 0.3 at $b\lesssim$30 kpc. Therefore, one can have a 
consistent picture where the covering factor of gas that can produce DLA absorption is $\sim$0.3 around galaxies for $b\le$30 kpc. Roughly sixty percent of these clouds have cold gas that can produce 
detectable 21-cm absorption. However, we caution that most of the QGPs discussed in this work are at $z\le$0.2 (Fig.~\ref{fig:hist}), while all except one of the low-$z$ DLAs are at $z\ge$0.2 
(Table~\ref{tab:dlagalaxy}). 

We can use the following relation between the geometrical cross-section of the absorbers and the absorber number density per unit redshift ($dN/dz$) to infer the spatial extent of absorbing gas around galaxies 
\citep[see e.g.][]{schaye2007,kacprzak2008},
\begin{equation}
 \frac{dN}{dz} = \frac{1}{2} \frac{c}{H_0} \frac{(1+z)^2}{\sqrt{\Omega_M (1+z)^3 + \Omega_\Lambda}} \Phi^* \Gamma(1 + \alpha +2\beta , \frac{L_{min}}{L^{*}}) C \pi R^{*2},
\end{equation}
where, $\Phi^*$ is the number density of $L^*$ galaxies, $\alpha$ is the faint-end slope of the Schechter galaxy luminosity function, $\beta$ is a parameter to scale the radius with luminosity 
($R/R^* = (L/L^*)^\beta$), $L_{min}$ is the minimum luminosity of galaxies contributing to absorption, $C$ is the covering factor of the absorber, and $R^*$ is the absorber radius for an $L^*$ 
galaxy. The factor $1/2$ is to account for the average projection of the physical cross-section area on the sky plane, assuming disc geometry (see Section~\ref{sec_orientation}). We adopt the 
Schechter function fit of $\Phi^*$ = 1.49 $\pm$ 0.04 $\times$ 10$^{-2}h^3$ Mpc$^{-3}$, $M^*$ $-$ 5log$_{10}h$ = $-$20.44 $\pm$ 0.01, $\alpha$ = $-$1.05 $\pm$ 0.01, estimated by \citet{blanton2003} 
for SDSS $r$-band data of galaxies at $z$ = 0.1. We assume a luminosity scaling of radius with $\beta$ = 0.4 as estimated from \hi\ sizes of galaxies (see Section~\ref{sec_scaledimpact}). We consider 
galaxies down to $L_{min}$ = 0.05$L^*$, though we note that for the values of $\alpha$ and $\beta$ considered here, the value of the incomplete gamma function in Eqn.~2 does not depend significantly 
on the choice of $L_{min}$. Then for $dN/dz$ of DLAs, $n_{DLA}$ = 0.045 $\pm$ 0.006, at $z$ = 0 as obtained by \citet{zwaan2005}, $C_{DLA}$ is unity for $R^*$ = 30 kpc. However, if $C_{DLA}$ were lower, then 
$R^*$ could be larger than 30 kpc. For example if $C_{DLA}$ = 0.3 as obtained above, then $R^*$ $\sim$ 60 kpc.

There exists no measurement of the covering factor and extent of gas that can produce DLAs around low-$z$ galaxies in literature. Though we note that host galaxies of $z<$ 1 DLAs have been detected 
upto $b\sim$70 kpc with median $b\sim$ 17 kpc \citep{rao2011}. \citet{richter2011} estimated the mean projected covering factor of $z<$ 0.5 \caii\ absorbers around galaxies as 0.33 and the characteristic 
radial extent of partly neutral gas clouds (HVCs) with log~\nhi\ $\ge$ 17.4 around $L\ge0.05L^*$ galaxies as $\sim$55 kpc. Note that they have not assumed any luminosity scaling of galaxy sizes, i.e. 
$\beta$ = 0, and have considered a spherical distribution of the neutral gas clouds. Under these assumptions (i.e. $\beta$ = 0 and not considering the factor $1/2$ in Eqn.~2), we obtain $R^*$ $\sim$ 30 kpc 
for $C_{DLA}$ = 0.3 and $R^*$ $\sim$ 15 kpc for $C_{DLA}$ = 1. 
\citet{kacprzak2008} estimated a covering factor of $\sim$0.5 and $R^*$ of $\sim$120 kpc for \mgii\ absorbers at $z$ = 0.5 around $L\ge0.05L^*$ galaxies, for $\beta$ = 0.2 and spherical halo. 
For comparison, we obtain $R^*$ $\sim$ 37 kpc for $C_{DLA}$ = 0.3 and $R^*$ $\sim$ 20 kpc for $C_{DLA}$ = 1, under the same assumptions. The larger covering factor and halo radius of the \mgii\ 
absorbers is expected since \mgii\ absorbers can trace partly ionized gas with lower \nhi\ (sub-DLAs and Lyman limit systems) which extends further out from the galaxy.
\begin{table} 
\caption{List of $z<$ 1 DLAs from the literature with \hi\ \21\ absorption measurement and host galaxy information.}
\centering
\begin{tabular}{cccccc}
\hline
Quasar & \zabs\ & \taudv\ & $b$   & Ref. \\
       &        & (\kms)  & (kpc) &      \\
(1)    & (2)    & (3)     & (4)   & (5)  \\
\hline
0235$+$164     & 0.5243 & 13.1      & 13.1       & 1,2 \\ 
0454$+$039     & 0.8596 & $\le$0.10 & 5.5        & 1,2 \\
0738$+$313     & 0.0912 & 1.05      & $<$3.4$^a$ & 2,3 \\
0738$+$313     & 0.2212 & 0.49      & 19.2       & 1,2 \\
0827$+$243     & 0.5247 & 0.23      & 32.8       & 1,2 \\
0952$+$179     & 0.2378 & 0.12      & $<$4.2$^a$ & 1,2 \\
1122$-$168     & 0.6819 & $\le$0.06 & 25.2$^b$   & 1,2 \\
1127$-$145     & 0.3127 & 3.02      & 45.6       & 1,2 \\
1229$-$021     & 0.3950 & 1.36      & 7.5$^a$    & 1,2 \\
1243$-$072     & 0.4367 & 0.82      & 12.5       & 4   \\
1331$+$305     & 0.6922 & 0.91      & 6.4$^a$    & 2,5 \\
1629$+$120     & 0.5318 & 0.49      & 17.1$^b$   & 1,6 \\
2149$+$212     & 0.9115 & $\le$0.10 & 13.3$^a$   & 1,2 \\
\hline
\end{tabular}
\label{tab:dlagalaxy}
\begin{flushleft}
Column 1: Quasar name. Column 2: Absorption redshift. Column 3: Integrated \hi\ \21\ optical depth in case of detections or 3$\sigma$ upper limit with data smoothed to 10\,\kms\ in case of non-detections.
Column 4: Projected separation or impact parameter (kpc) between radio sightline and centre of galaxy. Columns 5 : References $-$ 1: \citet{rao2011}; 2: \citet{kanekar2014}; 3: \citet{turnshek2001}; 
4: \citet{kanekar2002}; 5: \citet{lebrun2000}; 6: \citet{kanekar2003}. \\
$^a$ based on proximity $^b$ based on photometric redshift 
\end{flushleft}
\end{table} 
\subsection{Redshift evolution of \hi\ \21\ absorbers}
\label{sec_dndz}
\begin{figure}
\includegraphics[width=0.35\textwidth, angle=90]{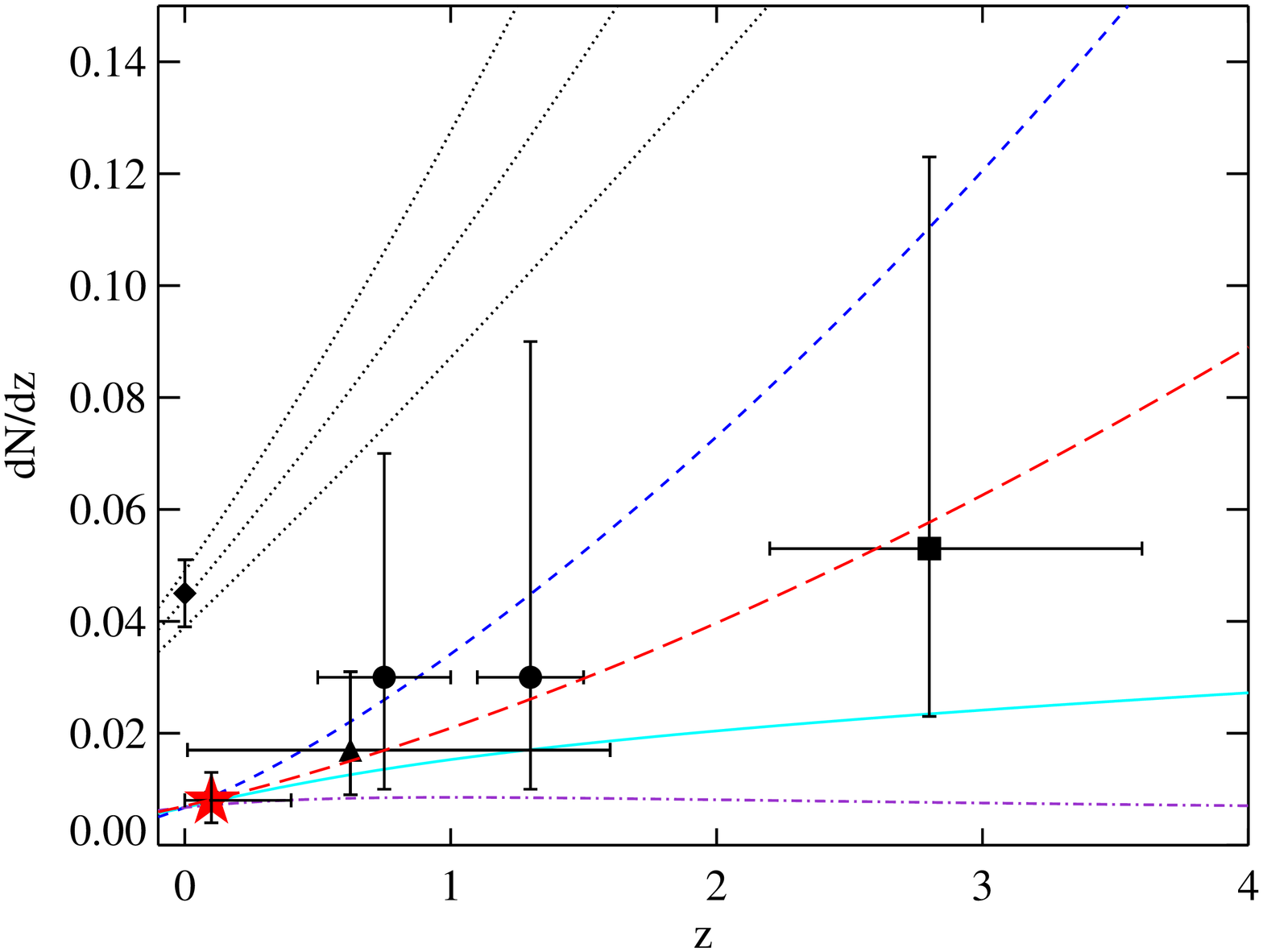}
\caption{Number of \hi\ \21\ absorbers per unit redshift range for \t0\ = 0.3\,\kms. The red star is the $n_{21}$ estimated in this work. The circles and the square are the $n_{21}$ estimates based on 
\mgii\ and DLAs, respectively \citep{gupta2012}. Also shown for reference are $n_{DLA}$ at $z$ = 0 estimated by \citet{zwaan2005} (diamond), $n_{DLA}$ at $z$ = 0.6 estimated by \citet{neeleman2016} (triangle) 
and its redshift evolution estimated by \citet{rao2006} with errors (dotted black lines). The long dashed red line is a fit by eye to the $n_{21}$ estimates, $n_{21}(z) = n_0(1 + z)^\gamma$ with $n_0$ = 0.007 
and $\gamma$ = 1.58. The solid cyan curve represents no intrinsic evolution in the product of space density and cross-section of \hi\ \21\ absorbers, normalized at $z$ = 0.1. The short dashed blue curve 
represents the redshift evolution of $n_{21}$ for the observed redshift evolution of $L^*$ and $\Phi^*$ of galaxies (and no evolution in \c21). The dotted-dashed purple curve represents $n_{21}(z)$ accounting 
for the redshift evolution of sizes of galaxies in addition to that of $L^*$ and $\Phi^*$, and constant \c21.
}
\label{fig:dndz}
\end{figure}

We can use Eqn.~2 to estimate the $dN/dz$ of \hi\ \21\ absorbers ($n_{21}$) at low redshift. We assume that the distribution of \hi\ \21\ absorbers is planar, and \c21\ is constant ($\sim$0.2) within 
$R^*$ with no luminosity dependence. We consider $R^*$ for \hi\ \21\ absorbers around low-$z$ galaxies to be $\sim$30 kpc, where we have measured \c21. Under these assumptions, we obtain the $n_{21}$ 
at $z$ = 0.1 as 0.008$^{+0.005}_{-0.004}$. In Fig.~\ref{fig:dndz}, we show our $n_{21}$ estimate along with those made by \citet{gupta2012} based on search for \hi\ \21\ absorption in \mgii\ absorbers 
and DLAs. We also show for comparison $n_{DLA}$ at $z$ = 0 as estimated by \citet{zwaan2005}, at $z$ = 0.6 as estimated by \citet{neeleman2016}, and its redshift evolution as estimated by \citet{rao2006}, 
i.e. $n_{DLA}(z) = n_0(1 + z)^\gamma$ with $n_0$ = 0.044 $\pm$ 0.005 and $\gamma$ = 1.27 $\pm$ 0.11.
Note that the $n_{21}$ measurements at different redshifts, while obtained for an uniform sensitivity of \t0\ = 0.3\,\kms, are based on different kinds of systems, possibly having different \nhi\ 
thresholds used in the target selection. For example, all the \mgii\ systems at $z\sim1.3$ used to derive $n_{21}$ need not be DLAs and all the QGPs in our sample need not produce \mgii\ absorption 
with rest equivalent width greater than 1\,\AA. Therefore, the correct mapping of $n_{21}(z)$ has to wait till the advent of large blind \hi\ \21\ absorption surveys using SKA pathfinders. Here we make 
some generic statements based on the measurements in hand. It is interesting to note that the $n_{21}$ we infer from our QGP observations is lower than what has been found from the high-$z$ \mgii\ 
absorbers and DLAs. We try to understand this by using Eqn.~2 and the measured redshift evolution of $L^*$ and $\Phi^*$ of galaxies.

Firstly, the expected $n_{21}(z)$ for no intrinsic redshift evolution in all of the parameters in Eqn.~2 is shown in Fig.~\ref{fig:dndz} (solid cyan curve).
Next, from the measured redshift evolution of the galaxy luminosity function in the optical bands, it can be seen that $L^*$ evolves as $\sim (1+z)^{2.7}$ and $\Phi^*$ evolves as $\sim (1+z)^{-1}$
\citep{gabasch2004,ilbert2005,dahlen2007}. The redshift evolution of $L^*$ comes into Eqn.~2 in two ways. First is through the dependence of $R^*$ on $L^*$ ($R^*$ $\sim$ $L^{*~0.4}$), and second is through 
the incomplete gamma function. For the simplistic calculation here we neglect any redshift evolution in the value of the incomplete gamma function. If we assume that the scaling relation between galaxy 
luminosity and \hi\ size, as well as \c21, do not evolve with redshift, then it can be seen from Fig.~\ref{fig:dndz} that $n_{21}$ is expected to increase with redshift (short dashed blue curve). 
However, galaxies of all morphological types have been observed to have much smaller sizes at high redshifts, with the effective optical radius of disc galaxies decreasing as $\sim (1+z)^{-1}$ 
\citep{trujillo2007,buitrago2008,vanderwel2014}. Assuming that $R^*$ of \hi\ \21\ absorbers follows the redshift evolution of the optical radius of disc galaxies and factoring in its decrease with redshift,
we find that $n_{21}$ is expected to decrease with redshift for constant \c21\ (Fig.~\ref{fig:dndz}; dotted-dashed purple curve). On the other hand, observations suggest that $n_{21}$ is increasing with 
redshift, i.e. $n_{21}(z) = 0.007~(1 + z)^{1.58}$ (Fig.~\ref{fig:dndz}; long dashed red curve). Further, the covering factor of cold gas is expected to decrease with redshift, with $C_{21}^{DLA}$ at $z<$1 
being $\sim$0.6 compared to $\sim$0.2 at 2.0$\le z\le$3.5 \citep{srianand2012}. This basically implies that if we take into account the evolution of size and luminosity of galaxies with redshift, the radius 
of the cold \hi\ gas around a galaxy that gives rise to \hi\ \21\ absorption will be much higher at high-$z$ than what is seen at low-$z$ for a galaxy with same optical luminosity. 

As noted above, the interpretation of the redshift evolution of \hi\ \21\ absorbers based on the present measurements is difficult due to the large errors and different selection techniques used at different 
redshifts. Accurate (upto $\sim$10\%) and uniform determination of the redshift evolution of $n_{21}$ and hence that of the cold gas fraction in galaxies at $z\le$ 1.5 will be possible with the upcoming blind 
\hi\ \21\ absorption line surveys, e.g. the MeerKAT Absorption Line Survey, the Search for \hi\ absorption with AperTIF, and the ASKAP First Large Absorption Survey in HI \citep[see e.g.][]{allison2016}.
Preliminary results of blind \hi\ \21\ absorption searches from the ALFALFA survey \citep{darling2011,wu2015} and ASKAP \citep{allison2015} are already beginning to demonstrate the ability of future 
blind \hi\ \21\ absorption surveys.
%
%
%=========================== SUMMARY =======================================================================================
%
\section{Summary}  
\label{sec_summary} 
We have presented in this work the results from our survey of \hi\ \21\ absorption of a large sample (55) of low-$z$ ($z<$0.4) galaxies at small impact parameters to background radio sources using the GMRT, 
WSRT and VLA. Our primary sample of 40 QGPs spans a range of impact parameters (5$-$34 kpc) and galaxy types (log~$M_*$ $\sim$7.8$-$11.0 $M_{\odot}$, log~$L_B$ $\sim$8.5$-$10.6 $L_{\odot}$, $g-r$ $\sim$ 0.1$-$1.6). 
The \hi\ \21\ spectral line measurements from our survey have increased the number of existing sensitive  (i.e \taudvl\ $\sim$ 0.3 \kms) \hi\ \21\ optical depth measurements at low-$z$ by a factor $\sim$3. 
We have detected \hi\ \21\ absorption from seven QGPs in our primary sample, two of which are reported here for the first time. The \hi\ \21\ detections from our survey have increased the existing number of 
detections from $z<$0.4 QGPs by almost a factor of two. Combining our primary sample with measurements from the literature having similar optical depth sensitivity, we have quantified the \hi\ \21\ optical 
depth and covering factor of \hi\ \21\ absorbers as a function of spatial location around low-$z$ galaxies. Below we list our main results.

\begin{itemize}
 \item The \hi\ \21\ optical depth and impact parameter are weakly anti-correlated. Performing survival analysis by including the upper limits as censored data points, we obtain, $r_k$ = $-$0.20, $P(r_k)$ = 0.01, 
       $S(r_k)$ = 2.42$\sigma$. Similar anti-correlation is also present between \taudv\ and the radial distance along the galaxy's major axis, as well the impact parameter and the radial distance scaled with 
       the effective \hi\ radius of the galaxies.
 \item We find that the covering factor of \hi\ \21\ absorbers is 0.24$^{+0.12}_{-0.08}$ at $b\le$ 15 kpc and decreases to 0.06$^{+0.09}_{-0.04}$ at $b=$ 15$-$35 kpc when we use \t0\ = 0.3 \kms. The average \c21\ 
       within $b$ = 30 kpc is 0.16$^{+0.07}_{-0.05}$. \c21\ shows similar declining trend with increasing scaled $b$, $R$ and scaled $R$.
 \item The \hi\ \21\ optical depth shows weak anti-correlation ($r_k$ = $-$0.17, $P(r_k)$ = 0.07, $S(r_k)$ = 1.79$\sigma$) with the azimuthal orientation of the radio sightline with respect to the 
       galaxy's major axis. There is similar weak correlation between \taudv\ and $sin(i)cos(\phi)$, i.e. \taudv\ shows mild increase with increasing galaxy inclination and decreasing orientation.
       However, these could be driven by the stronger and more significant anti-correlation of \taudv\ with $b$, since these parameters are correlated with $b$ in our sample.
 \item We find that \c21\ is 0.20$^{+0.14}_{-0.09}$ near the galaxy major axis ($\phi <$ 45\textdegree) and 0.04$^{+0.09}_{-0.03}$ near the minor axis ($\phi \ge$ 45\textdegree), and this declining trend 
       of \c21\ with increasing $\phi$ is also seen at $b<$ 15 kpc. This tentative indication that most of the \hi\ \21\ absorbers could be co-planar with that of the \hi\ disk is supported by the fact that 
       the only two \hi\ \21\ detections at $b\ge$ 15 kpc in the Combined sample arise from face-on galaxies. Further, \c21\ is maximum (0.31$^{+0.24}_{-0.15}$) for sightlines that pass near the major axis of 
       edge-on galaxies ($i\ge$58\textdegree). If this is true even for high-$z$ galaxies, then it will have important implications for the detection rate of \hi\ \21\ absorption towards \mgii\ absorbers, which 
       are found to be typically tracing the halo gas and not the extended \hi\ disks. However, high-$z$ galaxies are not expected to have such well-formed \hi\ disks, but to be more irregular. Hence, the dependence 
       of \hi\ \21\ absorption on galaxy orientation needs to be probed over a larger redshift range. 
 \item No significant dependence of \taudv\ and \c21\ on galaxy luminosity, stellar mass, colour and surface star formation rate density is found from the present data. However, our results suggest
       that the \hi\ \21\ absorbing gas cross-section may be larger for the brighter galaxies. A correlation between \taudv\ and \c21\ and properties associated with star formation in galaxies are expected 
       from the models of ISM as discussed in Section~\ref{sec_introduction}. Hence, it appears that most of our sightlines are outside the stellar disks of galaxies and probably tracing the outer quiescent 
       regions that are not affected by the ongoing star formation activities.
 \item No correlation is found between \hi\ \21\ optical depth and $W_r$(\caii) or $W_r$(\nai). We also find that $W_r$(\caii) and the ratio $W_r$(\nai)$/W_r$(\caii) suggests that most of the \hi\ \21\ 
       absorbers observed around low-$z$ galaxies are not tracing the dusty star-forming disks.
 \item From the data available in the literature, we find that the covering factor of \hi\ \21\ absorption from host galaxies of $z<$ 1 DLAs (0.62$^{+0.30}_{-0.21}$\%) is a factor of $\sim$4 times higher
       compared to that from the galaxy-selected QGPs. There is also no correlation between \taudv\ and $b$ for DLAs. This result is analogous to the finding that the covering factor of \mgii\ gaseous halos 
       around galaxies is close to 1 when one searches for galaxies around \mgii\ absorbers \citep[see][]{bergeron1991,steidel1995}, and is lower when one searches for \mgii\ absorption along sightlines 
       close to known galaxies \citep[see][]{bechtold1992,tripp2005}. Broadly, we can conclude that the \hi\ distribution around galaxies that can contribute to the DLA population is patchy (i.e with a covering 
       factor of about 30\%) and about 60\% of the DLAs have cold gas that can produce detectable \hi\ \21\ absorption. Such a picture will be consistent with the observed covering factors of DLAs and \hi\ \21\ 
       absorbers around galaxies and fraction of DLAs producing \hi\ \21\ absorption. 
 \item We estimate $n_{21}$ at $z$ = 0.1 as 0.008$^{+0.005}_{-0.004}$ from our QGP observations. This is lower than what has been found from the high-$z$ \mgii\ absorbers and DLAs. This along with the 
       observed redshift evolution of galaxy size and luminosity may suggest an evolution in the correlation between optical luminosity and the extent of the \hi\ gas around galaxies with redshift. \\
\end{itemize}
%
%========================================= Acknowledgment ==================================================================
%

\noindent \textbf{ACKNOWLEDGEMENTS} \newline

\noindent 
We thank the staff at GMRT, VLA, WSRT and SALT for their help during the observations. 
GMRT is run by the National Centre for Radio Astrophysics of the Tata Institute of Fundamental Research. 
The VLA is run by the National Radio Astronomy Observatory (NRAO). The NRAO is a facility of the National Science Foundation operated under cooperative agreement by Associated Universities, Inc. 
The WSRT is operated by the ASTRON (Netherlands Foundation for Research in Astronomy) with support from the Netherlands Foundation for Scientific Research (NWO). 
Some of the observations reported in this paper were obtained with the Southern African Large Telescope (SALT).
NG, PN, RS and PPJ acknowledge the support from Indo-Fench centre for the promotion of Advanced Research (IFCPAR) under Project No. 5504-2.
We thank the referee for his/her detailed and useful comments.
This research has made use of the NASA/IPAC Extragalactic Database (NED) which is operated by the Jet Propulsion Laboratory, California Institute of Technology, 
under contract with the National Aeronautics and Space Administration. 
Funding for SDSS-III has been provided by the Alfred P. Sloan Foundation, the Participating Institutions, the National Science Foundation, and the U.S. Department of Energy Office of Science. 
The SDSS-III web site is http://www.sdss3.org/. SDSS-III is managed by the Astrophysical Research Consortium for the Participating Institutions of the SDSS-III Collaboration including the 
University of Arizona, the Brazilian Participation Group, Brookhaven National Laboratory, Carnegie Mellon University, University of Florida, the French Participation Group, the German Participation 
Group, Harvard University, the Instituto de Astrofisica de Canarias, the Michigan State/Notre Dame/JINA Participation Group, Johns Hopkins University, Lawrence Berkeley National Laboratory, Max Planck 
Institute for Astrophysics, Max Planck Institute for Extraterrestrial Physics, New Mexico State University, New York University, Ohio State University, Pennsylvania State University, University of 
Portsmouth, Princeton University, the Spanish Participation Group, University of Tokyo, University of Utah, Vanderbilt University, University of Virginia, University of Washington, and Yale University. 
%
%======================================== Bibliography =====================================================================
% 
\def\aj{AJ}%
\def\actaa{Acta Astron.}%
\def\araa{ARA\&A}%
\def\apj{ApJ}%
\def\apjl{ApJ}%
\def\apjs{ApJS}%
\def\ao{Appl.~Opt.}%
\def\apss{Ap\&SS}%
\def\aap{A\&A}%
\def\aapr{A\&A~Rev.}%
\def\aaps{A\&AS}%
\def\azh{A$Z$h}%
\def\baas{BAAS}%
\def\bac{Bull. astr. Inst. Czechosl.}%
\def\caa{Chinese Astron. Astrophys.}%
\def\cjaa{Chinese J. Astron. Astrophys.}%
\def\icarus{Icarus}%
\def\jcap{J. Cosmology Astropart. Phys.}%
\def\jrasc{JRASC}%
\def\mnras{MNRAS}%
\def\memras{MmRAS}%
\def\na{New A}%
\def\nar{New A Rev.}%
\def\pasa{PASA}%
\def\pra{Phys.~Rev.~A}%
\def\prb{Phys.~Rev.~B}%
\def\prc{Phys.~Rev.~C}%
\def\prd{Phys.~Rev.~D}%
\def\pre{Phys.~Rev.~E}%
\def\prl{Phys.~Rev.~Lett.}%
\def\pasp{PASP}%
\def\pasj{PASJ}%
\def\qjras{QJRAS}%
\def\rmxaa{Rev. Mexicana Astron. Astrofis.}%
\def\skytel{S\&T}%
\def\solphys{Sol.~Phys.}%
\def\sovast{Soviet~Ast.}%
\def\ssr{Space~Sci.~Rev.}%
\def\zap{$Z$Ap}%
\def\nat{Nature}%
\def\iaucirc{IAU~Circ.}%
\def\aplett{Astrophys.~Lett.}%
\def\apspr{Astrophys.~Space~Phys.~Res.}%
\def\bain{Bull.~Astron.~Inst.~Netherlands}%
\def\fcp{Fund.~Cosmic~Phys.}%
\def\gca{Geochim.~Cosmochim.~Acta}%
\def\grl{Geophys.~Res.~Lett.}%
\def\jcp{J.~Chem.~Phys.}%
\def\jgr{J.~Geophys.~Res.}%
\def\jqsrt{J.~Quant.~Spec.~Radiat.~Transf.}%
\def\memsai{Mem.~Soc.~Astron.~Italiana}%
\def\nphysa{Nucl.~Phys.~A}%
\def\physrep{Phys.~Rep.}%
\def\physscr{Phys.~Scr}%
\def\planss{Planet.~Space~Sci.}%
\def\procspie{Proc.~SPIE}%
\let\astap=\aap
\let\apjlett=\apjl
\let\apjsupp=\apjs
\let\applopt=\ao
\bibliographystyle{mn}
\bibliography{mybib}

\begin{thebibliography}{152}
\expandafter\ifx\csname natexlab\endcsname\relax\def\natexlab#1{#1}\fi

\bibitem[{{Alam} {et~al.}(2015){Alam}, {Albareti}, {Allende Prieto}, {Anders},
  {Anderson}, {Anderton}, {Andrews}, {Armengaud}, {Aubourg}, {Bailey}, \&
  et~al.}]{alam2015}
{Alam}, S., {Albareti}, F.~D., {Allende Prieto}, C., {et~al.}, 2015, \apjs,
  219, 12

\bibitem[{{Allison} {et~al.}(2015){Allison}, {Sadler}, {Moss}, {Whiting},
  {Hunstead}, {Pracy}, {Curran}, {Croom}, {Glowacki}, {Morganti}, {Shabala},
  {Zwaan}, {Allen}, {Amy}, {Axtens}, {Ball}, {Bannister}, {Barker}, {Bell},
  {Bock}, {Bolton}, {Bowen}, {Boyle}, {Braun}, {Broadhurst}, {Brodrick},
  {Brothers}, {Brown}, {Bunton}, {Cantrall}, {Chapman}, {Cheng}, {Chippendale},
  {Chung}, {Cooray}, {Cornwell}, {DeBoer}, {Diamond}, {Edwards}, {Ekers},
  {Feain}, {Ferris}, {Forsyth}, {Gough}, {Grancea}, {Gupta}, {Guzman},
  {Hampson}, {Harvey-Smith}, {Haskins}, {Hay}, {Hayman}, {Heywood}, {Hotan},
  {Hoyle}, {Humphreys}, {Indermuehle}, {Jacka}, {Jackson}, {Jackson},
  {Jeganathan}, {Johnston}, {Joseph}, {Kendall}, {Kesteven}, {Kiraly},
  {Koribalski}, {Leach}, {Lenc}, {Lensson}, {Mackay}, {Macleod}, {Marquarding},
  {Marvil}, {McClure-Griffiths}, {McConnell}, {Mirtschin}, {Norris}, {Neuhold},
  {Ng}, {O'Sullivan}, {Pathikulangara}, {Pearce}, {Phillips}, {Popping},
  {Qiao}, {Reynolds}, {Roberts}, {Sault}, {Schinckel}, {Serra}, {Shaw},
  {Shields}, {Shimwell}, {Storey}, {Sweetnam}, {Troup}, {Turner}, {Tuthill},
  {Tzioumis}, {Voronkov}, {Westmeier}, \& {Wilson}}]{allison2015}
{Allison}, J.~R., {Sadler}, E.~M., {Moss}, V.~A., {et~al.}, 2015, \mnras, 453,
  1249

\bibitem[{{Allison} {et~al.}(2016){Allison}, {Zwaan}, {Duchesne}, \&
  {Curran}}]{allison2016}
{Allison}, J.~R., {Zwaan}, M.~A., {Duchesne}, S.~W., \& {Curran}, S.~J., 2016,
  \mnras, 462, 1341

\bibitem[{{Argence} \& {Lamareille}(2009)}]{argence2009}
{Argence}, B. \& {Lamareille}, F., 2009, \aap, 495, 759

\bibitem[{{Bagetakos} {et~al.}(2011){Bagetakos}, {Brinks}, {Walter}, {de Blok},
  {Usero}, {Leroy}, {Rich}, \& {Kennicutt}}]{bagetakos2011}
{Bagetakos}, I., {Brinks}, E., {Walter}, F., {de Blok}, W.~J.~G., {Usero}, A.,
  {Leroy}, A.~K., {Rich}, J.~W., \& {Kennicutt}, Jr., R.~C., 2011, \aj, 141, 23

\bibitem[{{Bahcall} \& {Ekers}(1969)}]{bahcall1969}
{Bahcall}, J.~N. \& {Ekers}, R.~D., 1969, \apj, 157, 1055

\bibitem[{{Bechtold} \& {Ellingson}(1992)}]{bechtold1992}
{Bechtold}, J. \& {Ellingson}, E., 1992, \apj, 396, 20

\bibitem[{{Begum} {et~al.}(2008){Begum}, {Chengalur}, {Karachentsev},
  {Sharina}, \& {Kaisin}}]{begum2008}
{Begum}, A., {Chengalur}, J.~N., {Karachentsev}, I.~D., {Sharina}, M.~E., \&
  {Kaisin}, S.~S., 2008, \mnras, 386, 1667

\bibitem[{{Bergeron} \& {Boiss{\'e}}(1991)}]{bergeron1991}
{Bergeron}, J. \& {Boiss{\'e}}, P., 1991, \aap, 243, 344

\bibitem[{{Best} {et~al.}(2003){Best}, {Arts}, {R{\"o}ttgering}, {Rengelink},
  {Brookes}, \& {Wall}}]{best2003}
{Best}, P.~N., {Arts}, J.~N., {R{\"o}ttgering}, H.~J.~A., {Rengelink}, R.,
  {Brookes}, M.~H., \& {Wall}, J., 2003, \mnras, 346, 627

\bibitem[{{Biggs} {et~al.}(2016){Biggs}, {Zwaan}, {Hatziminaoglou},
  {P{\'e}roux}, \& {Liske}}]{biggs2016}
{Biggs}, A.~D., {Zwaan}, M.~A., {Hatziminaoglou}, E., {P{\'e}roux}, C., \&
  {Liske}, J., 2016, \mnras, 462, 2819

\bibitem[{{Bigiel} {et~al.}(2010){Bigiel}, {Leroy}, {Walter}, {Blitz},
  {Brinks}, {de Blok}, \& {Madore}}]{bigiel2010}
{Bigiel}, F., {Leroy}, A., {Walter}, F., {Blitz}, L., {Brinks}, E., {de Blok},
  W.~J.~G., \& {Madore}, B., 2010, \aj, 140, 1194

\bibitem[{{Blanton} \& {Berlind}(2007)}]{blanton2007b}
{Blanton}, M.~R. \& {Berlind}, A.~A., 2007, \apj, 664, 791

\bibitem[{{Blanton} {et~al.}(2003){Blanton}, {Hogg}, {Bahcall}, {Brinkmann},
  {Britton}, {Connolly}, {Csabai}, {Fukugita}, {Loveday}, {Meiksin}, {Munn},
  {Nichol}, {Okamura}, {Quinn}, {Schneider}, {Shimasaku}, {Strauss}, {Tegmark},
  {Vogeley}, \& {Weinberg}}]{blanton2003}
{Blanton}, M.~R., {Hogg}, D.~W., {Bahcall}, N.~A., {et~al.}, 2003, \apj, 592,
  819

\bibitem[{{Blanton} \& {Roweis}(2007)}]{blanton2007a}
{Blanton}, M.~R. \& {Roweis}, S., 2007, \aj, 133, 734

\bibitem[{{Bohlin} {et~al.}(1978){Bohlin}, {Savage}, \& {Drake}}]{bohlin1978}
{Bohlin}, R.~C., {Savage}, B.~D., \& {Drake}, J.~F., 1978, \apj, 224, 132

\bibitem[{{Boisse} {et~al.}(1988){Boisse}, {Dickey}, {Kazes}, \&
  {Bergeron}}]{boisse1988}
{Boisse}, P., {Dickey}, J.~M., {Kazes}, I., \& {Bergeron}, J., 1988, \aap, 191,
  193

\bibitem[{{Bordoloi} {et~al.}(2014){Bordoloi}, {Lilly}, {Kacprzak}, \&
  {Churchill}}]{bordoloi2014}
{Bordoloi}, R., {Lilly}, S.~J., {Kacprzak}, G.~G., \& {Churchill}, C.~W., 2014,
  \apj, 784, 108

\bibitem[{{Borthakur}(2016)}]{borthakur2016}
{Borthakur}, S., 2016, \apj, 829, 128

\bibitem[{{Borthakur} {et~al.}(2015){Borthakur}, {Heckman}, {Tumlinson},
  {Bordoloi}, {Thom}, {Catinella}, {Schiminovich}, {Dav{\'e}}, {Kauffmann},
  {Moran}, \& {Saintonge}}]{borthakur2015}
{Borthakur}, S., {Heckman}, T., {Tumlinson}, J., {et~al.}, 2015, \apj, 813, 46

\bibitem[{{Borthakur} {et~al.}(2014){Borthakur}, {Momjian}, {Heckman}, {York},
  {Bowen}, {Yun}, \& {Tripp}}]{borthakur2014}
{Borthakur}, S., {Momjian}, E., {Heckman}, T.~M., {York}, D.~G., {Bowen},
  D.~V., {Yun}, M.~S., \& {Tripp}, T.~M., 2014, \apj, 795, 98

\bibitem[{{Borthakur} {et~al.}(2011){Borthakur}, {Tripp}, {Yun}, {Bowen},
  {Meiring}, {York}, \& {Momjian}}]{borthakur2011}
{Borthakur}, S., {Tripp}, T.~M., {Yun}, M.~S., {Bowen}, D.~V., {Meiring},
  J.~D., {York}, D.~G., \& {Momjian}, E., 2011, \apj, 727, 52

\bibitem[{{Borthakur} {et~al.}(2010){Borthakur}, {Tripp}, {Yun}, {Momjian},
  {Meiring}, {Bowen}, \& {York}}]{borthakur2010}
{Borthakur}, S., {Tripp}, T.~M., {Yun}, M.~S., {Momjian}, E., {Meiring}, J.~D.,
  {Bowen}, D.~V., \& {York}, D.~G., 2010, \apj, 713, 131

\bibitem[{{Bouch{\'e}} {et~al.}(2012){Bouch{\'e}}, {Hohensee}, {Vargas},
  {Kacprzak}, {Martin}, {Cooke}, \& {Churchill}}]{bouche2012}
{Bouch{\'e}}, N., {Hohensee}, W., {Vargas}, R., {Kacprzak}, G.~G., {Martin},
  C.~L., {Cooke}, J., \& {Churchill}, C.~W., 2012, \mnras, 426, 801

\bibitem[{{Briggs} \& {Wolfe}(1983)}]{briggs1983}
{Briggs}, F.~H. \& {Wolfe}, A.~M., 1983, \apj, 268, 76

\bibitem[{{Broeils} \& {Rhee}(1997)}]{broeils1997}
{Broeils}, A.~H. \& {Rhee}, M.-H., 1997, \aap, 324, 877

\bibitem[{{Brookes} {et~al.}(2008){Brookes}, {Best}, {Peacock},
  {R{\"o}ttgering}, \& {Dunlop}}]{brookes2008}
{Brookes}, M.~H., {Best}, P.~N., {Peacock}, J.~A., {R{\"o}ttgering}, H.~J.~A.,
  \& {Dunlop}, J.~S., 2008, \mnras, 385, 1297

\bibitem[{{Brookes} {et~al.}(2006){Brookes}, {Best}, {Rengelink}, \&
  {R{\"o}ttgering}}]{brookes2006}
{Brookes}, M.~H., {Best}, P.~N., {Rengelink}, R., \& {R{\"o}ttgering},
  H.~J.~A., 2006, \mnras, 366, 1265

\bibitem[{{Buitrago} {et~al.}(2008){Buitrago}, {Trujillo}, {Conselice},
  {Bouwens}, {Dickinson}, \& {Yan}}]{buitrago2008}
{Buitrago}, F., {Trujillo}, I., {Conselice}, C.~J., {Bouwens}, R.~J.,
  {Dickinson}, M., \& {Yan}, H., 2008, \apjl, 687, L61

\bibitem[{{Carilli} \& {van Gorkom}(1992)}]{carilli1992}
{Carilli}, C.~L. \& {van Gorkom}, J.~H., 1992, \apj, 399, 373

\bibitem[{{Catinella} \& {Cortese}(2015)}]{catinella2015}
{Catinella}, B. \& {Cortese}, L., 2015, \mnras, 446, 3526

\bibitem[{{Chen} {et~al.}(2001){Chen}, {Lanzetta}, {Webb}, \&
  {Barcons}}]{chen2001}
{Chen}, H.-W., {Lanzetta}, K.~M., {Webb}, J.~K., \& {Barcons}, X., 2001, \apj,
  559, 654

\bibitem[{{Chen} {et~al.}(2010){Chen}, {Wild}, {Tinker}, {Gauthier}, {Helsby},
  {Shectman}, \& {Thompson}}]{chen2010}
{Chen}, H.-W., {Wild}, V., {Tinker}, J.~L., {Gauthier}, J.-R., {Helsby}, J.~E.,
  {Shectman}, S.~A., \& {Thompson}, I.~B., 2010, \apjl, 724, L176

\bibitem[{{Chung} {et~al.}(2009){Chung}, {van Gorkom}, {Kenney}, {Crowl}, \&
  {Vollmer}}]{chung2009}
{Chung}, A., {van Gorkom}, J.~H., {Kenney}, J.~D.~P., {Crowl}, H., \&
  {Vollmer}, B., 2009, \aj, 138, 1741

\bibitem[{{Churchill} {et~al.}(2013){Churchill}, {Nielsen}, {Kacprzak}, \&
  {Trujillo-Gomez}}]{churchill2013}
{Churchill}, C.~W., {Nielsen}, N.~M., {Kacprzak}, G.~G., \& {Trujillo-Gomez},
  S., 2013, \apjl, 763, L42

\bibitem[{{Corbelli} \& {Schneider}(1990)}]{corbelli1990}
{Corbelli}, E. \& {Schneider}, S.~E., 1990, \apj, 356, 14

\bibitem[{{Crawford} {et~al.}(2010){Crawford}, {Still}, {Schellart}, {Balona},
  {Buckley}, {Dugmore}, {Gulbis}, {Kniazev}, {Kotze}, {Loaring}, {Nordsieck},
  {Pickering}, {Potter}, {Romero Colmenero}, {Vaisanen}, {Williams}, \&
  {Zietsman}}]{crawford2010}
{Crawford}, S.~M., {Still}, M., {Schellart}, P., {et~al.}, 2010, in \procspie,
  Vol. 7737, Observatory Operations: Strategies, Processes, and Systems III, p.
  773725

\bibitem[{{Curran} {et~al.}(2005){Curran}, {Murphy}, {Pihlstr{\"o}m}, {Webb},
  \& {Purcell}}]{curran2005}
{Curran}, S.~J., {Murphy}, M.~T., {Pihlstr{\"o}m}, Y.~M., {Webb}, J.~K., \&
  {Purcell}, C.~R., 2005, \mnras, 356, 1509

\bibitem[{{Curran} {et~al.}(2016){Curran}, {Reeves}, {Allison}, \&
  {Sadler}}]{curran2016}
{Curran}, S.~J., {Reeves}, S.~N., {Allison}, J.~R., \& {Sadler}, E.~M., 2016,
  \mnras, 459, 4136

\bibitem[{{Curran} {et~al.}(2010){Curran}, {Tzanavaris}, {Darling}, {Whiting},
  {Webb}, {Bignell}, {Athreya}, \& {Murphy}}]{curran2010}
{Curran}, S.~J., {Tzanavaris}, P., {Darling}, J.~K., {Whiting}, M.~T., {Webb},
  J.~K., {Bignell}, C., {Athreya}, R., \& {Murphy}, M.~T., 2010, \mnras, 402,
  35

\bibitem[{{Dahlen} {et~al.}(2007){Dahlen}, {Mobasher}, {Dickinson}, {Ferguson},
  {Giavalisco}, {Kretchmer}, \& {Ravindranath}}]{dahlen2007}
{Dahlen}, T., {Mobasher}, B., {Dickinson}, M., {Ferguson}, H.~C., {Giavalisco},
  M., {Kretchmer}, C., \& {Ravindranath}, S., 2007, \apj, 654, 172

\bibitem[{{Darling} {et~al.}(2011){Darling}, {Macdonald}, {Haynes}, \&
  {Giovanelli}}]{darling2011}
{Darling}, J., {Macdonald}, E.~P., {Haynes}, M.~P., \& {Giovanelli}, R., 2011,
  \apj, 742, 60

\bibitem[{{de Blok} {et~al.}(2008){de Blok}, {Walter}, {Brinks},
  {Trachternach}, {Oh}, \& {Kennicutt}}]{deblok2008}
{de Blok}, W.~J.~G., {Walter}, F., {Brinks}, E., {Trachternach}, C., {Oh},
  S.-H., \& {Kennicutt}, Jr., R.~C., 2008, \aj, 136, 2648

\bibitem[{{DeBoer} {et~al.}(2009){DeBoer}, {Gough}, {Bunton}, {Cornwell},
  {Beresford}, {Johnston}, {Feain}, {Schinckel}, {Jackson}, {Kesteven},
  {Chippendale}, {Hampson}, {O'Sullivan}, {Hay}, {Jacka}, {Sweetnam}, {Storey},
  {Ball}, \& {Boyle}}]{deboer2009}
{DeBoer}, D.~R., {Gough}, R.~G., {Bunton}, J.~D., {et~al.}, 2009, IEEE
  Proceedings, 97, 1507

\bibitem[{{Dutta} {et~al.}(2014){Dutta}, {Chengalur}, {Roy}, {Goss},
  {Arjunwadkar}, {Minter}, {Brogan}, \& {Lazio}}]{dutta2014}
{Dutta}, P., {Chengalur}, J.~N., {Roy}, N., {Goss}, W.~M., {Arjunwadkar}, M.,
  {Minter}, A.~H., {Brogan}, C.~L., \& {Lazio}, T.~J.~W., 2014, \mnras, 442,
  647

\bibitem[{{Dutta} {et~al.}(2016){Dutta}, {Gupta}, {Srianand}, \&
  {O'Meara}}]{dutta2016}
{Dutta}, R., {Gupta}, N., {Srianand}, R., \& {O'Meara}, J.~M., 2016, \mnras,
  456, 4209

\bibitem[{{Elmegreen} \& {Hunter}(2015)}]{elmegreen2015}
{Elmegreen}, B.~G. \& {Hunter}, D.~A., 2015, \apj, 805, 145

\bibitem[{{Epinat} {et~al.}(2008){Epinat}, {Amram}, {Marcelin}, {Balkowski},
  {Daigle}, {Hernandez}, {Chemin}, {Carignan}, {Gach}, \&
  {Balard}}]{epinat2008}
{Epinat}, B., {Amram}, P., {Marcelin}, M., {et~al.}, 2008, \mnras, 388, 500

\bibitem[{{Fern{\'a}ndez} {et~al.}(2016){Fern{\'a}ndez}, {Gim}, {van Gorkom},
  {Yun}, {Momjian}, {Popping}, {Chomiuk}, {Hess}, {Hunt}, {Kreckel}, {Lucero},
  {Maddox}, {Oosterloo}, {Pisano}, {Verheijen}, {Hales}, {Chung}, {Dodson},
  {Golap}, {Gross}, {Henning}, {Hibbard}, {Jaff{\'e}}, {Donovan Meyer},
  {Meyer}, {Sanchez-Barrantes}, {Schiminovich}, {Wicenec}, {Wilcots},
  {Bershady}, {Scoville}, {Strader}, {Tremou}, {Salinas}, \&
  {Ch{\'a}vez}}]{fernandez2016}
{Fern{\'a}ndez}, X., {Gim}, H.~B., {van Gorkom}, J.~H., {et~al.}, 2016, \apjl,
  824, L1

\bibitem[{{Field}(1959)}]{field1959}
{Field}, G.~B., 1959, \apj, 129, 536

\bibitem[{{Gabasch} {et~al.}(2004){Gabasch}, {Bender}, {Seitz}, {Hopp},
  {Saglia}, {Feulner}, {Snigula}, {Drory}, {Appenzeller}, {Heidt}, {Mehlert},
  {Noll}, {B{\"o}hm}, {J{\"a}ger}, {Ziegler}, \& {Fricke}}]{gabasch2004}
{Gabasch}, A., {Bender}, R., {Seitz}, S., {et~al.}, 2004, \aap, 421, 41

\bibitem[{{Gehrels}(1986)}]{gehrels1986}
{Gehrels}, N., 1986, \apj, 303, 336

\bibitem[{{Greisen} {et~al.}(2009){Greisen}, {Spekkens}, \& {van
  Moorsel}}]{greisen2009}
{Greisen}, E.~W., {Spekkens}, K., \& {van Moorsel}, G.~A., 2009, \aj, 137, 4718

\bibitem[{{Gupta} {et~al.}(2010){Gupta}, {Srianand}, {Bowen}, {York}, \&
  {Wadadekar}}]{gupta2010}
{Gupta}, N., {Srianand}, R., {Bowen}, D.~V., {York}, D.~G., \& {Wadadekar}, Y.,
  2010, \mnras, 408, 849

\bibitem[{{Gupta} {et~al.}(2013){Gupta}, {Srianand}, {Noterdaeme}, {Petitjean},
  \& {Muzahid}}]{gupta2013}
{Gupta}, N., {Srianand}, R., {Noterdaeme}, P., {Petitjean}, P., \& {Muzahid},
  S., 2013, \aap, 558, A84

\bibitem[{{Gupta} {et~al.}(2012){Gupta}, {Srianand}, {Petitjean}, {Bergeron},
  {Noterdaeme}, \& {Muzahid}}]{gupta2012}
{Gupta}, N., {Srianand}, R., {Petitjean}, P., {Bergeron}, J., {Noterdaeme}, P.,
  \& {Muzahid}, S., 2012, \aap, 544, A21

\bibitem[{{Gupta} {et~al.}(2009){Gupta}, {Srianand}, {Petitjean}, {Noterdaeme},
  \& {Saikia}}]{gupta2009}
{Gupta}, N., {Srianand}, R., {Petitjean}, P., {Noterdaeme}, P., \& {Saikia},
  D.~J., 2009, \mnras, 398, 201

\bibitem[{{Haschick} \& {Burke}(1975)}]{haschick1975}
{Haschick}, A.~D. \& {Burke}, B.~F., 1975, \apjl, 200, L137

\bibitem[{{Haynes} {et~al.}(1979){Haynes}, {Giovanelli}, \&
  {Roberts}}]{haynes1979}
{Haynes}, M.~P., {Giovanelli}, R., \& {Roberts}, M.~S., 1979, \apj, 229, 83

\bibitem[{{Heiles} \& {Troland}(2003)}]{heiles2003}
{Heiles}, C. \& {Troland}, T.~H., 2003, \apj, 586, 1067

\bibitem[{{Hwang} \& {Chiou}(2004)}]{hwang2004}
{Hwang}, C.-Y. \& {Chiou}, S.-H., 2004, \apj, 600, 52

\bibitem[{{Ianjamasimanana} {et~al.}(2012){Ianjamasimanana}, {de Blok},
  {Walter}, \& {Heald}}]{ianjamasimanana2012}
{Ianjamasimanana}, R., {de Blok}, W.~J.~G., {Walter}, F., \& {Heald}, G.~H.,
  2012, \aj, 144, 96

\bibitem[{{Ilbert} {et~al.}(2005){Ilbert}, {Tresse}, {Zucca}, {Bardelli},
  {Arnouts}, {Zamorani}, {Pozzetti}, {Bottini}, {Garilli}, {Le Brun}, {Le
  F{\`e}vre}, {Maccagni}, {Picat}, {Scaramella}, {Scodeggio}, {Vettolani},
  {Zanichelli}, {Adami}, {Arnaboldi}, {Bolzonella}, {Cappi}, {Charlot},
  {Contini}, {Foucaud}, {Franzetti}, {Gavignaud}, {Guzzo}, {Iovino},
  {McCracken}, {Marano}, {Marinoni}, {Mathez}, {Mazure}, {Meneux}, {Merighi},
  {Paltani}, {Pello}, {Pollo}, {Radovich}, {Bondi}, {Bongiorno}, {Busarello},
  {Ciliegi}, {Lamareille}, {Mellier}, {Merluzzi}, {Ripepi}, \&
  {Rizzo}}]{ilbert2005}
{Ilbert}, O., {Tresse}, L., {Zucca}, E., {et~al.}, 2005, \aap, 439, 863

\bibitem[{{Jiang} {et~al.}(2011){Jiang}, {Ge}, {Zhou}, {Wang}, \&
  {Wang}}]{jiang2011}
{Jiang}, P., {Ge}, J., {Zhou}, H., {Wang}, J., \& {Wang}, T., 2011, \apj, 732,
  110

\bibitem[{{Jonas}(2009)}]{jonas2009}
{Jonas}, J.~L., 2009, IEEE Proceedings, 97, 1522

\bibitem[{{Kacprzak} {et~al.}(2011){Kacprzak}, {Churchill}, {Barton}, \&
  {Cooke}}]{kacprzak2011}
{Kacprzak}, G.~G., {Churchill}, C.~W., {Barton}, E.~J., \& {Cooke}, J., 2011,
  \apj, 733, 105

\bibitem[{{Kacprzak} {et~al.}(2012){Kacprzak}, {Churchill}, \&
  {Nielsen}}]{kacprzak2012}
{Kacprzak}, G.~G., {Churchill}, C.~W., \& {Nielsen}, N.~M., 2012, \apjl, 760,
  L7

\bibitem[{{Kacprzak} {et~al.}(2008){Kacprzak}, {Churchill}, {Steidel}, \&
  {Murphy}}]{kacprzak2008}
{Kacprzak}, G.~G., {Churchill}, C.~W., {Steidel}, C.~C., \& {Murphy}, M.~T.,
  2008, \aj, 135, 922

\bibitem[{{Kanekar} {et~al.}(2002){Kanekar}, {Athreya}, \&
  {Chengalur}}]{kanekar2002}
{Kanekar}, N., {Athreya}, R.~M., \& {Chengalur}, J.~N., 2002, \aap, 382, 838

\bibitem[{{Kanekar} \& {Chengalur}(2003)}]{kanekar2003}
{Kanekar}, N. \& {Chengalur}, J.~N., 2003, \aap, 399, 857

\bibitem[{{Kanekar} {et~al.}(2014){Kanekar}, {Prochaska}, {Smette}, {Ellison},
  {Ryan-Weber}, {Momjian}, {Briggs}, {Lane}, {Chengalur}, {Delafosse}, {Grave},
  {Jacobsen}, \& {de Bruyn}}]{kanekar2014}
{Kanekar}, N., {Prochaska}, J.~X., {Smette}, A., {et~al.}, 2014, \mnras, 438,
  2131

\bibitem[{{Keeney} {et~al.}(2005){Keeney}, {Momjian}, {Stocke}, {Carilli}, \&
  {Tumlinson}}]{keeney2005}
{Keeney}, B.~A., {Momjian}, E., {Stocke}, J.~T., {Carilli}, C.~L., \&
  {Tumlinson}, J., 2005, \apj, 622, 267

\bibitem[{{Keeney} {et~al.}(2011){Keeney}, {Stocke}, {Danforth}, \&
  {Carilli}}]{keeney2011}
{Keeney}, B.~A., {Stocke}, J.~T., {Danforth}, C.~W., \& {Carilli}, C.~L., 2011,
  \aj, 141, 66

\bibitem[{{Kennicutt}(1998{\natexlab{a}})}]{kennicutt1998a}
{Kennicutt}, Jr., R.~C., 1998{\natexlab{a}}, \araa, 36, 189

\bibitem[{{Kennicutt}(1998{\natexlab{b}})}]{kennicutt1998b}
---, 1998{\natexlab{b}}, \apj, 498, 541

\bibitem[{{Kewley} {et~al.}(2004){Kewley}, {Geller}, \& {Jansen}}]{kewley2004}
{Kewley}, L.~J., {Geller}, M.~J., \& {Jansen}, R.~A., 2004, \aj, 127, 2002

\bibitem[{{Khare} {et~al.}(2012){Khare}, {vanden Berk}, {York}, {Lundgren}, \&
  {Kulkarni}}]{khare2012}
{Khare}, P., {vanden Berk}, D., {York}, D.~G., {Lundgren}, B., \& {Kulkarni},
  V.~P., 2012, \mnras, 419, 1028

\bibitem[{{Krogager} {et~al.}(2012){Krogager}, {Fynbo}, {M{\o}ller}, {Ledoux},
  {Noterdaeme}, {Christensen}, {Milvang-Jensen}, \& {Sparre}}]{krogager2012}
{Krogager}, J.-K., {Fynbo}, J.~P.~U., {M{\o}ller}, P., {Ledoux}, C.,
  {Noterdaeme}, P., {Christensen}, L., {Milvang-Jensen}, B., \& {Sparre}, M.,
  2012, \mnras, 424, L1

\bibitem[{{Kulkarni} \& {Heiles}(1988)}]{kulkarni1988}
{Kulkarni}, S.~R. \& {Heiles}, C., 1988, {Neutral hydrogen and the diffuse
  interstellar medium}, {Kellermann}, K.~I. \& {Verschuur}, G.~L., eds., pp.
  95--153

\bibitem[{{Lah} {et~al.}(2009){Lah}, {Pracy}, {Chengalur}, {Briggs}, {Colless},
  {de Propris}, {Ferris}, {Schmidt}, \& {Tucker}}]{lah2009}
{Lah}, P., {Pracy}, M.~B., {Chengalur}, J.~N., {et~al.}, 2009, \mnras, 399,
  1447

\bibitem[{{Le Brun} {et~al.}(2000){Le Brun}, {Smette}, {Surdej}, \&
  {Claeskens}}]{lebrun2000}
{Le Brun}, V., {Smette}, A., {Surdej}, J., \& {Claeskens}, J.-F., 2000, \aap,
  363, 837

\bibitem[{{Liszt}(2001)}]{liszt2001}
{Liszt}, H., 2001, \aap, 371, 698

\bibitem[{{McKee} \& {Ostriker}(1977)}]{mckee1977}
{McKee}, C.~F. \& {Ostriker}, J.~P., 1977, \apj, 218, 148

\bibitem[{{Mihos} {et~al.}(2012){Mihos}, {Keating}, {Holley-Bockelmann},
  {Pisano}, \& {Kassim}}]{mihos2012}
{Mihos}, J.~C., {Keating}, K.~M., {Holley-Bockelmann}, K., {Pisano}, D.~J., \&
  {Kassim}, N.~E., 2012, \apj, 761, 186

\bibitem[{{Miller} {et~al.}(2009){Miller}, {Hornschemeier}, \&
  {Mobasher}}]{miller2009}
{Miller}, N.~A., {Hornschemeier}, A.~E., \& {Mobasher}, B., 2009, \aj, 137,
  4436

\bibitem[{{Morganti} {et~al.}(2015){Morganti}, {Sadler}, \&
  {Curran}}]{morganti2015}
{Morganti}, R., {Sadler}, E.~M., \& {Curran}, S., 2015, Advancing Astrophysics
  with the Square Kilometre Array (AASKA14), 134

\bibitem[{{Muzahid} {et~al.}(2015){Muzahid}, {Srianand}, \&
  {Charlton}}]{muzahid2015}
{Muzahid}, S., {Srianand}, R., \& {Charlton}, J., 2015, \mnras, 448, 2840

\bibitem[{{Neeleman} {et~al.}(2016){Neeleman}, {Prochaska}, {Ribaudo},
  {Lehner}, {Howk}, {Rafelski}, \& {Kanekar}}]{neeleman2016}
{Neeleman}, M., {Prochaska}, J.~X., {Ribaudo}, J., {Lehner}, N., {Howk}, J.~C.,
  {Rafelski}, M., \& {Kanekar}, N., 2016, \apj, 818, 113

\bibitem[{{Nielsen} {et~al.}(2013){Nielsen}, {Churchill}, {Kacprzak}, \&
  {Murphy}}]{nielsen2013}
{Nielsen}, N.~M., {Churchill}, C.~W., {Kacprzak}, G.~G., \& {Murphy}, M.~T.,
  2013, \apj, 776, 114

\bibitem[{{Nielsen} {et~al.}(2015){Nielsen}, {Churchill}, {Kacprzak}, {Murphy},
  \& {Evans}}]{nielsen2015}
{Nielsen}, N.~M., {Churchill}, C.~W., {Kacprzak}, G.~G., {Murphy}, M.~T., \&
  {Evans}, J.~L., 2015, \apj, 812, 83

\bibitem[{{Noterdaeme} {et~al.}(2009){Noterdaeme}, {Petitjean}, {Ledoux}, \&
  {Srianand}}]{noterdaeme2009}
{Noterdaeme}, P., {Petitjean}, P., {Ledoux}, C., \& {Srianand}, R., 2009, \aap,
  505, 1087

\bibitem[{{Noterdaeme} {et~al.}(2010){Noterdaeme}, {Srianand}, \&
  {Mohan}}]{noterdaeme2010}
{Noterdaeme}, P., {Srianand}, R., \& {Mohan}, V., 2010, \mnras, 403, 906

\bibitem[{{Oosterloo} {et~al.}(2007){Oosterloo}, {Fraternali}, \&
  {Sancisi}}]{oosterloo2007}
{Oosterloo}, T., {Fraternali}, F., \& {Sancisi}, R., 2007, \aj, 134, 1019

\bibitem[{{P{\'e}roux} {et~al.}(2012){P{\'e}roux}, {Bouch{\'e}}, {Kulkarni},
  {York}, \& {Vladilo}}]{peroux2012}
{P{\'e}roux}, C., {Bouch{\'e}}, N., {Kulkarni}, V.~P., {York}, D.~G., \&
  {Vladilo}, G., 2012, \mnras, 419, 3060

\bibitem[{{Poznanski} {et~al.}(2012){Poznanski}, {Prochaska}, \&
  {Bloom}}]{poznanski2012}
{Poznanski}, D., {Prochaska}, J.~X., \& {Bloom}, J.~S., 2012, \mnras, 426, 1465

\bibitem[{{Prochaska} {et~al.}(2011){Prochaska}, {Weiner}, {Chen}, {Mulchaey},
  \& {Cooksey}}]{prochaska2011}
{Prochaska}, J.~X., {Weiner}, B., {Chen}, H.-W., {Mulchaey}, J., \& {Cooksey},
  K., 2011, \apj, 740, 91

\bibitem[{{Rafelski} {et~al.}(2011){Rafelski}, {Wolfe}, \&
  {Chen}}]{rafelski2011}
{Rafelski}, M., {Wolfe}, A.~M., \& {Chen}, H.-W., 2011, \apj, 736, 48

\bibitem[{{Rahmani} {et~al.}(2016){Rahmani}, {P{\'e}roux}, {Turnshek}, {Rao},
  {Quiret}, {Hamilton}, {Kulkarni}, {Monier}, \& {Zafar}}]{rahmani2016}
{Rahmani}, H., {P{\'e}roux}, C., {Turnshek}, D.~A., {et~al.}, 2016, \mnras,
  463, 980

\bibitem[{{Rao} {et~al.}(2011){Rao}, {Belfort-Mihalyi}, {Turnshek}, {Monier},
  {Nestor}, \& {Quider}}]{rao2011}
{Rao}, S.~M., {Belfort-Mihalyi}, M., {Turnshek}, D.~A., {Monier}, E.~M.,
  {Nestor}, D.~B., \& {Quider}, A., 2011, \mnras, 416, 1215

\bibitem[{{Rao} {et~al.}(2006){Rao}, {Turnshek}, \& {Nestor}}]{rao2006}
{Rao}, S.~M., {Turnshek}, D.~A., \& {Nestor}, D.~B., 2006, \apj, 636, 610

\bibitem[{{Reeves} {et~al.}(2015){Reeves}, {Sadler}, {Allison}, {Koribalski},
  {Curran}, \& {Pracy}}]{reeves2015}
{Reeves}, S.~N., {Sadler}, E.~M., {Allison}, J.~R., {Koribalski}, B.~S.,
  {Curran}, S.~J., \& {Pracy}, M.~B., 2015, \mnras, 450, 926

\bibitem[{{Reeves} {et~al.}(2016){Reeves}, {Sadler}, {Allison}, {Koribalski},
  {Curran}, {Pracy}, {Phillips}, {Bignall}, \& {Reynolds}}]{reeves2016}
{Reeves}, S.~N., {Sadler}, E.~M., {Allison}, J.~R., {et~al.}, 2016, \mnras,
  457, 2613

\bibitem[{{Reid} {et~al.}(1999){Reid}, {Kronberg}, \& {Perley}}]{reid1999}
{Reid}, R.~I., {Kronberg}, P.~P., \& {Perley}, R.~A., 1999, \apjs, 124, 285

\bibitem[{{Richter} {et~al.}(2011){Richter}, {Krause}, {Fechner}, {Charlton},
  \& {Murphy}}]{richter2011}
{Richter}, P., {Krause}, F., {Fechner}, C., {Charlton}, J.~C., \& {Murphy},
  M.~T., 2011, \aap, 528, A12

\bibitem[{{Rosenberg} \& {Schneider}(2002)}]{rosenberg2002}
{Rosenberg}, J.~L. \& {Schneider}, S.~E., 2002, \apj, 567, 247

\bibitem[{{Roy} {et~al.}(2006){Roy}, {Chengalur}, \& {Srianand}}]{roy2006}
{Roy}, N., {Chengalur}, J.~N., \& {Srianand}, R., 2006, \mnras, 365, L1

\bibitem[{{Roy} {et~al.}(2012){Roy}, {Minter}, {Goss}, {Brogan}, \&
  {Lazio}}]{roy2012}
{Roy}, N., {Minter}, A.~H., {Goss}, W.~M., {Brogan}, C.~L., \& {Lazio},
  T.~J.~W., 2012, \apj, 749, 144

\bibitem[{{Sancisi} {et~al.}(2008){Sancisi}, {Fraternali}, {Oosterloo}, \& {van
  der Hulst}}]{sancisi2008}
{Sancisi}, R., {Fraternali}, F., {Oosterloo}, T., \& {van der Hulst}, T., 2008,
  \aapr, 15, 189

\bibitem[{{Sardane} {et~al.}(2015){Sardane}, {Turnshek}, \&
  {Rao}}]{sardane2015}
{Sardane}, G.~M., {Turnshek}, D.~A., \& {Rao}, S.~M., 2015, \mnras, 452, 3192

\bibitem[{{Schaye} {et~al.}(2007){Schaye}, {Carswell}, \& {Kim}}]{schaye2007}
{Schaye}, J., {Carswell}, R.~F., \& {Kim}, T.-S., 2007, \mnras, 379, 1169

\bibitem[{{Schneider} \& {Corbelli}(1993)}]{schneider1993}
{Schneider}, S.~E. \& {Corbelli}, E., 1993, \apj, 414, 500

\bibitem[{{Schombert} {et~al.}(1997){Schombert}, {Pildis}, \&
  {Eder}}]{schombert1997}
{Schombert}, J.~M., {Pildis}, R.~A., \& {Eder}, J.~A., 1997, \apjs, 111, 233

\bibitem[{{Snow} \& {McCall}(2006)}]{snow2006}
{Snow}, T.~P. \& {McCall}, B.~J., 2006, \araa, 44, 367

\bibitem[{{Srianand} {et~al.}(2015){Srianand}, {Gupta}, {Momjian}, \&
  {Vivek}}]{srianand2015}
{Srianand}, R., {Gupta}, N., {Momjian}, E., \& {Vivek}, M., 2015, \mnras, 451,
  917

\bibitem[{{Srianand} {et~al.}(2012){Srianand}, {Gupta}, {Petitjean},
  {Noterdaeme}, {Ledoux}, {Salter}, \& {Saikia}}]{srianand2012}
{Srianand}, R., {Gupta}, N., {Petitjean}, P., {Noterdaeme}, P., {Ledoux}, C.,
  {Salter}, C.~J., \& {Saikia}, D.~J., 2012, \mnras, 421, 651

\bibitem[{{Srianand} {et~al.}(2008){Srianand}, {Gupta}, {Petitjean},
  {Noterdaeme}, \& {Saikia}}]{srianand2008}
{Srianand}, R., {Gupta}, N., {Petitjean}, P., {Noterdaeme}, P., \& {Saikia},
  D.~J., 2008, \mnras, 391, L69

\bibitem[{{Srianand} {et~al.}(2013){Srianand}, {Gupta}, {Rahmani}, {Momjian},
  {Petitjean}, \& {Noterdaeme}}]{srianand2013}
{Srianand}, R., {Gupta}, N., {Rahmani}, H., {Momjian}, E., {Petitjean}, P., \&
  {Noterdaeme}, P., 2013, \mnras, 428, 2198

\bibitem[{{Stanghellini} {et~al.}(1998){Stanghellini}, {O'Dea}, {Dallacasa},
  {Baum}, {Fanti}, \& {Fanti}}]{stanghellini1998}
{Stanghellini}, C., {O'Dea}, C.~P., {Dallacasa}, D., {Baum}, S.~A., {Fanti},
  R., \& {Fanti}, C., 1998, \aaps, 131, 303

\bibitem[{{Stanimirovi{\'c}} {et~al.}(2010){Stanimirovi{\'c}}, {Weisberg},
  {Pei}, {Tuttle}, \& {Green}}]{stanimirovic2010}
{Stanimirovi{\'c}}, S., {Weisberg}, J.~M., {Pei}, Z., {Tuttle}, K., \& {Green},
  J.~T., 2010, \apj, 720, 415

\bibitem[{{Steidel}(1995)}]{steidel1995}
{Steidel}, C.~C., 1995, in QSO Absorption Lines, {Meylan}, G., ed., p. 139

\bibitem[{{Stocke} {et~al.}(2010){Stocke}, {Keeney}, \&
  {Danforth}}]{stocke2010}
{Stocke}, J.~T., {Keeney}, B.~A., \& {Danforth}, C.~W., 2010, \pasa, 27, 256

\bibitem[{{Stocke} {et~al.}(2013){Stocke}, {Keeney}, {Danforth}, {Shull},
  {Froning}, {Green}, {Penton}, \& {Savage}}]{stocke2013}
{Stocke}, J.~T., {Keeney}, B.~A., {Danforth}, C.~W., {Shull}, J.~M., {Froning},
  C.~S., {Green}, J.~C., {Penton}, S.~V., \& {Savage}, B.~D., 2013, \apj, 763,
  148

\bibitem[{{Straka} {et~al.}(2013){Straka}, {Whichard}, {Kulkarni}, {Bishof},
  {Bowen}, {Khare}, \& {York}}]{straka2013}
{Straka}, L.~A., {Whichard}, Z.~L., {Kulkarni}, V.~P., {Bishof}, M., {Bowen},
  D., {Khare}, P., \& {York}, D.~G., 2013, \mnras, 436, 3200

\bibitem[{{Tamburro} {et~al.}(2009){Tamburro}, {Rix}, {Leroy}, {Mac Low},
  {Walter}, {Kennicutt}, {Brinks}, \& {de Blok}}]{tamburro2009}
{Tamburro}, D., {Rix}, H.-W., {Leroy}, A.~K., {Mac Low}, M.-M., {Walter}, F.,
  {Kennicutt}, R.~C., {Brinks}, E., \& {de Blok}, W.~J.~G., 2009, \aj, 137,
  4424

\bibitem[{{Toribio} {et~al.}(2011){Toribio}, {Solanes}, {Giovanelli}, {Haynes},
  \& {Martin}}]{toribio2011}
{Toribio}, M.~C., {Solanes}, J.~M., {Giovanelli}, R., {Haynes}, M.~P., \&
  {Martin}, A.~M., 2011, \apj, 732, 93

\bibitem[{{Tripp} \& {Bowen}(2005)}]{tripp2005}
{Tripp}, T.~M. \& {Bowen}, D.~V., 2005, in IAU Colloq. 199: Probing Galaxies
  through Quasar Absorption Lines, {Williams}, P., {Shu}, C.-G., \& {Menard},
  B., eds., pp. 5--23

\bibitem[{{Trujillo} {et~al.}(2007){Trujillo}, {Conselice}, {Bundy}, {Cooper},
  {Eisenhardt}, \& {Ellis}}]{trujillo2007}
{Trujillo}, I., {Conselice}, C.~J., {Bundy}, K., {Cooper}, M.~C., {Eisenhardt},
  P., \& {Ellis}, R.~S., 2007, \mnras, 382, 109

\bibitem[{{Tumlinson} {et~al.}(2013){Tumlinson}, {Thom}, {Werk}, {Prochaska},
  {Tripp}, {Katz}, {Dav{\'e}}, {Oppenheimer}, {Meiring}, {Ford}, {O'Meara},
  {Peeples}, {Sembach}, \& {Weinberg}}]{tumlinson2013}
{Tumlinson}, J., {Thom}, C., {Werk}, J.~K., {et~al.}, 2013, \apj, 777, 59

\bibitem[{{Turnshek} {et~al.}(2001){Turnshek}, {Rao}, {Nestor}, {Lane},
  {Monier}, {Bergeron}, \& {Smette}}]{turnshek2001}
{Turnshek}, D.~A., {Rao}, S., {Nestor}, D., {Lane}, W., {Monier}, E.,
  {Bergeron}, J., \& {Smette}, A., 2001, \apj, 553, 288

\bibitem[{{van der Hulst} {et~al.}(2001){van der Hulst}, {van Albada}, \&
  {Sancisi}}]{vanderhulst2001}
{van der Hulst}, J.~M., {van Albada}, T.~S., \& {Sancisi}, R., 2001, in
  Astronomical Society of the Pacific Conference Series, Vol. 240, Gas and
  Galaxy Evolution, {Hibbard}, J.~E., {Rupen}, M., \& {van Gorkom}, J.~H.,
  eds., p. 451

\bibitem[{{van der Wel} {et~al.}(2014){van der Wel}, {Franx}, {van Dokkum},
  {Skelton}, {Momcheva}, {Whitaker}, {Brammer}, {Bell}, {Rix}, {Wuyts},
  {Ferguson}, {Holden}, {Barro}, {Koekemoer}, {Chang}, {McGrath},
  {H{\"a}ussler}, {Dekel}, {Behroozi}, {Fumagalli}, {Leja}, {Lundgren},
  {Maseda}, {Nelson}, {Wake}, {Patel}, {Labb{\'e}}, {Faber}, {Grogin}, \&
  {Kocevski}}]{vanderwel2014}
{van der Wel}, A., {Franx}, M., {van Dokkum}, P.~G., {et~al.}, 2014, \apj, 788,
  28

\bibitem[{{van Dishoeck} \& {Black}(1989)}]{vandishoeck1989}
{van Dishoeck}, E.~F. \& {Black}, J.~H., 1989, \apj, 340, 273

\bibitem[{{Vanden Berk} {et~al.}(2001){Vanden Berk}, {Richards}, {Bauer},
  {Strauss}, {Schneider}, {Heckman}, {York}, {Hall}, {Fan}, {Knapp},
  {Anderson}, {Annis}, {Bahcall}, {Bernardi}, {Briggs}, {Brinkmann}, {Brunner},
  {Burles}, {Carey}, {Castander}, {Connolly}, {Crocker}, {Csabai}, {Doi},
  {Finkbeiner}, {Friedman}, {Frieman}, {Fukugita}, {Gunn}, {Hennessy},
  {Ivezi{\'c}}, {Kent}, {Kunszt}, {Lamb}, {Leger}, {Long}, {Loveday}, {Lupton},
  {Meiksin}, {Merelli}, {Munn}, {Newberg}, {Newcomb}, {Nichol}, {Owen}, {Pier},
  {Pope}, {Rockosi}, {Schlegel}, {Siegmund}, {Smee}, {Snir}, {Stoughton},
  {Stubbs}, {SubbaRao}, {Szalay}, {Szokoly}, {Tremonti}, {Uomoto}, {Waddell},
  {Yanny}, \& {Zheng}}]{vandenberk2001}
{Vanden Berk}, D.~E., {Richards}, G.~T., {Bauer}, A., {et~al.}, 2001, \aj, 122,
  549

\bibitem[{{Verheijen} {et~al.}(2007){Verheijen}, {van Gorkom}, {Szomoru},
  {Dwarakanath}, {Poggianti}, \& {Schiminovich}}]{verheijen2007}
{Verheijen}, M., {van Gorkom}, J.~H., {Szomoru}, A., {Dwarakanath}, K.~S.,
  {Poggianti}, B.~M., \& {Schiminovich}, D., 2007, \apjl, 668, L9

\bibitem[{{Verheijen} {et~al.}(2008){Verheijen}, {Oosterloo}, {van Cappellen},
  {Bakker}, {Ivashina}, \& {van der Hulst}}]{verheijen2008}
{Verheijen}, M.~A.~W., {Oosterloo}, T.~A., {van Cappellen}, W.~A., {Bakker},
  L., {Ivashina}, M.~V., \& {van der Hulst}, J.~M., 2008, in American Institute
  of Physics Conference Series, Vol. 1035, The Evolution of Galaxies Through
  the Neutral Hydrogen Window, {Minchin}, R. \& {Momjian}, E., eds., pp.
  265--271

\bibitem[{{Wakker} \& {Mathis}(2000)}]{wakker2000}
{Wakker}, B.~P. \& {Mathis}, J.~S., 2000, \apjl, 544, L107

\bibitem[{{Walter} {et~al.}(2008){Walter}, {Brinks}, {de Blok}, {Bigiel},
  {Kennicutt}, {Thornley}, \& {Leroy}}]{walter2008}
{Walter}, F., {Brinks}, E., {de Blok}, W.~J.~G., {Bigiel}, F., {Kennicutt},
  Jr., R.~C., {Thornley}, M.~D., \& {Leroy}, A., 2008, \aj, 136, 2563

\bibitem[{{Welty} {et~al.}(1996){Welty}, {Morton}, \& {Hobbs}}]{welty1996}
{Welty}, D.~E., {Morton}, D.~C., \& {Hobbs}, L.~M., 1996, \apjs, 106, 533

\bibitem[{{White} {et~al.}(1997){White}, {Becker}, {Helfand}, \&
  {Gregg}}]{white1997}
{White}, R.~L., {Becker}, R.~H., {Helfand}, D.~J., \& {Gregg}, M.~D., 1997,
  \apj, 475, 479

\bibitem[{{Wild} {et~al.}(2006){Wild}, {Hewett}, \& {Pettini}}]{wild2006}
{Wild}, V., {Hewett}, P.~C., \& {Pettini}, M., 2006, \mnras, 367, 211

\bibitem[{{Wolfe} {et~al.}(2005){Wolfe}, {Gawiser}, \& {Prochaska}}]{wolfe2005}
{Wolfe}, A.~M., {Gawiser}, E., \& {Prochaska}, J.~X., 2005, \araa, 43, 861

\bibitem[{{Wolfe} {et~al.}(2003){Wolfe}, {Prochaska}, \& {Gawiser}}]{wolfe2003}
{Wolfe}, A.~M., {Prochaska}, J.~X., \& {Gawiser}, E., 2003, \apj, 593, 215

\bibitem[{{Wolfire} {et~al.}(1995){Wolfire}, {Hollenbach}, {McKee}, {Tielens},
  \& {Bakes}}]{wolfire1995a}
{Wolfire}, M.~G., {Hollenbach}, D., {McKee}, C.~F., {Tielens}, A.~G.~G.~M., \&
  {Bakes}, E.~L.~O., 1995, \apj, 443, 152

\bibitem[{{Womble}(1993)}]{womble1993}
{Womble}, D.~S., 1993, \pasp, 105, 1043

\bibitem[{{Wu} {et~al.}(2012){Wu}, {Hao}, {Jia}, {Zhang}, \& {Peng}}]{wu2012}
{Wu}, X.-B., {Hao}, G., {Jia}, Z., {Zhang}, Y., \& {Peng}, N., 2012, \aj, 144,
  49

\bibitem[{{Wu} {et~al.}(2015){Wu}, {Haynes}, {Giovanelli}, {Zhu}, \&
  {Chen}}]{wu2015}
{Wu}, Z., {Haynes}, M.~P., {Giovanelli}, R., {Zhu}, M., \& {Chen}, R., 2015,
  \caa, 39, 466

\bibitem[{{York} {et~al.}(2000){York}, {Adelman}, {Anderson}, {Anderson},
  {Annis}, {Bahcall}, {Bakken}, {Barkhouser}, {Bastian}, {Berman}, {Boroski},
  {Bracker}, {Briegel}, {Briggs}, {Brinkmann}, {Brunner}, {Burles}, {Carey},
  {Carr}, {Castander}, {Chen}, {Colestock}, {Connolly}, {Crocker}, {Csabai},
  {Czarapata}, {Davis}, {Doi}, {Dombeck}, {Eisenstein}, {Ellman}, {Elms},
  {Evans}, {Fan}, {Federwitz}, {Fiscelli}, {Friedman}, {Frieman}, {Fukugita},
  {Gillespie}, {Gunn}, {Gurbani}, {de Haas}, {Haldeman}, {Harris}, {Hayes},
  {Heckman}, {Hennessy}, {Hindsley}, {Holm}, {Holmgren}, {Huang}, {Hull},
  {Husby}, {Ichikawa}, {Ichikawa}, {Ivezi{\'c}}, {Kent}, {Kim}, {Kinney},
  {Klaene}, {Kleinman}, {Kleinman}, {Knapp}, {Korienek}, {Kron}, {Kunszt},
  {Lamb}, {Lee}, {Leger}, {Limmongkol}, {Lindenmeyer}, {Long}, {Loomis},
  {Loveday}, {Lucinio}, {Lupton}, {MacKinnon}, {Mannery}, {Mantsch}, {Margon},
  {McGehee}, {McKay}, {Meiksin}, {Merelli}, {Monet}, {Munn}, {Narayanan},
  {Nash}, {Neilsen}, {Neswold}, {Newberg}, {Nichol}, {Nicinski}, {Nonino},
  {Okada}, {Okamura}, {Ostriker}, {Owen}, {Pauls}, {Peoples}, {Peterson},
  {Petravick}, {Pier}, {Pope}, {Pordes}, {Prosapio}, {Rechenmacher}, {Quinn},
  {Richards}, {Richmond}, {Rivetta}, {Rockosi}, {Ruthmansdorfer}, {Sandford},
  {Schlegel}, {Schneider}, {Sekiguchi}, {Sergey}, {Shimasaku}, {Siegmund},
  {Smee}, {Smith}, {Snedden}, {Stone}, {Stoughton}, {Strauss}, {Stubbs},
  {SubbaRao}, {Szalay}, {Szapudi}, {Szokoly}, {Thakar}, {Tremonti}, {Tucker},
  {Uomoto}, {Vanden Berk}, {Vogeley}, {Waddell}, {Wang}, {Watanabe},
  {Weinberg}, {Yanny}, {Yasuda}, \& {SDSS Collaboration}}]{york2000}
{York}, D.~G., {Adelman}, J., {Anderson}, Jr., J.~E., {et~al.}, 2000, \aj, 120,
  1579

\bibitem[{{York} {et~al.}(2006){York}, {Khare}, {Vanden Berk}, {Kulkarni},
  {Crotts}, {Lauroesch}, {Richards}, {Schneider}, {Welty}, {Alsayyad}, {Kumar},
  {Lundgren}, {Shanidze}, {Smith}, {Vanlandingham}, {Baugher}, {Hall},
  {Jenkins}, {Menard}, {Rao}, {Tumlinson}, {Turnshek}, {Yip}, \&
  {Brinkmann}}]{york2006}
{York}, D.~G., {Khare}, P., {Vanden Berk}, D., {et~al.}, 2006, \mnras, 367, 945

\bibitem[{{York} {et~al.}(2012){York}, {Straka}, {Bishof}, {Kuttruff}, {Bowen},
  {Kulkarni}, {Subbarao}, {Richards}, {Vanden Berk}, {Hall}, {Heckman},
  {Khare}, {Quashnock}, {Ghering}, \& {Johnson}}]{york2012}
{York}, D.~G., {Straka}, L.~A., {Bishof}, M., {et~al.}, 2012, \mnras, 423, 3692

\bibitem[{{Zwaan} {et~al.}(2015){Zwaan}, {Liske}, {P{\'e}roux}, {Murphy},
  {Bouch{\'e}}, {Curran}, \& {Biggs}}]{zwaan2015}
{Zwaan}, M.~A., {Liske}, J., {P{\'e}roux}, C., {Murphy}, M.~T., {Bouch{\'e}},
  N., {Curran}, S.~J., \& {Biggs}, A.~D., 2015, \mnras, 453, 1268

\bibitem[{{Zwaan} {et~al.}(2005){Zwaan}, {van der Hulst}, {Briggs},
  {Verheijen}, \& {Ryan-Weber}}]{zwaan2005}
{Zwaan}, M.~A., {van der Hulst}, J.~M., {Briggs}, F.~H., {Verheijen}, M.~A.~W.,
  \& {Ryan-Weber}, E.~V., 2005, \mnras, 364, 1467

\bibitem[{{Zwaan} {et~al.}(2001){Zwaan}, {van Dokkum}, \&
  {Verheijen}}]{zwaan2001}
{Zwaan}, M.~A., {van Dokkum}, P.~G., \& {Verheijen}, M.~A.~W., 2001, Science,
  293, 1800

\end{thebibliography}
%
%=========================== APPENDIX ======================================================================================
%
\begin{appendices}
\appendix
\setcounter{table}{0}
\renewcommand{\table}{A\arabic{table}}
\setcounter{figure}{0}
\renewcommand{\figure}{A\arabic{figure}}
\section{Sample Properties}
\label{appendix_sample}
Properties of the QGPs observed by us as described in Section~\ref{sec_sampleprop} are provided here.
\begin{landscape}
%\begin{table} 
%\caption{Details of the QGPs observed by us.} 
{\bf Table A1.} Details of the QGPs observed by us. \newline \newline
\centering
\begin{tabular}{ccccccccccccccc}
\hline
QGP & Radio  & \zrad\ & Type & Mor. & Galaxy & \zgal\ & \zgal\ & $g-r$ & log~$M_*$   & log~$L_B$   & $b$   & $i$           & $\phi$        \\
    & Source &        &      &      &        &        & Source &       & ($M_\odot$) & ($L_\odot$) & (kpc) & (\textdegree) & (\textdegree) \\
(1) & (2)    & (3)    & (4)  & (5)  & (6)    & (7)    & (8)    & (9)   & (10)        & (11)        & (12)  & (13)          & (14)          \\
\hline
\multicolumn{14}{c}{\it Primary Sample} \\
\hline
J0041$-$0143N   & J004126.11$-$014305.6 & 1.679        & Q & R & J004121.53$-$014257.0 & 0.01769 & 1 & 0.58 & 10.5 & 10.4 &   25.0 & 15 &--- \\ 
J0041$-$0143C   & J004126.01$-$014314.6 &              &   & R &                       &         &   &      &      &      &   25.0 &    &--- \\ 
J0041$-$0143S   & J004125.98$-$014324.6 &              &   & R &                       &         &   &      &      &      &   26.0 &    &--- \\ 
J0043$+$0024    & J004332.71$+$002459.9 & 1.127        & Q & C & J004333.64$+$002516.2 & 0.07975 & 2 & 0.39 &  9.6 &  9.9 &   32.3 & 27 &--- \\
J0134$+$0003    & J013412.70$+$000345.2 & 0.879        & Q & C & J013411.28$+$000327.5 & 0.05198 & 2 & 0.44 &  9.8 &  9.9 &   28.1 & 58 & 52 \\
J0802$+$1601    & J080240.89$+$160104.6 & 1.784        & Q & C & J080241.68$+$160035.2 & 0.01565 & 2 & 0.43 &  8.8 &  9.0 &   10.1 & 34 & 45 \\
J0817$+$1958    & J081705.49$+$195842.8 & 5.154        & Q & C & J081707.13$+$195907.9 & 0.04363 & 2 & 0.74 & 10.8 & 10.4 &   29.3 & 38 & 43 \\
J0821$+$5031    & J082153.82$+$503120.4 & 2.123        & Q & C & J082153.74$+$503125.7 & 0.1835  & 3 & 0.23 &  9.4 & 10.1 &   16.5 & 38 & 28 \\
J0849$+$5108\_1 & J084957.97$+$510829.0 & 0.584        & Q & C & J084957.49$+$510842.3 & 0.07342 & 2 & 0.46 & 10.5 & 10.6 &   19.6 & 46 & 16 \\
J0849$+$5108\_2 & J084957.97$+$510829.0 & 0.584        & Q & C & J084958.10$+$510826.6 & 0.3120  & 3 & 1.65 & 10.4 &  9.7 &   12.3 & 0  &--- \\
J0856$+$1020    & J085624.91$+$102017.0 & 3.701        & Q & C & J085625.41$+$102020.1 & 0.1043  & 4 & 0.45 & 10.1 & 10.2 &   15.3 & 36 & 51 \\
J0911$+$1958    & J091133.45$+$195814.1 & 1.635        & Q & C & J091134.38$+$195830.2 & 0.02293 & 5 & 0.39 &  8.7 &  9.0 &    9.6 & 58 & 84 \\
J1015$+$1637    & J101514.63$+$163740.4 & $\ge$0.5$^a$ & U & C & J101514.22$+$163758.3 & 0.05436 & 2 & 0.43 &  9.5 &  9.8 &   19.9 & 58 & 40 \\
J1110$+$0321E   & J111022.59$+$032133.5 & 0.965        & Q & R & J111025.09$+$032138.8 & 0.02996 & 2 & 0.49 &  9.7 &  9.8 &   15.4 & 39 & 47 \\
J1110$+$0321W   & J111022.97$+$032130.5 &              &   & R &                       &         &   &      &      &      &   22.7 &    & 53 \\
J1126$+$3802    & J112613.04$+$380216.4 & 0.515$^b$    & U & C & J112612.57$+$380228.6 & 0.06190 & 2 & 0.57 & 10.1 & 10.0 &   16.0 & 60 & 70 \\
J1133$+$0015\_1 & J113303.02$+$001549.0 & 1.171        & Q & C & J113302.01$+$001541.7 & 0.0762  & 4 & 0.49 & 10.0 & 10.0 &   24.3 & 28 &--- \\
J1133$+$0015\_2 & J113303.02$+$001549.0 & 1.171        & Q & C & J113303.96$+$001538.2 & 0.07584 & 2 & 0.76 & 10.6 & 10.2 &   25.6 & 60 & 64 \\
J1156$+$3412    & J115659.62$+$341216.9 & $\ge$0.5$^a$ & G & R & J115659.59$+$341221.7 & 0.13797 & 2 & 0.59 & 10.6 & 10.5 &   11.7 & 46 & 27 \\
J1209$+$4119    & J120922.78$+$411941.3 & 1.364        & Q & C & J120923.25$+$411949.3 & 0.17098 & 2 & 0.73 & 10.7 & 10.4 &   27.9 & 53 & 66 \\
J1228$+$3706    & J122847.42$+$370612.0 & 1.517        & Q & C & J122847.72$+$370606.9 & 0.13837 & 2 & 0.64 & 10.6 & 10.4 &   15.2 & 26 &--- \\
J1241$+$6332    & J124157.54$+$633241.6 & 2.613        & Q & C & J124157.26$+$633237.6 & 0.1430  & 6 & 0.37 & 10.3 & 10.6 &   11.1 & 66 & 19 \\
J1243$+$4043    & J124355.78$+$404358.4 & 1.520        & Q & C & J124357.17$+$404346.1 & 0.01693 & 2 & 0.47 &  9.6 &  9.6 &    6.9 & 77 & 26 \\
J1243$+$1622N   & J124357.79$+$162246.3 & 0.555        & Q & R & J124342.64$+$162335.9 & 0.00263 & 7 & 0.77 &  9.8 &  9.4 &   12.2 & 47 & 25 \\
J1243$+$1622C   & J124357.79$+$162252.3 &              &   & R &                       &         &   &      &      &      &   12.1 &    & 24 \\
J1243$+$1622S   & J124357.51$+$162301.3 &              &   & R &                       &         &   &      &      &      &   11.8 &    & 22 \\
J1254$+$0546    & J125422.19$+$054619.8 & 2.252        & Q & C & J125422.21$+$054610.4 & 0.0246  & 4 & 0.21 &  7.8 &  8.5 &    4.7 & 69 &  6 \\
J1300$+$2830    & J130028.53$+$283010.1 & 0.647        & Q & C & J130028.66$+$283012.1 & 0.2229  & 6 & 0.47 &  9.8 &  9.9 &    9.4 & 71 & 30 \\
J1300$-$0121    & J130050.09$-$012146.2 & $\ge$0.5$^a$ & U & R & J130049.15$-$012136.4 & 0.02275 & 2 & 0.29 &  8.3 &  9.1 &    7.9 & 70 & 60 \\
J1345$+$0347    & J134528.74$+$034719.7 & 0.825        & Q & C & J134528.42$+$034723.0 & 0.2622  & 4 & 0.77 & 11.0 & 10.5 &   23.5 & 70 & 25 \\
J1359$+$2935    & J135907.77$+$293531.8 & 0.225$^b$    & U & R & J135908.47$+$293541.5 & 0.14607 & 2 & 0.76 & 10.9 & 10.5 &   34.0 & 36 & 85 \\
J1422$+$0540    & J142253.22$+$054013.4 & 0.618$^b$    & G & C & J142253.99$+$053947.9 & 0.05422 & 2 & 0.79 & 10.4 &  9.9 &   29.5 & 70 & 55 \\
J1438$+$1758    & J143806.79$+$175805.4 & 2.031        & Q & C & J143806.76$+$175802.5 & 0.1468  & 6 & 0.34 &  9.0 &  9.3 &    7.5 & 76 &  6 \\
J1443$+$0214    & J144304.53$+$021419.3 & 1.820        & Q & C & J144304.53$+$021419.3 & 0.3714  & 6 & ---  & ---  &  --- & $<$5.1 &--- &--- \\	
J1457$+$0519    & J145722.53$+$051921.8 & 3.174        & Q & C & J145722.40$+$051918.8 & 0.2198  & 2 & 0.29 &  9.4 & 10.1 &   12.7 & 29 &--- \\
J1525$+$4201    & J152523.55$+$420117.0 & 1.195        & Q & R & J152522.99$+$420131.7 & 0.09094 & 4 & 0.68 & 10.7 & 10.4 &   27.1 & 38 & 60 \\
J1539$+$0534    & J153905.20$+$053438.3 & 1.509        & Q & R & J153905.16$+$053416.5 & 0.02594 & 2 & 0.70 & 10.7 & 10.4 &   11.4 & 18 &--- \\	
J1551$+$0713    & J155121.13$+$071357.7 & 0.675        & Q & C & J155121.06$+$071352.4 & 0.1007  & 4 & 0.79 & 10.8 & 10.3 &   10.0 & 24 &--- \\
\hline
\end{tabular}
\label{tab:sampleprop1}
%\end{table}
\end{landscape}
\begin{landscape}
%\setcounter{table}{0}
%\begin{table*} 
%\caption{Continued from previous page.}
{\bf Table A1.} Continued from previous page. \newline \newline
\centering
\begin{tabular}{ccccccccccccccc}
\hline 
QGP & Radio  & \zrad\ & Type & Mor. & Galaxy & \zgal\ & \zgal\ & $g-r$ & log~$M_*$   & log~$L_B$   & $b$   & $i$           & $\phi$        \\
    & Source &        &      &      &        &        & Source &       & ($M_\odot$) & ($L_\odot$) & (kpc) & (\textdegree) & (\textdegree) \\
(1) & (2)    & (3)    & (4)  & (5)  & (6)    & (7)    & (8)    & (9)   & (10)        & (11)        & (12)  & (13)          & (14)          \\
\hline                                                                                     
J1606$+$2717    & J160658.29$+$271705.5 & 0.933        & Q & R & J160659.13$+$271642.6 & 0.04620 & 2 & 0.35 &  9.3 &  9.7 &   23.1 & 66 & 71 \\
J1634$+$3900    & J163402.95$+$390000.5 & 1.083        & Q & C & J163403.30$+$390001.0 & 0.36620 & 8 & 0.64 & 10.2 & 10.0 &   20.9 & 49 & 69 \\
J1639$+$1127    & J163956.35$+$112758.7 & 0.994        & Q & C & J163956.37$+$112802.1 & 0.0790  & 6 & 0.49 &  9.8 &  9.9 &    5.1 & 65 &  3 \\
J1643$+$3547    & J164315.58$+$354729.7 & 0.224$^b$    & G & R & J164315.10$+$354701.6 & 0.03121 & 2 & 0.65 &  9.9 &  9.7 &   17.9 & 55 & 23 \\
J1648$+$2429    & J164802.37$+$242916.9 & 0.791        & Q & C & J164804.38$+$242935.3 & 0.03583 & 2 & 0.75 & 10.0 &  9.6 &   23.5 & 38 & 46 \\
J1748$+$7005    & J174832.84$+$700550.8 & 0.770        & Q & C & J174926.43$+$700839.7 & 0.00008 & 7 & ---  & ---  &  9.2 &   10.3 & 70 & 44 \\
J2240$-$0836    & J224013.41$-$083628.7 & 0.581        & Q & C & J224013.49$-$083633.6 & 0.2111  & 4 & 0.84 & 10.8 & 10.3 &   17.4 & 62 & 84 \\
\hline
\multicolumn{14}{c}{\it Supplementary Sample} \\                                                                                                
\hline                                                                                                                                            
J0820$+$0826    & J082015.85$+$082612.6 & 0.686        & Q & C & J082015.88$+$082610.5 & 0.07750 & 9 & 1.63 &  8.5 &  7.5 &    3.1 &--- &--- \\ 
J0919$+$3400E   & J091917.07$+$340049.5 & 0.108$^b$    & U & R & J091917.45$+$340029.4 & 0.02376 & 2 & 0.73 & 10.6 & 10.2 &    9.9 & 60 & 45 \\
J0919$+$3400W   & J091914.80$+$340051.4 &              &   & R &                       &         &   &      &      &      &   19.0 &    & 90 \\
J0947$+$1113    & J094745.85$+$111354.0 & 1.774        & U & C & J094745.85$+$111354.0 & 0.11240 & 9 & ---  & ---  & ---  &    4.1 &--- &--- \\
J0954$+$3733    & J095433.96$+$373336.3 &   ---        & U & C & J095432.89$+$373333.1 & 0.07282 & 2 & 0.45 & 10.1 & 10.2 &   18.2 & 50 &  1 \\
J0956$+$1614N   & J095604.27$+$161420.5 &   ---        & U & R & J095604.09$+$161427.1 & 0.03346 & 2 & 0.49 &  9.3 &  9.4 &    4.7 & 37 & 37 \\
J0956$+$1614S   & J095604.20$+$161417.0 &   ---        &   & R &                       &         &   &      &      &      &    6.8 &    & 23 \\
J1050$+$0002    & J105053.93$+$000213.6 & 0.662        & Q & C & J105053.93$+$000213.6 & 0.23380 & 9 & ---  & ---  & ---  &    7.4 &--- &--- \\
J1359$+$2708    & J135947.90$+$270828.3 & 1.832        & Q & C & J135947.90$+$270828.3 & 0.04320 & 9 & ---  & ---  & ---  &    1.7 &--- &--- \\
J1438$+$0920    & J143809.90$+$092016.9 &   ---        & U & R & J143810.20$+$092009.8 & 0.03029 & 2 & 0.81 & 11.1 & 10.6 &    5.1 & 29 &--- \\
J1624$+$2328    & J162434.18$+$232859.9 & 0.230        & Q & C & J162433.76$+$232912.3 & 0.09382 & 2 & 0.50 & 10.3 & 10.3 &   23.8 & 29 &--- \\
J2242$-$0019    & J224247.89$-$001944.9 & 0.405        & Q & C & J224247.65$-$001918.7 & 0.05815 & 2 & 0.79 & 10.7 & 10.3 &   29.8 & 75 & 46 \\
\hline                                                                                                                                              
\multicolumn{14}{c}{\it Miscellaneous Sample} \\                                                                                                    
\hline                                                                                                                                              
J0846$+$0704    & J084600.36$+$070424.6 & 0.342        & Q & R & J084600.52$+$070428.8 & 0.34120 & 2 & 0.79 & 10.4 &  9.9 &   23.5 &  9 &--- \\
J0942$+$0623    & J094221.98$+$062335.2 & 0.123        & G & C & J094221.94$+$062337.1 & 0.12390 & 2 & 0.54 & 10.5 & 10.4 &    4.4 & 53 & 41 \\
J1214$+$2931    & J121417.80$+$293143.4 & 0.063        & Q & R & J121418.25$+$293146.7 & 0.06326 & 2 & 0.78 & 11.0 & 10.6 &    8.2 & 47 &  7 \\
J1643$+$2419    & J164330.03$+$241951.8 & 0.087        & G & R & J164331.92$+$241939.6 & 0.08723 & 2 & 0.53 & 10.9 & 10.9 &   37.1 & 29 &--- \\
J2054$+$0041    & J205449.61$+$004153.0 & 0.284$^b$    & G & C & J205449.64$+$004149.8 & 0.20153 & 2 & 0.53 & 10.2 & 10.1 &   10.7 & 36 & 21 \\
\hline
\end{tabular}
\label{tab:sampleprop2}
\begin{flushleft}
Column 1: QGP name; in case of multiple radio sightlines for the same QGP, the name is suffixed with N, S, E, W or C (denoting north, south, east, west and centre, respectively) 
to indicate the radio sightline; in case of two galaxies around the same radio sightline, the name is suffixed with \_1 and \_2 to differentiate the two QGPs. 
Column 2: Radio source name. Column 3: Redshift of radio source. Column 4: Type of radio source $-$ quasar (Q), radio galaxy (G) or unknown (U). 
Column 5: Morphology of radio source at arcsecond-scale $-$ compact (C) or resolved (R). Column 6: Galaxy name. Column 7: Galaxy redshift. 
Column 8: Source of galaxy redshift $-$ 1: \citet{jonas2009}; 2: SDSS; 3: APO observations \citep{gupta2009}; 4: SALT observations (this work);
5: \citet{schombert1997}; 6: identified by us as GOTOQ; 7: \citet{epinat2008}; 8: \citet{womble1993}; 9: \caii\ absorption identified by us in SDSS quasar spectra.
Column 9: Rest-frame SDSS $g-r$ colour of galaxy. Column 10: Log of stellar mass ($M_\odot$) of galaxy from {\it kcorrect} \citep{blanton2007a}. 
Column 11: Log of $B$-band luminosity ($L_\odot$). Column 12: Projected separation or impact parameter (kpc) between radio sightline and centre of galaxy. 
Column 13: Galaxy inclination. Column 14: Orientation of radio sightline with respect to galaxy's major axis. \\
$^a$ based on SDSS and WISE colours (see text for details)  $^b$ photometric redshift 
\end{flushleft}
%\end{table*}
\end{landscape}
\begin{landscape}
\section{Literature Sample}
\label{appendix_literature}
The details of the QGPs from literature that are considered in this work (as described in Section~\ref{sec_litsample}) are provided here. \newline \newline \newline \newline
%
%\begin{table}
%\caption{Details of the QGPs searched for \hi\ \21\ absorption in the literature that are included in this work.}
{\bf Table B1.} Details of the QGPs searched for \hi\ \21\ absorption in the literature that are included in this work. \newline \newline
\centering
\begin{tabular}{ccccccccccccccc}
\hline
Radio Source & \zrad\ & Type & Mor. & Galaxy & \zgal\ & $g-r$ & log~$M_*$   & log~$L_B$   & $b$   & $i$           & $\phi$        & \taudvl\ & \taudv\ & Ref. \\
             &        &      &      &        &        &       & ($M_\odot$) & ($L_\odot$) & (kpc) & (\textdegree) & (\textdegree) & (\kms)   & (\kms)  &      \\
(1)          & (2)    & (3)  & (4)  & (5)    & (6)   & (7)    & (8)         & (9)         & (10)  & (11)          & (12)          & (13)     & (14)    & (15) \\
\hline
J010644.15$-$103410.6 & 0.468     & Q & R & J010643.93$-$103419.3 & 0.1460  & 0.39 &  9.9 & 10.2 & 23.7 & 57 & 83 & $\le$0.063 & ---      & 1  \\
J025134.53$+$431515.8 & 1.313     & Q & C & J025135.79$+$431511.7 & 0.0520  & 0.81 & 11.3 & 10.8 & 14.6 & 47 &  0 & $\le$0.054 & 0.26     & 2  \\
J044858.80$-$204446.5 & 1.894     & Q & C & J044858.62$-$204454.7 & 0.0661  & ---  & ---  & ---  & 15.6 & 77 &  9 & $\le$0.210 & ---      & 3  \\
J085521.37$+$575144.1 & 0.368$^a$ & U & C & J085519.05$+$575140.7 & 0.02600 & 0.48 &  9.0 &  9.2 &  9.9 & 49 & 32 & $\le$0.019 & 1.02     & 4  \\
J091011.01$+$463617.8 & 1.020     & Q & C & J091010.55$+$463633.4 & 0.0998  & 0.58 & 10.5 & 10.4 & 30.0 & 23 &--- & $\le$0.032 & ---      & 1  \\
J095820.94$+$322402.2 & 0.531     & Q & C & J095821.07$+$322211.4 & 0.00484 & 0.74 & 10.0 &  9.7 & 11.1 & 67 & 76 & $\le$0.009 & 0.13     & 5  \\
J102258.41$+$123429.7 & 1.724     & Q & R & J102257.92$+$123439.2 & 0.1253  & 0.87 & 10.7 & 10.2 & 33.2 & 74 & 22 & $\le$0.095 & ---      & 1  \\
J104257.58$+$074850.5 & 2.667     & Q & C & J104257.74$+$074751.3 & 0.03321 & 0.50 &  9.1 &  9.0 &  1.7 & 50 & 12 & $\le$0.057 & 0.19     & 1  \\
J115917.87$+$441216.3 & 1.211     & Q & R & J115856.84$+$441133.8 & 0.00233 & 0.48 &  8.0 &  8.0 & 10.9 & 67 & 87 & $\le$0.130 & ---      & 6  \\
J120321.93$+$041419.0 & 1.224     & Q & C & J120322.71$+$041347.2 & 0.02036 & 0.65 &  9.0 &  8.8 & 14.0 & 76 & 77 & $\le$0.047 & ---      & 6  \\
J121846.60$+$173817.2 & 1.809     & Q & C & J121841.78$+$174308.6 & 0.00232 & 0.54 &  7.6 &  8.2 & 13.9 &  0 &  0 & $\le$0.168 & ---      & 6  \\
J122106.87$+$454852.1 & 0.525     & Q & R & J122115.22$+$454843.2 & 0.0015  & 0.08 &  7.4 &  7.8 &  2.7 & 33 &  1 & $\le$1.900 & 1.39     & 7  \\
J122154.17$+$305146.4 & 0.776     & Q & C & J122216.72$+$305323.5 & 0.00185 & 0.53 &  7.1 &  7.1 & 12.0 & 27 &--- & $\le$0.085 & ---      & 6  \\
J122847.42$+$370612.0 & 1.517     & Q & C & J122828.08$+$371405.3 & 0.00132 & 0.28 &  7.0 &  7.6 & 11.0 & 13 &--- & $\le$0.136 & ---      & 6  \\
J125248.28$+$474044.0 & 0.922     & Q & R & J125249.22$+$474106.1 & 0.0321  & 0.89 & 11.0 & 10.5 & 18.3 & 50 & 68 & $\le$0.190 & ---      & 1  \\
J125757.23$+$322929.2 & 0.805     & Q & C & J125756.40$+$322853.9 & 0.02337 & 0.44 &  8.8 &  9.0 & 17.4 & 69 & 44 & $\le$0.070 & ---      & 6  \\
J132155.68$+$351037.2 & 1.918     & Q & C & J132227.58$+$351219.5 & 0.00215 & 0.15 &  6.1 &  6.9 & 19.7 & 54 & 75 & $\le$0.126 & ---      & 6  \\
J132840.56$+$622136.9 & 1.218     & Q & C & J132839.89$+$622136.0 & 0.0423  & 0.51 &  8.9 &  8.9 &  4.0 & 51 & 14 & $\le$0.126 & ---      & 1  \\
J132846.53$-$485638.7 & ---       & U &---& J132844.09$-$485500.5 & 0.00997 & ---  & ---  & 10.4 & 20.7 & 25 &--- & $\le$0.030 & 0.82     & 8  \\
J133007.70$-$205616.2 & 1.169     & Q & R & J133005.30$-$205558.7 & 0.0178  & ---  & ---  & 10.1 & 13.7 & 52 & 60 & $\le$0.007 & 0.24$^b$ & 9  \\
J135400.09$+$565004.9 & ---       & U & R & J135400.68$+$565000.3 & 0.09549 & 0.80 & 10.7 & 10.2 & 11.8 & 73 & 80 & $\le$0.038 & 3.03     & 4  \\
J145907.58$+$714019.9 & 0.905     & Q &---& J145745.71$+$714056.4 & 0.00149 & ---  & ---  &  8.6 & 12.0 & 54 & 48 & $\le$0.014 & ---      & 10 \\
J202346.20$-$365520.5 & 1.048     & Q &---& J202345.21$-$365505.7 & 0.02700 & ---  & ---  &  9.9 & 10.3 & 66 & 21 & $\le$0.049 & 0.22     & 3  \\
J225503.89$+$131333.9 & 0.543     & Q &---& J225503.10$+$131313.7 & 0.03253 & 0.75 & 11.0 & 10.6 & 15.1 & 41 & 44 & $\le$0.026 & ---      & 11 \\
\hline
\end{tabular}
\label{tab:litsample}
\begin{flushleft}
Column 1: Radio source name. Column 2: Redshift of radio source. Column 3: Type of radio source $-$ quasar (Q), radio galaxy (G) or unknown (U). 
Column 4: Morphology of radio source at arcsecond-scale $-$ compact (C) or resolved (R). Column 5: Galaxy name. Column 6: Galaxy redshift.
Column 7: Rest-frame SDSS $g-r$ colour of galaxy. Column 8: Log of stellar mass ($M_\odot$) of galaxy from {\it kcorrect} \citep{blanton2007a}. 
Column 9: Log of $B$-band luminosity ($L_\odot$). Column 10: Projected separation or impact parameter (kpc) between radio sightline and centre of galaxy. 
Columns 11: Galaxy inclination. Column 12: Orientation of radio sightline with respect to galaxy's major axis. 
Column 13: 3$\sigma$ upper limit on integrated \hi\ \21\ optical depth from spectra smoothed to 10\,\kms. 
Column 14: Integrated \hi\ \21\ optical depth in case of detections at the observed spectral resolution.
Column 15: References for the \hi\ \21\ optical depth measurements $-$ 1: \citet{borthakur2011}; 2: \citet{hwang2004}; 3: \citet{carilli1992}; 4: \citet{zwaan2015}; 
5: \citet{keeney2005}; 6: \citet{borthakur2016} ; 7: \citet{borthakur2014}; 8: \citet{reeves2016}; 9: \citet{keeney2011}; 10: \citet{haschick1975}; 11: \citet{corbelli1990}. \\
$^a$ photometric redshift $^b$ \taudv\ corresponds to stronger of the two absorption components separated by 250\,\kms
\end{flushleft}
%\end{table}
\end{landscape}
\section{SALT spectra}
\label{appendix_saltspectra}
The details of the redshifts obtained from the SALT spectra of galaxies are provided in Table~\ref{tab:saltgal} and the galaxy spectra are shown in Fig.~\ref{fig:saltspec1}.
\begin{table*} 
\caption{Details of the galaxy spectra obtained using SALT.}
\centering
\begin{tabular}{ccccc}
\hline
Galaxy & $z_g$ & Emission lines & Absorption lines \\
(1)    & (2)   & (3)            & (4)              \\
\hline
J085625.41$+$102020.1 & 0.1043 $\pm$ 0.0005 & H$\alpha$, H$\beta$, [O\,{\sc iii}], [N\,{\sc ii}], [S\,{\sc ii}] & Ca\,{\sc ii}                                   \\
J113302.01$+$001541.7 & 0.0762 $\pm$ 0.0003 & H$\alpha$, [O\,{\sc iii}], [N\,{\sc ii}], [S\,{\sc ii}]           & Ca\,{\sc ii}                                   \\
J125422.21$+$054610.4 & 0.0246 $\pm$ 0.0005 & H$\alpha$, H$\beta$, [O\,{\sc iii}], [N\,{\sc ii}], [S\,{\sc ii}] & ---                                            \\
J134528.42$+$034723.0 & 0.2622 $\pm$ 0.0005 & ---                                                               & H$\alpha$, H$\beta$, Na\,{\sc i}, Ca\,{\sc ii} \\
J145722.40$+$051918.8 & 0.2198 $\pm$ 0.0001 & H$\alpha$, [N\,{\sc ii}], [S\,{\sc ii}]                           & ---                                            \\
J155121.06$+$071352.4 & 0.1007 $\pm$ 0.0005 & ---                                                               & H$\alpha$, H$\beta$, Na\,{\sc i}, Ca\,{\sc ii} \\
J224013.49$-$083633.6 & 0.2111 $\pm$ 0.0005 & H$\alpha$, [N\,{\sc ii}], [S\,{\sc ii}]                           & Ca\,{\sc ii}                                   \\
\hline
\end{tabular}
\label{tab:saltgal}
\begin{flushleft}
Column 1: Galaxy name. Column 2: Galaxy redshift. Columns 3 and 4: Emission and absorption lines detected in the SALT spectrum, respectively.
\end{flushleft}
\end{table*}
\begin{figure*}
\subfloat{\includegraphics[width=0.6\textwidth, angle=90]{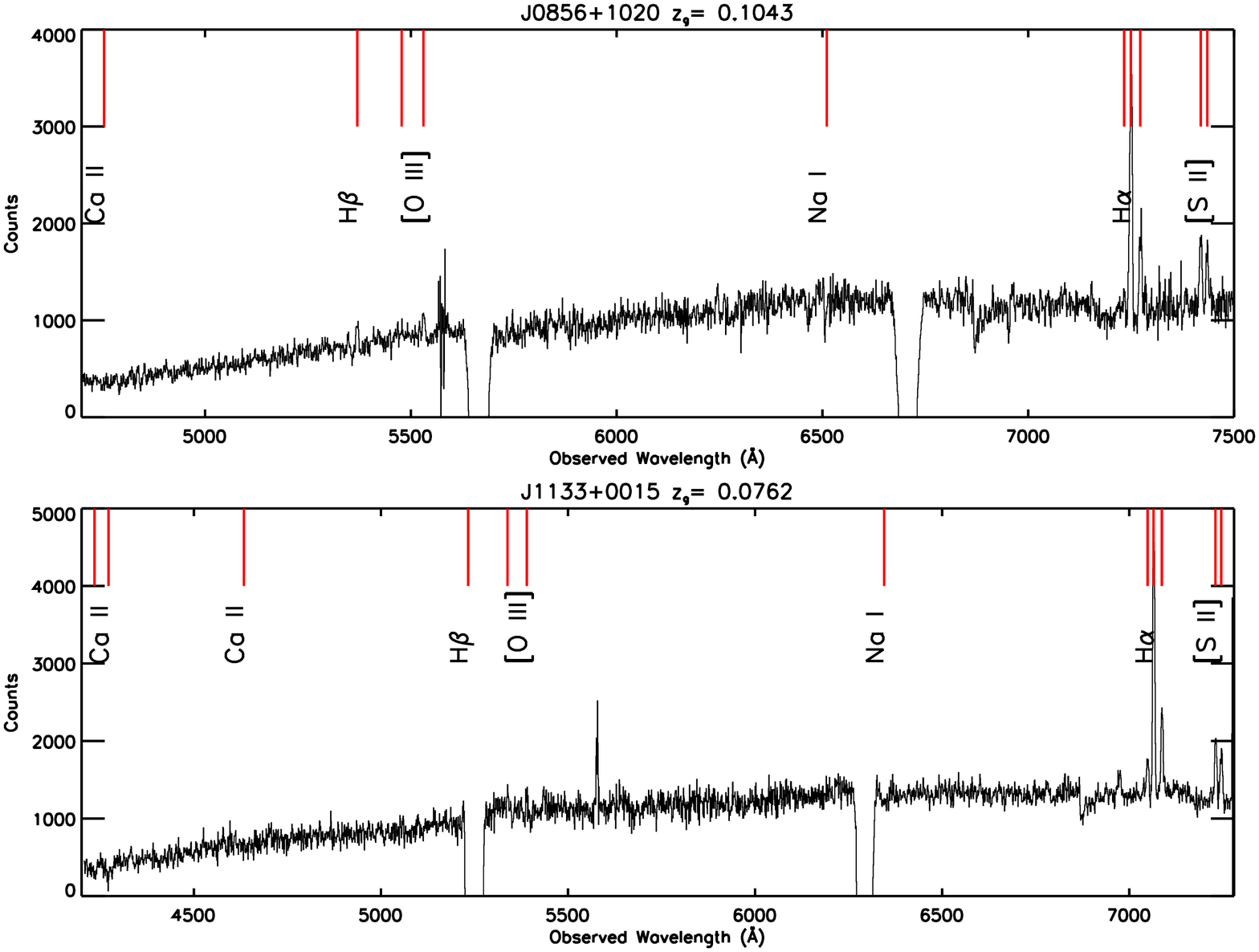} } \hspace{0.01cm}
\subfloat{\includegraphics[width=0.6\textwidth, angle=90]{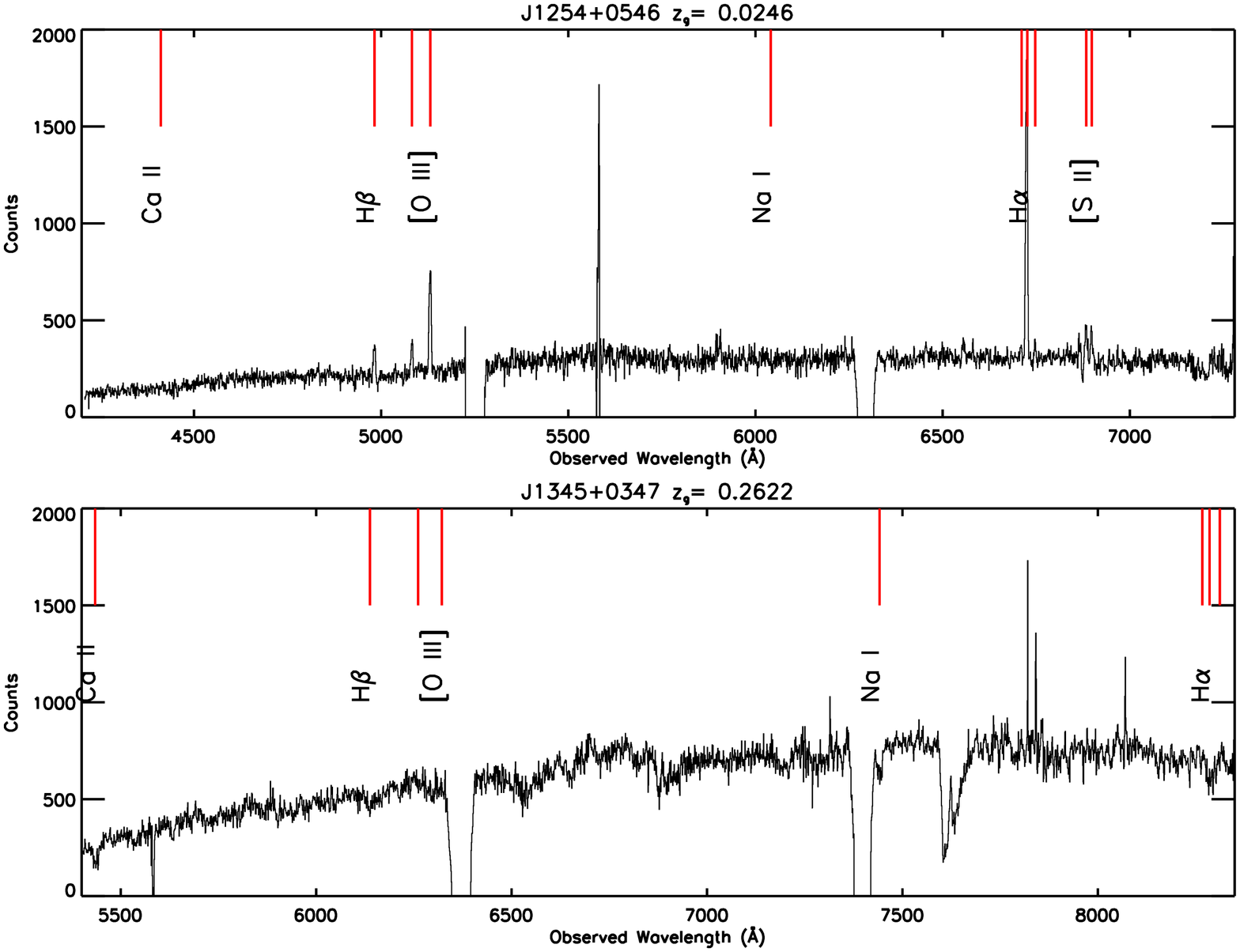} }  
\caption{SALT spectra for seven of the galaxies in our sample. The expected positions of the H$\alpha$, H$\beta$, [O\,{\sc iii}], [N\,{\sc ii}], and [S\,{\sc ii}] 
emission lines, and Na\,{\sc i} and Ca\,{\sc ii} absorption lines are marked by ticks.}
\label{fig:saltspec1}
\end{figure*}
\begin{figure*}
\ContinuedFloat
\subfloat{\includegraphics[width=0.6\textwidth, angle=90]{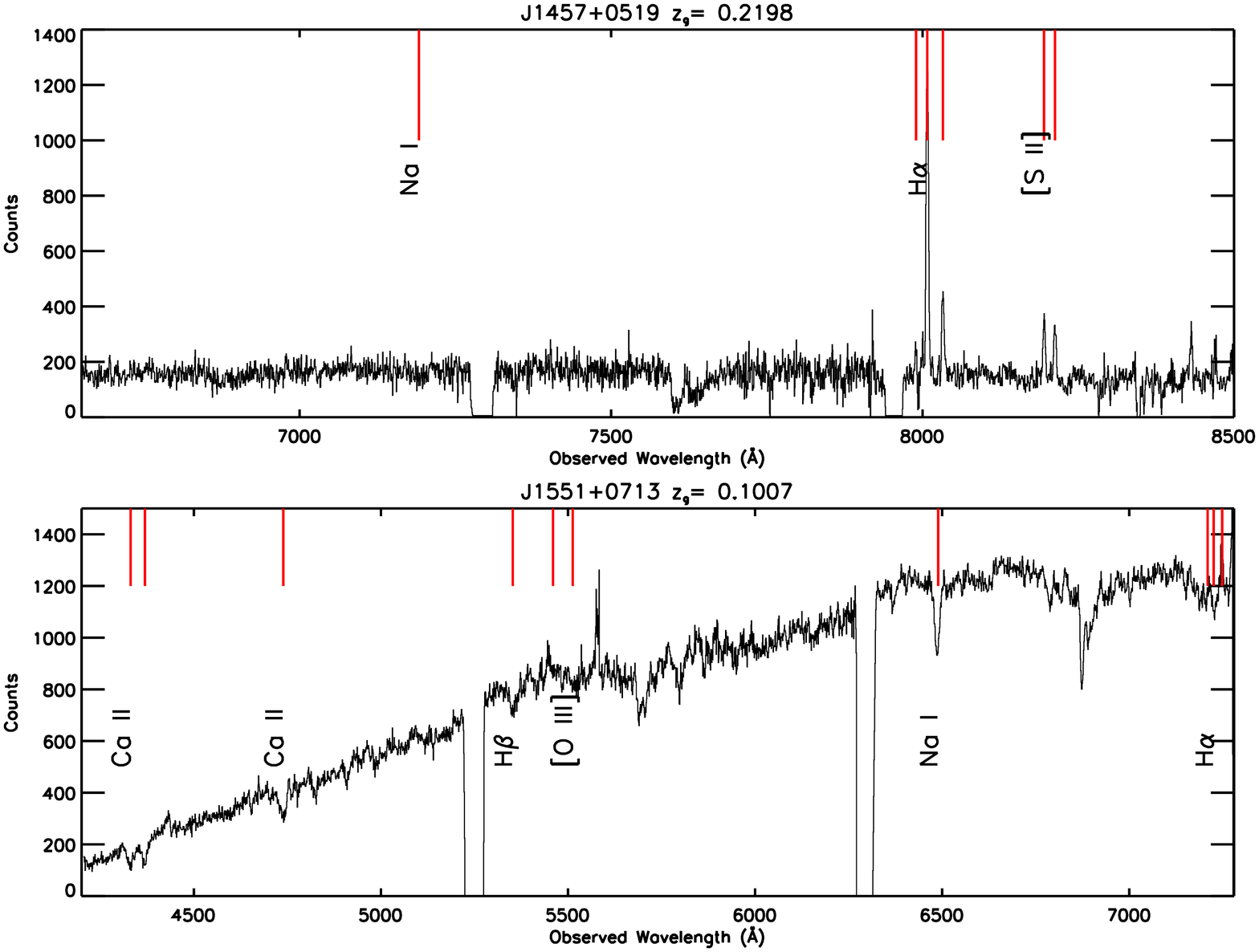} } \hspace{0.01cm}
\subfloat{\includegraphics[width=0.6\textwidth, angle=90]{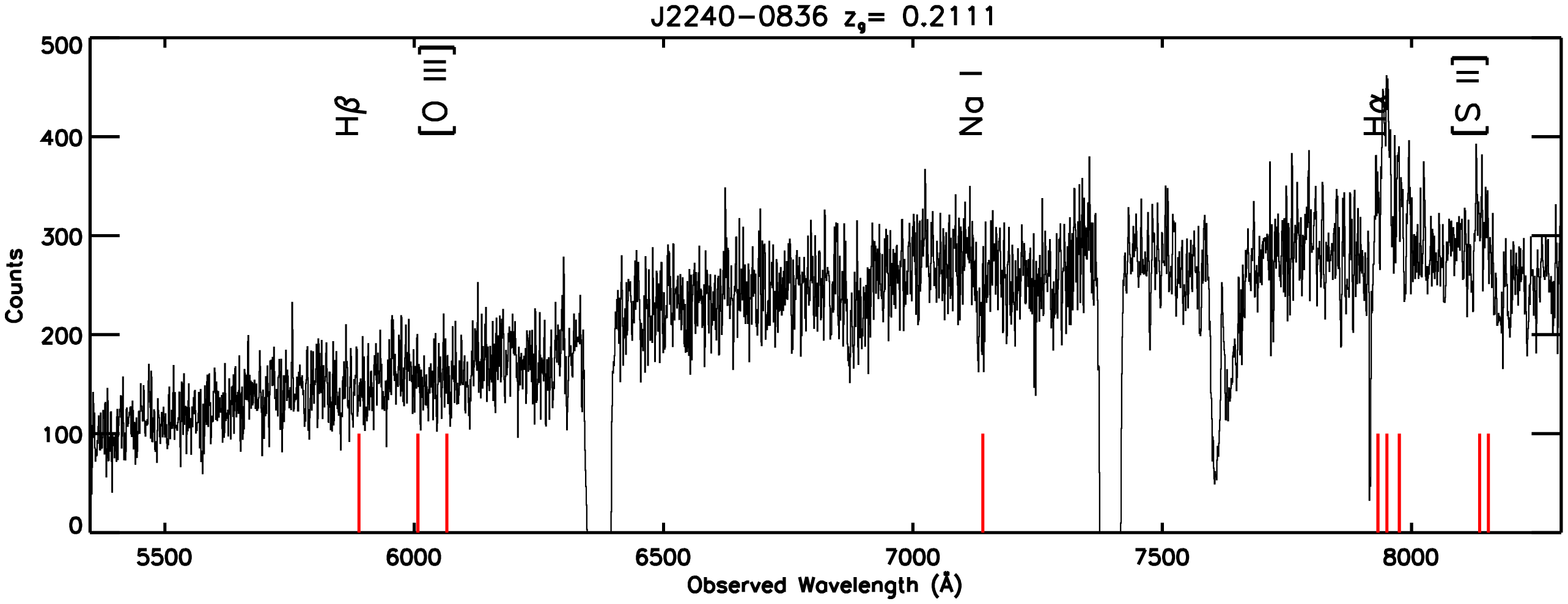} }  
\caption{Continued from previous page.}
\label{fig:saltspec2}
\end{figure*}
\section{Radio observation log}
\label{appendix_radiolog}
The log of the radio observations carried out by us is provided here. 
\begin{table*}
\caption{Radio observation log of the sources observed by us.} 
\centering
\begin{tabular}{ccccccccc}
\hline
Radio  & Galaxy & \zgal\ & Telescope & Date & $\delta v$ & Time on    & Beam size    & Spatial resolution \\
Source &        &        &           &      & (\kms)     & source (h) & (arcsec$^2$) & (kpc$^2$)          \\
(1)    & (2)    & (3)    & (4)       & (5)  & (6)        & (7)        & (8)          & (9)                \\
\hline
\multicolumn{9}{c}{\it Primary Sample} \\
\hline
J004126.01$-$014315.6 & J004121.53$-$014257.0 & 0.01769	& GMRT & 17 October 2014   & 1.8 & 2.6 &  3.4 $\times$  2.3 &  1.2 $\times$  0.8 \\    
J004332.71$+$002459.9 & J004333.64$+$002516.2 & 0.07975 & GMRT & 05 September 2014 & 1.9 & 2.7 &  2.4 $\times$  2.3 &  3.6 $\times$  3.4 \\	
J013412.70$+$000345.2 & J013411.28$+$000327.5 & 0.05198 & GMRT & 17 October 2014   & 1.8 & 2.4 &  2.9 $\times$  2.3 &  2.9 $\times$  2.4 \\       
J080240.89$+$160104.6 & J080241.68$+$160035.2 & 0.01565 & GMRT & 03 May 2014       & 1.7 & 4.2 &  2.8 $\times$  2.2 &  0.9 $\times$  0.7 \\       
J081705.49$+$195842.8 & J081707.13$+$195907.9 & 0.04363 & GMRT & 06 September 2014 & 1.8 & 2.4 &  2.4 $\times$  2.1 &  2.0 $\times$  1.8 \\       
J082153.82$+$503120.4 & J082153.74$+$503125.7 & 0.1835	& GMRT & 19 September 2007 & 3.9 & 8.0 &  3.5 $\times$  2.4 & 10.7 $\times$  7.3 \\
J084957.97$+$510829.0 & J084957.49$+$510842.3 & 0.07342	& GMRT & 19 June 2007      & 3.5 & 7.7 &  3.2 $\times$  2.5 &  4.5 $\times$  3.5 \\
J084957.97$+$510829.0 & J084958.10$+$510826.6 & 0.3120	& GMRT & 18 July 2011      & 2.3 & 2.7 &  3.4 $\times$  2.6 & 15.6 $\times$ 12.0 \\
                      &                       &         &      & 03 July 2012      &     & 3.5 &                    &                    \\
J085624.91$+$102017.0 & J085625.41$+$102020.1 & 0.1043  & GMRT & 25 April 2015     & 1.9 & 5.8 &  2.5 $\times$  2.3 &  4.8 $\times$  4.4 \\       
J091133.45$+$195814.1 & J091134.38$+$195830.2 & 0.02293 & GMRT & 18 October 2014   & 1.8 & 5.9 &  2.9 $\times$  2.2 &  1.3 $\times$  1.0 \\       
J101514.63$+$163740.4 & J101514.22$+$163758.3 & 0.05436 & GMRT & 16 July 2011      & 1.8 & 3.0 &  2.5 $\times$  2.2 &  2.6 $\times$  2.3 \\       
J111023.84$+$032136.1 & J111025.09$+$032138.8 & 0.02996	& GMRT & 01 March 2008     & 3.4 & 7.6 &  2.4 $\times$  1.9 &  1.4 $\times$  1.1 \\
J112613.04$+$380216.4 & J112612.57$+$380228.6 & 0.06190	& GMRT & 15 July 2011      & 1.8 & 4.6 &  2.3 $\times$  2.2 &  2.8 $\times$  2.6 \\       
J113303.02$+$001549.0 & J113302.01$+$001541.7 & 0.0762  & GMRT & 06 May 2014       & 1.9 & 2.2 &  3.2 $\times$  2.1 &  4.6 $\times$  3.0 \\       
J113303.02$+$001549.0 & J113303.96$+$001538.2 & 0.07584 & GMRT & 06 May 2014       & 1.9 & 2.2 &  3.2 $\times$  2.1 &  4.6 $\times$  3.0 \\       
J115659.62$+$341216.9 & J115659.59$+$341221.7 & 0.13797 & GMRT & 28 February 2015  & 1.9 & 5.0 &  2.9 $\times$  2.2 &  7.2 $\times$  5.5 \\
J120922.78$+$411941.3 & J120923.25$+$411949.3 & 0.17098 & GMRT & 25 July 2014      & 2.0 & 1.6 &  4.1 $\times$  2.3 & 12.0 $\times$  6.8 \\       
J122847.42$+$370612.0 & J122847.72$+$370606.9 & 0.13837	& GMRT & 02 March 2008     & 3.7 & 7.5 &  2.5 $\times$  1.9 &  6.1 $\times$  4.7 \\
J124157.54$+$633241.6 & J124157.26$+$633237.6 & 0.1430	& GMRT & 03 March 2008     & 3.8 & 7.7 &  4.0 $\times$  2.6 & 10.1 $\times$  6.5 \\
                      &                       &         &      & 13 June 2009      &     & 8.0 &                    &                    \\
                      &                       &         &      & 14 June 2009      &     & 5.4 &                    &                    \\
J124355.78$+$404358.4 & J124357.17$+$404346.1 & 0.01693 & GMRT & 01 July 2010      & 3.4 & 4.6 &  2.6 $\times$  2.1 &  0.9 $\times$  0.7 \\                     
J124357.65$+$162253.3 & J124342.64$+$162335.9 & 0.00263 & GMRT & 18 Ocotber 2014   & 1.7 & 1.6 &  2.8 $\times$  2.2 &  0.2 $\times$  0.1 \\       
                      &                       &         &      & 22 November 2014  &     & 2.6 &                    &                    \\                       
J125422.19$+$054619.8 & J125422.21$+$054610.4 & 0.0246  & GMRT & 26 May 2014       & 1.7 & 2.2 &  2.4 $\times$  2.1 &  1.2 $\times$  1.0 \\       
                      &                       &         &      & 29 August 2014    &     & 6.4 &                    &                    \\       
J130028.53$+$283010.1 & J130028.66$+$283012.1 & 0.2228  & VLA  & 21 June 2015      & 0.5 & 1.9 &  2.2 $\times$  1.8 &  7.9 $\times$  6.4 \\       
                      &                       &         &      & 26 June 2015      &     & 1.9 &                    &                    \\       
J130050.09$-$012146.2 & J130049.15$-$012136.4 & 0.02275 & GMRT & 08 November 2014  & 1.7 & 4.5 &  3.2 $\times$  2.8 &  1.5 $\times$  1.3 \\       
J134528.74$+$034719.7 & J134528.42$+$034723.0 & 0.2622  & GMRT & 30 January 2015   & 2.2 & 5.3 &  3.3 $\times$  3.2 & 13.5 $\times$ 12.8 \\       
J135907.77$+$293531.8 & J135908.47$+$293541.5 & 0.14607 & GMRT & 17 July 2011      & 2.0 & 3.0 &  2.9 $\times$  2.2 &  7.5 $\times$  5.8 \\       
J142253.22$+$054013.4 & J142253.99$+$053947.9 & 0.05422 & GMRT & 06 March 2015     & 1.8 & 3.4 &  2.6 $\times$  2.3 &  2.8 $\times$  2.4 \\       
J143806.79$+$175805.4 & J143806.76$+$175802.5 & 0.1468  & GMRT & 03 May 2014       & 2.0 & 5.1 &  3.1 $\times$  2.3 &  7.9 $\times$  6.0 \\       
J144304.53$+$021419.3 & J144304.53$+$021419.3 & 0.3714	& GMRT & 02 June 2012      & 2.3 & 5.7 &  3.9 $\times$  2.9 & 20.0 $\times$ 15.0 \\
J145722.53$+$051921.8 & J145722.40$+$051918.8 & 0.2198  & GMRT & 01 May 2014       & 2.1 & 3.0 &  3.9 $\times$  2.6 & 13.7 $\times$  9.2 \\       
J152523.55$+$420117.0 & J152522.99$+$420131.7 & 0.09094 & GMRT & 25 May 2014       & 1.9 & 2.2 &  3.0 $\times$  2.6 &  5.2 $\times$  4.4 \\       
J153905.20$+$053438.3 & J153905.16$+$053416.5 & 0.02594	& GMRT & 25 August 2013    & 1.8 & 3.0 &  3.2 $\times$  2.4 &  1.7 $\times$  1.3 \\
J155121.13$+$071357.7 & J155121.06$+$071352.4 & 0.1007  & GMRT & 06 May 2014       & 1.9 & 2.2 &  3.6 $\times$  2.1 &  6.6 $\times$  3.9 \\       
                      &                       &         &      & 22 November 2014  &     & 6.1 &                    &                    \\       
J160658.29$+$271705.5 & J160659.13$+$271642.6 & 0.04620 & GMRT & 25 July 2014      & 1.8 & 2.6 &  3.4 $\times$  2.3 &  3.1 $\times$  2.1 \\       
J163402.95$+$390000.5 & J163403.30$+$390001.0 & 0.36620 & GMRT & 23 November 2014  & 2.4 & 2.5 &  5.6 $\times$  2.7 & 28.6 $\times$ 13.6 \\       
J163956.35$+$112758.7 &	J163956.37$+$112802.1 & 0.0790	& GMRT & 02 July 2010      & 3.5 & 4.2 &                    &                    \\
                      &                       &         &      & 14 July 2010      & 0.9 & 3.8 & 2.8 $\times$  2.1  &  4.2 $\times$  3.1 \\
                      &                       &         &      & 16 July 2010      &     & 3.1 &                    &                    \\
J164315.58$+$354729.7 & J164315.10$+$354701.6 & 0.03121 & WSRT & 18 February 2011  & 1.1 & 9.7 & 27.0 $\times$ 13.2 & 16.9 $\times$  8.2 \\       
J164802.37$+$242916.9 & J164804.38$+$242935.3 & 0.03583 & GMRT & 24 July 2014      & 1.8 & 2.8 &  2.4 $\times$  2.0 &  1.7 $\times$  1.4 \\       
J174832.84$+$700550.8 & J174926.43$+$700839.7 & 0.00008 & GMRT & 19 October 2014   & 1.7 & 2.6 &  3.6 $\times$  2.4 &  0.1 $\times$  0.1 \\
J224013.41$-$083628.7 & J224013.49$-$083633.6 & 0.2111  & GMRT & 06 December 2014  & 2.1 & 2.4 &  3.3 $\times$  2.9 & 11.3 $\times$ 10.0 \\  
\hline
\multicolumn{9}{c}{\it Supplementary Sample} \\
\hline
J082015.85$+$082612.6 & J082015.88$+$082610.5 & 0.0775  & GMRT & 08 November 2014  & 1.8 & 3.5 &  2.7 $\times$  2.2 &  3.9 $\times$  3.3 \\
                      &                       &         &      & 26 April 2015     &     & 5.2 &                    &                    \\
J091917.06$+$340049.5 &	J091917.45$+$340029.4 & 0.02376 & GMRT & 15 July 2011      & 1.8 & 3.0 &  3.0 $\times$  1.8 &  1.5 $\times$  0.9 \\
J094745.85$+$111354.0 & J094745.85$+$111354.0 & 0.1124  & GMRT & 02 May 2014       & 1.9 & 3.0 &  4.7 $\times$  2.0 &  9.6 $\times$  4.0 \\
J095433.96$+$373336.3 &	J095432.89$+$373333.1 & 0.07282	& WSRT & 04 March 2011     & 1.1 & 6.0 & 37.4 $\times$ 13.6 & 51.8 $\times$ 18.9 \\
J095604.34$+$161418.5 &	J095604.09$+$161427.1 & 0.03346	& GMRT & 03 July 2010      & 3.4 & 2.3 &  4.0 $\times$  2.1 &  2.7 $\times$  1.4 \\
J105053.93$+$000213.6 & J105053.93$+$000213.6 & 0.2338  & GMRT & 22 May 2014       & 2.1 & 4.7 &  3.5 $\times$  2.6 & 12.8 $\times$  9.6 \\
J135947.90$+$270828.3 & J135947.90$+$270828.3 & 0.0432  & GMRT & 02 May 2014       & 1.8 & 3.0 &  2.9 $\times$  2.2 &  2.5 $\times$  1.8 \\
J143809.90$+$092016.9 &	J143810.20$+$092009.8 & 0.03029 & GMRT & 24 August 2014    & 1.8 & 2.0 &  4.0 $\times$  2.3 &  2.4 $\times$  1.4 \\	
J162434.18$+$232859.9 & J162433.76$+$232912.3 & 0.09382 & GMRT & 07 March 2015     & 1.9 & 2.4 &  2.7 $\times$  2.3 &  4.7 $\times$  4.0 \\
J224247.89$-$001944.9 & J224247.65$-$001918.7 & 0.05815 & GMRT & 19 October 2014   & 1.8 & 2.5 &  3.7 $\times$  2.1 &  4.2 $\times$  2.3 \\
\hline
\end{tabular}
\label{tab:radiolog1}
\end{table*}
\setcounter{table}{0}
\begin{table*}
\caption{Continued from previous page.} 
\centering
\begin{tabular}{ccccccccc}
\hline
Radio  & Galaxy & \zgal\ & Telescope & Date & $\delta v$ & Time on    & Beam size    & Spatial resolution \\
Source &        &        &           &      & (\kms)     & source (h) & (arcsec$^2$) & (kpc$^2$)          \\
(1)    & (2)    & (3)    & (4)       & (5)  & (6)        & (7)        & (8)          & (9)                \\
\hline
\multicolumn{9}{c}{\it Miscellaneous Sample} \\
\hline
J084600.36$+$070424.6 & J084600.52$+$070428.8 & 0.34235 & GMRT & 01 June 2014      & 2.3 & 2.2 & 4.3 $\times$ 3.0 & 21.2 $\times$ 14.8 \\
J094221.98$+$062335.2 & J094221.94$+$062337.1 & 0.12390 & WSRT & 03 March 2011     & 1.2 & 9.8 &                  &                    \\
                      &                       &         & GMRT & 16 July 2011      & 1.9 & 3.0 & 2.4 $\times$ 1.8 &  5.3 $\times$  4.1 \\
                      &                       &         & GMRT & 17 July 2011      &     & 3.0 &                  &                    \\
J121417.80$+$293143.4 & J121418.25$+$293146.7 & 0.06326 & GMRT & 24 December 2014  & 1.8 & 5.9 & 2.8 $\times$ 2.0 &  3.5 $\times$  2.4 \\
J164330.03$+$241951.8 & J164331.92$+$241939.6 & 0.08723 & GMRT & 15 July 2011      & 1.9 & 4.6 & 3.3 $\times$ 2.4 &  5.4 $\times$  3.9 \\
J205449.61$+$004153.0 & J205449.64$+$004149.8 & 0.20153 & VLA  & 20 June 2015      & 0.5 & 1.2 & 2.1 $\times$ 1.6 &  6.9 $\times$  5.3 \\
\hline
\end{tabular}
\label{tab:radiolog2}
\begin{flushleft}
Column 1: Radio source name. Column 2: Galaxy name. Column 3: Galaxy redshift. Column 4: Telescope used for radio observation. 
Column 5: Date of observation. Column 6: Spectral resolution in \kms\ per channel. Column 7: Time on source in h. 
Column 8: The size of the restoring beam of the continuum image in arcsec$^2$. 
Column 9: The spatial resolution corresponding to the beam size at the redshift of the galaxy in kpc$^2$.\\
\end{flushleft}
\end{table*}
\section{Results from \hi\ \21\ spectral line observations}
\label{appendix_21cmresults}
The parameters derived from our radio spectral line observations are provided here.
\begin{table*}
\caption{Parameters derived from radio observations.}  
\centering
\begin{tabular}{ccccccccc}
\hline
QGP & Peak Flux Density & $\delta v$ & Spectral rms & $\sigma_\tau$ & $\tau_p$ & \taudvl\ & \taudv\ & Ref. \\
    & (\mjb)            & (\kms)     & (\mjb)       &               &          & (\kms)   & (\kms)  &      \\
(1) & (2)               & (3)        & (4)          & (5)           & (6)      & (7)      & (8)     & (9)  \\
\hline
\multicolumn{9}{c}{\it Primary Sample} \\
\hline
J0041$-$0143N   & 354  & 1.8 & 2.3 & 0.006 & ---  & $\le$0.070 & ---              & 1 \\ 
J0041$-$0143C   & 67   & 1.8 & 2.1 & 0.033 & ---  & $\le$0.282 & ---              & 1 \\ 
J0041$-$0143S   & 215  & 1.8 & 2.1 & 0.010 & 0.08 & $\le$0.094 & 0.52 $\pm$ 0.07  & 1 \\ 
J0043$+$0024    & 128  & 1.9 & 1.9 & 0.015 & ---  & $\le$0.171 & ---              & 2 \\
J0134$+$0003    & 907  & 1.8 & 2.1 & 0.002 & ---  & $\le$0.030 & ---              & 2 \\
J0802$+$1601    & 74   & 1.7 & 1.6 & 0.022 & ---  & $\le$0.286 & ---              & 2 \\
J0817$+$1958    & 309  & 1.8 & 2.2 & 0.007 & ---  & $\le$0.104 & ---              & 2 \\
J0821$+$5031    & 48   & 3.9 & 0.9 & 0.019 & ---  & $\le$0.380 & ---              & 3 \\
J0849$+$5108\_1 & 248  & 3.5 & 1.0 & 0.004 & ---  & $\le$0.080 & ---              & 3 \\
J0849$+$5108\_2 & 233  & 2.3 & 2.0 & 0.009 & 0.09 & $\le$0.077 & 0.95 $\pm$ 0.06  & 4 \\
J0856$+$1020    & 54   & 1.9 & 1.3 & 0.024 & ---  & $\le$0.360 & ---              & 2 \\
J0911$+$1958    & 259  & 1.8 & 1.4 & 0.006 & ---  & $\le$0.084 & ---              & 2 \\
J1015$+$1637    & 88   & 1.8 & 1.8 & 0.020 & ---  & $\le$0.257 & ---              & 2 \\
J1110$+$0321E   & 155  & 3.4 & 1.1 & 0.007 & ---  & $\le$0.080 & ---              & 3 \\ 
J1110$+$0321W   & 223  & 3.4 & 0.9 & 0.004 & ---  & $\le$0.130 & ---              & 3 \\ 
J1126$+$3802    & 123  & 1.8 & 2.4 & 0.020 & ---  & $\le$0.254 & ---              & 2 \\
J1133$+$0015\_1 & 420  & 1.9 & 3.0 & 0.007 & ---  & $\le$0.095 & ---              & 2 \\
J1133$+$0015\_2 & 420  & 1.9 & 3.0 & 0.007 & ---  & $\le$0.095 & ---              & 2 \\  
J1156$+$3412    & 97   & 1.9 & 1.8 & 0.018 & ---  & $\le$0.202 & ---              & 2 \\
J1209$+$4119    & 178  & 2.0 & 2.5 & 0.014 & ---  & $\le$0.218 & ---              & 2 \\
J1228$+$3706    & 298  & 3.7 & 1.1 & 0.004 & ---  & $\le$0.070 & ---              & 3 \\
J1241$+$6332    & 68   & 3.8 & 0.7 & 0.009 & 0.16 & $\le$0.170 & 2.90 $\pm$ 0.16  & 3 \\
J1243$+$4043    & 196  & 3.4 & 1.5 & 0.008 & 0.18 & $\le$0.131 & 2.26 $\pm$ 0.10  & 2 \\
J1243$+$1622N   & 1257 & 1.7 & 1.6 & 0.001 & ---  & $\le$0.030 & ---              & 2 \\ 
J1243$+$1622C   & 419  & 1.7 & 1.3 & 0.003 & ---  & $\le$0.060 & ---              & 2 \\ 
J1243$+$1622S   & 645  & 1.7 & 1.4 & 0.002 & ---  & $\le$0.030 & ---              & 2 \\ 
J1254$+$0546    & 109  & 1.7 & 0.8 & 0.008 & ---  & $\le$0.250 & ---              & 2 \\
J1300$+$2830    & 119  & 0.5 & 3.9 & 0.033 & ---  & $\le$0.239 & ---              & 2 \\
J1300$-$0121    & 132  & 1.7 & 1.9 & 0.015 & ---  & $\le$0.205 & ---              & 2 \\
J1345$+$0347    & 91   & 2.2 & 1.1 & 0.013 & ---  & $\le$0.182 & ---              & 2 \\
J1359$+$2935    & 104  & 2.0 & 2.5 & 0.024 & ---  & $\le$0.316 & ---              & 2 \\
J1422$+$0540    & 220  & 1.8 & 2.0 & 0.009 & ---  & $\le$0.133 & ---              & 2 \\
J1438$+$1758    & 53   & 2.0 & 1.3 & 0.024 & 0.38 & $\le$0.254 & 4.89 $\pm$ 0.19  & 2 \\
J1443$+$0214    & 163  & 2.4 & 0.9 & 0.006 & 0.33 & $\le$0.072 & 3.40 $\pm$ 0.10  & 4 \\
J1457$+$0519    & 157  & 2.1 & 2.1 & 0.014 & ---  & $\le$0.135 & ---              & 2 \\
J1525$+$4201    & 93   & 1.9 & 1.9 & 0.021 & ---  & $\le$0.235 & ---              & 2 \\
J1539$+$0534    & 124  & 1.8 & 1.8 & 0.015 & ---  & $\le$0.179 & ---              & 2 \\
J1551$+$0713    & 68   & 1.9 & 1.0 & 0.015 & ---  & $\le$0.265 & ---              & 2 \\
J1606$+$2717    & 214  & 1.8 & 2.0 & 0.010 & ---  & $\le$0.140 & ---              & 2 \\
J1634$+$3900    & 1100 & 2.4 & 1.7 & 0.002 & ---  & $\le$0.021 & ---              & 2 \\
J1639$+$1127    & 155  & 0.9 & 1.6 & 0.012 & 0.71 & $\le$0.088 & 15.70 $\pm$ 0.13 & 5 \\
J1643$+$3547    & 126  & 1.1 & 1.4 & 0.011 & ---  & $\le$0.110 & ---              & 2 \\
J1648$+$2429    & 293  & 1.8 & 2.1 & 0.007 & ---  & $\le$0.111 & ---              & 2 \\
J1748$+$7005    & 977  & 1.7 & 3.8 & 0.004 & ---  & $\le$0.077 & ---              & 2 \\
J2240$-$0836    & 171  & 2.1 & 3.9 & 0.009 & ---  & $\le$0.309 & ---              & 2 \\
\hline
\multicolumn{9}{c}{\it Supplementary Sample} \\
\hline
J0820$+$0826    & 106  & 1.8 & 1.0 & 0.010 & --- & $\le$0.157 & --- & 2 \\ 
J0919$+$3400E   & 66   & 1.8 & 2.5 & 0.039 & --- & $\le$0.490 & --- & 2 \\
J0919$+$3400W   & 38   & 1.8 & 2.4 & 0.063 & --- & $\le$0.797 & --- & 2 \\
J0947$+$1113    & 227  & 1.9 & 2.7 & 0.012 & --- & $\le$0.162 & --- & 2 \\
J0954$+$3733    & 311  & 1.1 & 1.7 & 0.005 & --- & $\le$0.075 & --- & 2 \\
J0956$+$1614N   & 73   & 3.4 & 1.7 & 0.023 & --- & $\le$0.460 & --- & 2 \\
J0956$+$1614S   & 35   & 3.4 & 1.7 & 0.049 & --- & $\le$0.718 & --- & 2 \\
J1050$+$0002    & 59   & 2.1 & 1.2 & 0.021 & --- & $\le$0.293 & --- & 2 \\
J1359$+$2708    & 142  & 1.8 & 1.8 & 0.013 & --- & $\le$0.220 & --- & 2 \\
J1438$+$0920    & 59   & 1.8 & 2.1 & 0.036 & --- & $\le$0.442 & --- & 2 \\
J1624$+$2328    & 120  & 1.9 & 1.8 & 0.015 & --- & $\le$0.187 & --- & 2 \\
J2242$-$0019    & 291  & 1.8 & 2.6 & 0.009 & --- & $\le$0.114 & --- & 2 \\
\hline
\end{tabular}
\label{tab:radioparam1}
\end{table*}
\setcounter{table}{0}
\begin{table*}
\caption{Continued from previous page.}  
\centering
\begin{tabular}{ccccccccc}
\hline
QGP & Peak Flux Density & $\delta v$ & Spectral rms & $\sigma_\tau$ & $\tau_p$ & \taudvl\ & \taudv\ & Ref. \\
    & (\mjb)            & (\kms)     & (\mjb)       &               &          & (\kms)   & (\kms)  &      \\
(1) & (2)               & (3)        & (4)          & (5)           & (6)      & (7)      & (8)     & (9)  \\
\hline
\multicolumn{9}{c}{\it Miscellaneous Sample} \\
\hline
J0846$+$0704    & 155  & 2.3 & 3.7 & 0.024 & ---  & $\le$0.333 & ---              & 2 \\
J0942$+$0623    & 116  & 2.0 & 1.9 & 0.016 & 0.70 & $\le$0.215 & 49.90 $\pm$ 0.40 & 6 \\
J1214$+$2931    & 59   & 1.8 & 1.8 & 0.030 & ---  & $\le$0.390 & ---              & 2 \\
J1643$+$2419    & 199  & 1.9 & 1.8 & 0.009 & ---  & $\le$0.157 & ---              & 2 \\
J2054$+$0041    & 362  & 0.5 & 9.5 & 0.022 & 0.61 & $\le$0.174 & 29.92 $\pm$ 0.22 & 2 \\
\hline
\end{tabular}
\label{tab:radioparam2}
\begin{flushleft}
Column 1: QGP name as given in Table~A1. Column 2: Peak flux density in \mjb. Column 3: Spectral resolution in \kms\ per channel. 
Column 4: Spectral rms in \mjb. Column 5: Standard deviation of the \hi\ \21\ optical depth. Column 6: Maximum \hi\ \21\ optical depth in case of \hi\ \21\ detections. 
Column 7: 3$\sigma$ upper limit on integrated \hi\ \21\ optical depth from spectra smoothed to 10\,\kms. Column 8: Integrated \hi\ \21\ optical depth in case of \hi\ \21\ detections. 
Column 9: References $-$ 1: \citet{dutta2016}; 2: This work; 3: \citet{gupta2010}; 4: \citet{gupta2013}; 5: \citet{srianand2013}; 6: \citet{srianand2015}. \\
Note that the values in Columns 4, 5, 6 and 8 are at the spectral resolution specified in Column 3.
\end{flushleft}
\end{table*}
\section{\hi\ \21\ absorption spectra}
\label{appendix_21cmspectra}
The \hi\ \21\ absorption spectra towards all the sources observed by us are shown here. The QGP name as given in Table~A1 are provided for each spectra. 
\begin{figure*}
\subfloat{\includegraphics[height=0.5\textheight, angle=90]{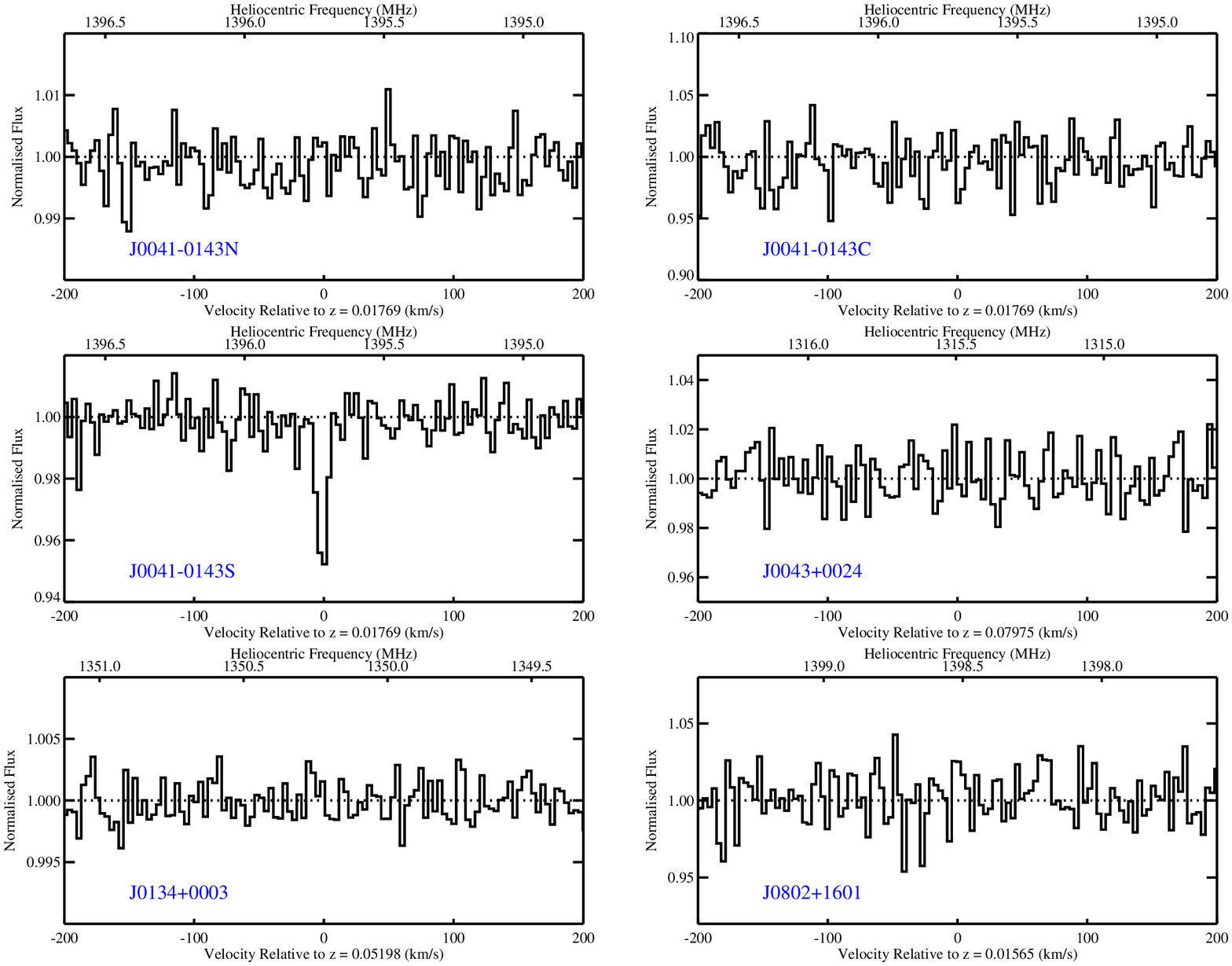} } \hspace{0.01cm}
\subfloat{\includegraphics[height=0.5\textheight, angle=90]{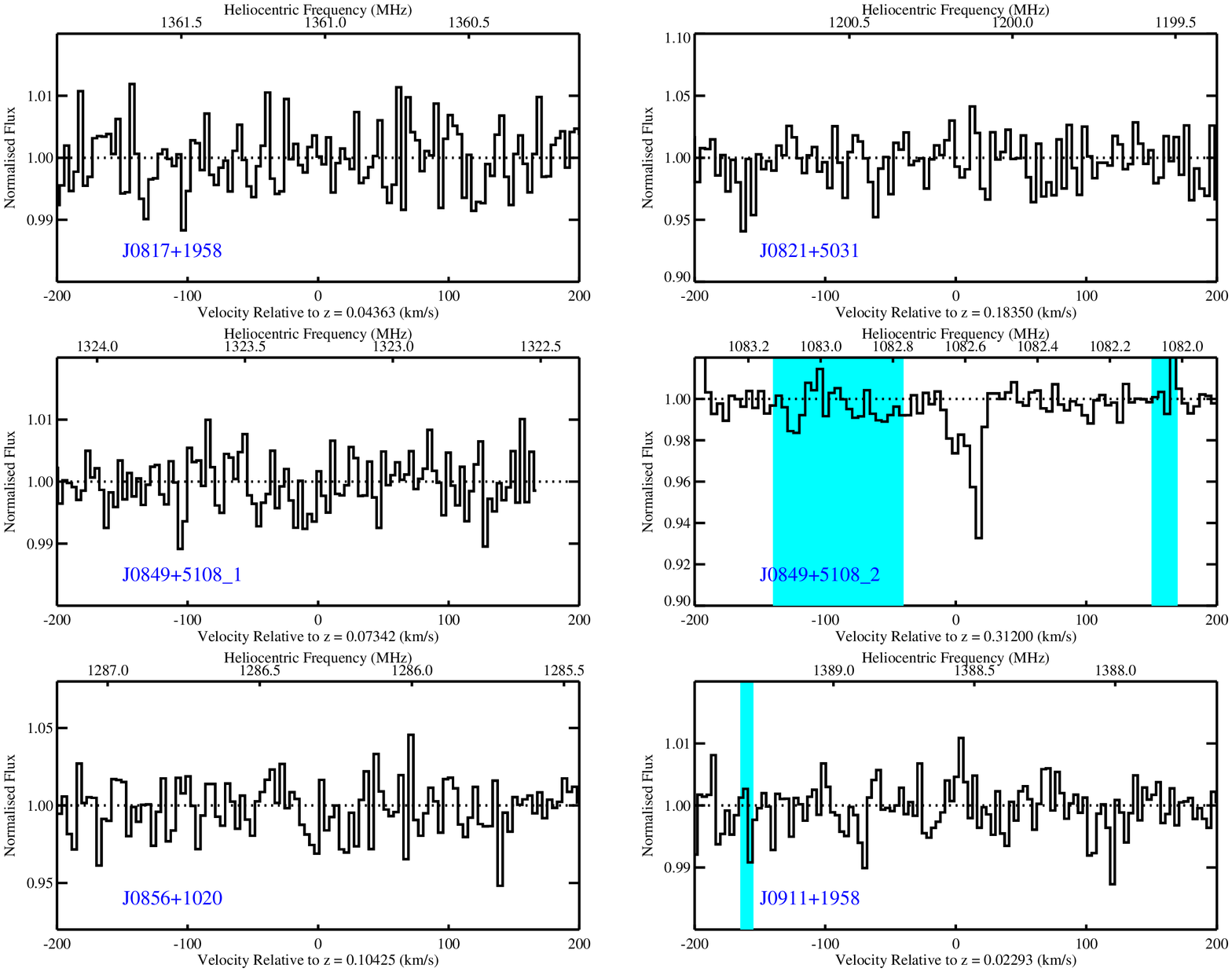} }
\caption{\hi\ 21-cm absorption spectra towards the radio sources in our primary sample, smoothed to $\sim$4 \kms. The shaded regions are affected by RFI.}
\label{fig:21cmspectra1}
\end{figure*}
\begin{figure*}
\ContinuedFloat
\subfloat{\includegraphics[height=0.5\textheight, angle=90]{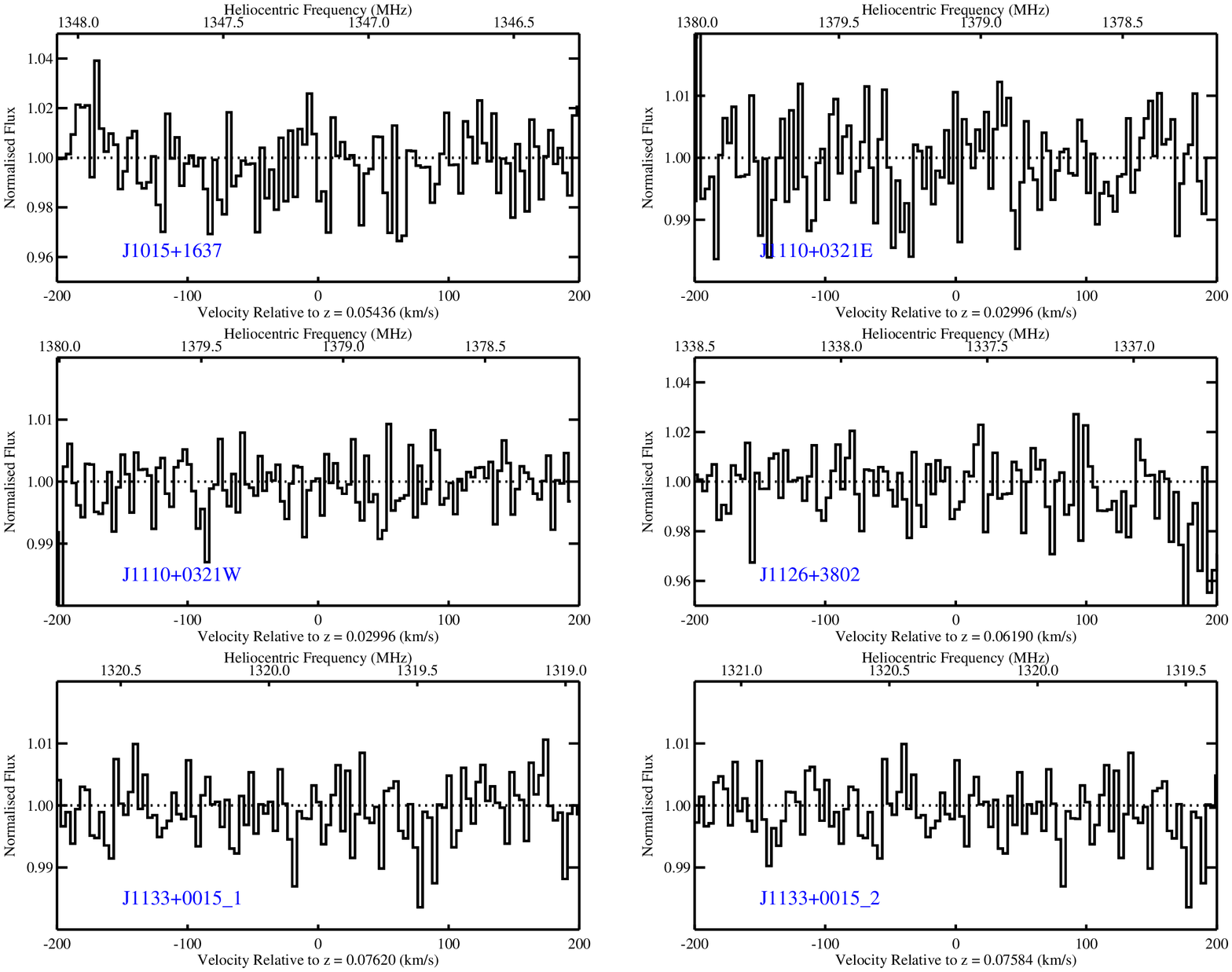} } \hspace{0.01cm}
\subfloat{\includegraphics[height=0.5\textheight, angle=90]{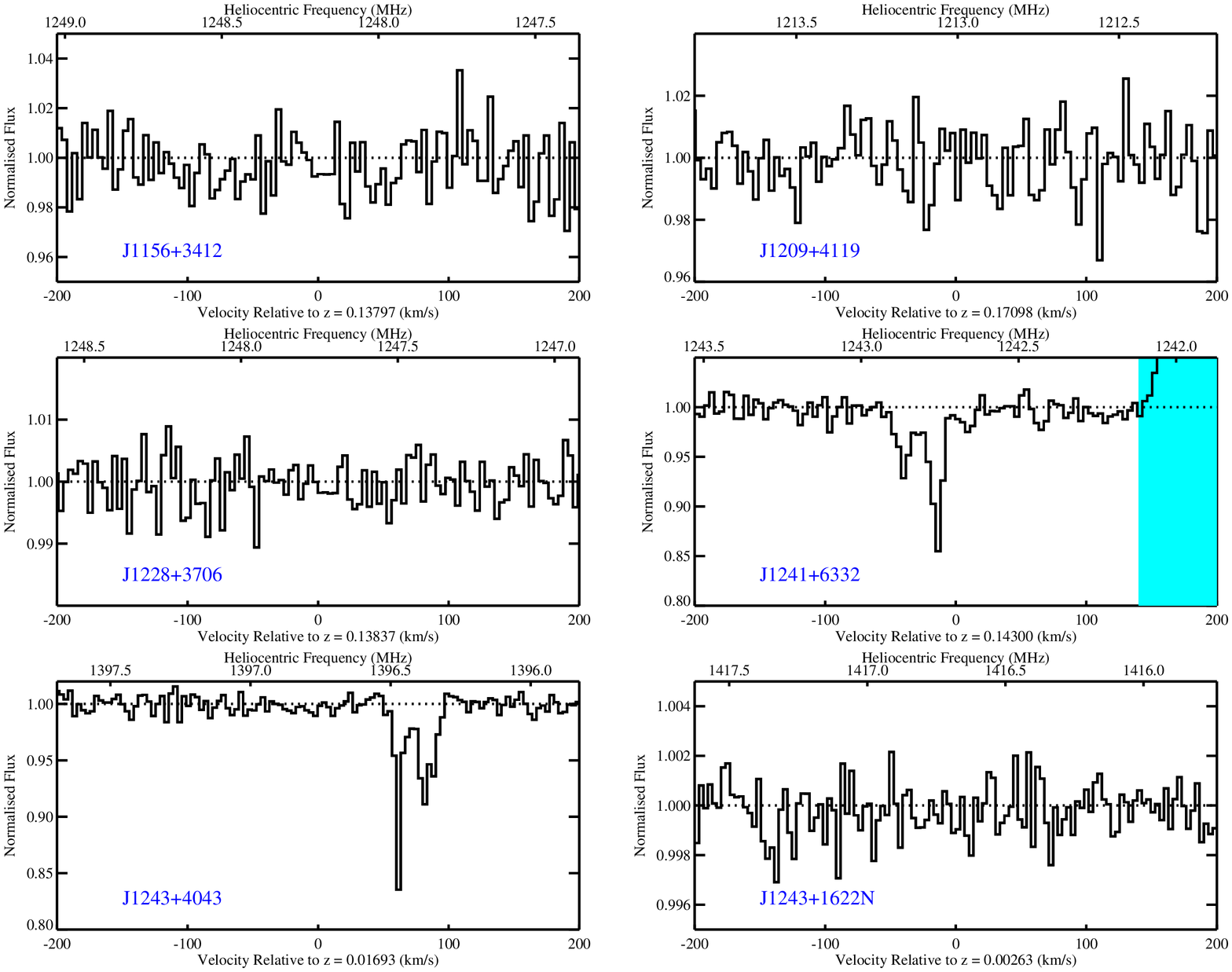} }
\caption{Continued from previous page.}
\label{fig:21cmspectra2}
\end{figure*}
\begin{figure*}
\ContinuedFloat
\subfloat{\includegraphics[height=0.5\textheight, angle=90]{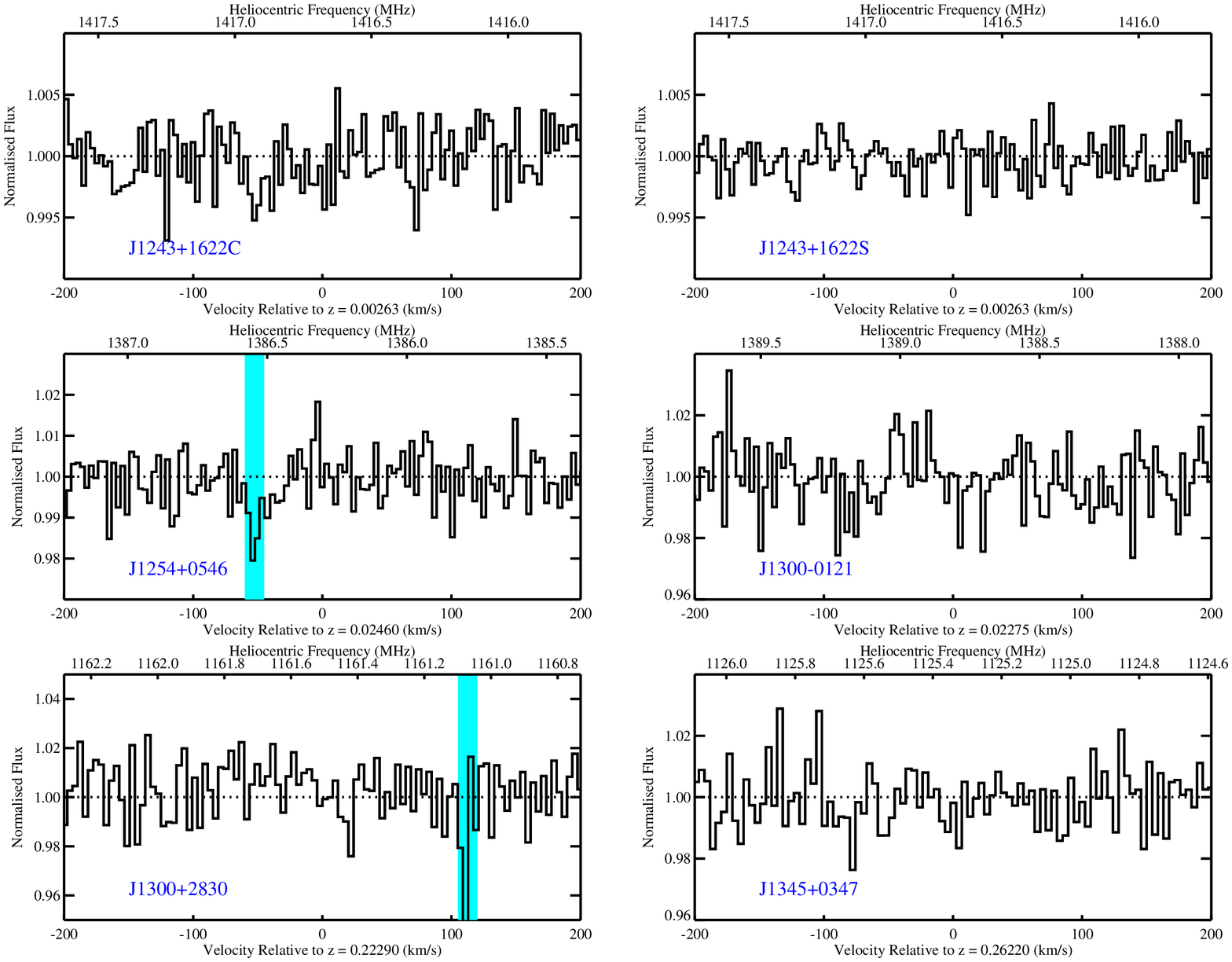} } \hspace{0.01cm}
\subfloat{\includegraphics[height=0.5\textheight, angle=90]{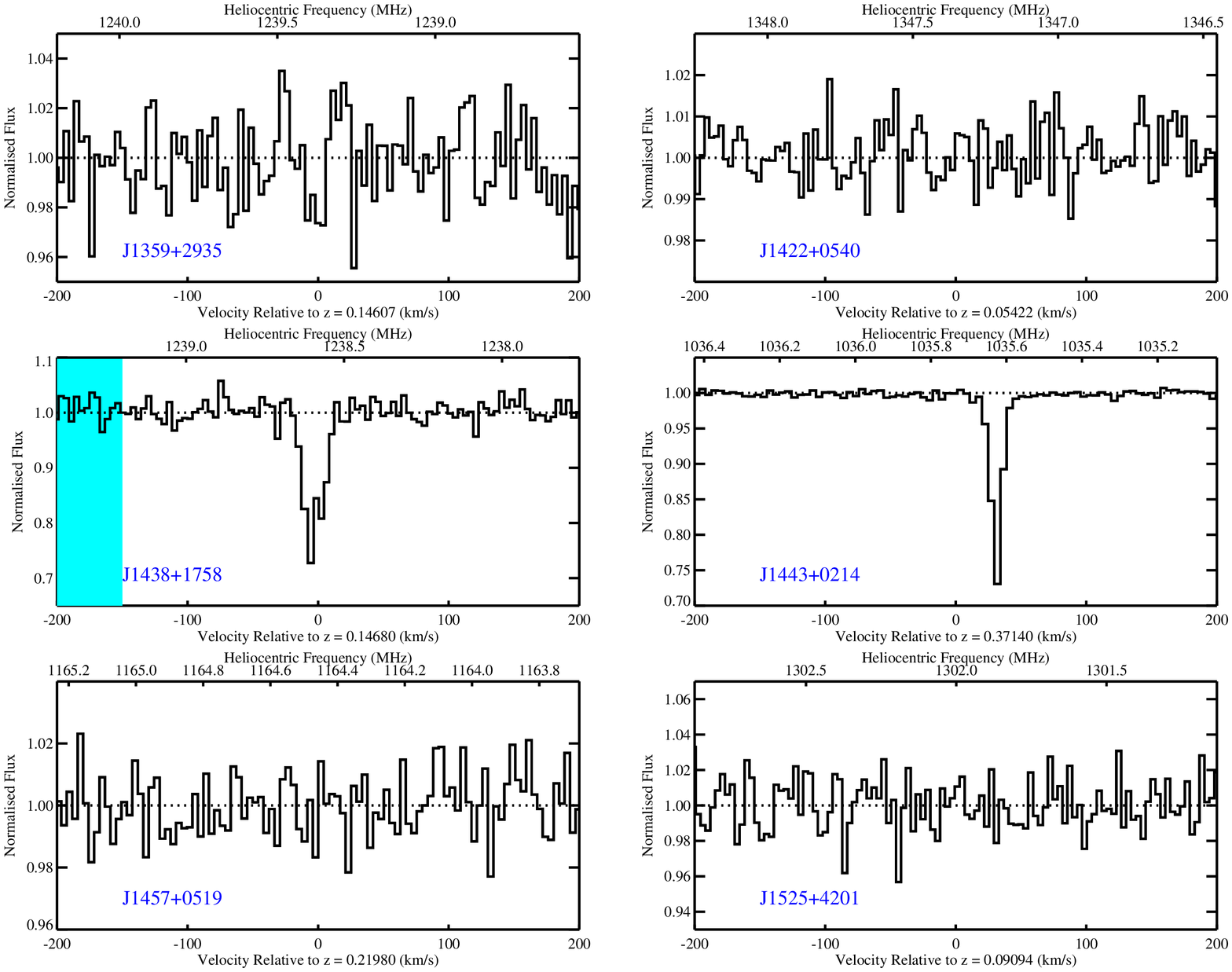} }
\caption{Continued from previous page.}
\label{fig:21cmspectra3}
\end{figure*}
\begin{figure*}
\ContinuedFloat
\subfloat{\includegraphics[height=0.5\textheight, angle=90]{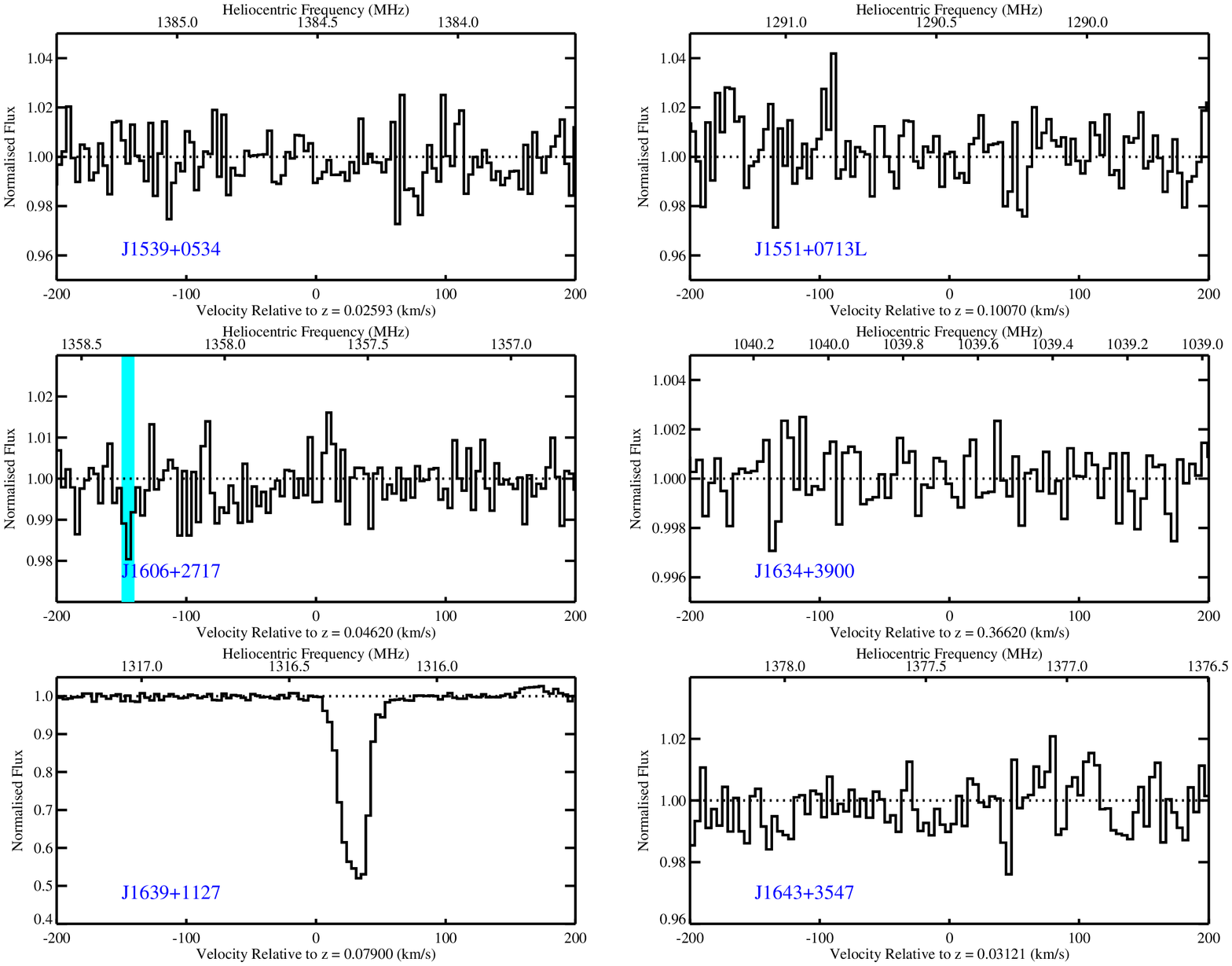} } \hspace{0.01cm}
\subfloat{\includegraphics[height=0.5\textheight, angle=90]{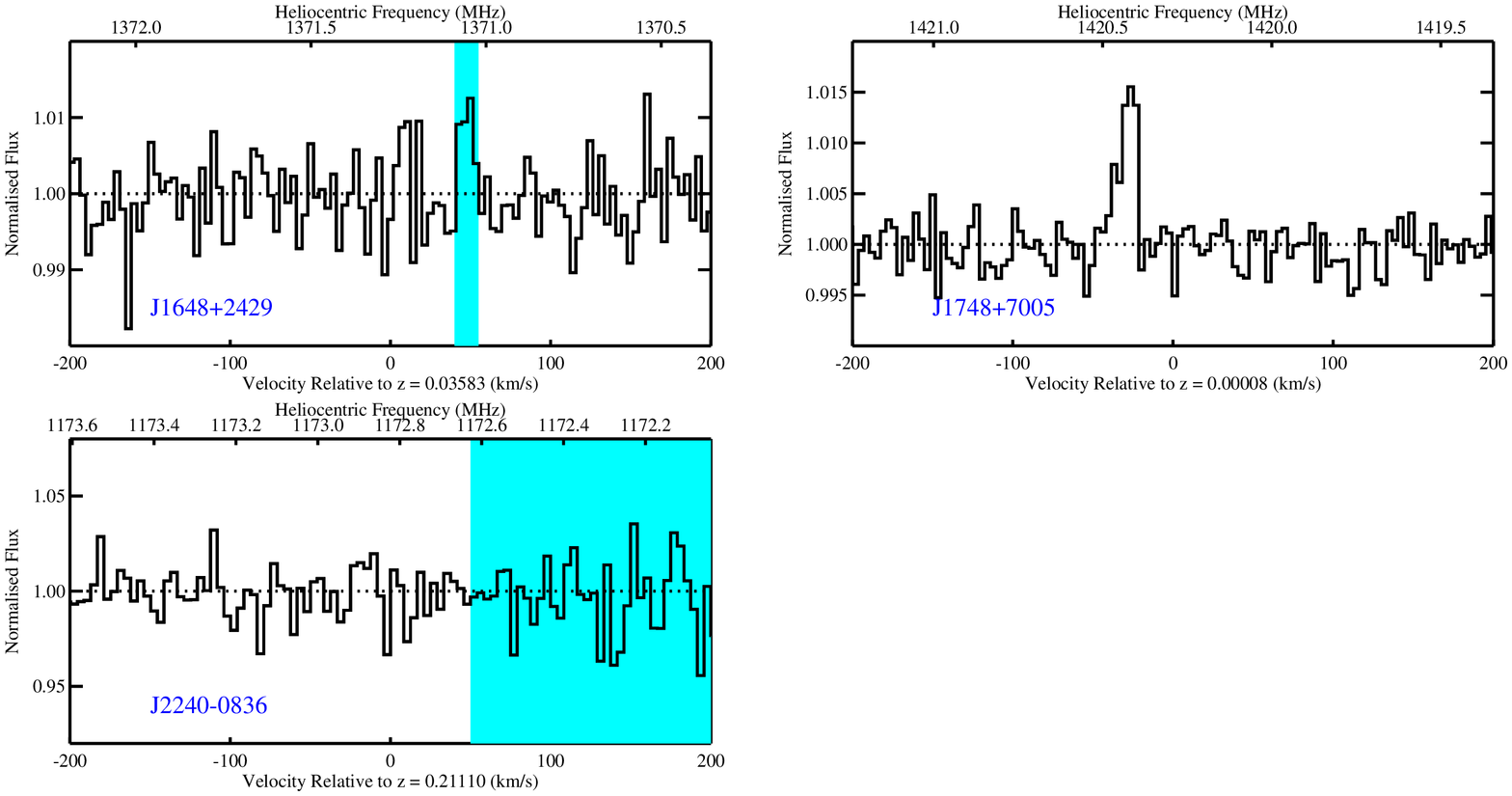} }
\caption{Continued from previous page.}
\label{fig:21cmspectra4}
\end{figure*}
\begin{figure*}
\subfloat{\includegraphics[height=0.5\textheight, angle=90]{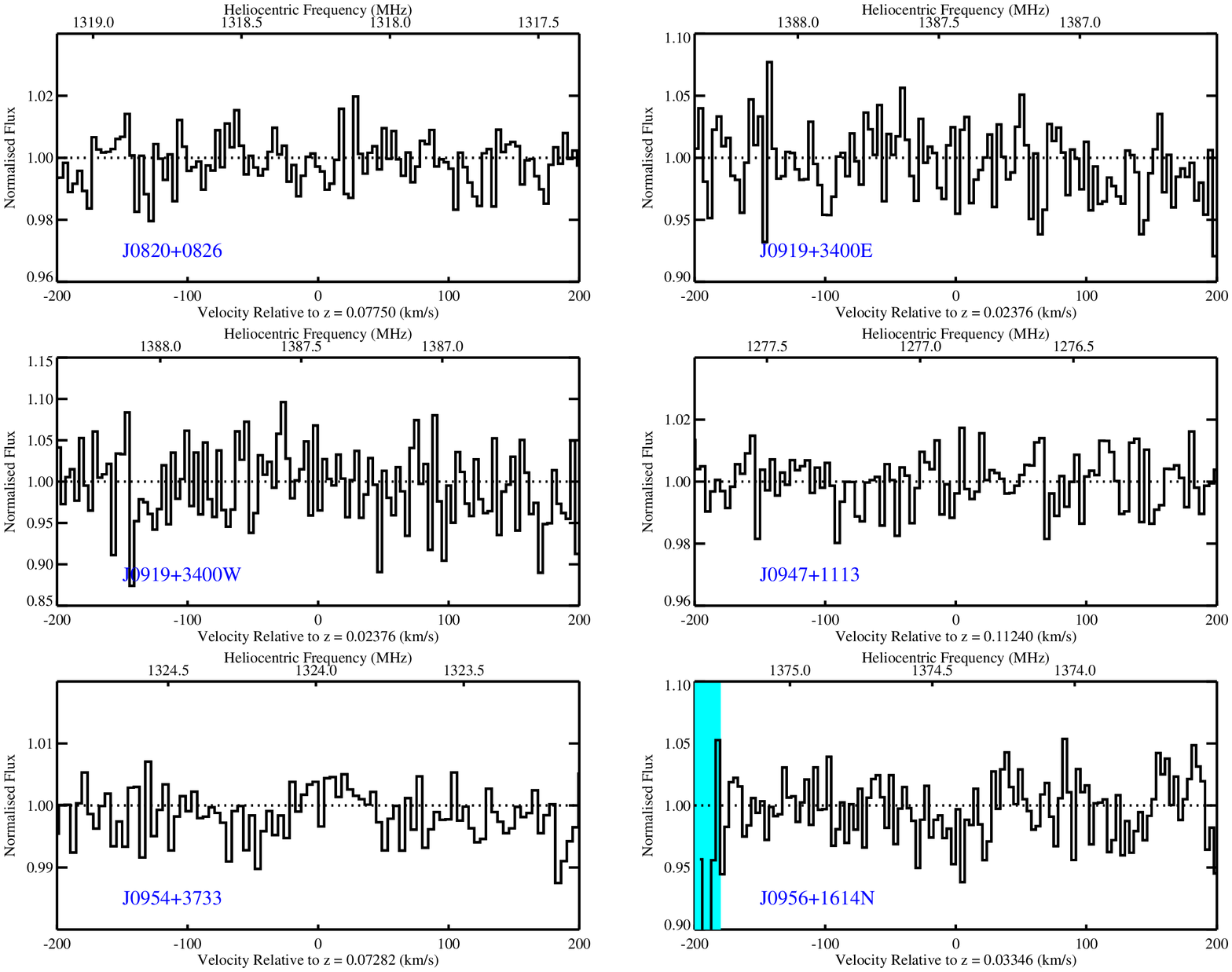} } \hspace{0.01cm}
\subfloat{\includegraphics[height=0.5\textheight, angle=90]{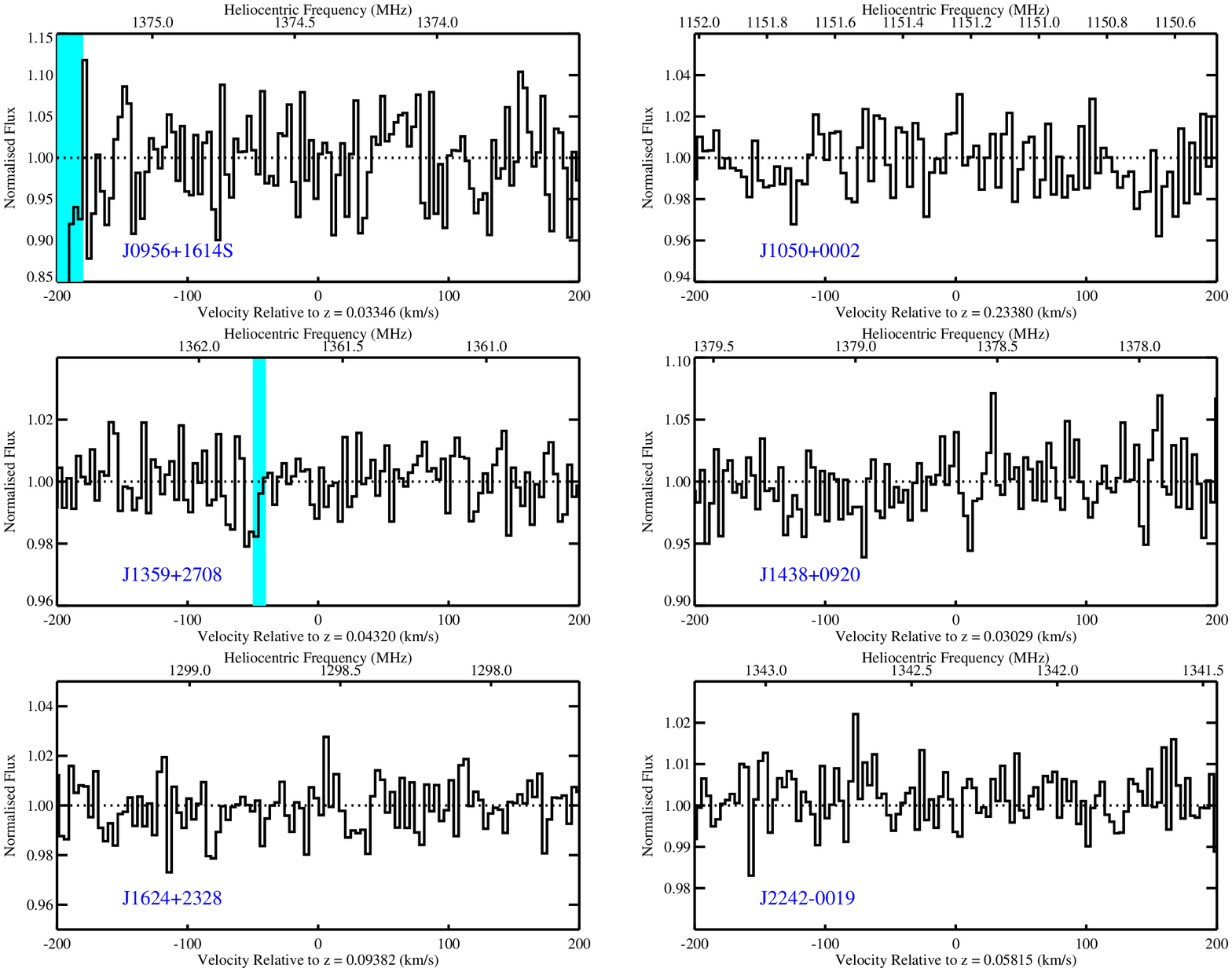} }
\caption{Same as in Fig.~\ref{fig:21cmspectra1} for the radio sources in the supplementary sample.}
\label{fig:nonstat_spectra}
\end{figure*}
\begin{figure*}
\subfloat{\includegraphics[height=0.5\textheight, angle=90]{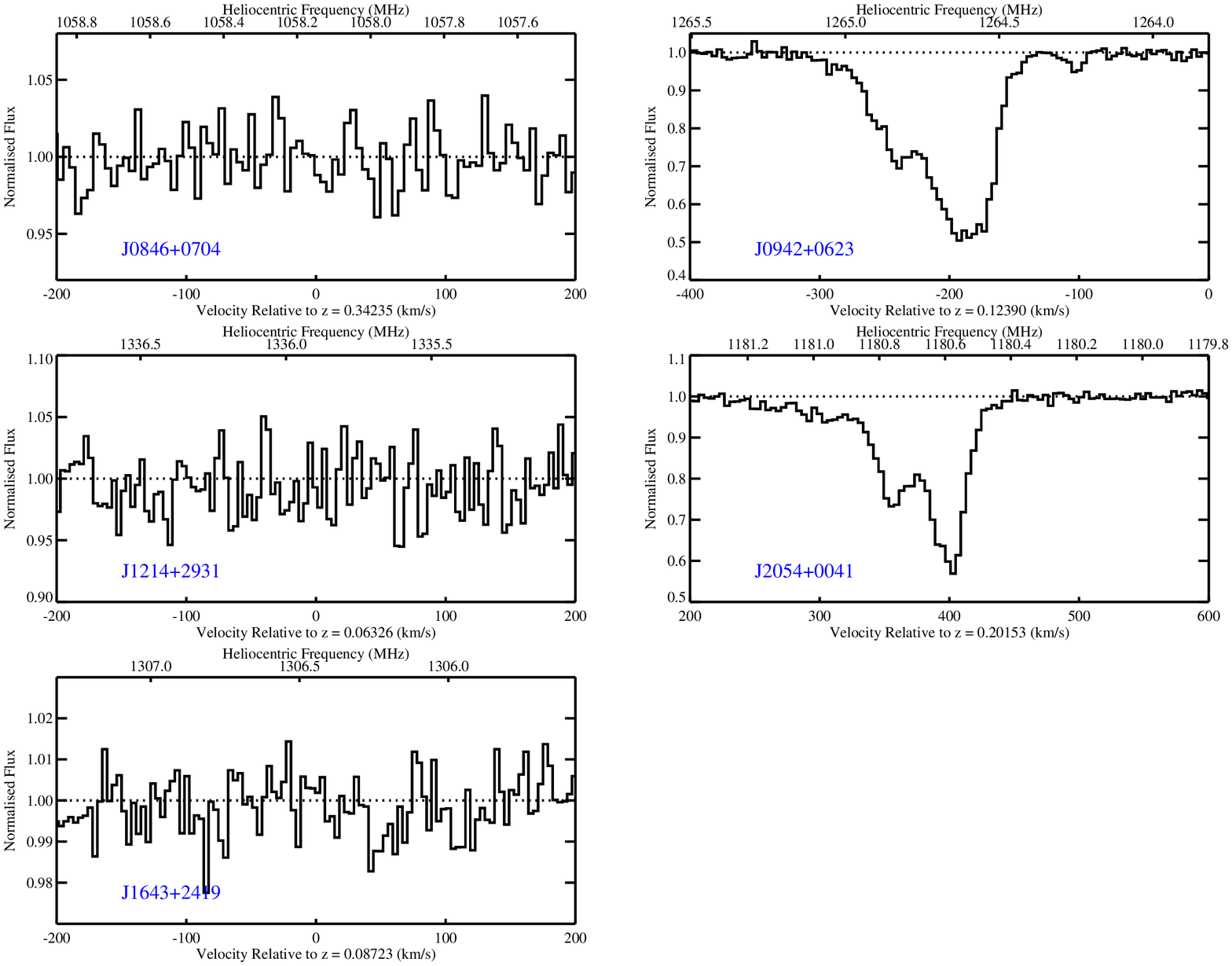} } 
\caption{Same as in Fig.~\ref{fig:21cmspectra1} for the radio sources in the miscellaneous sample.}
\label{fig:misc_spectra}
\end{figure*}
\end{appendices}
\end{document}